\documentclass[a4paper, 11pt]{nicearticle}
\usepackage{extrapackages}

\title{Lectures on the Bondi--Metzner--Sachs Group}
\subtitle{and related topics in infrared physics}
\shorttitle{Lectures on the BMS Group}
\date{\today}
\copyrightyear{2025}
\author[Aguiar Alves]{Níckolas de Aguiar Alves}
\orcid{0000-0002-0309-735X}
\institution{Center for Natural and Human Sciences, \href{https://ror.org/028kg9j04}{Federal University of ABC},\\Avenida dos Estados 5001, Bangú, Santo André, São Paulo 09280-560, Brazil}
\email{alves.nickolas@ufabc.edu.br}
\abstract{These are the extended lecture notes for a minicourse presented at the \href{https://graspschool.github.io/2024}{I São Paulo School on Gravitational Physics} discussing the Bondi--Metzner--Sachs (BMS) group, the group of symmetries at null infinity on asymptotically flat spacetimes. The BMS group has found many applications in classical gravity, quantum field theory in flat and curved spacetimes, and quantum gravity. These notes build the BMS group from its most basic prerequisites (such as group theory, symmetries in differential geometry, and asymptotic flatness) up to modern developments. These include its connections to the Weinberg soft graviton theorem, the memory effect, its use to construct Hadamard states in quantum field theory in curved spacetimes, and other ideas. Advanced sections briefly discuss the main concepts behind the infrared triangle in electrodynamics, superrotations, and the Dappiaggi--Moretti--Pinamonti group in expanding universes with cosmological horizons. New contributions by the author concerning asymptotic (conformal) Killing horizons are discussed at the end.}
\fullcitation{aguiaralves2026LecturesBondiMetzner}
\license{\href{https://creativecommons.org/licenses/by/4.0/}{Creative Commons Attribution 4.0 International} \faCreativeCommons\faCreativeCommonsBy}
\illustration{
\begin{multline*}
            \vcenter{\hbox{\begin{tikzpicture}
            \begin{feynman}
                \node[blob, pattern color=white] (c) at (0,0);
                \vertex (a1) at (180:2);
                \vertex (a2) at (135:2);
                \vertex (a3) at (225:2);
                \vertex (b1) at (30:2);
                \vertex (b2) at (330:2);
                \diagram*{
                    (a1) -- [fermion] (c),
                    (a2) -- [fermion] (c),
                    (a3) -- [fermion] (c),
                    (c) -- [fermion] (b1),
                    (c) -- [fermion] (b2),
                };
            \end{feynman}
        \end{tikzpicture}}} + \vcenter{\hbox{\begin{tikzpicture}
            \begin{feynman}
                \node[blob, pattern color=white] (c) at (0,0);
                \vertex (a1) at (180:2);
                \vertex (a2) at (135:2);
                \vertex (a3) at (225:2);
                \vertex (a4) at (180:1);
                \vertex (a5) at (225:1);
                \vertex (b1) at (30:2);
                \vertex (b2) at (330:2);
                \diagram*{
                    (a4) -- [graviton,accent] (a5),
                    (a1) -- [fermion] (c),
                    (a2) -- [fermion] (c),
                    (a3) -- [fermion] (c),
                    (c) -- [fermion] (b1),
                    (c) -- [fermion] (b2),
                };
            \end{feynman}
        \end{tikzpicture}}} \\ + \vcenter{\hbox{\begin{tikzpicture}
            \begin{feynman}
                \node[blob, pattern color=white] (c) at (0,0);
                \vertex (a1) at (180:2);
                \vertex (a2) at (135:2);
                \vertex (a3) at (225:2);
                \vertex (a4) at (180:1);
                \vertex (a5) at (225:1);
                \vertex (a6) at (135:1);
                \vertex (b1) at (30:2);
                \vertex (b2) at (330:2);
                \vertex (b3) at (30:0.8);
                \diagram*{
                    (a4) -- [graviton,accent] (a5),
                    (a6) -- [graviton,accent] (b3),
                    (a1) -- [fermion] (c),
                    (a2) -- [fermion] (c),
                    (a3) -- [fermion] (c),
                    (c) -- [fermion] (b1),
                    (c) -- [fermion] (b2),
                };
            \end{feynman}
        \end{tikzpicture}}} + \vcenter{\hbox{\begin{tikzpicture}
            \begin{feynman}
                \node[blob, pattern color=white] (c) at (0,0);
                \vertex (a1) at (180:2);
                \vertex (a2) at (135:2);
                \vertex (a3) at (225:2);
                \vertex (a4) at (180:1);
                \vertex (a5) at (225:1);
                \vertex (a6) at (135:1);
                \vertex (b1) at (30:2);
                \vertex (b2) at (330:2);
                \vertex (b3) at (30:0.8);
                \vertex (b4) at (30:1.5);
                \vertex (b5) at (330:1);
                \diagram*{
                    (b4) -- [graviton,accent] (b5),
                    (a4) -- [graviton,accent] (a5),
                    (a6) -- [graviton,accent] (b3),
                    (a1) -- [fermion] (c),
                    (a2) -- [fermion] (c),
                    (a3) -- [fermion] (c),
                    (c) -- [fermion] (b1),
                    (c) -- [fermion] (b2),
                };
            \end{feynman}
        \end{tikzpicture}}} + \cdots
        \end{multline*}
}
\coverinfo{Cover: perturbative diagrams used in the proof of Weinberg's soft graviton theorem \cite{weinberg1965InfraredPhotonsGravitons}, discussed on \cref{subsec: weinberg-soft-graviton}.}

\hypersetup{pdfauthor={Níckolas de Aguiar Alves}}

\begin{document}
\maketitle
\tableofcontents

\section{Introduction}
    The notion of symmetry is ubiquitous throughout physics. It is at the core of every physical theory and one could argue it is a necessity for one to even do physics at all. Symmetries allow us to greatly simplify the difficult problems we encounter in the Universe and approach them in a feasible manner, and they often reflect fundamental insights about the structure of nature. 

    When studying physical symmetries, the concept of group proves to be particularly useful \cite{zee2016GroupTheoryNutshell,hamermesh1989GroupTheoryIts,sundermeyer2014SymmetriesFundamentalPhysics,schwichtenberg2018PhysicsSymmetry}. Groups are abstract mathematical structures similar to, but simpler than, vector spaces. The ``collection'' of symmetries of a certain kind of a given physical system often constitutes a group. Hence, knowing the basic parlance of group theory is a powerful manner of studying symmetries and their consequences. 

    The \gls{BMS} group is the group of symmetries at infinity in asymptotically flat spacetimes. Consider the following scenario in \gls{GR}. One has a given distribution of matter in spacetime arranged in such a manner that, far away from all of this matter, spacetime is nearly flat in a suitable sense (to be defined on \cref{sec: asymptotically-flat-spacetimes}). Far from all these sources, spacetime resembles Minkowski spacetime, so we would expect that the symmetries at infinity should somehow resemble the symmetries of Minkowski spacetime---the latter form the so-called Poincaré group. Hence, it was a surprise when Bondi, Metzner, and Sachs discovered the correct group is much larger than the Poincaré group \cite{bondi1962GravitationalWavesGeneral,sachs1962GravitationalWavesGeneral,sachs1962AsymptoticSymmetriesGravitational}---in addition to translations, rotations, and boosts one also has an infinite number of ``supertranslations''. 

    This simple fact hides deep physical truths about \gls{GR}. Namely, it is a statement that \gls{GR} \emph{does not} reduce to special relativity at large distances. Rather, it reduces to a much more complex structure that arises only at infinity. This rich structure at infinity can be exploited, as done by \textcite{dappiaggi2017HadamardStatesLightlike}, to construct physically interesting states for quantum fields evolving on asymptotically flat backgrounds. There are also other physical implications. For example, the \gls{BMS} symmetries imply correlations between \(S\)-matrix elements that yield the so-called soft graviton theorem \cite{weinberg1965InfraredPhotonsGravitons,weinberg1995Foundations}, which is a fundamental ingredient for understanding the \gls{IR} structure of cross sections in scattering processes involving gravitons. \gls{BMS} transformations can also be related to the so-called memory effect---originally discovered by \textcite{zeldovich1974RadiationGravitationalWaves}---which predicts that after the passage of a \gls{GW} two nearby detectors will be permanently displaced. It is expected that this effect should be measurable in the near future \cite{favata2010GravitationalwaveMemoryEffect,grant2023OutlookDetectingGravitationalwave}.

    In these lecture notes, I provide an introduction to the \gls{BMS} group and some of the developments built upon it. They are an extremely extended version of the lecture notes for the ten-hour-long minicourse I presented at the \href{https://graspschool.github.io/2024}{I São Paulo School on Gravitational Physics}, which was entitled ``Infrared Symmetries of General Relativity''. The original course was aimed at advanced undergraduate students and early graduate students with previous knowledge of basic \gls{GR}, for example at the level of the textbook by \textcite{carroll2019SpacetimeGeometryIntroduction}. The course was then an introduction to group theory in special relativity (\cref{sec: symmetries-groups}), isometries and confomorphisms in differential geometry and relativity (\cref{sec: symmetries-curved-spacetime}), asymptotic flatness (some of \cref{sec: asymptotically-flat-spacetimes}), the \gls{BMS} group (some of \cref{sec: bms-group}), and some applications (some of \cref{sec: modern-developments}). More specifically, the course covered much of the material in \cref{sec: symmetries-groups,sec: symmetries-curved-spacetime,sec: asymptotically-flat-spacetimes,sec: bms-group}, with the exception of the effects of Lorentz transformations on the night sky, the Bondi--Sachs formalism, and \gls{CK} spacetimes. Due to time constraints, I had chosen to focus on the concepts and ideas rather than the detailed calculations available in the notes (this text). I briefly discussed the main ideas of \cref{sec: modern-developments}, although without presenting any calculations. 

    Due to the nature of the original introductory course, most of the material in this text tries to assume only undergraduate-level \gls{GR} and the basic jargon about differential geometry (such as what is a manifold). Some applications require previous knowledge about \gls{QFT} to be fully appreciated. Apart from these prerequisites, the text is essentially self-contained. For example, no previous experience with \gls{CFT} is assumed. \cref{sec: symmetries-groups} discusses basic facts about group theory, with a particular view toward special relativity. \cref{sec: symmetries-curved-spacetime} further develops this discussion for the study of isometry and conformal groups in relativistic spacetimes and other manifolds. This includes discussing maps between manifolds (including pullbacks and pushforwards), Lie derivatives, and the notion of (conformal) Killing vector fields. Some examples connected to other parts of the text are given. \cref{sec: asymptotically-flat-spacetimes} discusses the definition of asymptotic flatness at null infinity both in the conformal completion approach pioneered by Penrose and others and in the Bondi--Sachs formalism. I also comment a bit on \gls{CK} spacetimes, which have proven to be useful in the study of \gls{IR} structure. \cref{sec: bms-group} discusses Carrollian manifolds and uses these notions to derive the \gls{BMS} group in two different ways. Modern developments and applications of \gls{BMS} symmetries are discussed on \cref{sec: modern-developments}, and some of them will be better appreciated with previous knowledge of \gls{QFT}. \cref{sec: electrodynamics} consider analogous constructions in electrodynamics, by discussing the so-called large gauge transformations. \cref{sec: superrotations} discusses the notion of superrotation, which yields one possible extension of the \gls{BMS} group. \cref{sec: infrared-de-sitter} considers the \gls{DMP} group, which is the analog of the \gls{BMS} group for asymptotically de Sitter spacetimes. \cref{sec: sky-killing-horizon} discusses my work with Landulfo on viewing null infinity as a Killing horizon. I conclude on \cref{sec: conclusion}. 
    
    A few appendices are available as well. Appendix \ref{app: complex-analysis} discusses a few results on complex analysis that may be unfamiliar to some readers (although basic complex analysis is assumed). Appendix \ref{app: double-cover-lorentz} proves that \(\SL(2,\Comp)\) is the double cover of the Lorentz group \(\SO*(3,1)\), which is an important statement in understanding the conformal group of the sphere. This group, in turn, plays a prominent role in the derivation of the \gls{BMS} group. Appendix \ref{app: linearized-gravity} reviews some standard results about linearized gravity and is useful in understanding the \gls{GW} displacement memory effect on \cref{sec: modern-developments}. Appendix \ref{app: preserving-scs} discusses a possible critique of the derivations of the \gls{BMS} group. Appendix \ref{app: stationary-phase-approximation-sphere} considers the stationary phase approximation for certain integrals over the sphere which play a relevant role in calculations involving the Weinberg soft theorems. 
    
    I often use the same notation and conventions employed in the textbook by \textcite{wald1984GeneralRelativity}. This includes abstract index notation, metric signature \({-}{+}{+}{+}\), and geometric units with \(G = c = 1\). My sign conventions are \({+}{+}{+}\) in the Misner--Thorne--Wheeler classification \cite{misner2017Gravitation}. Unless stated otherwise, Latin indices \(a, b, \ldots\) represent abstract indices, Greek indices \(\mu, \nu, \ldots\) represent spacetime coordinate indices. Occasionally Latin indices \(i, j, \ldots\) represent spatial indices and capital Latin indices \(A, B, \ldots\) represent indices on a two-sphere, which should be clear from context.

\section{Symmetries and Groups}\label{sec: symmetries-groups}
    We begin by discussing what is the mathematical structure of the symmetries of a physical system. This will naturally lead us to the notion of group, which is a well-established and well-studied concept in mathematics. We shall then understand some of the basic ideas of group theory and in particular exploit them to investigate properties of relativistic spacetimes. 

    \subsection{Case Study: Rotations}
        Perhaps the paradigmatic example of a symmetry is rotational symmetry. This means that the physical system does not change when we rotate it about a specific point and is present in many interesting examples in fundamental physics. For example, the hydrogen atom model in \gls{QM} presents spherical symmetry. So does the Schwarzschild solution in \gls{GR} and many stellar solutions of interest. 

        Let us for a moment ponder about what is it that \emph{defines} a rotation as a rotation. To be concrete, we are thinking right now about rotations as operations on \(n\)-dimensional real vectors. We know that a generic rotation should be a linear transformation \(R \colon \R^n \to \R^n\). Indeed, \(R(\alpha \vb{v}) = \alpha R \vb{v}\), for rotations should not ``see'' the length of a vector---they only care about its direction. Furthermore, \(R(\vb{u} + \vb{v}) = R \vb{u} + R \vb{v}\), because the rotation occurs in a ``solid'' manner---both elements of the sum and the sum itself should all be rotated in precisely the same way. Hence, indeed, a general rotation should be a linear transformation. 

        This, however, is surely not enough to characterize what we mean by a rotation. There are far too many linear transformations! For example, given the canonical basis \(\qty{\vb{e}_i}\), we could pick the transformation defined by \(T\vb{e}_i = (1 + \delta_{i2})\vb{e}_i\), where \(\delta_{ij}\) is the Kronecker delta. This transformation simply stretches one of the coordinate axes, while keeping the remaining ones constant. This is surely a linear transformation, but it looks nothing like a rotation. 

        The extra property that characterizes rotations is that they preserve angles and norms. Hence, they do not impart any sort of stretching into the vectors they act on and they do not change the relative angle between them. Mathematically this is expressed in terms of the scalar product in \(\R^n\) as 
        \begin{equation}
            R\vb{u} \vdot R\vb{v} = \vb{u} \vdot \vb{v}.
        \end{equation}
        This can be shown to be equivalent to the requirement that
        \begin{equation}\label{eq: orthogonality-R}
            R^\intercal R = \Eins,
        \end{equation}
        where \(R^\intercal\) is the transpose of \(R\) and \(\Eins\) is the identity matrix. Matrices that satisfy \cref{eq: orthogonality-R} are said to be orthogonal. This invites us to define 
        \begin{equation}
            \Og(n) = \qty{R \in \Mat(n); R^\intercal R = \Eins},
        \end{equation}
        where \(\Mat(n)\) is the space of \(n \times n\) real matrices and the ``O'' in \(\Og(n)\) stands for ``orthogonal''. \(\Og(n)\) is the so-called orthogonal group in \(n\) dimensions.

        We would like to know which sort of structure represents collections of symmetries in general, so it is interesting for us to study some properties of \(\Og(n)\) to figure out an abstract definition. \(R^\intercal R = \Eins\) seems too strict, since this is specific to rotations, and not every symmetry is a rotation. In fact, even linearity might be a stretch, since complicated symmetries could be nonlinear in principle. 

        One of the most basic facts we can notice about \(\Og(n)\) is that it is endowed with a product. Namely, given two rotations \(R\) and \(S\), we can also define \(RS\), which stands for ``apply the rotation \(S\), and then the rotation \(R\)''. This constitutes a rotation, for one can show through direct calculation that \(RS \in \Og(n)\). This is our first axiom for a group: a group should be a set \(G\) endowed with a product \(\cdot \colon G \times G \to G\). Indeed, notice that if we apply two symmetry transformations in sequence, the result should also be a symmetry. This is due to the fact that a symmetry relates two physically equivalent configurations, and thus applying two symmetries in sequence yields a configuration that is physically equivalent to the original one. 

        Next, we notice that the product in \(\Og(n)\) is associative. Indeed, given rotations \(R\), \(S\), and \(T\), we have that \((RS)T = R(ST)\). We will also impose associativity as one of the axioms for a group. This is due both to the fact that it is mathematically interesting to do so---the resulting theory is quite rich---and due to the fact that it is even difficult to conceive how one could apply transformations to a physical system in a manner which is not associative. If we think of the transformation \((RS)T\) in terms of two consecutive transformations (\(T\) and then \(RS\)), it is not obvious that it is equivalent to \(R(ST)\) (\(ST\) and then \(R\)). However, it seems intuitive that these two transformation processes should be equivalent to the ``physical realization'' \(T\), then \(S\), then \(R\), which we would denote as \(RST\) (without parentheses). Hence, it seems these transformations should compose in an associative manner. A second argument could be that each transformation should be viewed as a mapping between configurations of a physical system, and the composition of mappings is always associative. 

        We then notice that ``doing nothing'' should be considered a symmetry. Indeed, a physical configuration is physically indistinguishable from itself. Hence, we expect there to be an identity in a group. And, correctly, we have that \(\Eins \in \Og(n)\), as one can promptly check. 

        Finally, undoing a symmetry transformation also constitutes a symmetry. If we perform a transformation in the system in such a way that the configuration after the transformation is indistinguishable from the configuration before the transformation, it is surely true that the statement also holds the other way around. Thus, the group should be populated with the inverses of every group element. As expected, for every \(R \in \Og(n)\), \(R^{-1} = R^\intercal \in \Og(n)\).

        We thus arrive at the following definition. 

        \begin{definition}[Group]
            A \emph{group} is a pair \((G,\cdot)\) where \(G\) is a set and \(\cdot \colon G \times G \to G\) is a mapping satisfying the following conditions:
            \begin{enumerate}
                \item \(\cdot\) is associative, so \((g_1 \cdot g_2) \cdot g_3 = g_1 \cdot (g_2 \cdot g_3)\) for every \(g_1, g_2, g_3 \in G\),
                \item \(\cdot\) has a neutral element \(e\), \ie, there is an element \(e \in G\) with \(e \cdot g = g \cdot e = g\) for all \(g \in G\),
                \item all elements in \(G\) have inverses, meaning that for each \(g \in G\) there is some \(g^{-1} \in G\) with \(g \cdot g^{-1} = g^{-1} \cdot g = e\).
            \end{enumerate}
            Groups which are also commutative (\(g_1 \cdot g_2 = g_2 \cdot g_1\) for all \(g_1, g_2 \in G\)) are said to be \emph{Abelian}. I will often omit \(\cdot\) when writing products when the product operation is understood: \(g_1 g_2 = g_1 \cdot g_2\).
        \end{definition}

        Our interest in groups is similar to one's interest in vector spaces. These are useful mathematical concepts that have been well-developed by mathematicians and that occur once and again in physical contexts. As such, it is useful to know a thing or two about groups because this helps us understanding physical phenomena better.  Refs. \citeonline{zee2016GroupTheoryNutshell,hamermesh1989GroupTheoryIts,sundermeyer2014SymmetriesFundamentalPhysics,schwichtenberg2018PhysicsSymmetry} discuss many applications of group theory to physics, but in the following we will restrict our attention to groups that are of interest for relativity (with a particular focus toward the \gls{BMS} group). 

        Just as one has the notion of a subspace of a vector space, we also have the notion of a subgroup of a group. 

        \begin{definition}[Subgroup]
            Let \((G,\cdot)\) be a group and \(H \subeq G\). We say \((H,\cdot)\) is a \emph{subgroup} of \((G, \cdot)\) if \((H,\cdot)\) is a group. Notice the product has to be inherited from the larger group.  
        \end{definition}

        As an example, consider 
        \begin{equation}
            \SO(n) = \qty{R \in \Mat(n); R^\intercal R = \Eins, \det R = +1}.
        \end{equation}
        Notice that \(\SO(n) \subeq \Og(n)\). In fact, every element \(R \in \Og(n)\) either lies in \(\SO(n)\) or has \(\det R = - 1\). One can show that \(\SO(n)\) is a subgroup of \(\Og(n)\), and it is \(\SO(n)\) that is often referred to as the group of rotations. The reason is that \(\Og(n)\) is essentially \(\SO(n)\) with the possibility of composing an element of \(\SO(n)\) with a reflection, so \(\SO(n)\) restricts our attention to the rotations that involve no reflections. The ``S'' in \(\SO(n)\) stands for ``special'', and refers to the fact that the matrices have unit determinant. 

        In order to understand further the structure of \(\SO(n)\) and \(\Og(n)\), it is useful to define a Lie group. 

        \begin{definition}[Lie Group]
            A \emph{Lie group} is a group \((G,\cdot)\) endowed with a smooth manifold structure in such a manner that the maps \(\mu \colon G \times G \to G\) and \(\iota \colon G \to G\) given by \(\mu(g_1,g_2) = g_1 \cdot g_2\) and \(\iota(g) = g^{-1}\) are smooth.
        \end{definition}

        Hence, a Lie group is a smooth group. This is useful because often in physics we encounter groups that have infinitely many elements, where ``infinitely many'' is meant in a continuous (or, more appropriately, smooth) way. This is the case of \(\Og(n)\) and \(\SO(n)\). We can define coordinates on these groups by employing the so-called Euler angles (see, \eg, Refs. \citeonline{goldstein2001ClassicalMechanics,lemos2018AnalyticalMechanics}). \(\SO(n)\) is composed of a connected manifold \cite{oneill1983SemiRiemannianGeometryApplications,onishchik1993FoundationsLieTheory}, which means it is a ``unique continuous piece''. \(\Og(n)\), on the other hand, has two connected components: there is one continuous piece with \(\det R = + 1\) (\(\SO(n)\)) and another with \(\det R = - 1\) \cite{oneill1983SemiRiemannianGeometryApplications,onishchik1993FoundationsLieTheory}, because one cannot smoothly change the sign of the determinant from \(+1\) to \(-1\). 

    \subsection{Case Study: Lorentz Group}
        Let us now move on to a more complicated example. Let us consider a spacetime \((\R^n,\tensor{\eta}{_a_b})\) with metric signature \((p,q)\) (\(p+q=n\)). Euclidean space is then the case with \(q=0\) and Minkowski spacetime is the case with \(q=1\). We would like to investigate the linear isometries of this spacetime. At this point we can think of isometries as coordinate transformations that do not change the components of the metric (I will give a more appropriate definition on \cref{sec: symmetries-curved-spacetime}). They can be understood as the fundamental symmetries of the spacetime. Linear isometries are a particular class of isometries that have the form 
        \begin{equation}
            \tensor{x}{^\mu} \to \tensor{x}{^\prime^\mu} = \tensor{\Lambda}{^\mu_\nu} \tensor{x}{^\nu} 
        \end{equation}
        for some matrix \(\tensor{\Lambda}{^\mu_\nu}\).

        For a general coordinate transformation to preserve the metric it should satisfy 
        \begin{equation}
            \tensor{\eta}{_\mu_\nu}\pdv{\tensor{x}{^\prime^\mu}}{\tensor{x}{^\rho}}\pdv{\tensor{x}{^\prime^\nu}}{\tensor{x}{^\sigma}} = \tensor{\eta}{_\rho_\sigma}.
        \end{equation}
        For a linear coordinate transformation, we can write this simply as
        \begin{equation}\label{eq: tensor-lambda-preserves-eta}
            \tensor{\eta}{_\mu_\nu}\tensor{\Lambda}{^\mu_\rho}\tensor{\Lambda}{^\nu_\sigma} = \tensor{\eta}{_\rho_\sigma}.
        \end{equation}
        In Cartesian coordinates, we can write the components \(\tensor{\eta}{_\mu_\nu}\) as the diagonal matrix
        \begin{equation}
            \eta = \mqty(\dmat{-1,-1,\ddots,+1,+1}),
        \end{equation}
        with \(p\) positive entries and \(q\) negative entries. \cref{eq: tensor-lambda-preserves-eta} can now be written simply as
        \begin{equation}
            \Lambda^{\intercal} \eta \Lambda = \eta.
        \end{equation}

        We thus define the group
        \begin{equation}
            \Og(p,q) = \qty{\Lambda \in \Mat(p+q); \Lambda^\intercal \eta \Lambda = \eta},
        \end{equation}
        known as the \((p,q)\)-pseudo-orthogonal group. ``\(p\)'' refers to the \(p\) positive signs in the Minkowski metric and ``\(q\)'' refers to the remaining negative signs. Notice this is a generalization of the orthogonal groups: \(\Og(n)\) could be defined as the group with matrices satisfying \(R^{\intercal} \Eins R = \Eins\), which is similar to the condition defining \(\Og(p,q)\).  

        Our main interest is in \(\Og(3,1)\), known as the Lorentz group in four-dimensions. It is composed of rotations, Lorentz boosts, and both spatial and time reflections. Notice that \(\Og(3)\) is a subgroup of \(\Og(3,1)\) formed by the elements with 
        \begin{equation}\label{eq: lambda-R-o31}
            \Lambda = \mqty(1 & 0 & 0 & 0 \\ 0 & & & \\ 0 & & R & \\ 0 & & &),
        \end{equation}
        where \(R\) is an orthogonal matrix. 

        One could argue, correctly, that technically (\ref{eq: lambda-R-o31}) is not an element of \(\Og(3)\). Indeed, \(\Og(3)\) is formed by \(3 \times 3\) matrices and (\ref{eq: lambda-R-o31}) is obviously a \(4 \times 4\) matrix. Yet, it seems odd to not consider (\ref{eq: lambda-R-o31}) an element of \(\Og(3)\), given that matrices of that form behave precisely as elements of \(\Og(3)\). Let us then give meaning to this.

        \begin{definition}[Homomorphism]
            Let \(G\) and \(H\) be groups. We say a function \(\phi \colon G \to H\) is a \emph{homomorphism} if, and only if, \(\phi(g_1)\phi(g_2) = \phi(g_1 g_2)\) for every \(g_1, g_2 \in G\).
        \end{definition}
        Notice I wrote simply \(G\) rather than the tuple \((G,\cdot)\). This is common in group theory, and perhaps in all of mathematics. Notice also that the definition of a homomorphism resembles that of a linear transformation between vector spaces---both of them are mappings between two algebraic structures of the same kind (either groups or vector spaces) that preserve the algebraic structure. Hence, they are natural maps between these sorts of structures. 

        Just as in linear algebra, we can now define an isomorphism.
        \begin{definition}[Isomorphism]
            Let \(G\) and \(H\) be groups and \(\phi \colon G \to H\) be a homomorphism. If \(\phi\) is bijective, it is said to be an isomorphism.
        \end{definition}
        The inverse of a homomorphism, when it exists, is automatically a homomorphism. Isomorphisms between groups are analogous to isomorphisms between vector spaces---they mean the two groups (or vector spaces) are algebraically ``equal'' as far as the structure they preserve is concerned. Two isomorphic groups can be understood as two copies of the same group. 

        Now we can give meaning to our previous idea: there is an injective homomorphism from \(\Og(3)\) into \(\Og(3,1)\). Or, alternatively, \(\Og(3,1)\) has a subgroup that is isomorphic to \(\Og(3)\). We often just make these statements briefly by saying \(\Og(3)\) is a subgroup of \(\Og(3,1)\), where the necessary homomorphisms are understood. 

        Just as we defined \(\SO(n)\) to get rid of the reflections of \(\Og(n)\), we can define \(\SO(p,q)\) to focus on a smaller subgroup of \(\Og(p,q)\). There is, however, a caveat: while \(\Og(n)\) has two connected components, \(\Og(p,q)\) has four. This is because in addition to spatial reflections we also have the independent time reflections (where ``time'' refers to the \(q\) directions with negative components in the metric). If we reverse time and apply a spatial reflection, we get a transformation with positive determinant, but which certainly is not a composition of ``pure'' rotations and boosts. Thus, we provide the following definitions\footnote{I focus on the Lorentzian case for simplicity, but it is possible to define \(\SO*(p,q)\) for arbitrary signature \cite{oneill1983SemiRiemannianGeometryApplications,hamilton2017MathematicalGaugeTheory}.}. 
        \begin{equation}
            \SO(n-1,1) = \qty{\Lambda \in \Mat(n); \Lambda^\intercal \eta \Lambda = \eta, \det\Lambda = +1}
        \end{equation}
        and
        \begin{equation}
            \SO*(n-1,1) = \qty{\Lambda \in \Mat(n); \Lambda^\intercal \eta \Lambda = \eta, \det\Lambda = +1, \tensor{\Lambda}{^0_0} > 0}.
        \end{equation}
        \(\tensor{\Lambda}{^0_0} > 0\) prevents time reflections, and when combined with \(\det\Lambda = 1\) we end up ruling out spatial reflections too. Hence, \(\SO*(n-1,1)\) is connected \cite{oneill1983SemiRiemannianGeometryApplications}. The property \(\tensor{\Lambda}{^0_0} > 0\) is preserved by matrix multiplication, so \(\SO*(n-1,1)\) is indeed a group \cite{lester1993OrthochronousSubgroups}. 

        Restricting our attention to \(\SO*(3,1)\) also has physical motivation. It is well-known that the weak interactions explicitly break parity (spatial reflection) and time reversal symmetries. See, \eg, Refs. \citeonline{schwartz2014QuantumFieldTheory,weinberg1995Foundations}. Hence, even in flat spacetime, \(\Og(3,1)\) is not a fundamental symmetry of nature, but \(\SO*(3,1)\) is. Of course, the group describing the spacetime symmetries is \(\Og(3,1)\), but mathematical simplicity also allows us to focus on \(\SO*(3,1)\). This is what I will do for the rest of these notes. 

        Finally, a comment about nomenclature is in place. \(\SO(3,1)\) is known as the proper Lorentz group. The group
        \begin{equation}
            \Og*(3,1) = \qty{\Lambda \in \Mat(4); \Lambda^\intercal \eta \Lambda = \eta, \tensor{\Lambda}{^0_0} > 0}
        \end{equation}
        is known as the orthochronous Lorentz group, since it forbids time reversals. Finally, 
        \begin{equation}
            \SO*(3,1) = \SO(3,1) \cap \Og*(3,1)
        \end{equation}
        is the proper orthochronous Lorentz group, also called the restricted Lorentz group. Through the remainder of this text, I will simply say ``Lorentz group'' when referencing the restricted Lorentz group. We will not be interested in the larger versions.

    \subsection{Case Study: Poincaré Group}
        When discussing the Lorentz group we restricted our attention to linear isometries. Apart from mathematical simplicity, there is no reason to make this restriction. Hence, we now consider the full group of isometries of Minkowski spacetime. These are the transformations \(\tensor{x}{^\mu} \to \tensor{x}{^\prime^\mu}\) such that
        \begin{equation}
            \tensor{\eta}{_\mu_\nu}\pdv{\tensor{x}{^\prime^\mu}}{\tensor{x}{^\rho}}\pdv{\tensor{x}{^\prime^\nu}}{\tensor{x}{^\sigma}} = \tensor{\eta}{_\rho_\sigma}.
        \end{equation}
        Notice this equation means the Jacobian \(\pdv*{\tensor{x}{^\prime^\mu}}{\tensor{x}{^\rho}}\) must be a Lorentz transformation. Hence, we have the differential equation
        \begin{equation}
            \pdv{\tensor{x}{^\prime^\mu}}{\tensor{x}{^\rho}} = \tensor{\Lambda}{^\mu_\rho}.
        \end{equation}
        Integrating this equation leads us to 
        \begin{equation}
            \tensor{x}{^\prime^\mu} = \tensor{\Lambda}{^\mu_\rho} \tensor{x}{^\rho} + \tensor{a}{^\mu},
        \end{equation}
        for an arbitrary vector \(\tensor{a}{^b} \in \R^n\) which appears as an integration constant. If we omit the indices, we can write the resulting transformation as
        \begin{equation}\label{eq: general-poincare-transf}
            x \to x' = \Lambda x + a.
        \end{equation}
        This is known as a Poincaré transformation, and we denote (\ref{eq: general-poincare-transf}) as \((\Lambda,a)\). Once we restrict attention to \(\Lambda \in \SO*(n-1,1)\), we get the group 
        \begin{equation}
            \ISO*(n-1,1) = \qty{(\Lambda,a); \Lambda \in \SO*(n-1,1), a \in \R^n}.
        \end{equation}
        This is known as the proper orthochronous Poincaré group, or restricted Poincaré group. As with the Lorentz group, I will simply call this the Poincaré group.

        To find the group product of the restricted Poincaré group we perform two Poincaré transformations in sequence. Define 
        \begin{equation}
            \left\lbrace\begin{aligned}
                x &\to x' = \Lambda x + a, \\
                x' &\to x'' = \Lambda' x' + a'.
            \end{aligned}\right.
        \end{equation}
        Then
        \begin{equation}
            x \to x'' = \Lambda' \Lambda x + \Lambda' a + a'.
        \end{equation}
        Hence, 
        \begin{equation}
            (\Lambda',a') \odot (\Lambda,a) = (\Lambda' \Lambda, \Lambda' a + a'),
        \end{equation}
        where \(\odot\) denotes the group product. 

        We have two particularly interesting subgroups of \(\ISO*(n-1,1)\). Notice that
        \begin{equation}\label{eq: subgroup-poincare-lorentz}
            \qty{(\Lambda,0); \Lambda \in \SO*(n-1,1)}
        \end{equation}
        is isomorphic to the restricted Lorentz group \(\SO*(n-1,1)\). Furthermore, 
        \begin{equation}\label{eq: subgroup-poincare-translations}
            \qty{(\Eins,a); a \in \R^n}
        \end{equation}
        is isomorphic to \((\R^n,+)\), the group of translations\footnote{Mathematically, this is the group of \(n\)-dimensional real vectors with the operation of addition as the group product.}. There are some interesting ideas connected to these two subgroups. 

        Firstly, notice that given any \((\Lambda,b) \in \ISO*(n-1,1)\) and \(a \in \R^n\) we have that
        \begin{subequations}
            \begin{align}
                (\Lambda,b)^{-1}\odot(\Eins,a)\odot(\Lambda,b) &= (\Lambda^{-1},-\Lambda^{-1}b)\odot(\Eins,a)\odot(\Lambda,b), \\ 
                &= (\Lambda^{-1},-\Lambda^{-1}b)\odot(\Lambda,a+b), \\
                &= (\Lambda^{-1} \Lambda, \Lambda^{-1}a + \Lambda^{-1}b -\Lambda^{-1}b), \\
                &= (\Eins, \Lambda^{-1}a) \in \R^n,
            \end{align}
        \end{subequations}
        where the inclusion in the last line implicitly assumes an isomorphism between (\ref{eq: subgroup-poincare-translations}) and \((\R^n,+)\). This means that \(R^n\) is not only a subgroup of \(\ISO*(n-1,1)\), but a normal subgroup. 

        \begin{definition}[Normal Subgroup]
            Let \(G\) be a group and \(N \subeq G\) be a subgroup. We say \(N\) is a \emph{normal} subgroup, and write \(N \nsub G\), if it holds that for any \(n \in N\) and any \(g \in G\) we have \(g^{-1} n g \in N\).
        \end{definition}

        Normal subgroups are a particularly important class of subgroups, as discussed in Ref. \citeonline{geroch1985MathematicalPhysics}, for example. For our purposes, it is useful to notice the following. Let \(g \in G\) (\(G\) a group) and \(H \subeq G\) be a subgroup. Then define
        \begin{equation}
            g^{-1} H g = \qty{g^{-1} h g; h \in H}.
        \end{equation}
        This is in general a new subgroup of \(G\) that is isomorphic to \(H\). Hence, we get many copies of \(H\) inside \(G\). For a normal subgroup, we have a ``uniqueness property'' in the sense that all of these copies coincide, for \(g^{-1} N g \subeq N\) due to the very definition of normal subgroup\footnote{This ``uniqueness property'' should be read with care. Consider \(G = (\R^2,+)\), for example. While \(N = (\R,+) \nsub G\) is a normal subgroup of \(G\), there are multiple copies of \(N\) inside \(G\) (one for each direction in the plane). However, it still holds that \(g^{-1} N g \subeq N\) for all \(g \in G\), and hence we cannot generate new copies of \(N\) by merely writing \(g^{-1} N g\).}. Some authors, especially in the physics literature, prefer to call normal subgroups ``invariant subgroups''. 

        Notice that \(\SO*(n-1,1)\) is not a normal subgroup. Indeed, 
        \begin{subequations}
            \begin{align}
                (\Lambda',a)^{-1}\odot(\Lambda,0)\odot(\Lambda',a) &= (\Lambda'^{-1},-\Lambda'^{-1}a)\odot(\Lambda,0)\odot(\Lambda',a), \\ &= (\Lambda'^{-1},-\Lambda'^{-1}a)\odot(\Lambda\Lambda',\Lambda a), \\
                &= (\Lambda'^{-1}\Lambda\Lambda',\Lambda'^{-1}\Lambda a -\Lambda'^{-1}a),
            \end{align}
        \end{subequations}
        and this is not in general an element of (\ref{eq: subgroup-poincare-lorentz}).

        It is also interesting to notice that if we understand \(\SO*(n-1,1)\) and \(\R^n\) as the subgroups given on (\ref{eq: subgroup-poincare-lorentz}) and (\ref{eq: subgroup-poincare-translations}), then \(\SO*(n-1,1) \cap \R^n = \qty{(\Eins,0)}\), which is the subgroup composed solely by the neutral element. Furthermore, notice that any element of \(\ISO*(n-1,1)\) can be written in the form
        \begin{equation}
            (\Lambda,a) = (\Lambda,0)\odot(\Eins,\Lambda^{-1}a).
        \end{equation}
        We thus write \(\ISO*(n-1,1) = \SO*(n-1,1) \odot \R^n\), meaning that every Poincaré transformation is a translation followed by a Lorentz transformation. Due to \(\R^n\) being normal, this can also be rewritten as \(\ISO*(n-1,1) = \R^n \odot \SO*(n-1,1)\) with 
        \begin{equation}
            (\Lambda,a) = (\Eins,a)\odot(\Lambda,0).
        \end{equation}

        This sort of structure has a particular name in group theory. 

        \begin{definition}[Semidirect Product of Groups]
            Let \(G\) be a group with neutral element \(e\), \(H, N \subeq G\) subgroups, and \(N \nsub G\). If
            \begin{equation}
                G = HN = \qty{hn; h \in H, n \in N}
            \end{equation}
            and \(H \cap N = \qty{e}\), then we say \(G\) is the semidirect product of \(N\) and \(H\) and write \(G = H \ltimes N\).
        \end{definition}

        The notation \(H \ltimes N\) makes reference to the facts that, as a set, \(G = H \times N\) and, as a group, \(N \nsub G\). 

        Semidirect products can also be defined for two given groups and used as a way of building a third new group \cite{robinson1996CourseTheoryGroups,robinson2022AbstractAlgebraIntroduction}, but for us this internal definition will be enough. 

        It is useful to introduce a last definition for future use: the notion of a direct product of groups. 
        \begin{definition}[Direct Product of Groups]
            Let \(G\) be a group with neutral element \(e\), \(H, N \subeq G\) subgroups, and \(H, N \nsub G\). If \(G = H \ltimes N\), we say \(G\) is the \emph{direct product} of \(H\) and \(N\) and write \(G = H \times N\).
        \end{definition}
        Notice the direct product is symmetric in \(H\) and \(N\), since both of them are normal subgroups. In particular, it holds that \(G = H \ltimes N = N \ltimes H\). Some authors, especially in the physics literature, denote the direct product of groups by \(\otimes\).

\section{Symmetries in Curved Spacetimes}\label{sec: symmetries-curved-spacetime}
    Now that we know how to characterize symmetries in terms of groups, the next step should be to adapt this discussion to curved spacetimes. This will require us to develop some new geometric language in order to appropriately formulate what constitutes a symmetry in a curved spacetime. Furthermore, we will be able to provide a geometric, coordinate-free construction of the isometry groups we previously discussed. Parts of this section are inspired by Appendix C of Ref. \citeonline{wald1984GeneralRelativity}.

    This section will rely considerably on a basic understanding of differential geometry as required for \gls{GR}. Examples of textbooks on differential geometry are Refs. \citeonline{tu2011IntroductionManifolds,lee2012IntroductionSmoothManifolds}, but basic differential geometry is also reviewed in \gls{GR} books such as Refs. \citeonline{hawking1973LargeScaleStructure,wald1984GeneralRelativity}.

    \subsection{Pullbacks and Pushforwards}
        We now take an active point of view on coordinate transformations, which should now be called diffeomorphisms. Rather than thinking about them as mere changes of coordinates, we will think of them as active transformations that take one manifold into another one and discuss how the structures on these manifolds transform. To understand this, we will have to discuss some properties about smooth mappings between manifolds. 

        Let \(M\) and \(N\) be manifolds. Given a point \(p \in M\), we denote the space of tangent vectors to \(M\) at \(p\) by \(\T[p]M\). Suppose we are given a smooth mapping \(\phi \colon M \to N\). This mapping gives us some structure to relate \(M\) and \(N\). Which sorts of relations can we derive from this? 

        Suppose first that we are given a smooth function \(f \colon N \to \R\). We can ``pullback'' \(f\) using \(\phi\) by defining a new function \(\phi_* f \colon M \to \R\) through \(\phi_* f = f \circ \phi\). This can be depicted in the diagram 
        \begin{center}
            \includestandalone{pullback}
        \end{center}
        We call \(\phi_* f\) the pullback of \(f\) through \(\phi\). This name comes from the fact that \(\phi\) is ``pulling back'' the function \(f\) from \(N\) to \(M\). I should remark that in the mathematics literature the pullback is often written as \(\phi^*\), while \(\phi_*\) denotes the pushforward (to be defined). I prefer to follow the conventions used by \textcite{wald1984GeneralRelativity}.

        Next suppose we have a vector \(v \in \T[p]M\) (I will sometimes omit the abstract indices to avoid cluttering the notation in what follows). While we could pull back a function, we can ``push forward'' this vector. Recall that a vector at \(p\) is a linear map \(v \colon \ck[\infty](p) \to \R\) (with \(\ck[\infty](p)\) being the space of functions which are smooth on some neighborhood of \(p\)) satisfying certain conditions so that it behaves as a directional derivative \cite{lee2012IntroductionSmoothManifolds,tu2011IntroductionManifolds}. We can define a new vector \(\phi^* v \in \T[\phi(p)]N\) by imposing that
        \begin{equation}
            \phi^* v (f) = v(\phi_* f) = v(f \circ \phi)
        \end{equation}
        for every \(f \in \ck[\infty](\phi(p))\). Diagrammatically this can be written as
        \begin{center}
            \includestandalone{pushforward}
        \end{center}
        where \(\phi_*\) is the pullback map defined as \(\phi_* f = f \circ \phi\).

        Notice that the pushforward \(\phi^*\) is a linear map \(\phi^* \colon \T[p]M \to \T[\phi(p)]N\) and can thus be interpreted as the derivative of \(\phi\) at \(p\) in the usual sense of multivariable calculus (although it is now generalized to manifolds). The pushforward of a smooth mapping \(\phi\) is also known as the differential or the tangent map of \(\phi\). 

        This process does not stop at the level of vectors. Let \(\omega \in \T*[\phi(p)]N\) be a one-form at \(\phi(p) \in N\). Recall that a one-form in \(\T*[\phi(p)]N\) is a linear map \(\omega \colon \T[\phi(p)]N \to \R\). We can pull back this one-form to a new one-form \(\phi_* \omega \in \T*[p]M\) by defining
        \begin{equation}
            \phi_* \omega(v) = \omega(\phi^* v)
        \end{equation}
        for every \(v \in \T[p]M\). We can write this in the diagram 
        \begin{center}
            \includestandalone{pullback-form}
        \end{center}
        where \(\phi^* \colon \T[p]M \to \T[\phi(p)]N\) is the pushforward. 

        Given a tensor \(T\) of type \((0,l)\) at \(p\), we can also pull-it-back by defining \(\phi_* T\) through
        \begin{equation}
            \phi_* T (v_1, \ldots, v_l) = T(\phi^* v_1, \ldots, \phi^* v_l).
        \end{equation}
        Analogously, for a tensor \(T\) of type \((k,0)\) we define the pushforward through
        \begin{equation}
            \phi^* T (\omega_1, \ldots, \omega_k) = T(\phi_* \omega_1, \ldots, \phi_* \omega_k).
        \end{equation}
        This is consistent with our previous definition of pushforward of a vector due to the definition of pullback of a one-form. 

        Notice we thus get a number of relations between different structures defined on \(M\) and \(N\). Nevertheless, we have quite some restrictions. We cannot push forward nor pull back a tensor of mixed type, since we do not know how to pull back contravariant tensors and we do not know how to push forward covariant tensors. 

        Suppose, however, that \(\phi \colon M \to N\) is not only smooth, but actually a diffeomorphism. This means that \(\phi\) is smooth, bijective, and has a smooth inverse\footnote{Not all smooth and bijective maps have a smooth inverse. For example, consider the function \(x \mapsto x^3\) over the real numbers. While it is smooth and bijective, the inverse \(x \mapsto \sqrt[3]{x}\) is not differentiable at \(x = 0\).}. In particular this implies that \(\dim M = \dim N\). In this case, we have the inverse mapping \(\phi^{-1}\) and we can exploit it to define the pullbacks and pushforwards that \(\phi\) cannot handle. Given a tensor \(T\) of type \((k,l)\) at \(p \in M\), we define 
        \begin{equation}
            \phi^* T (\omega_1, \ldots, \omega_k, v_1, \ldots, v_l) = T(\phi_* \omega_1, \ldots, \phi_* \omega_k, (\phi^{-1})^* v_1, \ldots, (\phi^{-1})^* v_l).
        \end{equation}
        The pullback \(\phi_*\) is defined analogously, but since it holds that \(\phi_* = (\phi^{-1})^*\) it suffices to work with the pushforward. 

        Notice that a diffeomorphism is analogous to an isomorphism, but for manifolds. It gives the natural equivalence relation between manifolds. If two manifolds admit a diffeomorphism, they are essentially the same as far as the differentiable structure go. Nevertheless, further structure added to the manifold---such as a metric---may not be invariant under a given diffeomorphism.

        If we consider a diffeomorphism \(\phi \colon M \to M\), we get an interesting structure. Given a tensor field \(T\), we can now consider the pushforward of the entire field, \(\phi^* T\). This leads us to a new tensor field that can be compared to the original tensor field \(T\). In general, these two tensor fields will not be the same, even though they are related by a diffeomorphism. In coordinate-parlance, the coordinate components of \(T\) are preserved, but they are taken to \(\phi(p)\) instead of being kept at \(p\). This is an ambiguity in the description, and hence it is understood as a gauge symmetry. We get physical symmetries in the particular case in which \(\phi^* T = T\), so that acting on the spacetime with a diffeomorphism keeps everything unchanged---this is not a mere redundancy, but an actual symmetry of the spacetime. The diffeomorphisms that keep the metric invariant, \ie, the diffeomorphisms with \(\phi^* g = g\), are called isometries. 

    \subsection{Lie Derivatives}
        An interesting structure we can consider when dealing with families of diffeomorphisms is a one-parameter group of diffeomorphisms. This is a map \(\phi \colon \R \times M \to M\) such that \(\phi_t \colon M \to M\) is a diffeomorphism for every \(t \in \R\) and the map \(t \to \phi_t\) is such that \(\phi_t \circ \phi_s = \phi_{t+s}\). Notice then that \(\qty{\phi_t}_{t\in\R}\) has a natural group structure, with the product being given by the composition of diffeomorphisms.

        Take a point \(p \in M\) and consider a one-parameter group of diffeomorphisms in \(M\). Notice that \(\gamma_p(t) = \phi_t(p)\) defines a curve in \(M\)---this is called an orbit of \(\phi_t\). If we differentiate \(\gamma_p\) at \(t = 0\), we get a vector \(v_p \in \T[p]M\). Through this process we get a vector field \(v\) associated to \(\phi_t\) which is everywhere parallel to the orbits of \(\phi_t\). Inversely, given a smooth vector field \(v\), it is always possible to find a one-parameter group of diffeomorphisms in \(M\) whose orbits are all parallel to \(v\). This is said to be the flow of \(v\). For details, see Refs. \citeonline{lee2012IntroductionSmoothManifolds,tu2011IntroductionManifolds,wald1984GeneralRelativity}, for example. I should mention that the one-parameter group of diffeomorphisms may not be defined for all parameter values \(t \in \R\), but rather be defined only on a smaller interval. In any case, it is always possible to find a local flow. I will ignore these subtleties in what follows.

        Suppose now we want to consider how a given tensor field \(T\) changes in spacetime. It is, of course, useful to have a derivative of \(T\) to analyze this variation. We would like to compute something like 
        \begin{equation}\label{eq: goal-lie-derivative}
            \eval{\dv{T}{t}}_p \sim \lim_{t \to 0} \frac{T(p + t) - T(p)}{t}.
        \end{equation}
        Of course, this equation makes no sense. There is no vector space structure on the manifold for \(p+t\) to make sense and we cannot compare tensors defined at different points, since they live in different tangent spaces. However, if we are given a one-parameter group of diffeomorphisms \(\qty{\phi_t}\) with tangent vector field \(v\) (or, equivalently, if we are given the vector field and consider the one-parameter group of diffeomorphisms) we can define 
        \begin{equation}\label{eq: definition-lie-derivative}
            \Lie[v]\tensor{T}{^{a_1}^\cdots^{a_k}_{b_1}_\cdots_{b_l}} = \lim_{t\to 0} \frac{\phi_{-t}^*\tensor{T}{^{a_1}^\cdots^{a_k}_{b_1}_\cdots_{b_l}} - \tensor{T}{^{a_1}^\cdots^{a_k}_{b_1}_\cdots_{b_l}}}{t}.
        \end{equation}
        Notice that \(\phi_{-t}^*T = \phi_{t*}T\) picks the value of the tensor field at ``\(p+t\)'' and pulls it back to \(p\). From a different perspective, we use \(\phi_{-t}\) to push forward the tensor at ``\(p+t\)'' to ``\(p+t-t=p\)''. Hence, \cref{eq: definition-lie-derivative} gives a precise mathematical meaning to \cref{eq: goal-lie-derivative}. \(\Lie[v]\) is known as the Lie derivative with respect to \(\tensor{v}{^a}\). Notice it is a linear map that takes smooth \((k,l)\)-tensor fields to smooth \((k,l)\)-tensor fields. The Lie derivative preserves contractions and it can be shown that the Lie derivative obeys the Leibnitz rule on tensor products,
        \begin{equation}
            \Lie[v](S \otimes T) = (\Lie[v]S) \otimes T + S \otimes (\Lie[v]T).
        \end{equation}
        Notice that \(\Lie[v]T = 0\) everywhere if, and only if, \(\phi_t\) is a symmetry of \(T\) for all \(t\), \ie, \(\phi_t^* T = T\) for all \(t\).

        At a point \(p\), notice that
        \begin{subequations}\label{eq: lie-derivative-t-derivative}
            \begin{align}
                \qty(\Lie[v]\tensor{T}{^{a_1}^\cdots^{a_k}_{b_1}_\cdots_{b_l}})_p &= \lim_{t\to 0} \frac{\qty(\phi_{-t}^*\tensor{T}{^{a_1}^\cdots^{a_k}_{b_1}_\cdots_{b_l}})_p - \qty(\tensor{T}{^{a_1}^\cdots^{a_k}_{b_1}_\cdots_{b_l}})_p}{t}, \\ 
                &= \eval{\dv{t}}_{t=0} \qty(\phi_{-t}^*\tensor{T}{^{a_1}^\cdots^{a_k}_{b_1}_\cdots_{b_l}})_p.
            \end{align}
        \end{subequations}
        In particular, consider the case of a scalar function. Then 
        \begin{subequations}
            \begin{align}
                (\Lie[v]f)_p &= \eval{\dv{t}}_{t=0} (\phi_{-t}^*f)_p, \\
                &= \eval{\dv{t}}_{t=0} (\phi_{t*}f)_p, \\
                &= \eval{\dv{t}}_{t=0} (f \circ \phi_{t})_p, \\
                &= v(f)_p,
            \end{align}
        \end{subequations}
        where the last step can be performed by choosing a coordinate system and employing the chain rule \cite{wald1984GeneralRelativity}. We thus learn that
        \begin{equation}
            \Lie[v]f = v(f).
        \end{equation}

        Now that we know how the Lie derivative acts on functions, the next simplest step is to learn how it acts on vector fields. To find that, introduce a coordinate system such that the parameter \(t\) along the integral lines of \(\tensor{v}{^a}\) is one of the coordinates, \(\tensor{x}{^1}\). In this manner, we have that
        \begin{equation}
            \tensor{v}{^a} = \tensor{\qty(\pdv{\tensor{x}{^1}})}{^a}.
        \end{equation}
        This corresponds to choosing the function \(\tensor{x}{^1}\) such that \(v(\tensor{x}{^1}) = 1\), which can always be done in neighborhoods in which \(\tensor{v}{^a}\) does not vanish. 

        In such a coordinate system, acting with \(\phi_{-t}\) is equivalent to performing the coordinate transformation \(\tensor{x}{^1} \to \tensor{x}{^1} + t\) while holding the remaining coordinates fixed. The matrix components of the pushforward \(\phi_{-t}^* \colon \T[p]M \to \T[\phi_{-t}(p)]M\) are then given in the coordinate basis by 
        \begin{equation}
            \tensor{(\phi_{-t}^*)}{^\mu_\nu} = \tensor{\delta}{^\mu_\nu}.
        \end{equation}
        Hence, this means that 
        \begin{equation}
            (\phi_{-t}^*\tensor{T}{^{\mu_1}^\cdots^{\mu_k}_{\nu_1}_\cdots_{\nu_l}})(\tensor{x}{^1},\ldots,\tensor{x}{^n}) = \tensor{T}{^{\mu_1}^\cdots^{\mu_k}_{\nu_1}_\cdots_{\nu_l}}(\tensor{x}{^1} + t,\ldots,\tensor{x}{^n}).
        \end{equation}
        Using this in the expression for the Lie derivative, \cref{eq: definition-lie-derivative}, we conclude that in this coordinate system
        \begin{equation}
            \Lie[v]\tensor{T}{^{\mu_1}^\cdots^{\mu_k}_{\nu_1}_\cdots_{\nu_l}} = \pdv{\tensor{T}{^{\mu_1}^\cdots^{\mu_k}_{\nu_1}_\cdots_{\nu_l}}}{\tensor{x}{^1}}.
        \end{equation}
        Notice that with this expression it is particularly simple to prove the Leibnitz rule.

        Using this general expression, we see that given a vector field \(\tensor{w}{^a}\) we have, in this coordinate system adapted to \(\tensor{v}{^a}\),
        \begin{equation}
            \Lie[v]\tensor{w}{^\mu} = \pdv{\tensor{w}{^\mu}}{\tensor{x}{^1}}.
        \end{equation}

        Similarly, we can compute the commutator\footnote{Recall that the commutator of two vector fields \(\tensor{v}{^a}\) and \(\tensor{w}{^a}\) is defined as the vector field \(\tensor{\comm{v}{w}}{^a}\) such that \(\comm{v}{w}(f) = v(w(f)) - w(v(f))\). Using a covariant derivative, we can write \(\tensor{\comm{v}{w}}{^a} = \tensor{v}{^b}\tensor{\nabla}{_b}\tensor{w}{^a} - \tensor{w}{^b}\tensor{\nabla}{_b}\tensor{v}{^a}\).} \(\tensor{\comm{v}{w}}{^\mu}\) and find that
        \begin{equation}
            \tensor{\comm{v}{w}}{^\mu} = \pdv{\tensor{w}{^\mu}}{\tensor{x}{^1}}
        \end{equation}
        too. Hence, we conclude that in this coordinate system we have 
        \begin{equation}
            \Lie[v]\tensor{w}{^\mu} = \tensor{\comm{v}{w}}{^\mu}.
        \end{equation}
        Since this equation is covariant and both quantities are defined in a coordinate-independent manner, we can conclude that 
        \begin{equation}
            \Lie[v]\tensor{w}{^a} = \tensor{\comm{v}{w}}{^a}
        \end{equation}
        as a tensor equality. 

        Having the expressions for the Lie derivatives of scalars and vector fields, we can derive all other cases. These formulae are more easily expressed in terms of covariant derivatives, so that is what we will do. As an example, let us consider the Lie derivative for one-forms. Given some one-form \(\tensor{\mu}{_a}\) and a vector field \(\tensor{w}{^a}\), we have that 
        \begin{subequations}
            \begin{align}
                \Lie[v](\tensor{\mu}{_a}\tensor{w}{^a}) &= v(\tensor{\mu}{_a}\tensor{w}{^a}), \\ 
                &= \tensor{v}{^b}\tensor{\nabla}{_b}(\tensor{\mu}{_a}\tensor{w}{^a}), \\ 
                &= \tensor{v}{^b}\tensor{w}{^a}\tensor{\nabla}{_b}\tensor{\mu}{_a} + \tensor{v}{^b}\tensor{\mu}{_a}\tensor{\nabla}{_b}\tensor{w}{^a}.
            \end{align}
        \end{subequations}
        This follows from using the expression of the Lie derivative for scalar functions. However, if we instead used the Leibnitz rule and the expression for the Lie derivative of a vector field we would have found
        \begin{subequations}
            \begin{align}
                \Lie[v](\tensor{\mu}{_a}\tensor{w}{^a}) &= \tensor{w}{^a}\Lie[v]\tensor{\mu}{_a} + \tensor{\mu}{_a}\Lie[v]\tensor{w}{^a}, \\
                &= \tensor{w}{^a}\Lie[v]\tensor{\mu}{_a} + \tensor{\mu}{_a}\tensor{\comm{v}{w}}{^a}, \\
                &= \tensor{w}{^a}\Lie[v]\tensor{\mu}{_a} + \tensor{\mu}{_a}\tensor{v}{^b}\tensor{\nabla}{_b}\tensor{w}{^a} - \tensor{\mu}{_a}\tensor{w}{^b}\tensor{\nabla}{_b}\tensor{v}{^a}.
            \end{align}
        \end{subequations}
        Bringing everything together and solving for \(\tensor{w}{^a}\Lie[v]\tensor{\mu}{_a}\) we find that
        \begin{equation}
            \tensor{w}{^a}\Lie[v]\tensor{\mu}{_a} = \tensor{v}{^b}\tensor{w}{^a}\tensor{\nabla}{_b}\tensor{\mu}{_a} + \tensor{\mu}{_b}\tensor{w}{^a}\tensor{\nabla}{_a}\tensor{v}{^b},
        \end{equation}
        with \(\tensor{w}{^a}\) arbitrary. Therefore, it follows that
        \begin{equation}
            \Lie[v]\tensor{\mu}{_a} = \tensor{v}{^b}\tensor{\nabla}{_b}\tensor{\mu}{_a} + \tensor{\mu}{_b}\tensor{\nabla}{_a}\tensor{v}{^b}.
        \end{equation}

        The general expression for an arbitrary tensor field can be obtained by induction. One finds that
        \begin{equation}\label{eq: general-expression-lie-derivative}
            \Lie[v]\tensor{T}{^{a_1}^\cdots^{a_k}_{b_1}_\cdots_{b_l}} = \tensor{v}{^c}\tensor{\nabla}{_c}\tensor{T}{^{a_1}^\cdots^{a_k}_{b_1}_\cdots_{b_l}} - \sum_{i=1}^{k} \tensor{T}{^{a_1}^\cdots^c^\cdots^{a_k}_{b_1}_\cdots_{b_l}}\tensor{\nabla}{_c}\tensor{v}{^{a_i}} + \sum_{j=1}^l \tensor{T}{^{a_1}^\cdots^{a_k}_{b_1}_\cdots_c_\cdots_{b_l}}\tensor{\nabla}{_{b_j}}\tensor{v}{^c}.
        \end{equation}
        This expression holds for \emph{any} covariant derivative (see Sec. 3.1 of Ref. \citeonline{wald1984GeneralRelativity} for a definition), not only for the covariant derivative \(\tensor{\nabla}{_a}\) with \(\tensor{\nabla}{_a}\tensor{g}{_b_c} = 0\) (known as the Levi-Civita connection). 

    \subsection{Conformal Killing Vector Fields}
        At this stage, we are ready to discuss an important class of concepts: Killing vector fields and conformal Killing vector fields. 

        We begin by giving meaning to the term ``conformal''. Let \((M,\tensor{g}{_a_b})\) and \((N,\tensor*{g}{^\prime_a_b})\) be pseudo-Riemannian manifolds. A conformal transformation is a transformation that preserves angles. Within \gls{GR} and differential geometry, a conformal transformation is a smooth map \(\phi \colon M \to N\) such that \(\phi_* \tensor*{g}{^\prime_a_b} = \Omega^2 \tensor*{g}{_a_b} \equiv \tensor*{\tilde{g}}{_a_b}\) for some (smooth) function \(\Omega > 0\). This transformation preserves angles because 
        \begin{equation}
            \frac{\tensor{g}{_a_b}\tensor{\xi}{^a}\tensor{\psi}{^b}}{\sqrt{(\tensor{g}{_c_d}\tensor{\xi}{^c}\tensor{\xi}{^d})(\tensor{g}{_e_f}\tensor{\psi}{^e}\tensor{\psi}{^f})}} = \frac{\tensor{\tilde{g}}{_a_b}\tensor{\xi}{^a}\tensor{\psi}{^b}}{\sqrt{(\tensor{\tilde{g}}{_c_d}\tensor{\xi}{^c}\tensor{\xi}{^d})(\tensor{\tilde{g}}{_e_f}\tensor{\psi}{^e}\tensor{\psi}{^f})}}.
        \end{equation}
        In particular, \(\tensor{g}{_a_b}\tensor{k}{^a}\tensor{k}{^b} = 0\) if, and only if, \(\tensor{\tilde{g}}{_a_b}\tensor{k}{^a}\tensor{k}{^b} = 0\). Therefore, under a conformal transformation of the metric, null vectors are preserved and, as a consequence, so are the lightcones. This means that the causal structure of spacetime is preserved under conformal transformations.

        A diffeomorphism \(\phi \colon M \to N\) such that \(\phi_* \tensor*{g}{^\prime_a_b} = \tensor{g}{_a_b}\) is called an isometry. There is a second interesting class of diffeomorphisms that ``almost preserve'' the metric, known as confomorphisms\footnote{This is the nomenclature used by \textcite{choquet-bruhat2009GeneralRelativityEinstein}. Ref.  \textcite{wald1984GeneralRelativity}, for example, uses the term ``conformal isometry''. I prefer to use confomorphism to make it clear that it is not necessarily an isometry.}. These are the diffeomorphisms such that \(\phi_* \tensor*{g}{^\prime_a_b} = \Omega^2 \tensor{g}{_a_b}\) for some smooth function \(\Omega > 0\). Hence, a confomorphism is a diffeomorphism that is also a conformal transformation. While confomorphisms do not need to preserve the metric, they always preserve the so-called conformal structure, which is the equivalence class of pseudo-Riemannian manifolds related by confomorphisms.

        Suppose now we are given a one-parameter group of confomorphisms \(\qty{\phi_t}\) everywhere parallel to the vector field \(\tensor{v}{^a}\). Isometries will be just a limiting case. Then, for small values of \(t\), we can use a Taylor expansion to find that
        \begin{subequations}
            \begin{align}
                \phi_{t*} \tensor{g}{_a_b} &= \phi^*_{-t} \tensor{g}{_a_b}, \\ 
                &= \phi^*_{0} \tensor{g}{_a_b} + t \eval{\dv{t}}_{t=0}\qty(\phi^*_{-t} \tensor{g}{_a_b}) + \order{t^2}, \\
                &= \tensor{g}{_a_b} + t \Lie[v] \tensor{g}{_a_b} + \order{t^2},
            \end{align}
        \end{subequations}
        where I used \cref{eq: lie-derivative-t-derivative} and the fact that \(\phi^*_0 = \mathrm{id}\) for a one-parameter group of diffeomorphisms. Since \(\qty{\phi_t}\) is a one-parameter group of confomorphisms, it must hold that \(\phi_{t*}\tensor{g}{_a_b} = \Omega_t^2 \tensor{g}{_a_b}\) for all \(t\), with \(\Omega_t\) a smooth positive function for each value of \(t\). Since \(\phi^*_0 = \mathrm{id}\), we know \(\Omega_0 = 1\). Using this we find that
        \begin{subequations}\label{eq: CKV-lie-derivative-from-pullback}
            \begin{align}
                \phi_{t*} \tensor{g}{_a_b} &= \Omega_t^2 \tensor{g}{_a_b}, \\
                \tensor{g}{_a_b} + t \Lie[v] \tensor{g}{_a_b} + \order{t^2} &= \tensor{g}{_a_b} + t \eval{\dv{\Omega_t^2}{t}}_{t=0} \tensor{g}{_a_b} + \order{t^2}, \\
                \Lie[v] \tensor{g}{_a_b} &= \eval{\dv{\Omega_t^2}{t}}_{t=0} \tensor{g}{_a_b}.
            \end{align}
        \end{subequations}
        Defining the function \(\lambda\) as
        \begin{equation}
            \lambda \equiv \eval{\dv{\Omega_t^2}{t}}_{t=0},
        \end{equation}
        we conclude that
        \begin{equation}\label{eq: conformal-killing-lie-derivative}
            \Lie[v]\tensor{g}{_a_b} = \lambda \tensor{g}{_a_b}.
        \end{equation}
        Notice that for an isometry we have \(\Omega_t^2 = 1\) for all \(t\), and therefore \(\lambda = 0\) in this case.

        Using \cref{eq: general-expression-lie-derivative} for a covariant derivative with \(\tensor{\nabla}{_a}\tensor{g}{_b_c} = 0\), we find that
        \begin{equation}
            \Lie[v]\tensor{g}{_a_b} = \tensor{\nabla}{_a}\tensor{v}{_b} + \tensor{\nabla}{_b}\tensor{v}{_a}
        \end{equation}
        and \cref{eq: conformal-killing-lie-derivative} becomes
        \begin{equation}
            \tensor{\nabla}{_a}\tensor{v}{_b} + \tensor{\nabla}{_b}\tensor{v}{_a} = \lambda \tensor{g}{_a_b}.
        \end{equation}
        For general \(\lambda\), this is known as the conformal Killing equation. For \(\lambda = 0\) (corresponding to isometries rather than confomorphisms) this is known as the Killing equation.  

        \(\lambda\) is uniquely fixed by the vector field \(\tensor{v}{^a}\), which has all the information about the one-parameter group of confomorphisms and therefore has all the information about the original conformal factor \(\Omega_t\). Assuming spacetime is \(n\)-dimensional and contracting both sides of the conformal Killing equation with \(\tensor{g}{^a^b}\) leads to
        \begin{equation}
            2 \tensor{\nabla}{_a}\tensor{v}{^a} = n \lambda,
        \end{equation}
        which establishes the value of \(\lambda\). Hence, the conformal Killing equation becomes
        \begin{equation}
            \tensor{\nabla}{_a}\tensor{v}{_b} + \tensor{\nabla}{_b}\tensor{v}{_a} = \frac{2}{n} \qty(\tensor{\nabla}{_c}\tensor{v}{^c}) \tensor{g}{_a_b}.
        \end{equation}

        Notice that Killing vector fields are infinitesimal generators of isometries, so they capture---in an infinitesimal sense---what are the symmetries of the metric. Conformal Killing vector fields are a bit more generous and consider the conformal symmetries of the metric. 

        In a curved spacetime, symmetries correspond to the integral lines of complete Killing vector fields, where ``complete'' means that their integral lines (\ie, their flow) is defined for all times. The symmetry group of the spacetime is then a subgroup of the group composed by all diffeomorphisms, \(\Diff(M)\) (the group product is the composition of maps). As an example, we say that a spacetime is spherically symmetric when the group of isometries has an \(\SO(3)\) subgroup and the orbits of this subgroup trace out spacelike two-spheres. This is discussed in more detail in Chap. 6 of Ref. \citeonline{wald1984GeneralRelativity} and Appendix B of Ref. \citeonline{hawking1973LargeScaleStructure}.
    
    \subsection{Case Study: Confomorphisms on a Sphere}\label{subsec: conformal-isometries-sphere}
        As a case study, let us consider what are the isometries and the confomorphisms on the two-sphere \(\Sph^2\). This will be a convenient discussion for later. 

\begin{figure}
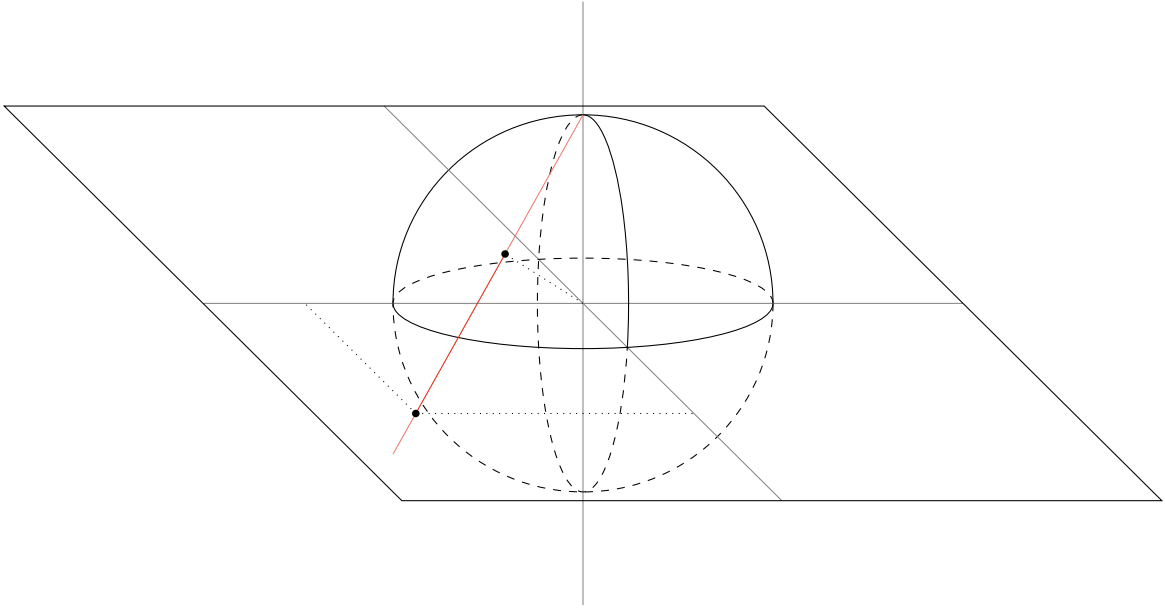

            \centering
            \includestandalone{stereographic}
            \caption{Diagram for construction of stereographic coordinates. One picks a point on the sphere and connects it to the north pole. One follows the line determined in this manner until it intercepts the \(xy\)-plane, at which point one records the coordinates \(x\) and \(y\) of the intersection. The coordinates attributed to the point on the sphere are then \(\zeta = x + iy\) and its conjugate \(\bar{\zeta}\). This corresponds to \(\zeta = e^{i\phi}\cot(\frac{\theta}{2})\).}
            \label{fig: stereographic-coordinates}
        \end{figure}

        To perform the calculations, it will be useful to employ some coordinate system. Spherical coordinates\footnote{In my conventions, \(\theta\) is the polar angle and \(\phi\) is the azimuthal angle.} \((\theta,\phi)\) are one of the options, but we will employ a different choice. Namely, we will work with stereographic coordinates. This corresponds to defining a complex coordinate
        \begin{equation}\label{eq: def-stereographic}
            \zeta = e^{i\phi} \cot(\frac{\theta}{2}) = \frac{z + r}{x - i y}.
        \end{equation}
        Geometrically this construction is illustrated on \cref{fig: stereographic-coordinates}. One draws the sphere as the unit sphere in three dimensions and traces lines from the north pole to different points of the sphere. Given a point on the sphere, the line from it to the north pole can be extended until it crosses the \(xy\)-plane, marking the coordinate \(\zeta = x + iy\) at the crossing point. In this manner, each point on the sphere is mapped to a point on the plane, and the north pole is mapped to infinity. It is worth mentioning that this is a way of viewing the sphere as the Riemann sphere \(\Comp \cup \qty{\infty}\).

        The round metric on the sphere is given by
        \begin{equation}
            \dd{\Sph}^2 = \dd{\theta}^2 + \sin^2{\theta}\dd{\phi}^2.
        \end{equation}
        In stereographic coordinates, we get
        \begin{subequations}\label{eq: sphere-metric-stereographic}
            \begin{gather}
                \dd{\Sph}^2 = 2 \tensor{\gamma}{_{\zeta}_{\bar\zeta}} \dd{\zeta} \dd{\bar{\zeta}},
                \intertext{with}
                \tensor{\gamma}{_{\zeta}_{\bar{\zeta}}} = \frac{2}{(1 + \zeta \bar{\zeta})^2}.
            \end{gather}
        \end{subequations}
        The Christoffel symbols for this metric in stereographic coordinates are
        \begin{equation}\label{eq: christoffel-sphere-stereographic}
            \tensor{\Gamma}{^{\zeta}_{\zeta}_{\zeta}} = - \frac{2 \bar{\zeta}}{1 + \zeta \bar{\zeta}} \qq{and} \tensor{\Gamma}{^{\bar{\zeta}}_{\bar{\zeta}}_{\bar{\zeta}}} = - \frac{2 \zeta}{1 + \zeta \bar{\zeta}}.
        \end{equation}

        Let us then consider the conformal Killing equation. A vector field \(\tensor{Y}{^a}\) is a conformal Killing vector field on the sphere if it satisfies
        \begin{equation}
            \tensor{\nabla}{_a}\tensor{Y}{_b} + \tensor{\nabla}{_b}\tensor{Y}{_a} = \tensor{\nabla}{_c}\tensor{Y}{^c} \tensor{\gamma}{_a_b},
        \end{equation}
        where \(\tensor{\gamma}{_a_b}\) is the round metric. It is convenient to write this in the form 
        \begin{equation}
            \tensor{\gamma}{_b_c}\tensor{\nabla}{_a}\tensor{Y}{^c} + \tensor{\gamma}{_a_c}\tensor{\nabla}{_b}\tensor{Y}{^c} = \tensor{\nabla}{_c}\tensor{Y}{^c} \tensor{\gamma}{_a_b}.
        \end{equation}
        In components, this expression is given by\footnote{Notice that for a vector field \(\tensor{Y}{^a}\) to be real, it is necessary that \(\tensor{Y}{^{\bar{\zeta}}} = \overline{\tensor{Y}{^{\zeta}}}\).} 
        \begin{subequations}\label{eq: ckv-sphere}
            \begin{align}
                \tensor{\partial}{_{\zeta}}\tensor{Y}{^{\bar{\zeta}}} &= 0, \label{subeq: ckv-sphere-1} \\
                \tensor{\partial}{_{\bar{\zeta}}}\tensor{Y}{^{\zeta}} &= 0, \label{subeq: ckv-sphere-2} \\
                \tensor{\partial}{_{\bar{\zeta}}}\tensor{Y}{^{\bar{\zeta}}} + \tensor{\Gamma}{^{\bar{\zeta}}_{\bar{\zeta}}_{\bar{\zeta}}}\tensor{Y}{^{\bar{\zeta}}} + \tensor{\partial}{_{\zeta}}\tensor{Y}{^{\zeta}} + \tensor{\Gamma}{^{\zeta}_{\zeta}_{\zeta}}\tensor{Y}{^{\zeta}} &= \tensor{\nabla}{_c}\tensor{Y}{^c}. \label{subeq: ckv-sphere-trivial}
            \end{align}
        \end{subequations}
        \cref{subeq: ckv-sphere-trivial} is trivial. The derivatives in \cref{subeq: ckv-sphere-1,subeq: ckv-sphere-2} should be understood as Wirtinger derivatives, which are defined and discussed on Appendix \ref{app: complex-analysis}. In summary, for \(\zeta = x + iy\) they are the linear operators
        \begin{equation}\label{eq: def-wirtinger-derivatives}
            \pdv{\zeta} = \frac{1}{2}\qty(\pdv{x} - i \pdv{y}) \qq{and} \pdv{\bar{\zeta}} = \frac{1}{2}\qty(\pdv{x} + i \pdv{y})
        \end{equation}
        and, as a consequence, the equation
        \begin{equation}
            \pdv{f}{\bar{\zeta}} = 0
        \end{equation}
        is equivalent to the Cauchy--Riemann equations. It follows that \(f\) is a holomorphic function of \(\zeta\) if, and only if, it is real-differentiable on the plane and \(\pdv*{f}{\bar{\zeta}} = 0\). Hence, \cref{subeq: ckv-sphere-1,subeq: ckv-sphere-2} imply that \(\tensor{Y}{^{\zeta}}\) (resp. \(\tensor{Y}{^{\bar{\zeta}}}\)) is a holomorphic function of \(\zeta\) (\(\bar{\zeta}\)). 

        It follows that the local Killing vector fields are characterized by linear combinations of fields with the form \(\tensor{Y}{^{\zeta}} = \alpha \zeta^n\) for \(\alpha \in \Comp\) and \(n \geq 0\) an integer. There is, however, a caveat: some of these fields may not be global Killing vector fields (as we would need for a complete Killing vector field). This is because the stereographic coordinate we are working with is only defined away from the north pole, since \(\zeta = \infty\) \emph{at} the north pole. To avoid this, we define a new stereographic coordinate \(\xi\) through
        \begin{equation}\label{eq: def-stereographic-antipodal}
            \xi = -e^{i\phi}\tan(\frac{\theta}{2}) = \frac{z - r}{x - i y}.
        \end{equation}
        \(\xi\) is then the coordinate antipodally related to \(\zeta\). These two coordinate systems are related by
        \begin{equation}
            \xi = - \frac{1}{\bar{\zeta}},
        \end{equation}
        with an analogous expression for the conjugates. The line element in the \(\xi\) coordinate is identical to the expression in terms of the \(\zeta\) coordinate, and hence so are the results of the Killing equation: \(\tensor{Y}{^\xi}\) must be a holomorphic function of \(\xi\).

        Consider now the vector field defined by \(\tensor{Y}{^\zeta} = \alpha \zeta^n\). When we change coordinates from \(\zeta\) to \(\xi\), we find that the same vector field has the component
        \begin{equation}
            \tensor{Y}{^\xi} = \frac{(-1)^n \bar{\alpha}}{\xi^{n-2}}.
        \end{equation}
        This is only a holomorphic function of \(\xi\) for \(n \leq 2\). 

        Therefore, we find that there is a six-dimensional real space of possible conformal Killing vector fields, which is spanned by \(\tensor{Y}{^{\zeta}} = 1, \zeta, \zeta^2, i, i\zeta, i\zeta^2\). Are these vector fields complete? 

        Let us consider some curve that is everywhere parallel to the most general vector field with
        \begin{equation}\label{eq: general-CKV-sphere}
            \tensor{Y}{^{\zeta}} = \alpha + \beta \zeta + \gamma \zeta^2
        \end{equation}
        for \(\alpha, \beta, \gamma \in \Comp\). This curve is characterized by the differential equation
        \begin{equation}
            \dv{\zeta}{t} = \alpha + \beta \zeta + \gamma \zeta^2.
        \end{equation}
        The full solution has the form 
        \begin{equation}
            \zeta(t) = \frac{a(t) \zeta(0) + b(t)}{c(t) \zeta(0) + d(t)},
        \end{equation}
        where the specific functional form of \(a(t)\), \(b(t)\), \(c(t)\), and \(d(t)\) depends on the values of \(\alpha, \beta, \gamma\), and the general case is particularly complicated. One can choose their normalization such that \(ad - bc = 1\). Such a transformation \(\zeta(0) \to \zeta(t)\) is known as a Möbius transformation, and these transformations correspond precisely to the conformal transformations on the Riemann sphere. 

        Notice that a Möbius transformation
        \begin{equation}
            \zeta \to \frac{a \zeta + b}{c \zeta + d}
        \end{equation}
        with \(ad - bc = 1\) has precisely one pole at \(\zeta = - d/c\) and one zero at \(\zeta = - b/a\). They are meromorphic in the complex plane and are a diffeomorphism of the Riemann sphere onto itself.  

        The (non-conformal) Killing vector fields on the sphere are the conformal Killing vector fields which satisfy the extra condition \(\tensor{\nabla}{_a}\tensor{Y}{^a} = 0\). Imposing this condition on a vector field with the form (\ref{eq: general-CKV-sphere}) we conclude that a (non-conformal) Killing vector field has the form (\ref{eq: general-CKV-sphere}) with the extra conditions that \(\beta + \bar{\beta} = 0\) and \(\gamma = \bar{\alpha}\). These are three real constraints, so we get a three-dimensional real space of Killing vector fields. 
        
        What is the group of confomorphisms on the sphere? To answer this question, we must study how two Möbius transformations compose. Consider two consecutive Möbius transformations
        \begin{equation}
            \zeta \to \zeta' = \frac{a \zeta + b}{c \zeta + d} \qq{and} \zeta' \to \zeta'' = \frac{a' \zeta' + b'}{c' \zeta' + d'}.
        \end{equation}
        We want to express the coefficients of the resulting transformation
        \begin{equation}
            \zeta \to \zeta'' = \frac{a'' \zeta + b''}{c'' \zeta + d''}
        \end{equation}
        in terms of the original coefficients. We assume that
        \begin{equation}
            ad-bc = a'd'-b'c' = 1.
        \end{equation}
        One can show that
        \begin{equation}
            \mqty(a'' & b'' \\ c'' & d'') = \mqty(a' & b' \\ c' & d')\mqty(a & b \\ c & d),
        \end{equation}
        with 
        \begin{equation}
            a'' d'' - b'' c'' = 1.
        \end{equation}
        This is the product of two \(2 \times 2\) complex matrices with unit determinant. Furthermore, notice that the Möbius transformation corresponding to the matrix \(L\) is the same Möbius transformation corresponding to the matrix \(-L\). Hence, the group is the group \(\SL(2,\Comp)/\Z_2\): the group of \(2 \times 2\) complex matrices with unit determinant up to sign. This may seem complicated, but it turns out that there is an isomorphism
        \begin{equation}\label{eq: lorentz-double-cover}
            \SL(2,\Comp)/\Z_2 \cong \SO*(3,1),
        \end{equation}
        which is proven on Appendix \ref{app: double-cover-lorentz}. Hence, the group of confomorphisms on the two-sphere is just the restricted Lorentz group!

        It can also be shown that the group of isometries on the two-sphere is \(\SO(3)\), as one would expect. This is done by proving the matrices representing the Möbius transformations are now unitary, \(L^\dagger L = \Eins\). As a consequence, they now live in the group \(\SU(2)\) of \(2 \times 2\) unitary matrices. Up to sign, we get \(\SU(2)/\Z_2 \cong \SO(3)\) through arguments analogous to those of Appendix \ref{app: double-cover-lorentz}.

    \subsection{Case Study: Isometries on Minkowski Spacetime}
        It is instructive to consider as a second case study the isometries in Minkowski spacetime. We should, of course, recover the Poincaré transformations. We work in \(n\)-dimensional Minkowski spacetime, for generality. 

        The Killing equation is 
        \begin{equation}
            \tensor{\nabla}{_a}\tensor{\xi}{_b} + \tensor{\nabla}{_b}\tensor{\xi}{_a} = 0.
        \end{equation}
        In globally inertial Cartesian coordinates this expression becomes
        \begin{equation}\label{eq: killing-eq-minkowski-cartesian}
            \tensor{\partial}{_\mu}\tensor{\xi}{_\nu} + \tensor{\partial}{_\nu}\tensor{\xi}{_\mu} = 0.
        \end{equation}

        Differentiate \cref{eq: killing-eq-minkowski-cartesian} and apply it a few times to see that
        \begin{subequations}
            \begin{align}
                \tensor{\partial}{_\rho}\tensor{\partial}{_\mu}\tensor{\xi}{_\nu} &= - \tensor{\partial}{_\rho}\tensor{\partial}{_\nu}\tensor{\xi}{_\mu}, \\
                &= - \tensor{\partial}{_\nu}\tensor{\partial}{_\rho}\tensor{\xi}{_\mu}, \\
                &= + \tensor{\partial}{_\nu}\tensor{\partial}{_\mu}\tensor{\xi}{_\rho}, \\
                &= + \tensor{\partial}{_\mu}\tensor{\partial}{_\nu}\tensor{\xi}{_\rho}, \\
                &= - \tensor{\partial}{_\mu}\tensor{\partial}{_\rho}\tensor{\xi}{_\nu}, \\
                &= - \tensor{\partial}{_\rho}\tensor{\partial}{_\mu}\tensor{\xi}{_\nu},
            \end{align}
        \end{subequations}
        and thus 
        \begin{equation}
            \tensor{\partial}{_\rho}\tensor{\partial}{_\mu}\tensor{\xi}{_\nu} = 0,
        \end{equation}
        meaning \(\tensor{\xi}{_\mu}\) is at most linear in the coordinates. 

        The most general expression for \(\tensor{\xi}{^a}\) at this stage is
        \begin{equation}
            \tensor{\xi}{_\mu} = \tensor{\alpha}{_\mu} + \tensor{\beta}{_\mu_\nu} \tensor{x}{^\nu}.
        \end{equation}
        \cref{eq: killing-eq-minkowski-cartesian} now yields
        \begin{equation}
            \tensor{\beta}{_\nu_\mu} + \tensor{\beta}{_\mu_\nu} = 0,
        \end{equation}
        establishing that \(\tensor{\beta}{_\mu_\nu}\) is antisymmetric.
        
        We thus see that the most general Killing vector field for Minkowski spacetime is
        \begin{equation}
            \tensor{\xi}{^a} = (\tensor{\alpha}{^\mu} + \tensor{\beta}{^\mu_\nu} \tensor{x}{^\nu}) \tensor{\qty(\pdv{\tensor{x}{^\mu}})}{^a}.
        \end{equation}
        This field is defined everywhere.

        Let us compute the flow of this Killing vector field. In Cartesian coordinates, it is given by the system of differential equations
        \begin{equation}
            \dv{\tensor{x}{^\mu}}{\lambda} = \tensor{\alpha}{^\mu} + \tensor{\beta}{^\mu_\nu} \tensor{x}{^\nu}.
        \end{equation}
        This can be written in matrix form as
        \begin{equation}
            \dv{x}{\lambda} = \alpha + \beta x,
        \end{equation}
        and one can then promptly integrate the system to get, in terms of matrix exponentials,
        \begin{equation}\label{eq: poincare-from-killing}
            x(\lambda) = e^{\lambda \beta}x(0) + e^{\lambda \beta} \int_0^{\lambda} e^{-\lambda' \beta} \dd{\lambda'} \alpha.
        \end{equation}
        The second term on the \gls{RHS} is an element of \(\R^n\), and hence this transformation is quite similar to the general transformation rule for the Poincaré group, \cref{eq: general-poincare-transf}. There is, however, an important question: is the matrix \(e^{\lambda\beta}\) an element of the Lorentz group \(\SO*(n-1,1)\)?

        To prove it is, first notice that 
        \begin{subequations}
            \begin{align}
                \tensor{\beta}{_\mu_\nu} + \tensor{\beta}{_\nu_\mu} &=0, \\
                \tensor{\beta}{^\rho_\nu}\tensor{\eta}{_\rho_\mu} + \tensor{\eta}{_\nu_\sigma}\tensor{\beta}{^\sigma_\mu} &=0, \\
                \beta^\intercal \eta + \eta \beta &= 0,
            \end{align}
        \end{subequations}
        which translates the antisymmetry properties of \(\tensor{\beta}{_\mu_\nu}\) to the matrix \(\beta\) (which has entries \(\tensor{\beta}{^\mu_\nu}\)). With this expression in mind, we notice that
        \begin{subequations}
            \begin{align}
                \qty(e^{\lambda \beta})^\intercal \eta \qty(e^{\lambda \beta}) &= \sum_{n=0}^{+\infty} \frac{\lambda^n}{n!} (\beta^{\intercal})^n \eta \qty(e^{\lambda \beta}), \\
                &= \sum_{n=0}^{+\infty} \frac{(-1)^n \lambda^n}{n!} \eta \beta^n \qty(e^{\lambda \beta}), \\
                &= \eta e^{-\lambda \beta} e^{\lambda \beta}, \\
                &= \eta.
            \end{align}
        \end{subequations}
        Therefore, \(e^{\lambda \beta} \in \Og(n-1,1)\). Since \(e^{0 \beta} = \Eins \in \SO*(n-1,1)\) and \(e^{\lambda \beta}\) is continuous in \(\lambda\), we know \(e^{\lambda \beta} \in \SO*(n-1,1)\) for all \(\lambda\).

        Hence, we conclude (\ref{eq: poincare-from-killing}) represents a Poincaré transformation. By carefully selecting \(\alpha\) and \(\beta\) one can show that all Poincaré transformations are of the form (\ref{eq: poincare-from-killing}). This is, of course, exactly what we expected.

        One may notice that we have \(n(n+1)/2\) parameters in between \(\alpha\) and \(\beta\). This is the dimensionality of \(\ISO*(n-1,1)\) as a Lie group (\ie, its manifold dimension). We have an equal number of linearly independent Killing vector fields, and a convenient basis is
        \begin{equation}\label{eq: poincare-generators}
            \tensor*{P}{^a_\mu} = \tensor{\qty(\pdv{\tensor{x}{^\mu}})}{^a} \qq{and} \tensor*{J}{^a_\mu_\nu} = \tensor{x}{_\mu}\tensor{\qty(\pdv{\tensor{x}{^\nu}})}{^a} - \tensor{x}{_\nu}\tensor{\qty(\pdv{\tensor{x}{^\mu}})}{^a}.
        \end{equation}
        The vectors \(\tensor*{P}{^a_\mu}\) generate translations, the \(\tensor*{J}{^a_i_j}\) generate rotations, and the \(\tensor*{J}{^a_i_t}\) generate boosts.

\section{Asymptotically Flat Spacetimes}\label{sec: asymptotically-flat-spacetimes}
    Our goal in these lectures is to understand the asymptotic symmetries of asymptotically flat spacetimes. For that end, we must have a good understanding of what we mean by ``asymptotic'' and what we mean by ``asymptotically flat''. The goal of this section is to give attention to these terms and discuss how they are currently understood in \gls{GR}.

    Asymptotic flatness and the \gls{BMS} group depend crucially on the spacetime dimension \cite{ashtekar1997AsymptoticStructureSymmetryreduced,hollands2005AsymptoticFlatnessBondi,hollands2017BMSSupertranslationsMemory}. As a consequence, from this point onward I will focus on the four-dimensional case.

    \subsection{Infinity in Minkowski Spacetime}\label{subsec: infinity-minkowski}
        We shall begin by giving meaning to ``infinity'' in Minkowski spacetime. Our approach is similar in nature to the discussion in Chap. 11 of Ref. \citeonline{wald1984GeneralRelativity}. 

        To begin our discussion, let us choose a system of coordinates to better visualize what we are doing. Spherical coordinates turn out to be convenient, so we write the Minkowski line element as 
        \begin{equation}
            \dd{s}^2 = - \dd{t}^2 + \dd{r}^2 + r^2 \dd{\Sph}^2,
        \end{equation}
        where \(\dd{\Sph}^2\) stands for the line element of the round metric on the sphere. 

        When we talk about what happens ``at infinity'' in Minkowski spacetime, there are three possibilities we could be considering. Namely, we could be thinking about taking the limit \(t \to \pm \infty\) or taking the limit \(r \to + \infty\). This, however, is not a clear enough picture of what infinity is. For instance, we do not know if taking \(r \to + \infty\) toward different angular directions will lead to the same result. Furthermore, we could take \(r \to + \infty\) while holding \(t\) constant or while holding \(u = t - r\) constant, for example, and these options turn out to lead to very different results. Therefore, we must be more careful. 

        For concreteness, let us assume we want to take the limit \(r \to +\infty\) at constant \(u = t-r\). In this case, it makes sense to change coordinates so we use \(u\) rather than \(t\) when writing the line element. Doing so leads us to the line element
        \begin{equation}
            \dd{s}^2 = - \dd{u}^2 - 2 \dd{u}\dd{r} + r^2 \dd{\Sph}^2.
        \end{equation}
        At this point we could attempt to take the limit \(r \to + \infty\) in order to understand what is the geometry of infinity in Minkowski spacetime. Nevertheless, there is a difficulty in doing so: this limit leads to a divergence in the metric. This divergence is not due to a bad choice of coordinates, because it has geometrical (and hence physical) implications. Namely, the area of a sphere with radius \(r\) diverges as we take \(r \to + \infty\). Hence, it is not possible to take this limit for this metric. 

        Despite this difficulty, we would still like to understand the structure of infinity. To do so, we will make a new change of coordinates. We define a new coordinate \(l = 1/r\), which turns the line element into 
        \begin{equation}\label{eq: minkowski-u-l}
            \dd{s}^2 = - \dd{u}^2 + 2 l^{-2} \dd{u} \dd{l} + l^{-2} \dd{\Sph}^2.
        \end{equation}
        A few comments are in order. Firstly, notice that the physical region of Minkowski spacetime corresponds to \(l > 0\)---the limiting case \(l=0\) would in principle be infinity (in the sense that \(r \to + \infty\) as \(l \to 0^+\)) and \(l < 0\) is non-physical. Secondly, this coordinate patch does not cover the points with \(r = 0\), since \(l\) diverges there. Thirdly, notice that this was merely a coordinate change, and taking the limit \(l \to 0^+\) is not any more possible than taking the limit \(r \to + \infty\) was before. 

        Nevertheless, this new coordinate system seems to be ``better behaved'' at infinity. Namely, we know infinity sits at \(l=0\), which we can barely grasp. The only problem we face now is the presence of the \(l^{-2}\) factors in \cref{eq: minkowski-u-l}. To deal with them, we will take an apparently \emph{ad hoc} approach. We will define a new, unphysical metric \(\tensor{\tilde{g}}{_a_b}\) through
        \begin{equation}
            \tensor{\tilde{g}}{_a_b} = l^2 \tensor{\eta}{_a_b},
        \end{equation}
        where \(\tensor{\eta}{_a_b}\) is the Minkowski metric. The line element for this new unphysical metric is 
        \begin{equation}
            \dd{\tilde{s}}^2 = - l^2 \dd{u}^2 + 2 \dd{u} \dd{l} + \dd{\Sph}^2.
        \end{equation}
        This new unphysical metric has an advantage: it is well-behaved in the \(l \to 0^+\) limit. We can thus describe the geometry of infinity by plugging in \(l = 0\) and getting the induced metric
        \begin{equation}\label{eq: minkowski-induced-metric-nullfut}
            \dd{\tilde{\sigma}}^2 = 0 \dd{u}^2 + \dd{\Sph}^2,
        \end{equation}
        where we write \(0 \dd{u}^2\) explicitly as a reminder that this is a metric in a three-dimensional manifold. I wrote \(\dd{\tilde{\sigma}}^2\) instead of \(\dd{\tilde{s}}^2\) because this is the induced metric. We can now consider the behavior of tensor fields at infinity by mapping them from the original spacetime with metric \(\tensor{\eta}{_a_b}\) to the unphysical spacetime with metric \(\tensor{\tilde{g}}{_a_b}\) and subsequently evaluating them at \(l=0\). In this sense, infinity becomes a set of regular points. 

        Let me summarize what is the trick we just employed and notice some of the ideas that went in performing it. We had an issue with the metric \(\tensor{\eta}{_a_b}\) because it diverged in the region we wanted to consider---namely, infinity. Nevertheless, we found a function \(l\) that conveniently vanishes precisely at the region in which the metric diverges. We then noticed that the metric \(\tensor{\tilde{g}}{_a_b} = l^2 \tensor{\eta}{_a_b}\) is finite in the region we wanted to study. Hence, we decided to let go of \(\tensor{\eta}{_a_b}\) and work with \(\tensor{\tilde{g}}{_a_b}\) instead, at least while we are considering the behavior at infinity. By making appropriate transformations of tensor fields so they can be understood as being defined in the spacetime given by \(\tensor{\tilde{g}}{_a_b}\), we can analyze their behavior at infinity by simply evaluating them at infinity, which became a set of regular points in the new unphysical spacetime. 

        It is important to emphasize that \(\tensor{\tilde{g}}{_a_b}\) is an unphysical metric. It is not the metric that describes the spacetime geometry as measured in experiments. It is defined as a mathematical construction that allows us to ``bring in infinity'' and consider it as a regular point. In general, \(\tensor{\tilde{g}}{_a_b}\) will not even satisfy the Einstein field equations, so it is not an accurate model of the bulk spacetime\footnote{By ``bulk'' I mean the interior of the spacetime, or the set of all ``finite'' points. This contrasts with infinity, which would be the ``boundary'' of spacetime.} we are trying to describe. Nevertheless, the price we pay to study infinity is that we are forced to work with \(\tensor{\tilde{g}}{_a_b}\) rather than with the physical metric \(\tensor{\eta}{_a_b}\).

        This procedure is called a conformal compactification. It is conformal because we have a transformation of the form \(\tensor{\eta}{_a_b} \to \Omega^2 \tensor{\eta}{_a_b}\) for some function \(\Omega\) which is positive in the bulk spacetime. It is a compactification because it transforms a spacetime that is not compact into a spacetime that is compact by adding the points at infinity. There are, however, better choices of compactification. In our construction, we defined a function \(\Omega = l\) which was not defined at some points of spacetime---namely those with \(r = 0\). We can, however, define a conformal compactification that takes proper account of all points in Minkowski spacetime. 

\begin{figure}
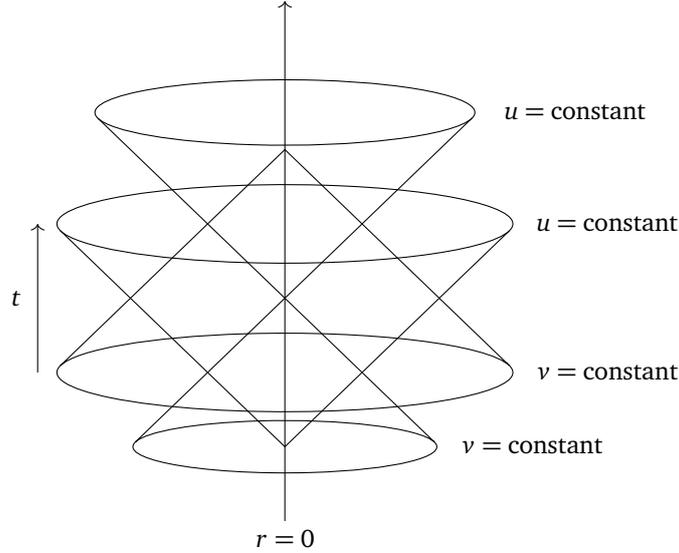

            \centering
            \includestandalone{lightcones}
            \caption{Illustration, with one dimension suppressed, of the physical meaning of the null coordinates \(u\) and \(v\). Surfaces of constant \(u\) are outgoing spherical ``waves'', while surfaces of constant \(v\) are their incoming analogues. This figure is based on Fig. 12.i of Ref. \citeonline{hawking1973LargeScaleStructure}.}
            \label{fig: minkowskinullcoordinates}
        \end{figure}

        To that end, we will make some more changes of coordinates. We begin by defining the advanced time coordinate \(v = t + r\). This is in contrast to the retarded time coordinate \(u = t - r\). Both of these coordinates are depicted on \cref{fig: minkowskinullcoordinates}. In terms of retarded and advanced times and angular coordinates the Minkowski line element is written as 
        \begin{equation}\label{eq: minkowski-u-v}
            \dd{s}^2 = - \dd{u}\dd{v} + \frac{1}{4}(v-u)^2 \dd{\Sph^2}.
        \end{equation}
        It is convenient to work with null coordinates---\ie, coordinates that are constant on a hypersurface normal to a null vector---because we can compactify each of these coordinates separately and still get null coordinates. This will allow us at the end to draw a diagram in which all (radial) null geodesics are at \(\ang{45}\) angles, which is convenient for reading the causal structure of spacetime. 

        We will now compactify the coordinates of \cref{eq: minkowski-u-v}. Our goal is to take the coordinates \(u\) and \(v\) which obey 
        \begin{equation}
            - \infty < u \leq v < + \infty
        \end{equation}
        and map them to coordinates with range in a finite interval. This can be done by means of a function \(f \colon (a,b) \to \R\) where \(-\infty < a < b < + \infty\) and such that \(f\) is injective. Two examples are the functions
        \begin{gather}
            \artanh \colon (-1,1) \to \R \\
            \intertext{and}
            \tan \colon \qty(-\frac{\pi}{2},+\frac{\pi}{2}) \to \R,
        \end{gather}
        but one could also work with other options. We will use the tangent function. We define new compactified null coordinates \(U\) and \(V\) through
        \begin{equation}
            u = \tan U \qq{and} v = \tan V.
        \end{equation}
        Notice that this maps the infinite range of \(u\) and \(v\) to a finite range of \(U\) and \(V\). Namely, \(U\) and \(V\) satisfy
        \begin{equation}
            - \frac{\pi}{2} < U \leq V < + \frac{\pi}{2}.
        \end{equation}
        The middle inequality holds because \(\tan\) is crescent and \(u \leq v\). 

        In our new choice of coordinates, we have the line element
        \begin{equation}\label{eq: minkowski-U-V}
            \dd{s}^2 = \sec^2 U \sec^2 V \qty[- \dd{U}\dd{V} + \frac{1}{4}\sin[2](V-U)\dd{\Sph}^2].
        \end{equation}
        At this stage, our coordinates are already compactified. We just need to get rid of the divergences of the metric at infinity. To do so, we perform a conformal transformation. As \cref{eq: minkowski-U-V} suggests, we take 
        \begin{equation}
            \Omega = 2 \cos U \cos V
        \end{equation}
        so that we define \(\tensor{\tilde{g}}{_a_b} = \Omega^2 \tensor{\eta}{_a_b}\) and get
        \begin{equation}
            \dd{\tilde{s}}^2 = - 4 \dd{U}\dd{V} + \sin[2](V-U)\dd{\Sph}^2.
        \end{equation}
        For convenience, we finally introduce coordinates \(T\) and \(R\) through
        \begin{equation}
            T = U + V \qq{and} R = V - U
        \end{equation}
        which mimic (up to normalization) \(u = t -r\) and \(v = t + r\). This finally leads to the unphysical metric 
        \begin{equation}\label{eq: einstein-metric}
            \dd{\tilde{s}}^2 = - \dd{T}^2 + \dd{R}^2 + \sin^2 R\dd{\Sph}^2.
        \end{equation}
        The physical region corresponds to the limits 
        \begin{equation}\label{eq: physical-region-compactified-minkowski}
            - \pi < T - R \leq T + R < + \pi,
        \end{equation}
        which in particular imply \(R \geq 0\).

        \cref{eq: einstein-metric} is the line element for the Einstein static universe (see, \eg, Sec. VII.2.1 of Ref. \citeonline{choquet-bruhat2015IntroductionGeneralRelativity}). This spacetime has topology \(\R \times \Sph^3\) and it is a solution to the Einstein field equations, but for a perfect fluid stress tensor with a cosmological constant component. This is, of course, a very different scenario from the Minkowski metric, which is a vacuum solution. This greatly exemplifies that the unphysical metric is really unphysical. 

\begin{figure}[tbp]
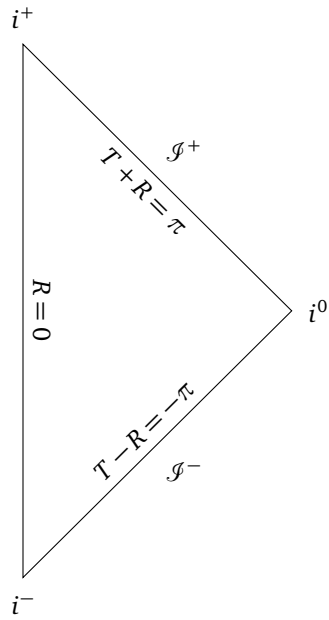

            \centering
            \includestandalone{minkowski-penrose}
            \caption{Carter--Penrose diagram of Minkowski spacetime.}
            \label{fig: minkowskipenrose}
        \end{figure}

\begin{figure}
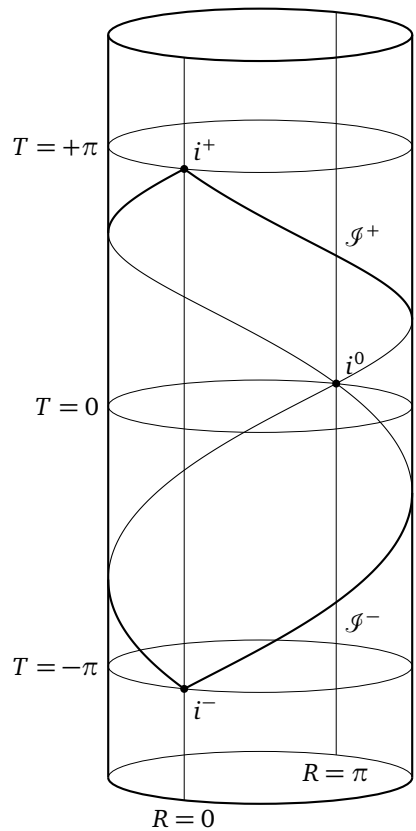

            \centering\tdplotsetmaincoords{80}{90}
            \includestandalone{einstein-cylinder}
            \caption{Minkowski spacetime embedded in the Einstein static universe (two spatial dimensions are suppressed). Since the Einstein static universe has the topology \(\R \times \Sph^3\), it is often depicted as a cylinder. The Carter--Penrose diagram for Minkowski spacetime is wrapped around the Einstein cylinder.}
            \label{fig: minkowski-einstein-cylinder}
        \end{figure}

        We can make a plot of the region given on \cref{eq: physical-region-compactified-minkowski}. Plotting \(R\) on the horizontal axis and \(T\) on the vertical axis, we get the diagram shown in \cref{fig: minkowskipenrose}. This is known as the Carter--Penrose diagram for Minkowski spacetime. It is a finite drawing of all of Minkowski spacetime, with each point representing a sphere \(\Sph^2\) (except for the points on the \(R=0\) line and the vertices of the diagram, which represent a single point each). We can ``double'' the diagram by letting each point be a hemisphere, and this version of the diagram is interesting because it can be ``wrapped around the Einstein static universe'', as shown in \cref{fig: minkowski-einstein-cylinder}.

        The region \(T+R = + \pi\) of the Carter--Penrose diagram corresponds to the limits \(r \to +\infty\) at constant \(u\). Each angular direction leads to a different point and each different value of \(u\) leads to a different point as well. In total, we get a three-dimensional manifold \(\nullfut\) with topology \(\R \times \Sph^2\): \(\R\) corresponds to the values of \(u\) and \(\Sph^2\) to the possible angular directions. Similarly, \(T-R = - \pi\) is the region \(\nullpas\) of limits \(r \to + \infty\) under constant \(v\) and it also has topology \(\R \times \Sph^2\) for analogous reasons. \(\nullfut\) is known as the future null infinity and the symbol \(\nullfut\) is pronounced ``scri-plus''. \(\nullpas\) (``scri-minus'') is the past null infinity. 

        The point \(T=0\) with \(R = \pi\) is at the boundary between \(\nullfut\) and \(\nullpas\). Notice this is indeed a point---\(R = \pi\) is one of the poles of the Einstein universe's three-sphere, which is thus a single point. This point, denoted \(\spatinf\), is known as spatial infinity and it corresponds to the limits \(r \to + \infty\) at constant \(t\). Notice that, since this is a single point, limits in different angular directions all coincide. 

        We still have two interesting points. \(T = \pi\) with \(R = 0\) is at the future of \(\nullfut\). This is located at the other pole of the three-sphere and thus it also corresponds to a single point, denoted \(\timefut\) and known as future timelike infinity. This is the limit \(t \to + \infty\) at constant \(r\), and there is no angular direction-dependence either. Similar comments are in order for the past timelike infinity \(\timepas\), located at \(T = -\pi\) with \(R = 0\). 

\begin{figure}[tbp]
            \centering
            \null\hfill
            \includestandalone[width=0.2\textwidth]{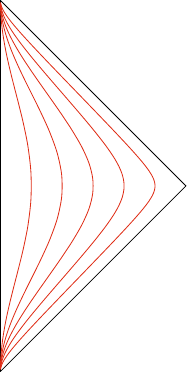}
            \hfill
            \includestandalone[width=0.2\textwidth]{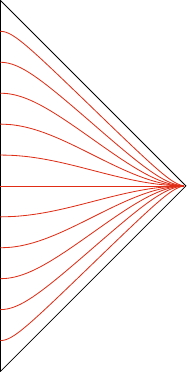}
            \hfill
            \includestandalone[width=0.2\textwidth]{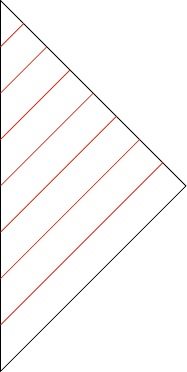}
            \hfill
            \includestandalone[width=0.2\textwidth]{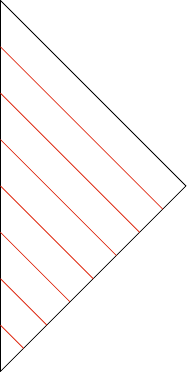}
            \hfill\null            
            \caption{Different curves represented in the Carter--Penrose diagram for Minkowski spacetime. Vertical curves (far-left) are curves with constant \(r\). Horizontal curves (center-left) are curves with constant \(t\). Diagonal lines have constant \(u\) (outgoing, center-right) or \(v\) (incoming, far-right).}
            \label{fig: minkowski-penrose-const}
        \end{figure}

        All timelike curves start at \(\timepas\) and end at \(\timefut\). All null curves start at \(\nullpas\) and end at \(\nullfut\). All spacelike curves start and end at \(\spatinf\). \cref{fig: minkowski-penrose-const} illustrates some curves of interest on the Carter--Penrose diagram for Minkowski spacetime.

        Conformal compactifications were introduced in \gls{GR} by \textcite{penrose1963AsymptoticPropertiesFields,penrose1964ConformalTreatmentInfinity,penrose1965ZeroRestmassFields}, who also presented the first versions of conformal diagrams, which were similar to \cref{fig: minkowski-einstein-cylinder}. \textcite{carter1966CompleteAnalyticExtension} later adapted them to the modern two-dimensional version. We shall extensively use these techniques in the rest of these notes. 

        I should mention an important property of the conformal transformations we are employing to find an unphysical spacetime. While it is true that \(\tensor{\tilde{g}}{_a_b}\) is unphysical it is important to recall that a vector \(\tensor{k}{^a}\) will satisfy \(\tensor{\tilde{g}}{_a_b}\tensor{k}{^a}\tensor{k}{^b} = 0\) for \(\tensor{\tilde{g}}{_a_b} = \Omega^2 \tensor{g}{_a_b}\) if, and only if, \(\tensor{g}{_a_b}\tensor{k}{^a}\tensor{k}{^b} = 0\). As a consequence, conformal transformations preserve lightcones, and thus they preserve the causal structure of spacetime. This means that Carter--Penrose diagrams are useful tools for analyzing the causal structure of spacetime. 

        Notice also that in the Carter--Penrose diagram for Minkowski spacetime all radial light rays move at \(\ang{45}\) angles. This allows us to read the causal structure of Minkowski spacetime directly from its Carter--Penrose diagram. This is the reason why these sorts of diagrams are popular in \gls{GR}. 

    \subsection{Asymptotically Flat Spacetimes}
        At this point we would like to adapt our discussion of infinity to more general spacetimes. This is related to defining what is an asymptotically flat spacetime. 

        In a general spacetime, the structure of infinity could be much different from that of Minkowski spacetime. Some spacetimes might not even get to infinity. Consider a dust-filled and spatially closed \gls{FLRW} universe, for example. Such a universe begins at a Big Bang, ends at a Big Crunch, and has compact spatial sections (it is spatially a sphere). Therefore, we never really get to infinity. We either reach the north or south pole of the spatial spheres or we reach the singularities at the beginning or end of spacetime. 

        We want to consider spacetimes that asymptotically look like Minkowski spacetime. Hence, their behavior at infinity should somehow resemble that of Minkowski spacetime. We know Minkowski spacetime has five different infinities, so there are five different meanings we can give to ``asymptotically flat'', plus combinations of them. 

        What are reasonable conditions we can impose? It makes sense to consider spacetimes in which the gravitational field falls off away from a central distribution of matter, so we have an asymptotically flat spacetime at null and spatial infinities. There is less interest in spacetimes which are asymptotically flat at timelike infinities, though, because this means the spacetime becomes flat at late or early times, so the matter content has dispersed. While there are situations in which this can be interesting, they are not what we are looking for at this stage. Hence, one often defines asymptotic flatness at spatial and null infinities. 

        Our definition will actually be even simpler than that. For our purposes, a single null infinity will typically be enough. Hence, we will discuss asymptotically flat spacetimes at future null infinity (the past case is analogous). We follow Chap. 2 of Ref. \citeonline{dappiaggi2017HadamardStatesLightlike}, which also discusses the timelike case. For the more complete case involving null and spatial infinity, see the Chap. 11 of Ref. \citeonline{wald1984GeneralRelativity}.

        I will first discuss in a handwaving way each of the axioms that we will employ and then state the final definition. The goal is to motivate the definition. We will find a definition by mimicking and generalizing the construction of conformal infinity we performed for Minkowski spacetime.

        We start with a spacetime \((M,\tensor{g}{_a_b})\). We want to enforce conditions on this spacetime so that it can be considered asymptotically flat. Our construction should be coordinate-independent in nature, so that it actually captures the physical aspects of spacetime rather than spurious coordinate behaviors.

        Our first demand will be that there is an unphysical spacetime \((\tilde{M},\tensor{\tilde{g}}{_a_b})\) which conformally extends \((M,\tensor{g}{_a_b})\). This unphysical spacetime will play the same role that the Einstein static universe played for Minkowski spacetime. 

        We need a way of injecting \(M\) into \(\tilde{M}\). Before, we had a natural way of viewing Minkowski spacetime as a submanifold of the Einstein static universe, which was given by the coordinate restrictions on the Einstein universe. In our more general case, this role is played by a function \(\psi \colon M \to \tilde{M}\) that should satisfy some convenient properties. In practice, it is convenient to choose \(\psi\) to be an embedding. Let us define this by following \textcite{tu2011IntroductionManifolds}. 

        \begin{definition}[Immersion]
            Let \(\psi \colon M \to \tilde{M}\) be a smooth map between the manifolds \(M\) and \(\tilde{M}\). We say \(\psi\) is an immersion if, and only if, the pushforward \(\psi^* \colon \T[p]M \to \T[\psi(p)]\tilde{M}\) is injective for all \(p \in M\).
        \end{definition}

        \begin{definition}[Embedding]
            Let \(M\) and \(\tilde{M}\) be smooth manifolds. An immersion \(\psi \colon M \to \tilde{M}\) is called an embedding if, and only if, it is injective and \(\psi\colon M \to \psi(M)\) is continuous with a continuous inverse. 
        \end{definition}

        \begin{definition}[Embedded Submanifold]
            Let \(M\) and \(\tilde{M}\) be smooth manifolds. If \(\psi \colon M \to \tilde{M}\) is an embedding, we say that \(\psi(M)\) is an embedded submanifold of \(\tilde{M}\).
        \end{definition}

        Embeddings are diffeomorphisms onto their ranges. Hence, we can think of an embedding as being a generalization of diffeomorphism that does not need to be surjective. It makes a copy of \(M\) in \(\tilde{M}\).

        We thus have the next condition we want to impose on \((M,\tensor{g}{_a_b})\): there is an embedding \(\psi \colon M \to \tilde{M}\), and we also require as a technical condition that \(\psi(M)\) is an open subset of \(\tilde{M}\).

        Having a smooth embedding, we have a way of connecting the two manifolds \(M\) and \(\tilde{M}\). Nevertheless, we still have not imposed any restrictions on the metrics \(\tensor{g}{_a_b}\) and \(\tensor{\tilde{g}}{_a_b}\). To do so, we notice that \(\psi \colon M \to \psi(M)\) is a diffeomorphism, and hence we can use it to perform both pullbacks and pushforwards of tensor fields in \(M\) and \(\psi(M)\). We will then ask that there is a function \(\Omega \in \ck[\infty](\psi(M))\) such that \(\Omega > 0\) and 
        \begin{equation}
            \tensor{\tilde{g}}{_a_b}|_{\psi(M)} = \Omega^2 \psi^* \tensor{g}{_a_b}.
        \end{equation}
        This ensures that \(\psi\) represents a conformal transformation between the spacetimes \((M,\tensor{g}{_a_b})\) and \((\tilde{M},\tensor{\tilde{g}}{_a_b})\).

        Given these objects, we define future null infinity as the boundary \(\nullfut = \bound\psi(M)\). \(\nullfut\) is an embedded submanifold of \(\tilde{M}\) and it is not to the past of any points of \(\psi(M)\). In symbols, we write
        \begin{equation}
            \nullfut \cap \causalp(\psi(M);\tilde{M}) = \varnothing.
        \end{equation}
        The causal past of a set \(S\), \(\causalp(S)\), is defined as follows.

        \begin{definition}[Causal Past]
            Let \((M,\tensor{g}{_a_b})\) be a spacetime. Consider a set \(S \subeq M\). The causal past of \(S\) in \(M\), denoted \(\causalp(S;M)\) or simply \(\causalp(S)\) when \(M\) is understood, is defined as the set of all points in \(M\) from which one can reach \(S\) by means of a future-directed causal curve. See Chap. 8 of Ref. \citeonline{wald1984GeneralRelativity} for further details.
        \end{definition}

        Hence, \(\nullfut\) is defined in this general construction as a submanifold of \(\tilde{M}\) satisfying a couple of special properties. In particular, it is a future boundary in the sense that is is not in the past of any points in the physical spacetime.

        Near infinity we expect the spacetime to be reasonably well-behaved, so that it actually looks like Minkowski spacetime. Thus, we demand it enjoys reasonably good causal properties. We cannot allow for closed timelike curves near infinity, for example. We enforce this behavior by demanding that strong causality holds in a neighborhood of infinity.

        \begin{definition}[Strong Causality]
            Consider a spacetime \((M,\tensor{g}{_a_b})\). We say \((M,\tensor{g}{_a_b})\) is strongly causal if for any point \(p \in M\) and every neighborhood \(U\) of \(p\) there is a neighborhood \(V\) of \(p\) with \(p \in V \subeq U\) such that no causal curve intersects \(V\) more than once.
        \end{definition}

        As discussed in Chap. 8 of Ref. \citeonline{wald1984GeneralRelativity}, strong causality essentially means that no causal curve comes arbitrarily close to intersecting itself. This prevents causality violations upon small perturbations of the metric. There are stronger causality impositions one could consider, but we shall assume only strong causality. 

        We need to impose that \(\Omega\) brings infinity in. This is done by imposing that \(\Omega\) vanishes on \(\nullfut\). However, we defined \(\Omega\) only on \(\psi(M)\). Hence, we ask that \(\Omega\) extends to a function \(\Omega \in \ck[\infty](\tilde{M})\) such that
        \begin{equation}
            \Omega|_{\nullfut} = 0.
        \end{equation}
        Furthermore, we want \(\nullfut\) to be precisely the submanifold with \(\Omega = 0\), so we must also require that
        \begin{equation}
            \tensor{\tilde{\nabla}}{_a}\Omega|_{\nullfut} \neq 0.
        \end{equation}
        This ensures \(\Omega = 0\) is indeed a hypersurface and ensures there is a non-vanishing vector normal to the surface. Hence, we get a single ``sheet'' with \(\Omega = 0\), as opposed to a plateau.

        We want to obtain the full range of limits \(r \to + \infty\) at constant \(u\)'s. To ensure we get all points at infinity, we define the vector field \(\tensor{n}{^a} = \tensor{\tilde{g}}{^a^b}\tensor{\tilde{\nabla}}{_b}\Omega\). We then demand the existence of a function \(\omega \in \ck[\infty](\tilde{M})\) with \(\omega > 0\) on \(\psi(M) \cup \nullfut\) such that 
        \begin{equation}
            \tensor{\tilde{\nabla}}{_a}(\omega^4 \tensor{n}{^a})|_{\nullfut} = 0
        \end{equation}
        and such that the integral curves of \(\omega^{-1}\tensor{n}{^a}\) are complete on \(\nullfut\). This technical condition is meant to ensure that \(\nullfut \cong \R \times \Sph^2\).

        Finally, we need the Riemann tensor to fall off sufficiently quickly near infinity. Thus, we ask that \((M,\tensor{g}{_a_b})\) satisfies the vacuum Einstein field equations on some neighborhood of the boundary of \(\psi(M)\). This can be weakened to requiring that the vacuum Einstein field equations hold asymptotically \cite{wald1984GeneralRelativity}. The point is that at infinity we should get a vacuum solution, so that we are far away from any matter sources.

        We now bring everything together to get to the following definition.

        \begin{definition}[Asymptotically Flat Spacetime at Future Null Infinity]
            Consider a spacetime \((M,\tensor{g}{_a_b})\). Suppose that there are
            \begin{enumerate}
                \item an unphysical spacetime \((\tilde{M},\tensor{\tilde{g}}{_a_b})\),
                \item a smooth embedding \(\psi \colon M \to \tilde{M}\) such that \(\psi(M)\) is open in \(\tilde{M}\),
                \item and a smooth function \(\Omega \colon \psi(M) \to \R\) with \(\Omega > 0\) and
                \begin{equation}
                    \tensor{\tilde{g}}{_a_b}|_{\psi(M)} = \Omega^2 \psi^* \tensor{g}{_a_b}.
                \end{equation}
            \end{enumerate}
            Furthermore, suppose that these objects are such that the following conditions are met:
            \begin{enumerate}
                \item \(\psi(M)\) is the interior of a manifold with boundary \(\nullfut = \bound\psi(M)\), where \(\nullfut\) is an embedded three-manifold of \(\tilde{M}\) and it holds that \(\nullfut \cap \causalp(\psi(M);\tilde{M}) = \varnothing\).
                \item Strong causality holds in \((\tilde{M},\tensor{\tilde{g}}{_a_b})\) at least on a neighborhood of \(\nullfut\).
                \item \(\Omega\) can be extended to a smooth function \(\Omega \colon \tilde{M} \to \R\) with \(\Omega|_{\nullfut} = 0\) and \(\tensor{\tilde{\nabla}}{_a}\Omega|_{\nullfut} \neq 0\).
                \item Denoting \(\tensor{n}{^a} = \tensor{\tilde{g}}{^a^b}\tensor{\tilde{\nabla}}{_b}\Omega\), there is a smooth function \(\omega \colon \tilde{M} \to \R\) with \(\omega > 0\) such that \(\tensor{\tilde{\nabla}}{_a}(\omega^4 \tensor{n}{^a})|_{\nullfut} = 0\) and such that the integral lines of \(\omega^{-1} \tensor{n}{^a}\) are complete. 
                \item The vacuum Einstein field equations hold for \((M,\tensor{g}{_a_b})\) on a neighborhood of infinity, or at least asymptotically as one approaches infinity. 
            \end{enumerate}
            If all of these conditions are met, we say that \((M,\tensor{g}{_a_b})\) is asymptotically flat at future null infinity. 
        \end{definition}

    \subsection{Case Study: Schwarzschild Spacetime}
        As a case study, let us show that Schwarzschild spacetime is asymptotically flat at future null infinity (and actually at past null infinity too). We follow  \textcite{schmidt1983AnalyticConformalExtensions}. I will not make an effort to prove all of the conditions of asymptotic flatness, but rather show the conformal compactification procedure. 

        We want to get to future null infinity, so it makes sense to employ a retarded time coordinate instead of the more usual Schwarzschild time coordinate. In this scenario, \(u\) is known as the retarded Eddington--Finkelstein coordinate. The Schwarzschild metric is written as 
        \begin{equation}
            \dd{s}^2 = - \qty(1 - \frac{2M}{r}) \dd{u}^2 - 2 \dd{u}\dd{r} + r^2 \dd{\Sph}^2.
        \end{equation}
        \(M\) is the black hole's mass, and the coordinate ranges are \(u \in \R\) and \(r > 0\). Notice that \(r = 0\) is \emph{not} a part of spacetime, since it is a curvature singularity. 

        We proceed as with Minkowski spacetime. Define a new coordinate \(l = 1/r\), which is now defined on the entire spacetime, since \(r > 0\) strictly. Using this new coordinate we get the line element
        \begin{equation}
            \dd{s}^2 = - \qty(1 - 2 M l) \dd{u}^2 + 2 l^{-2} \dd{u} \dd{l} + l^{-2} \dd{\Sph}^2.
        \end{equation}
        As with Minkowski spacetime, we have a divergence at \(l \to 0^+\) because the area of spheres tends to infinity. We solve this by multiplying the metric by \(l^2\) (which means our conformal factor is \(\Omega = l\)) to get to
        \begin{equation}
            \dd{\tilde{s}}^2 = - l^2 \qty(1 - 2 M l) \dd{u}^2 + 2 \dd{u} \dd{l} + \dd{\Sph}^2.
        \end{equation}
        This unphysical metric can now be extended so that \(l \in \R\). Through this procedure we get the unphysical spacetime \((\tilde{M},\tensor{\tilde{g}}{_a_b})\). Since \(r = 0\) was never a point in spacetime, we did not ``give up'' on that point when using the new coordinate \(l\).        

    \subsection{Case Study: Poincaré Transformations at Null Infinity}
        At this point, we can do a simple calculation that will help us to understand the results in the next section. Namely, we will now compute what the Poincaré transformations look like at null infinity.

        We start by considering Lorentz transformations. These are the vector fields \(\tensor*{J}{^a_\mu_\nu}\) given on \cref{eq: poincare-generators}. These vectors induce vectors on \(\nullfut\). These induced vectors can be computed by considering the expressions on \cref{eq: poincare-generators}, changing coordinates from \((t,x,y,z)\) to \((u,l,\zeta,\bar{\zeta})\) (\(l = 1/r\)), and then taking the \(l \to 0^+\) limit\footnote{Where did the pushforward go? It is hidden inside the new coordinates. Since the physical spacetime \(M\) is embedded in the unphysical spacetime \(\tilde{M}\), we can use the same coordinates on \(M\) and on \(\psi(M)\). The coordinate system \((u,l,\zeta,\bar{\zeta})\) is useful here because it holds on a larger region of \(\tilde{M}\), while \((t,x,y,z)\) or even \((u,r,\zeta,\bar{\zeta})\) would break down at null infinity.}.
        
        Let me give an example. Take \(\tensor*{J}{^a_x_t}\) as defined on \cref{eq: poincare-generators}. This vector generates boosts along the \(x\)-direction. In Cartesian coordinates it is given by
        \begin{equation}
            \tensor*{J}{^a_x_t} = x \tensor{\qty(\pdv{t})}{^a} + t \tensor{\qty(\pdv{x})}{^a}.
        \end{equation}
        Next we convert to \((u,l,\zeta,\bar{\zeta})\) coordinates. We can then write
        \begin{equation}
            \tensor*{J}{^a_x_t} = - \frac{u (\zeta + \bar{\zeta})}{1 + \zeta \bar{\zeta}} \tensor{\qty(\pdv{u})}{^a} - \frac{l(1+lu) (\zeta + \bar{\zeta})}{1 + \zeta \bar{\zeta}} \tensor{\qty(\pdv{l})}{^a} + \frac{(1+lu) (1-\zeta^2)}{2} \tensor{\qty(\pdv{\zeta})}{^a} + \cc,
        \end{equation}
        where \(\cc\) stands for the complex conjugate of the angular component. By taking \(l \to 0^+\) we then find
        \begin{equation}
            \tensor*{\mathcal{J}}{^a_x_t} = - \frac{u (\zeta + \bar{\zeta})}{1 + \zeta \bar{\zeta}} \tensor{\qty(\pdv{u})}{^a} + \frac{1-\zeta^2}{2} \tensor{\qty(\pdv{\zeta})}{^a} + \cc.
        \end{equation}
        Notice that the angular part is a conformal Killing vector field on the sphere (see \cref{subsec: conformal-isometries-sphere}).
        
        Through this procedure, the generators of Lorentz transformations take the general form
        \begin{equation}
            \tensor{\mathcal{J}}{^a} = \tensor{Y}{^a} + \frac{u}{2}\tensor{D}{_b}\tensor{Y}{^b} \tensor{\qty(\pdv{u})}{^a},
        \end{equation}
        where \(\tensor{Y}{^a}\) is some conformal Killing vector field on the sphere. The detailed expressions are
        \begin{subequations}\label{eq: poincare-generators-infinity}
        \begin{align}
            \tensor*{\mathcal{J}}{^a_y_z} &= \frac{i(1-\zeta^2)}{2} \tensor{\qty(\pdv{\zeta})}{^a} + \cc, \\
            \tensor*{\mathcal{J}}{^a_z_x} &= -\frac{1+\zeta^2}{2} \tensor{\qty(\pdv{\zeta})}{^a} + \cc, \\
            \tensor*{\mathcal{J}}{^a_x_y} &= i\zeta \tensor{\qty(\pdv{\zeta})}{^a} + \cc, \\
            \tensor*{\mathcal{J}}{^a_x_t} &= - \frac{u (\zeta + \bar{\zeta})}{1 + \zeta \bar{\zeta}} \tensor{\qty(\pdv{u})}{^a} + \frac{1-\zeta^2}{2} \tensor{\qty(\pdv{\zeta})}{^a} + \cc, \\
            \tensor*{\mathcal{J}}{^a_y_t} &= \frac{i u (\zeta - \bar{\zeta})}{1 + \zeta \bar{\zeta}} \tensor{\qty(\pdv{u})}{^a} + \frac{i(1+\zeta^2)}{2} \tensor{\qty(\pdv{\zeta})}{^a} + \cc, \\
            \tensor*{\mathcal{J}}{^a_z_t} &= \frac{u (1 - \zeta \bar{\zeta})}{1 + \zeta \bar{\zeta}} \tensor{\qty(\pdv{u})}{^a} + \zeta \tensor{\qty(\pdv{\zeta})}{^a} + \cc.
        \end{align}

        We still have to discuss the role of translations. These are the fields \(\tensor*{P}{_\mu^a}\) given on \cref{eq: poincare-generators}. We can compute the induced vectors on future null infinity. The results are
        \begin{align}
            \tensor*{\mathcal{P}}{^a_t} = \tensor{\qty(\pdv{u})}{^a} &= \sqrt{4 \pi} \tensor{Y}{_0^0}(\zeta,\bar\zeta) \tensor{\qty(\pdv{u})}{^a}, \\
            \tensor*{\mathcal{P}}{^a_x} = - \frac{\zeta + \bar{\zeta}}{1 + \zeta \bar{\zeta}} \tensor{\qty(\pdv{u})}{^a} &= \sqrt{\frac{2\pi}{3}} \qty(\tensor{Y}{_1^1}(\zeta,\bar\zeta) - \tensor{Y}{_1^{-1}}(\zeta,\bar\zeta)) \tensor{\qty(\pdv{u})}{^a}, \\
            \tensor*{\mathcal{P}}{^a_y} = \frac{i(\zeta - \bar{\zeta})}{1 + \zeta \bar{\zeta}} \tensor{\qty(\pdv{u})}{^a} &= -i\sqrt{\frac{2\pi}{3}} \qty(\tensor{Y}{_1^1}(\zeta,\bar\zeta) + \tensor{Y}{_1^{-1}}(\zeta,\bar\zeta)) \tensor{\qty(\pdv{u})}{^a}, \\
            \tensor*{\mathcal{P}}{^a_z} = \frac{1 - \zeta \bar{\zeta}}{1 + \zeta \bar{\zeta}}\tensor{\qty(\pdv{u})}{^a} &= - \sqrt{\frac{4\pi}{3}} \tensor{Y}{_1^0}(\zeta,\bar\zeta) \tensor{\qty(\pdv{u})}{^a}.
        \end{align}
        \end{subequations}
        Notice the coefficients are all linear combinations of spherical harmonics \(\tensor{Y}{_l^m}(\zeta,\bar{\zeta})\) with \(l \leq 1\). This hints on how supertranslations will occur on \cref{sec: bms-group}: they will admit values with \(l \geq 2\). Similarly, on \cref{sec: superrotations} we will see how superrotations generalize the coefficients of the \(\tensor*{\mathcal{J}}{^a_\mu_\nu}\) vectors to allow for other powers of \(\zeta\) and \(\bar{\zeta}\).

    \subsection{The Night Sky as the Riemann Sphere}\label{subsec: night-sky}
        In \cref{subsec: conformal-isometries-sphere} we saw the conformal group of the two-dimensional sphere is the Lorentz group. At that point, this seemed a curious result with no obvious physical interpretation. \cref{eq: poincare-generators-infinity}, however, gives us some intuition on what is going on: we should look at what is happening in the sphere at infinity. We follow Sec. 1.3 of Ref. \citeonline{penrose1984TwoSpinorCalculusRelativistic}.

        We have been working mostly with future null infinity \(\nullfut\), and \cref{eq: poincare-generators-infinity} is written at \(\nullfut\), but for this discussion it is best to consider past null infinity, \(\nullpas\). In this way, we can picture the sphere at infinity---often called the celestial sphere---as what an observer sees as the night sky. After all, for each fixed advanced time \(v=t+r\), the observer sees the sphere \(\Sph^2 \subeq \nullpas\) with that fixed value of \(v\). 

        Consider then two observers, Alice and Bob, who observe the night sky at the same event. Their motion is related by a Lorentz transformation. For simplicity, we assume the Lorentz transformation is only a composition of a rotation and a boost along the same direction, \(\vu{z}\). This is an example of a four-screw \cite{synge1955RelativitySpecialTheory}, and it holds that any Lorentz transformation can either be written as a four-screw along some axis or as another type of transformation known as null rotation. I will ignore the null rotations, which are discussed in Sec. 1.3 of Ref. \citeonline{penrose1984TwoSpinorCalculusRelativistic}.

        The Lorentz matrix for a rotation by an angle \(\psi\) about the \(\vu{z}\) axis is
        \begin{equation}
            \Lambda_R(\psi) = \mqty(1 & 0 & 0 & 0 \\ 0 & \cos\psi & \sin\psi & 0 \\ 0 & -\sin\psi & \cos\psi & 0 \\ 0 & 0 & 0 & 1).
        \end{equation}
        The Lorentz matrix for a boost with velocity \(\tanh\chi\) is given by
        \begin{equation}
            \Lambda_B(\chi) = \mqty(\cosh\chi & 0 & 0 & -\sinh\chi \\ 0 & 1 & 0 & 0 \\ 0 & 0 & 1 & 0 \\ -\sinh\chi & 0 & 0 & \cosh\chi).
        \end{equation}
        The parameter \(\chi\) is known as ``rapidity''. Notice the boost and the rotation commute. A four-screw about the \(\vu{z}\) axis is then represented by
        \begin{equation}
            \Lambda(\psi,\chi) = \mqty(\cosh\chi & 0 & 0 & -\sinh\chi \\ 0 & \cos\psi & \sin\psi & 0 \\ 0 & -\sin\psi & \cos\psi & 0 \\ -\sinh\chi & 0 & 0 & \cosh\chi).
        \end{equation}

        If an event is labeled by Alice as \(\tensor{x}{^\mu} = (t,x,y,z)\), then Bob labels it as 
        \begin{equation}
            \mqty(t' \\ x' \\ y' \\ z') = \mqty(\cosh\chi & 0 & 0 & -\sinh\chi \\ 0 & \cos\psi & \sin\psi & 0 \\ 0 & -\sin\psi & \cos\psi & 0 \\ -\sinh\chi & 0 & 0 & \cosh\chi)\mqty(t \\ x \\ y \\ z).
        \end{equation}

        Define now, for Alice, a new coordinate system \((v,l,\zeta,\bar{\zeta})\). The stereographic coordinates are given in terms of the Cartesian coordinates by 
        \begin{equation}\label{eq: stereographic-from-Cartesian}
            \zeta = \frac{z + r}{x - i y},
        \end{equation}
        with inverse,
        \begin{equation}\label{eq: Cartesian-from-stereographic}
            x = r \frac{\zeta + \bar{\zeta}}{1 + \zeta\bar{\zeta}}\qc y = -ir \frac{\zeta - \bar{\zeta}}{1 + \zeta\bar{\zeta}}, \qq{and} z = r \frac{-1 + \zeta\bar{\zeta}}{1 + \zeta\bar{\zeta}}.
        \end{equation}
        As usual, \(v=t+r\) and \(l=1/r\). We define \((v',l',\zeta',\bar{\zeta}')\) similarly for Bob. 

        Using the previous equations, we can find the relation between \((v',\zeta',\bar{\zeta}')\) and \((v,\zeta,\bar{\zeta})\) in the \(l \to 0^+\) limit (meaning we are interested in the celestial sphere, not on a finite sphere). We have the relations 
        \begin{align}
            v' &= \frac{e^{\chi} (1 + \zeta \bar{\zeta})}{1 + e^{2\chi} \zeta \bar{\zeta}} v, \\
            \intertext{and}
            \zeta' &= e^{\chi - i \psi} \zeta. \label{eq: lorentz-zeta}
        \end{align}
        
        Physically, these relations answer the following question. Suppose Alice observes, at a given event, a light-ray incoming from the celestial sphere at past null infinity with angular coordinates \((\zeta,\bar{\zeta})\) at advanced time \(v\). At the same event, from which direction \((\zeta',\bar{\zeta}')\) does Bob see the light-ray incoming and at which advanced time was the light-ray sent in?

        It is particularly interesting to analyze how the Lorentz transformations affect, say, the positions of the stars on the night sky. This is answered by \cref{eq: lorentz-zeta}. 

        Consider first the case of a pure boost (\(\psi = 0\)). In this case, the transformation is \(\zeta' = e^{\chi} \zeta\), meaning that Bob sees larger absolute values of the stereographic coordinate in comparison to Alice. Since \(\zeta = \infty\) is the North pole and \(\zeta = 0\) is the South pole, this means Bob sees the stars in the sky closer to the North pole in comparison to Alice. By ``North pole'' I mean the direction in the sky to which Bob is moving relative to Alice. 

        The second case is that of a pure rotation (\(\chi = 0\)). In this case, the transformation is \(\zeta' = e^{-i\psi}\zeta\), which is just a phase shift. One may recall from \cref{eq: def-stereographic} that the phase of \(\zeta\) is the azimuthal angle, and therefore a rotation simply shifts \(\phi \to \phi' = \phi - \psi\). This is just a rotation of the night sky about the same axis in which Bob is rotated relative to the \(\vu{z}\) axis---as expected. 

\begin{figure}
            \centering
            \null\hfill\includegraphics[width=0.35\linewidth]{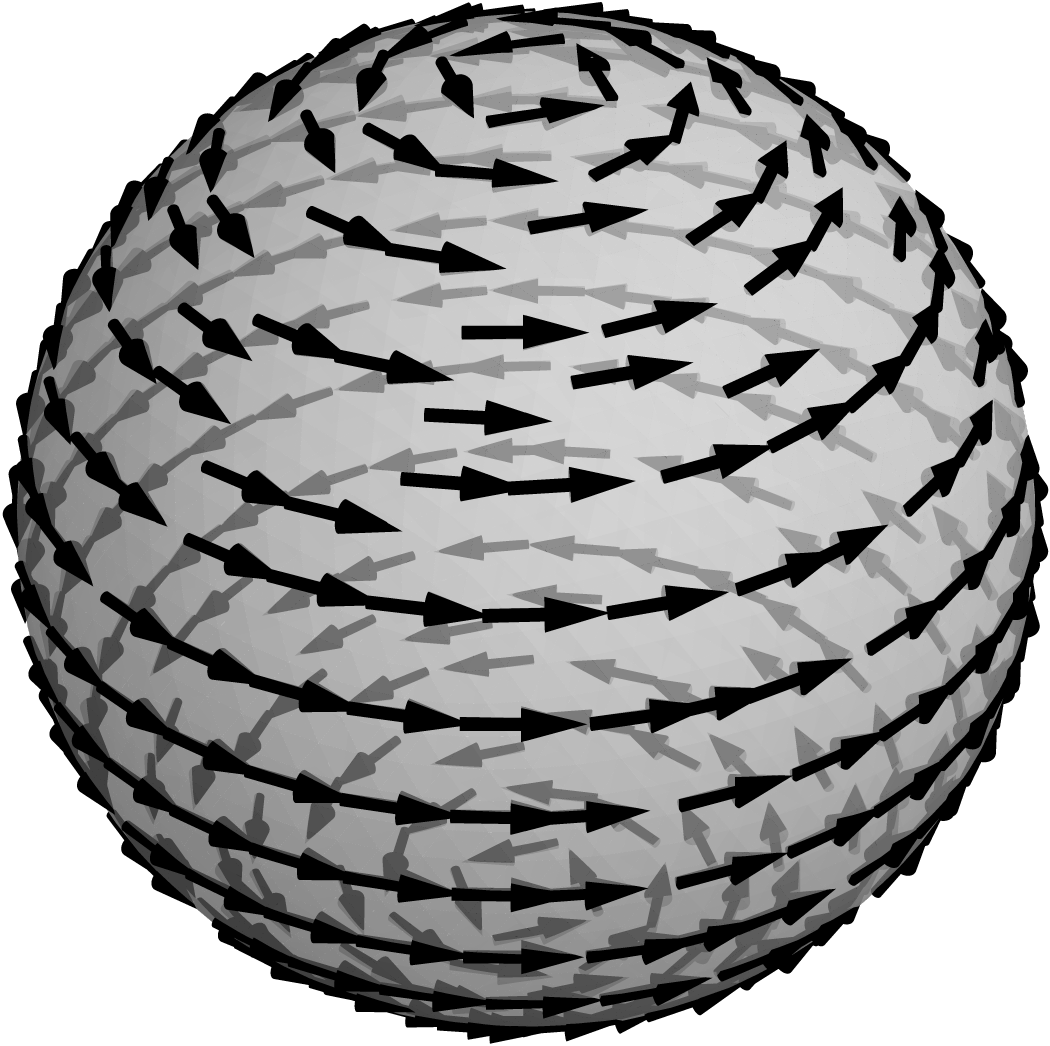}\hfill\includegraphics[width=0.35\linewidth]{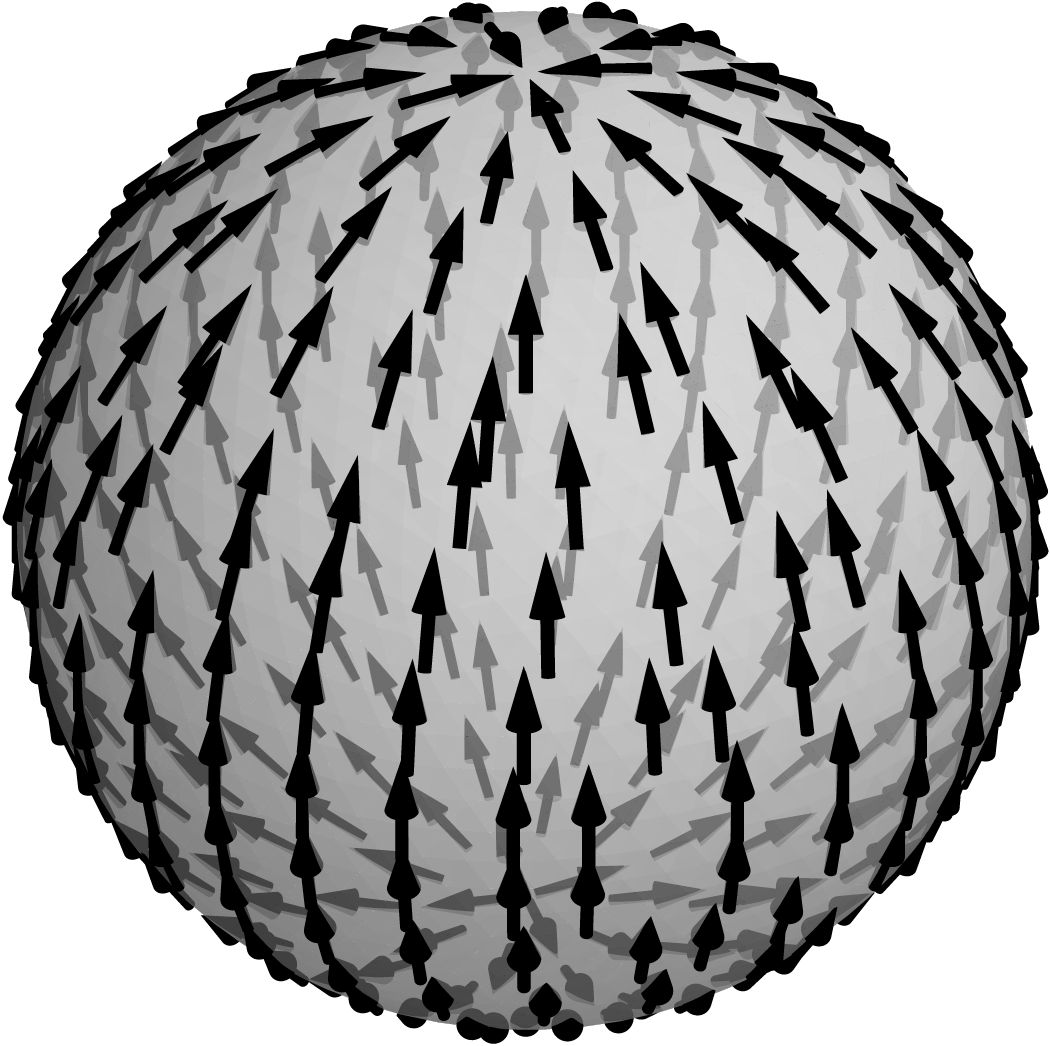}\hfill\null

            \null\hfill\includegraphics[width=0.35\linewidth]{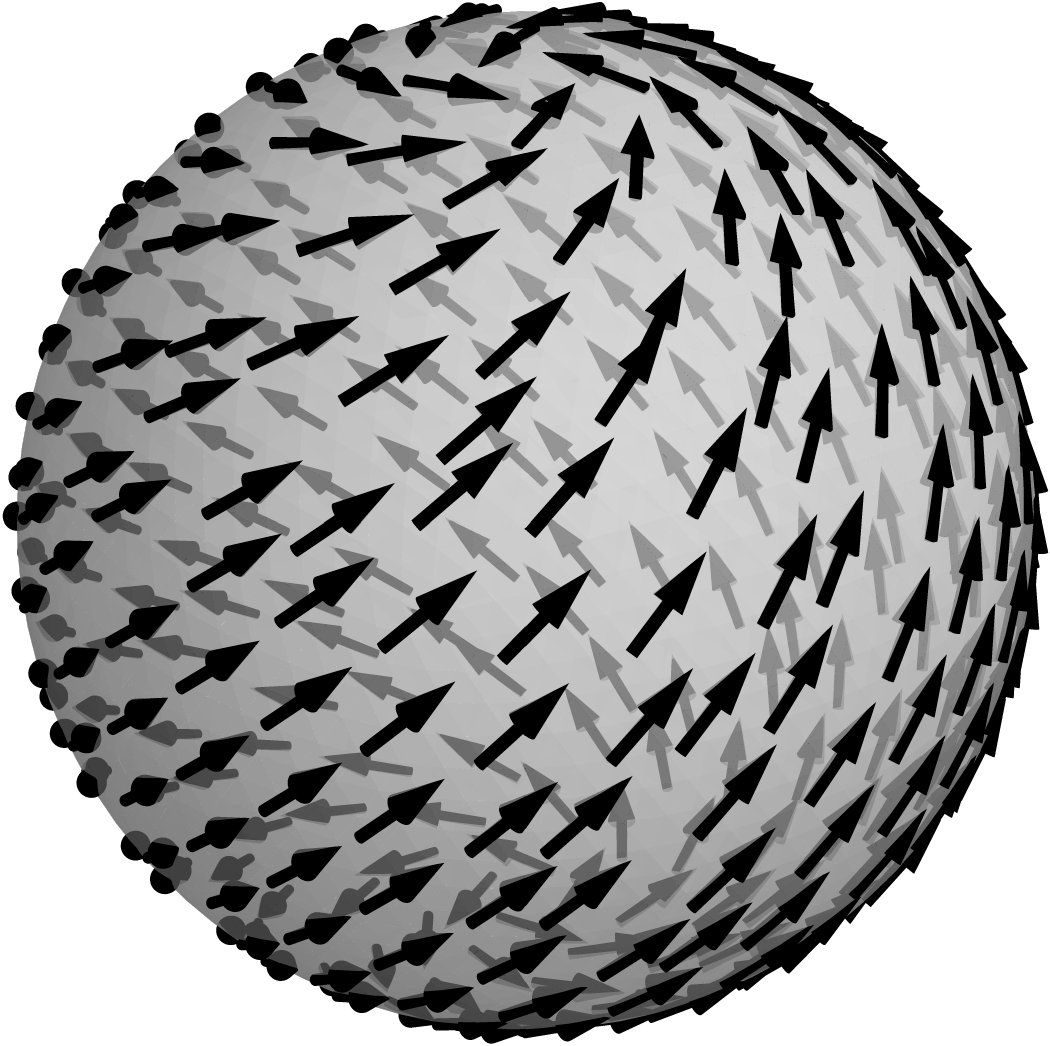}\hfill\includegraphics[width=0.35\linewidth]{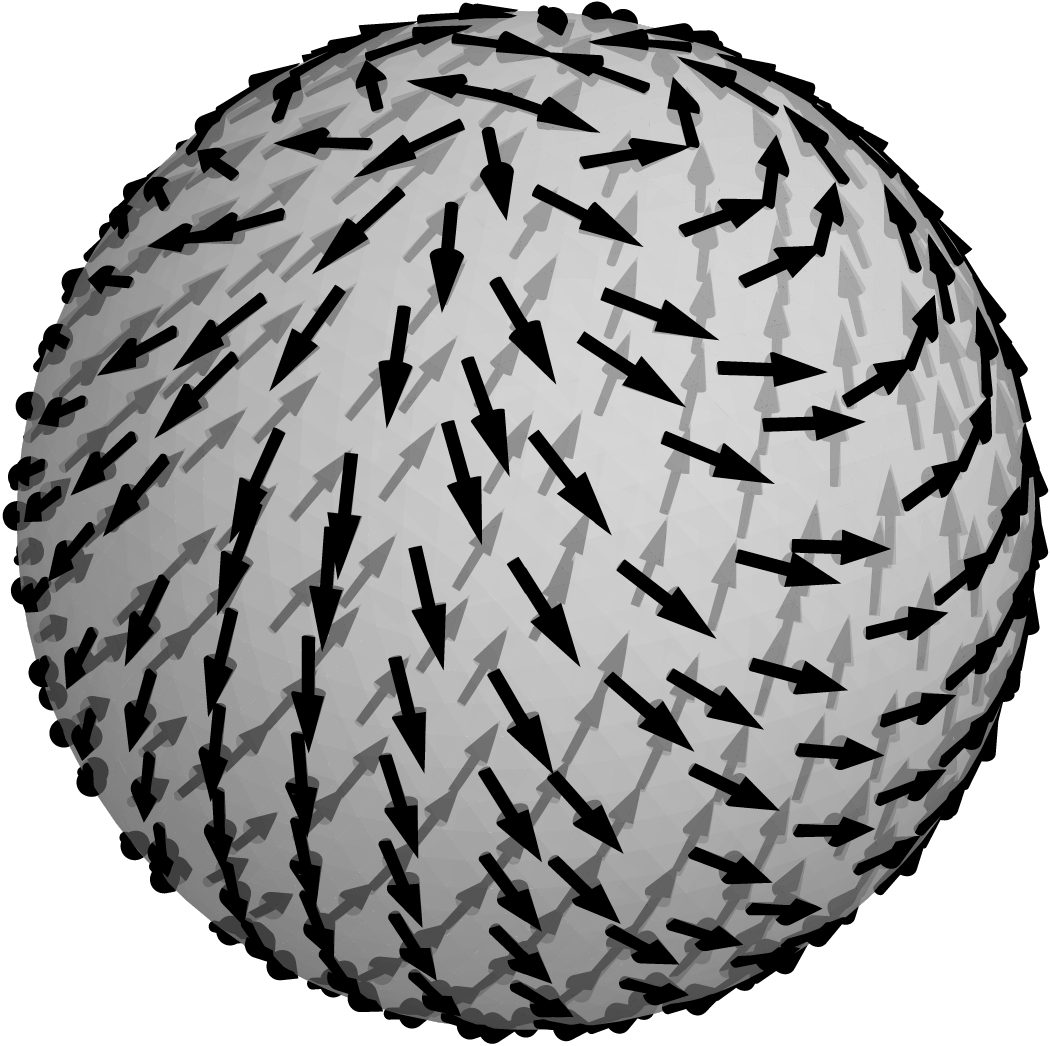}\hfill\null
            \caption{Effects of Lorentz transformations on the celestial sphere. If Alice saw a star in the sky at some point in the sphere, Bob (who is related to Alice by a Lorentz transformation) sees the star moved along the direction of the arrows. Top left: Bob is rotated relative to Alice. Top right: Bob is boosted relative to Alice (Alice sees Bob moving toward the North pole). Bottom left: Bob is related to Alice by a four-screw (a combination of the two previous cases). Bottom right: Bob is related to Alice by a null rotation. The pictures are drawn with \texttt{Mathematica} \cite{wolframresearch2024Mathematica140} and are based on the figures on Sec. 1.3 of the book by \textcite{penrose1984TwoSpinorCalculusRelativistic}.}
            \label{fig: conformal-sphere}
        \end{figure}

        For a general four-screw, the effect is a combination of the two previous ones. All of these effects, and the separate case of a null rotation, are illustrated on \cref{fig: conformal-sphere}. The lesson we take is that the night sky seen by Alice and the night sky seen by Bob---two observers in four-dimensional Minkowski spacetime related by a Lorentz transformation---are related by a conformal transformation. This is a physical interpretation for \(\SO*(3,1)\) being the conformal group on the sphere.

    \subsection{Bondi--Sachs Formalism}\label{subsec: bondi-sachs-formalism}
        There are some important remarks to be made concerning the definitions I used so far. I chose to discuss asymptotic flatness in the language of conformal compactifications, which is a coordinate-free approach and finds many uses in relativity. Nevertheless, this is not the only way of thinking about asymptotic flatness. In fact, the original works of \textcite{bondi1962GravitationalWavesGeneral,sachs1962GravitationalWavesGeneral,sachs1962AsymptoticSymmetriesGravitational} thought of asymptotic flatness in terms of the existence of a particular coordinate system in which the metric components fall-off sufficiently fast. This approach is employed in many modern references on the \gls{BMS} group and its extensions \cite{strominger2018LecturesInfraredStructure,compere2019AdvancedLecturesGeneral,pasterski2019ImplicationsSuperrotations,aneesh2022CelestialHolographyLectures}.

        There is a good reason for avoiding the use of conformal compactifications: there are physically interesting spacetimes that can be understood as asymptotically flat in a broader sense, but do not admit smooth conformal compactifications \cite{christodoulou2002GlobalInitialValue,dafermos2013FormationBlackHoles,friedrich2018PeelingNotPeeling,valientekroon2023ConformalMethodsGeneral}. This is due to the fact that the smoothness properties at null infinity are related to the decay properties of the Riemann tensor, which may be complicated in dynamical scenarios. 

        It is important to point out that there are very simple examples in which this sort of problem occurs. For instance, in Ref. \citeonline{christodoulou2002GlobalInitialValue}, Christodoulou considers the case of a system of \(n\) massive bodies interacting gravitationally in linearized gravity. In the far past the bodies are assumed to interact with each other through Newtonian interactions and to have asymptotically constant velocities. For convenience, a short review of linearized gravity is available in Appendix \ref{app: linearized-gravity}.

        For such a system, the quadrupole moment in the far past is given by
        \begin{equation}
            \tensor{Q}{_i_j} = \sum_{\substack{p\\\text{incoming}}} m_p \tensor*{x}{^{(p)}_i}\tensor*{x}{^{(p)}_j},
        \end{equation}
        where \(m_p\) is the mass of the \(p\)-th massive body and \(\tensor*{x}{^{(p)}_i}\) are the Cartesian coordinates of its position. In the far past, we know \(\tensor*{\dot{x}}{^{(p)}_i} = \tensor*{v}{^{(p)}_i}\), which are the constant incoming velocities (the dot denotes differentiation with respect to coordinate Cartesian time). In particular, notice that in the far past it holds that \(\tensor{\vb{x}}{^{(p)}} = \tensor{\vb{v}}{^{(p)}} u\), where \(u=t-r\) is retarded time\footnote{This follows from the fact that \(\dv*{\tensor{\vb{x}}{^{(p)}}}{u} = \dv*{\tensor{\vb{x}}{^{(p)}}}{t}\), since \(\tensor{\vb{x}}{^{(p)}}(t)\) is a function of time only.}. For the second derivative, Newton's law of gravitation yields
        \begin{equation}
            \tensor*{\ddot{x}}{^{(p)}_i} = - \sum_{\substack{q\neq p\\\text{incoming}}} \frac{m_q \qty(\tensor*{x}{^{(p)}_i} - \tensor*{x}{^{(q)}_i})}{\norm{\tensor{\vb{x}}{^{(p)}} - \tensor{\vb{x}}{^{(q)}}}^3}.
        \end{equation}
        For future convenience, it is useful to notice that
        \begin{equation}
            \tensor*{\dddot{x}}{^{(p)}_i} = - \sum_{\substack{q\neq p\\\text{incoming}}} \qty[\frac{m_q \qty(\tensor*{\dot{x}}{^{(p)}_i} - \tensor*{\dot{x}}{^{(q)}_i})}{\norm{\tensor{\vb{x}}{^{(p)}} - \tensor{\vb{x}}{^{(q)}}}^3} - \frac{3m_q \qty(\tensor*{x}{^{(p)}_i} - \tensor*{x}{^{(q)}_i})\qty(\tensor*{x}{^{(p)}_j} - \tensor*{x}{^{(q)}_j})\qty(\tensor*{\dot{x}}{^{(p)}^j} - \tensor*{\dot{x}}{^{(q)}^j})}{\norm{\tensor{\vb{x}}{^{(p)}} - \tensor{\vb{x}}{^{(q)}}}^5}].
        \end{equation}

        Using \(\tensor{\vb{x}}{^{(p)}} = \tensor{\vb{v}}{^{(p)}} u\), we find that in the far past the above expressions become
        \begin{subequations}
            \begin{align}
                \tensor*{x}{^{(p)}_i} &= \tensor*{v}{^{(p)}_i} u, \\
                \tensor*{\dot{x}}{^{(p)}_i} &= \tensor*{v}{^{(p)}_i}, \\
                \tensor*{\ddot{x}}{^{(p)}_i} &= - \sum_{\substack{q\neq p\\\text{incoming}}} \frac{m_q \qty(\tensor*{v}{^{(p)}_i} - \tensor*{v}{^{(q)}_i})}{u^2\norm{\tensor{\vb{v}}{^{(p)}} - \tensor{\vb{v}}{^{(q)}}}^3}, \\
                \tensor*{\dddot{x}}{^{(p)}_i} &= - \sum_{\substack{q\neq p\\\text{incoming}}} \qty[\frac{m_q \qty(\tensor*{v}{^{(p)}_i} - \tensor*{v}{^{(q)}_i})}{u^3 \norm{\tensor{\vb{v}}{^{(p)}} - \tensor{\vb{v}}{^{(q)}}}^3} - \frac{3m_q \qty(\tensor*{v}{^{(p)}_i} - \tensor*{v}{^{(q)}_i})}{u^3\norm{\tensor{\vb{v}}{^{(p)}} - \tensor{\vb{v}}{^{(q)}}}^3}].
            \end{align}
        \end{subequations}
        
        Using these expressions, one finds that
        \begin{subequations}
            \begin{align}
                \tensor*{\dddot{Q}}{_i_j} &= \sum_{\substack{p\\\text{incoming}}} m_p \qty[\tensor*{\dddot{x}}{^{(p)}_i}\tensor*{x}{^{(p)}_j} + 3 \tensor*{\ddot{x}}{^{(p)}_i}\tensor*{\dot{x}}{^{(p)}_j} + 3 \tensor*{\dot{x}}{^{(p)}_i}\tensor*{\ddot{x}}{^{(p)}_j} + \tensor*{x}{^{(p)}_i}\tensor*{\dddot{x}}{^{(p)}_j}], \\
                &= -\sum_{\substack{q\neq p\\\text{incoming}}} m_p m_q \frac{\qty[\qty(\tensor*{v}{^{(p)}_i} - \tensor*{v}{^{(q)}_i})\tensor*{v}{^{(p)}_j} + \tensor*{v}{^{(p)}_i}\qty(\tensor*{v}{^{(p)}_j} - \tensor*{v}{^{(q)}_j})]}{u^2 \norm{\tensor{\vb{v}}{^{(p)}} - \tensor{\vb{v}}{^{(q)}}}^3}.
            \end{align}
        \end{subequations}

        As attested by \cref{eq: power-GWs}, this result means there is energy being radiated to null infinity in the form of \glspl{GW} at early times (\(u \to -\infty\)), with the power emitted being suppressed by a factor of \(u^{4}\). While this suppression may seem considerable, it is not sufficient to ensure asymptotic flatness in the sense of a conformal compactification! The reason is that this emission of energy to infinity makes it so that the Riemann tensor decays too slowly at infinity. Let \(U\) be the null coordinate labeling outgoing null geodesics and \(V\) be the null coordinate labeling incoming null geodesics (these coordinates are meant in the full metric, not in the background Minkowski metric). Then a spacetime with a smooth conformal boundary has
        \begin{equation}
            \tensor{R}{_\mu_U_V_U} = \order{r^{-4}}
        \end{equation}
        as \(r \to +\infty\) with constant \(U\). This is a subcase of the so-called peeling property \cite{wald1984GeneralRelativity,valientekroon2023ConformalMethodsGeneral,friedrich2018PeelingNotPeeling,dafermos2013FormationBlackHoles}. In the scattering discussed above, one actually has \cite{christodoulou2002GlobalInitialValue}
        \begin{equation}
            \tensor{R}{_\mu_U_V_U} = \order{r^{-4} \log r},
        \end{equation}
        meaning the decay is too slow for the spacetime to admit a conformal compactification. 

        Let us then take a different approach to asymptotic flatness, which was originally taken by \textcite{bondi1962GravitationalWavesGeneral,sachs1962GravitationalWavesGeneral,sachs1962AsymptoticSymmetriesGravitational} and takes the name of Bondi--Sachs formalism \cite{madler2016BondiSachsFormalism}. This method will be deeply dependent of a particular coordinate system, but it will not require the introduction of a conformal factor or an unphysical spacetime. 
        
        We begin by defining the Bondi gauge \cite{bondi1960GravitationalWavesGeneral} following \textcite{compere2019AdvancedLecturesGeneral}. Assume an arbitrary spacetime. We consider some family of null hypersurfaces covering all of spacetime and labeled by a parameter \(u\). Since these hypersurfaces are null and they are normal to the one-form \(\tensor{\qty(\dd{u})}{_a}\), it follows that \(\tensor{g}{^u^u} = 0\). This fixes one of the components of the metric. 

        Next, we must fix the spatial coordinates. There will be a radial coordinate \(r\) and two angular coordinates \(\tensor{x}{^A}\) (\eg, \(A = \zeta, \bar{\zeta}\)). The angular coordinates will be taken to be orthogonal to the hypersurfaces of constant \(u\), which amounts to imposing 
        \begin{equation}
            \tensor{\qty(\dd{u})}{_a}\tensor{g}{^a^b}\tensor{\partial}{_b}\tensor{x}{^A} = 0,
        \end{equation}
        and hence \(\tensor{g}{^u^A} = 0\).

        So far we have three conditions on the coordinates: \(\tensor{g}{^u^u} = \tensor{g}{^u^A} = 0\). We can still impose one more. We do so by declaring that the third spatial coordinate, \(r\), is the areal distance, meaning\footnote{Notice that the factor of \(r^2\) is inside the determinant in \cref{eq: areal-radius}. This is crucial.} 
        \begin{equation}\label{eq: areal-radius}
            \tensor{\partial}{_r}\qty(\det(\frac{\tensor{g}{_A_B}}{r^2})) = 0.
        \end{equation}
        Hence, \(r\) is defined by the formula \(A \propto r^2\), where \(A\) is the area of the topological sphere with constant \(u\) and \(r\). Many authors refer to this coordinate as the ``luminosity distance'', but this term is also used in cosmology with a different meaning \cite{baumann2022Cosmology,weinberg2008Cosmology}. I thus prefer ``areal distance''. 
        
        The original \gls{BMS} analysis worked with the areal distance \cite{bondi1962GravitationalWavesGeneral,sachs1962AsymptoticSymmetriesGravitational,sachs1962GravitationalWavesGeneral}. This is the choice understood in the term ``Bondi gauge'' \cite{bondi1960GravitationalWavesGeneral}, and we will work with it in the following. Notice, though, that it is also possible to employ other choices of radial coordinate---after all, a choice of gauge should not (and does not) affect the physical results. For example, the Newman--Unti gauge \cite{newman1962BehaviorAsymptoticallyFlat} employs an affine parameter instead and reproduces the same results \cite{barnich2013AsymptoticSymmetriesNull,barnich2020BMSCurrentAlgebra}. In fact, there are also more general analyses that completely relax the condition on the radial coordinate and keep only the conditions \(\tensor{g}{^u^u} = \tensor{g}{^u^A} = 0\), allowing to treat the Bondi and Newman--Unti gauges in a unified way \cite{geiller2022PartialBondiGauge}.

        In matrix notation, we get the contravariant components
        \begin{equation}
            \tensor{g}{^\mu^\nu} = \mqty(0 & \tensor{g}{^u^r} & 0 \\ \tensor{g}{^u^r} & \tensor{g}{^r^r} & \tensor{g}{^r^A} \\ 0 & \tensor{g}{^r^A} & \tensor{g}{^A^B}),
        \end{equation}
        and inverting this block matrix will show that \(\tensor{g}{_r_r} = \tensor{g}{_r_A} = 0\). Hence, in Bondi gauge, a general Lorentzian metric has the form \cite{bondi1960GravitationalWavesGeneral,bondi1962GravitationalWavesGeneral,sachs1962GravitationalWavesGeneral}
        \begin{equation}\label{eq: arbitrary-bondi-gauge}
            \dd{s}^2 = - \frac{V e^{2\beta}}{r} \dd{u}^2 - 2 e^{2\beta} \dd{u}\dd{r} + r^2 \tensor{h}{_A_B} \qty(\dd{\tensor{x}{^A}} - \tensor{U}{^A}\dd{u})\qty(\dd{\tensor{x}{^B}} - \tensor{U}{^B}\dd{u}),
        \end{equation}
        where \(V\), \(\beta\), \(\tensor{U}{^A}\), and \(\tensor{h}{_A_B}\) take arbitrary values.

        At this stage, we have only chosen a specific coordinate system---the Bondi gauge---in an arbitrary spacetime. The next step is to introduce asymptotic flatness. To do so, first recall that in Bondi gauge the Minkowski metric is written as
        \begin{equation}
            \dd{s}^2 = - \dd{u}^2 - 2 \dd{u}\dd{r} + r^2 \tensor{\gamma}{_A_B} \dd{\tensor{x}{^A}} \dd{\tensor{x}{^B}}.
        \end{equation}
        Sachs \cite{sachs1962GravitationalWavesGeneral} then assumed that it is possible to expand the metric coefficients in powers of \(r\) near infinity with at most a finite pole at \(r = \infty\). We do not need analyticity, but rather only a couple of polynomial terms. Under this assumption, matching with Minkowski spacetime at large \(r\) leads to the asymptotic expansions
        \begin{subequations}
            \begin{align}
                V &\sim r + 2 m + \cdots, \\
                e^{2\beta} &\sim 1 + \frac{\alpha}{r} + \frac{\delta}{r^2} + \cdots, \\
                \tensor{U}{^A} &\sim \frac{\tensor{X}{^A}}{r^2} + \frac{\tensor{Y}{^A}}{r^3} + \cdots, \\
                \tensor{h}{_A_B} &\sim \tensor{\gamma}{_A_B} + \frac{\tensor{C}{_A_B}}{r} + \frac{\tensor{D}{_A_B}}{r^2} + \cdots.
            \end{align}
        \end{subequations}
        We may now start imposing restrictions on each of the terms in these asymptotic expansions. 

        First and foremost, we have the Bondi gauge imposition that \cref{eq: areal-radius} holds, which means that
        \begin{equation}
            \det(\tensor{h}{_A_B}) = \det(\tensor{\gamma}{_A_B})
        \end{equation}
        This expression holds exactly, and as a consequence it will impose a number of restrictions on the coefficients of the asymptotic expansion of \(\tensor{h}{_A_B}\). To see which sort of restrictions we get, notice that
        \begin{subequations}\label{eq: large-r-expansion-angular-metric}
            \begin{align}
                \det(\tensor{h}{_A_B}) &= \det(\tensor{\gamma}{_A_B} + \frac{\tensor{C}{_A_B}}{r} + \frac{\tensor{D}{_A_B}}{r^2}+\cdots), \\
                &= \det(\tensor{\gamma}{_A_C}\qty(\tensor{\delta}{^C_B} + \frac{\tensor{\gamma}{^C^D}\tensor{C}{_D_B}}{r} + \frac{\tensor{\gamma}{^C^D}\tensor{D}{_D_B}}{r^2}+\cdots)), \\
                &= \det(\tensor{\gamma}{_A_C})\det(\tensor{\delta}{^C_B} + \frac{\tensor{C}{^C_B}}{r} + \frac{\tensor{D}{^C_B}}{r^2}+\cdots), \\
                &= \det(\tensor{\gamma}{_A_C})\exp(\log(\det(\tensor{\delta}{^C_B} + \frac{\tensor{C}{^C_B}}{r} + \frac{\tensor{D}{^C_B}}{r^2}+\cdots))), \\
                &= \det(\tensor{\gamma}{_A_C})\exp(\tr(\log(\tensor{\delta}{^C_B} + \frac{\tensor{C}{^C_B}}{r} + \frac{\tensor{D}{^C_B}}{r^2}+\cdots))), \\
                &= \det(\tensor{\gamma}{_A_C})\exp(\tr(\frac{\tensor{C}{^C_B}}{r} + \frac{\tensor{D}{^C_B}}{r^2}- \frac{\tensor{C}{^C_D}\tensor{C}{^D_B}}{2r^2} +\cdots)), \\
                &= \det(\tensor{\gamma}{_A_C})\exp(\frac{\tensor{C}{^B_B}}{r} + \frac{\tensor{D}{^B_B}}{r^2}- \frac{\tensor{C}{^B_D}\tensor{C}{^D_B}}{2r^2} +\cdots), \\
                &= \det(\tensor{\gamma}{_A_C})\qty(1 + \frac{\tensor{C}{^B_B}}{r} + \frac{\tensor{D}{^B_B}}{r^2}- \frac{\tensor{C}{^B_D}\tensor{C}{^D_B}}{2r^2} + \frac{\tensor{C}{^B_B}\tensor{C}{^C_C}}{2r^2} +\cdots).
            \end{align}
        \end{subequations}
        We need all the subleading terms to vanish. This implies, at \(\order{1/r}\), that
        \begin{equation}\label{eq: CAB-traceless}
            \tensor{\gamma}{^A^B}\tensor{C}{_A_B} = 0.
        \end{equation}
        With this knowledge, we see the next order demands
        \begin{equation}\label{eq: DAB-trace-CAB-CAB}
            2\tensor{\gamma}{^A^B}\tensor{D}{_A_B} = \tensor{\gamma}{^A^B}\tensor{C}{_A_D}\tensor{C}{^D_B}.
        \end{equation}

        To fix the remaining components one needs to work with the Einstein equations. These calculations are lengthy, but see Ref. \citeonline{barnich2010AspectsBMSCFT} for details. At the end of the calculation, one concludes that an asymptotically flat metric should be written in Bondi gauge as \cite{compere2019AdvancedLecturesGeneral} 
        \begin{multline}\label{eq: bondi-gauge-AF}
            \dd{s}^2 = - \dd{u}^2 - 2 \dd{u}\dd{r} + r^2 \tensor{\gamma}{_{A}_{B}}\dd{\tensor{x}{^A}}\dd{\tensor{x}{^B}} \\ + \frac{2 m}{r}\dd{u}^2 + r \tensor{C}{_{A}_{B}}\dd{\tensor{x}{^A}}\dd{\tensor{x}{^B}} + \tensor{D}{^{B}}\tensor{C}{_{A}_{B}}\dd{u}\dd{\tensor{x}{^A}} \\ + \frac{1}{r}\qty(\frac{4}{3}\qty(\tensor{N}{_A} + u \tensor{\partial}{_A}m) - \frac{1}{8}\tensor{\partial}{_A}\qty(\tensor{C}{^{B}^{C}}\tensor{C}{_{B}_{C}}))\dd{u}\dd{\tensor{x}{^A}} \\ + \frac{1}{16r^2}\tensor{C}{_A_B}\tensor{C}{^A^B}\dd{u}\dd{r} + \frac{1}{4}\tensor{\gamma}{_A_B}\tensor{C}{_C_D}\tensor{C}{^C^D}\dd{\tensor{x}{^A}}\dd{\tensor{x}{^B}} + \cdots.
        \end{multline}
        The dots denote subleading terms suppressed by higher powers of \(r\). Capital Latin letters denote coordinates on the sphere. \(\tensor{D}{_A}\) is the covariant derivative on the sphere with the round metric \(\tensor{\gamma}{_{A}_{B}}\). This is the standard definition used in many modern texts on the \gls{BMS} group, particularly with a view toward high-energy physics \cite{compere2019AdvancedLecturesGeneral,strominger2018LecturesInfraredStructure,pasterski2019ImplicationsSuperrotations,aneesh2022CelestialHolographyLectures}.
        
        The tensors \(m\), \(\tensor{C}{_A_B}\), and \(\tensor{N}{_A}\) depend on retarded time \(u\) and on the angular variables, but not on \(r\). They have direct physical interpretations.
        \begin{enumerate}
            \item \(m\) is known as the Bondi mass aspect and its average over the unit sphere leads to the (Bondi--Sachs) mass at retarded time \(u\), while higher spherical harmonics lead to the spatial momentum.
            \item The shear \(\tensor{C}{_A_B}\) is symmetric and traceless (\cref{eq: CAB-traceless}), and hence it carries two degrees of freedom. It describes \glspl{GW} propagating to infinity. More specifically, \(\tensor{N}{_A_B} \equiv \tensor{\partial}{_u}\tensor{C}{_A_B}\) is known as the Bondi news tensor and its square yields the power emitted to infinity per solid angle at retarded time \(u\).
            \item \(\tensor{N}{_A}\) is known as the angular momentum aspect and it plays a role similar to that of the Bondi mass aspect, but with angular momentum. If one contracts it with the generator of rotations and integrates over the sphere, it yields that component of the angular momentum measured about the origin with \(r=0\) (in the above coordinates).
        \end{enumerate}

        \Cref{eq: bondi-gauge-AF} is not really the most general Ansatz we could work with. In fact, we implicitly imposed that
        \begin{equation}
            \tensor{D}{_A_B} = \frac{1}{4}\tensor{\gamma}{_A_B}\tensor{C}{_C_D}\tensor{C}{^C^D}
        \end{equation}
        to prevent a \(\log{r}\) factor from appearing in the metric components. This was done in the original analysis by \textcite{sachs1962GravitationalWavesGeneral}, who assumed the metric components to be written as a series in powers of \(1/r\) and anticipated that this could be too strong of a requirement. Indeed, this assumption leads to the peeling property mentioned above, and thus may fail to capture interesting scenarios. For discussions, see, \eg, Refs. \citeonline{barnich2010AspectsBMSCFT,winicour1985LogarithmicAsymptoticFlatness,geiller2024SymmetriesGravitationalScattering,madler2016BondiSachsFormalism}. We will proceed without the logarithm, but it is important to know it can be there in the most general case.

        One can compute the Einstein tensor for the metric \eqref{eq: bondi-gauge-AF} and write down the Einstein equations. In a large \(r\) expansion, the leading-order equation coming from \(\tensor{G}{_u_u} = 8 \pi \tensor{T}{_u_u}\) is
        \begin{subequations}\label{eq: BS-EFE-uu}
        \begin{equation}
            \tensor{\partial}{_u}m = \frac{1}{4}\tensor{D}{^A}\tensor{D}{^B}\tensor{N}{_A_B} - \tensor{\mathcal{T}}{_u_u},
        \end{equation}
        where \(\tensor{N}{_A_B}\) is the Bondi news tensor and we defined
        \begin{equation}
            \tensor{\mathcal{T}}{_u_u} = \frac{1}{8}\tensor{N}{_A_B}\tensor{N}{^A^B} + 4 \pi \lim_{r\to\infty} (r^2 \tensor{T}{_u_u}).
        \end{equation}
        \end{subequations}
        \(\tensor{T}{_u_u}\) is the \(uu\)-component of the matter stress tensor, while \(\tensor{\mathcal{T}}{_u_u}\) is the \(uu\)-component of an effective stress tensor including the contributions due to \glspl{GW} emitted to infinity. These contributions are precisely the terms that are quadratic in the fields. Notice that only contributions of massless fields to \(\tensor{T}{_u_u}\) will survive at infinity. 

        Similarly, we can consider the leading order equation coming from \(\tensor{G}{_u_A} = 8 \pi \tensor{T}{_u_A}\) and find
        \begin{subequations}\label{eq: BS-EFE-uA}
        \begin{equation}
            \tensor{\partial}{_u}\tensor{N}{_A} = -\frac{1}{4}\tensor{D}{^B}\qty(\tensor{D}{_B}\tensor{D}{^C}\tensor{C}{_A_C}-\tensor{D}{_A}\tensor{D}{^C}\tensor{C}{_B_C}) - u\tensor{\partial}{_A}\tensor{\partial}{_u}m - \tensor{\mathcal{T}}{_u_A},
        \end{equation}
        where \(\tensor{\partial}{_u}m\) is already determined by \cref{eq: BS-EFE-uu} and \(\tensor{\mathcal{T}}{_u_A}\) is given by 
        \begin{equation}
            \tensor{\mathcal{T}}{_u_A} = -\frac{1}{4}\tensor{\partial}{_A}(\tensor{C}{_B_C}\tensor{N}{^B^C}) + \frac{1}{4}\tensor{D}{_B}(\tensor{C}{^B^C}\tensor{N}{_C_A}) - \frac{1}{2}\tensor{C}{_A_B}\tensor{D}{_C}\tensor{N}{^B^C} + 8 \pi \lim_{r\to\infty} (r^2 \tensor{T}{_u_A}).
        \end{equation}
        \end{subequations}
        Once again, only quadratic contributions occur in \(\tensor{\mathcal{T}}{_u_A}\).

        From \cref{eq: BS-EFE-uu,eq: BS-EFE-uA}, we see that specifying the values of \(m\), \(\tensor{C}{_A_B}\), and \(\tensor{N}{_A}\) at early times in \(\nullfut\) (\ie, as \(u \to - \infty\)) and the values of \(\tensor{N}{_A_B}\) throughout \(\nullfut\) are sufficient to describe the gravitational field at \(\nullfut\) (I am assuming the stress tensor is also known). Hence, at the order we are working with in the large \(r\) expansion\footnote{The angular momentum aspect enters at subleading polynomial order, but all other fields enter at leading order. As we consider other orders in the large \(r\) expansion, more fields are necessary.}, the collection
        \begin{equation}
            \qty{\left.m(\zeta,\bar{\zeta})\right|_{\nullfut_-}, \left.\tensor{C}{_A_B}(\zeta,\bar{\zeta})\right|_{\nullfut_-}, \tensor{N}{_A_B}(u,\zeta,\bar{\zeta}), \left.\tensor{N}{_A}(\zeta,\bar{\zeta})\right|_{\nullfut_-}}
        \end{equation}
        is sufficient to determine the gravitational field throughout \(\nullfut\), assuming the stress tensor is known. Above, I wrote \(\nullfut_-\) to mean the value of the tensor fields in the limit \(u \to - \infty\). I am also assuming implicitly that the Bondi news tensor falls off sufficiently fast at \(u \to \pm \infty\), since this is necessary for the Bondi mass aspect and angular momentum aspect to be finite after integrating \cref{eq: BS-EFE-uu,eq: BS-EFE-uA}. I will comment more about this on \cref{subsec: CK-spacetimes}.

        We can further simplify the initial data concerning \(\tensor{C}{_A_B}\). We must assume \(\tensor{N}{_A_B}\) vanishes at early times (\(u \to -\infty\)) for \cref{eq: BS-EFE-uu,eq: BS-EFE-uA} to be well-posed. Assuming further that there is no energy flux due to matter at early times, we conclude \(\tensor{\partial}{_u}m\) vanishes at early times. Let us suppose the same is valid for \(\tensor{\partial}{_u}\tensor{N}{_A}\). In this case, it follows from \cref{eq: BS-EFE-uu,eq: BS-EFE-uA} that, at early times, 
        \begin{equation}\label{eq: initial-condition-shear-tensor}
            \left.\tensor{D}{^B}\pqty{\tensor{D}{_B}\tensor{D}{^C}\tensor{C}{_A_C}-\tensor{D}{_A}\tensor{D}{^C}\tensor{C}{_B_C}}\right|_{\nullfut_-} = 0.
        \end{equation}
        
        To solve \cref{eq: initial-condition-shear-tensor}, it is useful to decompose \(\tensor{C}{_A_B}\) in the form\footnote{Recall that symmetrization and antisymmetrization are often denoted by parentheses and square brackets. Specifically, \(\tensor{T}{_(_a_b_)} = \frac{1}{2}\qty(\tensor{T}{_a_b} + \tensor{T}{_b_a})\) and \(\tensor{T}{_[_a_b_]} = \frac{1}{2}\qty(\tensor{T}{_a_b} - \tensor{T}{_b_a})\).}
        \begin{equation}\label{eq: shear-decomposition}
            \tensor{C}{_A_B} = \qty(\tensor{\gamma}{_A_B}\tensor{D}{_C}\tensor{D}{^C} - 2 \tensor{D}{_A}\tensor{D}{_B})C + 2\tensor{\epsilon}{_C_(_A}\tensor{D}{_B_)}\tensor{D}{^C}\Psi,
        \end{equation}
        for suitable scalar fields \(C\) and \(\Psi\). Following \textcite{compere2019AdvancedLecturesGeneral}, we will refer to \(C\) as the memory field. \(\tensor{\epsilon}{_A_B}\) denotes the Levi-Civita tensor, with \(\tensor{\epsilon}{_1_2} = \sqrt{\det \tensor{\gamma}{_A_B}}\) in any particular coordinate system. This can be done for any symmetric traceless tensor field on the sphere. A proof for general compact Riemannian manifolds (not only the sphere) is given on Proposition 2.2 of Ref. \citeonline{ishibashi2004DynamicsNonGloballyHyperbolicStatic}. \cref{eq: shear-decomposition} is actually simpler than the general case because
        \begin{enumerate}
            \item there are no symmetric, traceless, divergenceless rank-\(2\) tensors on the \(2\)-sphere, 
            \item all divergenceless vectors on the \(2\)-sphere can be written as \(\tensor{V}{^A} = \tensor{\epsilon}{^A^B}\tensor{D}{_B}\Psi\) for a suitable scalar field \(\Psi\).
        \end{enumerate}
        These simplifications are specific to \(\Sph^2\), and thus we would need to consider a more general decomposition in higher spacetime dimensions \cite{satishchandran2019AsymptoticBehaviorMassless}.
        
        To illustrate how this decomposition is possible, we can take \cref{eq: shear-decomposition} as an Ansatz\footnote{Notice that taking \cref{eq: shear-decomposition} as an Ansatz is not proving the decomposition always holds. See Ref. \citeonline{ishibashi2004DynamicsNonGloballyHyperbolicStatic} for an actual proof.} and notice that
        \begin{subequations}
            \begin{align}
                \tensor{D}{_(_A}\tensor{D}{^C}\tensor{C}{_B_)_C} &= - \tensor{D}{_A}\tensor{D}{_B}\qty(\tensor{D}{_C}\tensor{D}{^C}+ 2)C + \tensor{\epsilon}{_D_(_B}\tensor{D}{_A_)}\tensor{D}{^D}\qty(\tensor{D}{_C}\tensor{D}{^C}+ 2)\Psi, \\
                \intertext{and}
                \tensor{D}{_[_A}\tensor{D}{^C}\tensor{C}{_B_]_C} &= \tensor{\epsilon}{_D_[_B}\tensor{D}{_A_]}\tensor{D}{^D}\qty(\tensor{D}{_C}\tensor{D}{^C}+ 2)\Psi.
            \end{align}
        \end{subequations}
        From the symmetric part we can conclude that
        \begin{equation}\label{eq: DA-DB-CAB}
            \tensor{D}{^A}\tensor{D}{^B}\tensor{C}{_A_B} = - \tensor{D}{_A}\tensor{D}{^A}\qty(\tensor{D}{_B}\tensor{D}{^B} + 2) C.
        \end{equation}
        This is a differential equation to determine \(C\), and it is just an iterated version of Helmholtz's equation. As for the antisymmetric part, we find 
        \begin{equation}
            \tensor{\epsilon}{^A^B}\tensor{D}{_A}\tensor{D}{^C}\tensor{C}{_B_C} = \tensor{D}{_A}\tensor{D}{^A}\qty(\tensor{D}{_B}\tensor{D}{^B} + 2) \Psi,
        \end{equation}
        which once again is an iterated Helmholtz equation. In both cases, we can find solutions for \(C\) and \(\Psi\) by employing Green functions twice\footnote{Since Green functions are the inverse of a differential operator and the Helmholtz operator has a nontrivial kernel, we need to ensure the source terms on the differential equations do not lie on this kernel. This can be done, as shown by \textcite{ishibashi2004DynamicsNonGloballyHyperbolicStatic}.}. The homogeneous solutions, in both cases, solve
        \begin{equation}
            \tensor{D}{_A}\tensor{D}{^A}\qty(\tensor{D}{_B}\tensor{D}{^B} + 2) Y = 0,
        \end{equation}
        and are thus spherical harmonics \(Y_{lm}\) with \(l=0\) or \(l=1\). In any case, these harmonics are annihilated when inserted in \cref{eq: shear-decomposition}, and therefore do not contribute to \(\tensor{C}{_A_B}\).
        
        Using this decomposition, we may now compute \cref{eq: initial-condition-shear-tensor} and find that it reduces to
        \begin{equation}
            \left.\tensor{\epsilon}{^B_A}\tensor{D}{_B}\tensor{D}{_C}\tensor{D}{^C}\qty(\tensor{D}{_D}\tensor{D}{^D} + 2)\Psi\right|_{\nullfut_-} = 0.
        \end{equation}
        This can be rewritten as 
        \begin{equation}
            \left.\tensor{\partial}{_A}\qty(\tensor{D}{_C}\tensor{D}{^C}\qty(\tensor{D}{_D}\tensor{D}{^D} + 2)\Psi)\right|_{\nullfut_-} = 0,
        \end{equation}
        and hence
        \begin{equation}
            \left.\tensor{D}{_C}\tensor{D}{^C}\qty(\tensor{D}{_D}\tensor{D}{^D} + 2)\Psi\right|_{\nullfut_-} = c,
        \end{equation}
        for some constant \(c\). Now the \gls{RHS} is a spherical harmonic with \(l=0\), while the \gls{LHS} cannot contain any spherical harmonics with \(l < 1\). Hence, \(c = 0\). By decomposing \(\Psi\) in spherical harmonics and considering the equation for each coefficient, we find all coefficients with \(l \geq 2\) vanish. The ones with \(l < 2\) are superfluous, and hence we conclude we can take \(\Psi|_{\nullfut_-} = 0\).

        Finally, we conclude that the data
        \begin{equation}\label{eq: initial-data-BS-AF}
            \qty{\left.m(\zeta,\bar{\zeta})\right|_{\nullfut_-}, \left.C(\zeta,\bar{\zeta})\right|_{\nullfut_-}, \tensor{N}{_A_B}(u,\zeta,\bar{\zeta}), \left.\tensor{N}{_A}(\zeta,\bar{\zeta})\right|_{\nullfut_-}}
        \end{equation}
        is sufficient to fully characterize the gravitational field at \(\nullfut\) at the orders we are working with. All remaining information can be obtained by considering \cref{eq: shear-decomposition} and integrating \cref{eq: BS-EFE-uu,eq: BS-EFE-uA} and the definition of the Bondi news tensor \(\tensor{\partial}{_u}\tensor{C}{_A_B} = \tensor{N}{_A_B}\), which is done with previous knowledge of the matter stress-energy tensor. We are implicitly assuming that there is no stress-energy reaching null infinity at early times and that the angular momentum aspect is constant at early times. We also assume the Bondi news tensor to decay sufficiently fast at early and late times.

        In these lectures, I chose the conformal compactification approach because it gives a direct and coordinate-independent construction of null infinity, which I find pedagogical. Also, it is the most popular approach among relativists, despite its limitations. In practice, this difference will not play a significant role in our discussions, because the notion of null infinity can be defined even in the absence of a conformal compactification \cite{christodoulou1993GlobalNonlinearStability}. It will, however, be useful to have two definitions we can work with, depending on what we are doing. Indeed, some applications we will consider in \cref{sec: modern-developments} are formulated using conformal compactifications. 

    \subsection{\texorpdfstring{\glsfmtlong{CK}}{Christodoulou--Klainerman} Spacetimes}\label{subsec: CK-spacetimes}
        An important development in the theory of asymptotically flat spacetimes came with the work of \textcite{christodoulou1989NonlinearStabilityMinkowski,christodoulou1990GlobalNonlinearStability,christodoulou1993GlobalNonlinearStability}, who were originally considering the study of spacetimes globally (not only asymptotically) similar to Minkowski spacetime. Their interest was mostly in proving the nonlinear stability of Minkowski spacetime, meaning that spacetimes that are similar to Minkowski at one specific instant of time are still similar to Minkowski after evolution.
        
        To write their definition, let \(\Sigma_t\) be a spacelike surface in the spacetime \(M\). ``Spacelike'' means this surface has a timelike normal vector \(\tensor{n}{^a}\) at every point (which we take to be unitary). We define the induced metric on each \(\Sigma_t\) to be\footnote{On a more formal level, given a submanifold \(\Sigma\) of a pseudo-Riemannian manifold \(M\), we can always define the induced metric by considering the identity map \(\imath \colon \Sigma \to M\) and considering the pullback \(\imath_* \tensor{g}{_a_b}\), where \(\tensor{g}{_a_b}\) is the metric on \(M\). In practice, this yields the same result as \cref{eq: induced-metric-Sigma-CK}.} 
        \begin{equation}\label{eq: induced-metric-Sigma-CK}
            \tensor{\bar{g}}{_a_b} = \tensor{g}{_a_b} + \tensor{n}{_a}\tensor{n}{_b}.
        \end{equation}
        Notice that \(\tensor{\bar{g}}{_a_b}\tensor{n}{^a} = 0\), and hence \(\tensor{\bar{g}}{_a_b}\) is a three-dimensional metric. 
        
        We also define the extrinsic curvature (or second fundamental form) of \(\Sigma\) as
        \begin{equation}
            \tensor{K}{_a_b} = \frac{1}{2}\Lie[n]\tensor{\bar{g}}{_a_b},
        \end{equation}
        which also has \(\tensor{K}{_a_b}\tensor{n}{^a} = 0\).
        
        \(\tensor{\bar{g}}{_i_j}\) tells us the intrinsic geometry of \(\Sigma_t\), while \(\tensor{K}{_i_j}\) tells us how \(\Sigma\) is embedded in the spacetime manifold \(M\). If \(\tensor{K}{_i_j}\) satisfies certain constraints, one can use the triple \((\Sigma,\tensor{\bar{g}}{_i_j},\tensor{K}{_i_j})\) as initial conditions and solve the Einstein equations starting from them \cite{choquet-bruhat2009GeneralRelativityEinstein,wald1984GeneralRelativity}. This is the basic idea behind numerical relativity \cite{baumgarte2010NumericalRelativitySolving,baumgarte2021NumericalRelativityStarting,gourgoulhon20123+1FormalismGeneral}.

        Asymptotically flat spacetimes in the sense of Christodoulou and Klainerman (from now on, \gls{CK} spacetimes) are characterized by the existence, in an initial value surface \(\Sigma_t\), of some coordinate system \((x,y,z)\) near infinity such that, as \(r = \sqrt{x^2 + y^2 + z^2}\) goes to infinity, 
        \begin{subequations}
            \begin{align}
                \tensor{\bar{g}}{_i_j} &= \qty(1 + \frac{2M}{r})\tensor{\delta}{_i_j} + o_4\qty(r^{-\frac{3}{2}}), \\
                \tensor{K}{_i_j} &= o_3\qty(r^{-\frac{5}{2}}).
            \end{align}
        \end{subequations}
        The symbol \(o_n(r^{-k})\) stands for any function \(f\) such that there are always constants \(C_l\) such that \(\abs{f^{(l)}} \leq C_l r^{-k-l}\) for \(0 \leq l \leq n\), where \(f^{(l)}\) is a shorthand for any \(l\)-order partial derivative. Any \gls{CK} spacetime satisfying a ``global smallness condition'' \cite{christodoulou1989NonlinearStabilityMinkowski,christodoulou1990GlobalNonlinearStability,christodoulou1993GlobalNonlinearStability} is a geodesically complete solution. In this sense, Minkowski spacetime is nonlinearly stable---perturbing the initial conditions for Minkowski spacetime will still lead to a similar spacetime. 

        One can get to other definitions of asymptotic flatness by considering other decay properties. For example, \textcite{bieri2007ExtensionStabilityTheorem,bieri2009ExtensionStabilityTheorem,bieri2009ExtensionsStabilityTheorem} considered the asymptotic falloffs
        \begin{subequations}
            \begin{align}
                \tensor{\bar{g}}{_i_j} &= \tensor{\delta}{_i_j} + o_3\qty(r^{-\frac{1}{2}}), \\
                \intertext{and}
                \tensor{K}{_i_j} &= o_2\qty(r^{-\frac{3}{2}}),
            \end{align}
        \end{subequations}
        and obtained results similar to those of Christodoulou and Klainerman. The basic similarities and differences between these two definitions---and a third one also by \textcite{bieri2024RadiationAsymptoticsSpacetimes}---are reviewed by \textcite{bieri2024GravitationalWaveDisplacement}. Whenever I speak of \gls{CK} spacetimes I mean one of these definitions, without paying too much attention to which one I am using. 
        
        These sorts of definitions lead to a number of results concerning the falloff properties of the spacetime metric \(\tensor{g}{_a_b}\) near null infinity, which can be relaxed enough to allow situations such as the scattering of massive bodies discussed at \cref{subsec: bondi-sachs-formalism}. While these expressions are originally written in terms of components of the Riemann tensor near infinity, they can be translated to yield constraints on the values of the tensors \(m\), \(\tensor{C}{_A_B}\), and \(\tensor{N}{_A}\) of \cref{eq: bondi-gauge-AF} as \(u \to \pm \infty\). For instance, a key result due to Christodoulou and Klainerman is that the asymptotic behavior of the Bondi news tensor is \cite{christodoulou1990GlobalNonlinearStability,christodoulou1991NonlinearNatureGravitation,christodoulou1993GlobalNonlinearStability,bieri2024GravitationalWaveDisplacement}
        \begin{equation}
            \tensor{N}{_A_B} = \order{\abs{u}^{-\frac{3}{2}}},
        \end{equation}
        which is sufficient for integration of \cref{eq: BS-EFE-uu,eq: BS-EFE-uA}.
    
\section{The \texorpdfstring{\glsfmtlong{BMS}}{Bondi--Metzner--Sachs} Group}\label{sec: bms-group}
    We are almost ready to derive the \gls{BMS} group. The last prerequisite we should cover is the basic theory of Carrollian manifolds, which is the sort of structure present at null infinity. This language will provide the natural arena for us to discuss asymptotic symmetries.

    \subsection{Carrollian Structures}
        By going back to the examples we provided of Minkowski and Schwarzschild spacetimes, one can notice that the line element for the induced metric at \(\nullfut\) appears to have the form given in \cref{eq: minkowski-induced-metric-nullfut}, 
        \begin{equation}
            \dd{\tilde{\sigma}}^2 = 0 \dd{u}^2 + \dd{\Sph}^2.
        \end{equation}
        Notice this is not a pseudo-Riemannian metric, because it is degenerate. If we denote this metric by \(\tensor{\tilde{h}}{_a_b}\), we know there is a vector \(\tensor{n}{^a} \neq 0\) such that
        \begin{equation}
            \tensor{\tilde{h}}{_a_b}\tensor{n}{^a} = 0.
        \end{equation}
        Namely, in the coordinates we used before, we have
        \begin{equation}
            \tensor{n}{^a} = \tensor{\qty(\pdv{u})}{^a}
        \end{equation}
        for both Minkowski and Schwarzschild spacetimes.

        Spacetimes with degenerate metrics can be understood as limiting cases of Lorentzian spacetimes. Namely, consider the Minkowski metric 
        \begin{equation}
            \dd{s}^2 = - c^2 \dd{t}^2 + \dd{x}^2 + \dd{y}^2 + \dd{z}^2,
        \end{equation}
        where I intentionally wrote the \(c\) factors explicitly. Notice now that in the unusual limit \(c \to 0\) we get 
        \begin{equation}
            \dd{s}^2 = - 0 \dd{t}^2 + \dd{x}^2 + \dd{y}^2 + \dd{z}^2,
        \end{equation}
        which is a degenerate metric. Hence, the sort of manifold we are interested in can be understood as the \(c \to 0\) limit of a Lorentzian manifold. 

        These sorts of limits were originally considered by \textcite{levy-leblond1965NouvelleLimiteNonrelativiste,sengupta1966AnalogueGalileiGroup}, who were interested in studying the \(c \to 0\) limit of the Poincaré group. Notice that in this limit the lightcones ``close'' around the time axis, and there is no propagation. Lévy-Leblond associated this strange causal behavior to some scenes in Lewis Carroll's books \emph{Alice's Adventures in Wonderland} and \emph{Through the Looking Glass and What Alice Found There}. Of particular notice are Chap. 7 of the former (in which the days of the month pass faster than the hours of the day) and Chap. 2 of the latter (in which the Red Queen states that ``Here, you see, it takes all the running you can do to keep in the same place''). This prompts the name ``Carrollian group'' for the \(c \to 0\) limit of the Poincaré group. Similarly, we get ``Carrollian manifold'' for manifolds endowed with a metric that is degenerate along a specific direction---one cannot move, even when moving at the speed of light \(c=0\).

        The work of \textcite{levy-leblond1965NouvelleLimiteNonrelativiste} was mostly focused on the Carroll group, which is the \(c \to 0\) limit of the Poincaré group. Carrollian structures in differential geometry were defined and studied by \textcite{duval2014CarrollNewtonGalilei,duval2014ConformalCarrollGroups,duval2014ConformalCarrollGroupsa}. Following their work, I shall now review the basic ideas and definitions necessary for the study of \gls{BMS} symmetries. 

        When describing a pseudo-Riemannian manifold, we typically provide a pair \((M,\tensor{g}{_a_b})\). \(M\) is the underlying smooth manifold on which we consider a smooth metric tensor \(\tensor{g}{_a_b}\). By analogy, one would in principle expect that for Carrollian manifolds we should specify a pair \((\nullfut,\tensor{\tilde{h}}{_a_b})\), where \(\nullfut\) is the underlying smooth manifold and \(\tensor{\tilde{h}}{_a_b}\) is the degenerate metric tensor field. This, however, is insufficient. Due to \(\tensor{\tilde{h}}{_a_b}\) being degenerate, we also specify the non-vanishing vector \(\tensor{n}{^a}\) such that \(\tensor{\tilde{h}}{_a_b}\tensor{n}{^a} = 0\). This vector has the role of defining the kernel of the metric tensor. A triple \((\nullfut,\tensor{\tilde{h}}{_a_b},\tensor{n}{^a})\) is known as a weak Carrollian structure, or simply a Carrollian structure. 

        The adjective ``weak'' makes reference to the fact that one can strengthen the definition of a Carrollian structure by further specifying a covariant derivative. To understand why, recall that in a pseudo-Riemannian manifold the metric singles out a particular choice of covariant derivative by imposing that the Christoffel symbols are given by
        \begin{equation}
            \tensor{\Gamma}{^\rho_\mu_\nu} = \frac{1}{2}\tensor{g}{^\rho^\sigma}\qty(\tensor{\partial}{_\mu}\tensor{g}{_\nu_\sigma} + \tensor{\partial}{_\nu}\tensor{g}{_\sigma_\mu} - \tensor{\partial}{_\sigma}\tensor{g}{_\mu_\nu}).
        \end{equation}
        Nevertheless, this is not applicable for a Carrollian metric. Since Carrollian metrics are degenerate, the inverse metric tensor \(\tensor{g}{^a^b}\) does not exist. Hence, imposing that the metric is parallelly transported by the covariant derivative does not single out a covariant derivative. \textcite{duval2014CarrollNewtonGalilei} then go further and ask that \(\tensor{n}{^a}\) is also parallelly transported, but this is also not enough to specify a single covariant derivative. Thus, one has to specify a covariant derivative in addition to a weak Carrollian structure. The resulting quadruple \((\nullfut,\tensor{\tilde{h}}{_a_b},\tensor{n}{^a},\tensor{\tilde{\nabla}}{_a})\) is then known as a strong Carrollian structure. 

        \begin{definition}[Carrollian Structures]
            A weak Carrollian structure, or simply a Carrollian structure, is a triple \((\nullfut,\tensor{\tilde{h}}{_a_b},\tensor{n}{^a})\) where \(\nullfut\) is a smooth manifold with dimension \(d\), \(\tensor{\tilde{h}}{_a_b}\) is a positive semi-definite symmetric tensor with matrix rank  \(d-1\), and \(\tensor{n}{^a}\) is a non-vanishing vector with \(\tensor{\tilde{h}}{_a_b}\tensor{n}{^a} = 0\). 

            A strong Carrollian structure is a quadruple \((\nullfut,\tensor{\tilde{h}}{_a_b},\tensor{n}{^a},\tensor{\tilde{\nabla}}{_a})\) where \((\nullfut,\tensor{\tilde{h}}{_a_b},\tensor{n}{^a})\) is a weak Carrollian structure and \(\tensor{\tilde{\nabla}}{_a}\) is a covariant derivative with \(\tensor{\tilde{\nabla}}{_a}\tensor{\tilde{h}}{_b_c} = 0\) and \(\tensor{\tilde{\nabla}}{_a}\tensor{n}{^b} = 0\).
        \end{definition}

        Now suppose we are given an asymptotically flat spacetime \((M,\tensor{g}{_a_b})\) with unphysical conformal extension \((\tilde{M},\tensor{\tilde{g}}{_a_b})\) and conformal factor \(\Omega\). Then future null infinity has a natural Carrollian structure \((\nullfut,\tensor{\tilde{h}}{_a_b},\tensor{n}{^a})\). Namely, \(\nullfut\) is just future null infinity itself, \(\tensor{\tilde{h}}{_a_b}\) is the metric induced on \(\nullfut\) by \(\tensor{\tilde{g}}{_a_b}\), and \(\tensor{n}{^a} = \tensor{\tilde{g}}{^a^b}\tensor{\tilde{\nabla}}{_b}\Omega\). 

        To proceed, notice that there is a gauge freedom in the definition of future null infinity. Namely, if \((M,\tensor{g}{_a_b})\) is asymptotically flat with conformal factor \(\Omega\), then the conformal factor \(\omega \Omega\) for \(\omega \in \ck[\infty](\tilde{M})\) and \(\omega > 0\) would work just as well. Using this freedom, the Carrollian structure of null infinity can be made into a strong Carrollian structure. As discussed in Chap. 11 of Ref. \citeonline{wald1984GeneralRelativity}, one can use this gauge freedom to impose that \(\tensor{\tilde{\nabla}}{_a}\tensor{n}{^b} = 0\). Notice that this condition is equivalent to imposing that \(\tensor{\tilde{\nabla}}{_a}\tensor{\tilde{\nabla}}{_b}\Omega = 0\), which does not need to hold in general. However, one can always find a choice of \(\Omega\) for which this holds. 

        The strong Carrollian structure ends up being too strong for our purposes. Hence, we will work mostly with the weak Carrollian structure of null infinity and only comment later on the role the strong Carrollian structure can play. 

    \subsection{Symmetries at Null Infinity}\label{subsec: BMS-derivation-global-geometric}
        We are now ready to discuss the symmetry group of future null infinity. To do so, consider the gauge freedom mentioned at the end of the last section, which consists of exchanging the conformal factor \(\Omega\) according to \(\Omega \to \omega \Omega\) for smooth \(\omega > 0\). Using that \(\tensor{\tilde{g}}{_a_b} = \Omega^2 \tensor{g}{_a_b}\) (I will omit the pushforward for simplicity), that \(\tensor{\tilde{h}}{_a_b}\) is the metric induced by \(\tensor{\tilde{g}}{_a_b}\) on \(\nullfut\), and that \(\tensor{n}{^a} = \tensor{\tilde{g}}{^a^b}\tensor{\tilde{\nabla}}{_b}\Omega\) we get that under \(\Omega \to \omega \Omega\) the Carrollian structure at null infinity transforms according to 
        \begin{equation}
            \nullfut \to \nullfut\qc \tensor{\tilde{h}}{_a_b} \to \omega^2 \tensor{\tilde{h}}{_a_b}, \qq{and} \tensor{n}{^a} \to \omega^{-1} \tensor{n}{^a}.
        \end{equation}

        There is no \emph{a priori} preference on whether we should choose \(\Omega\) or \(\omega \Omega\) when performing the conformal compactification. After all, this procedure is unphysical. Therefore, we must consider the two scenarios as being physically equivalent. We thus get an equivalence relation between Carrollian structures
        \begin{equation}\label{eq: equivalence-relation-carrollian-structures}
            (\nullfut, \tensor{\tilde{h}}{_a_b}, \tensor{n}{^a}) \sim (\nullfut, \omega^2 \tensor{\tilde{h}}{_a_b}, \omega^{-1} \tensor{n}{^a}).
        \end{equation}
        The transformations that take \((\nullfut, \tensor{\tilde{h}}{_a_b}, \tensor{n}{^a})\) to an equivalent structure \((\nullfut, \omega^2 \tensor{\tilde{h}}{_a_b}, \omega^{-1} \tensor{n}{^a})\) should be regarded as symmetries of future null infinity. The transformations that preserve \(\nullfut\) are diffeomorphisms, the transformations that preserve \((\nullfut, \tensor{\tilde{h}}{_a_b})\) up to gauge transformation are confomorphisms, and the transformations that preserve \((\nullfut, \tensor{\tilde{h}}{_a_b}, \tensor{n}{^a})\) up to the equivalence relation (\ref{eq: equivalence-relation-carrollian-structures}) are a subclass of confomorphisms that preserve the so-called strong conformal geometry (see, \eg, Refs.\citeonline{penrose1974RelativisticSymmetryGroups,penrose1986SpinorTwistorMethods}).

        We shall begin our discussion by taking a coordinate approach. Using the gauge freedom in the definition of future null infinity one can always choose a coordinate system near \(\nullfut\) in which (see, \eg, Chap. 11 of Ref. \citeonline{wald1984GeneralRelativity})
        \begin{equation}
            \dd{\tilde{s}}^2 = 2 \dd{\Omega} \dd{u} + \dd{\Sph^2} + \order{\Omega},
        \end{equation}
        where \(\Omega\) is the conformal factor used in the definition of null infinity. The induced metric \(\tensor{\tilde{h}}{_a_b}\) then has the line element
        \begin{equation}
            \dd{\tilde{\sigma}}^2 = \dd{\Sph}^2
        \end{equation}
        while the normal vector \(\tensor{n}{^a}\) is given by
        \begin{equation}
            \tensor{n}{^a} = \tensor{\qty(\pdv{u})}{^a}.
        \end{equation}
        Notice this choice of coordinates does not restrict the generality of our analysis. It merely provides a ``standard'' triple
        \begin{equation}
            (\nullfut, \tensor{\tilde{h}}{_a_b}, \tensor{n}{^a}) = \qty(\R \times \Sph^2, \dd{\Sph}^2, \tensor{\qty(\pdv{u})}{^a}).
        \end{equation}
        This can be thought of as a canonical choice of null infinity which we use to compare with other possible choices. Notice that these other possible choices are just as physical as this canonical choice.

        We will find the confomorphisms that preserve the Carrollian structure by finding the transformations that preserve \(\nullfut\), then restrict them to the transformations that preserve \((\nullfut, \tensor{\tilde{h}}{_a_b})\) up to gauge transformation, and finally restricting these to the transformations that preserve \((\nullfut, \tensor{\tilde{h}}{_a_b}, \tensor{n}{^a})\) up to equivalence. 

        The transformations that preserve \(\nullfut\) are diffeomorphisms. This is still extremely general. We can restrict this large group further by imposing that \((\nullfut, \tensor{\tilde{h}}{_a_b})\) is preserved up to gauge transformation. Since the gauge transformations map \((\nullfut, \tensor{\tilde{h}}{_a_b})\) to \((\nullfut, \omega^2 \tensor{\tilde{h}}{_a_b})\) we see that this is the group of confomorphisms of \((\nullfut, \tensor{\tilde{h}}{_a_b})\). In the canonical triplet, we have
        \begin{equation}
            (\nullfut, \tensor{\tilde{h}}{_a_b}) = (\nullfut, \dd{\Sph}^2).
        \end{equation}
        Hence, we want transformations that are confomorphisms of the sphere (since we have the metric of a sphere) and change \(u\) in an arbitrary way. We know from \cref{subsec: conformal-isometries-sphere} that the confomorphisms on the sphere are the Möbius transformations, which in stereographic coordinates can be written as
        \begin{equation}
            \zeta \to \frac{a \zeta + b}{c \zeta + d}
        \end{equation}
        with 
        \begin{equation}
            \mqty(a & b \\ c & d) \in \SL(2, \Comp)/\Z_2 \cong \SO*(3,1). 
        \end{equation}
        Furthermore, we have to assign a general transformation law for \(u\). Since \(u\) does not occur in the metric, we can preserve the conformal structure of the metric by perfoming any transformation of the form
        \begin{equation}
            u \to F(u,\zeta,\bar{\zeta})
        \end{equation}
        with \(u \mapsto F(u,\zeta,\bar{\zeta})\) being a diffeomorphism. 

        We thus get the general coordinate transformations
        \begin{equation}
            \left\lbrace
            \begin{aligned}
                \zeta &\to \frac{a \zeta + b}{c \zeta + d}, \\
                u &\to F(u,\zeta,\bar{\zeta}).
            \end{aligned}
            \right.
        \end{equation}
        This collection of transformations form a group known as the Newman--Unti group \cite{newman1962BehaviorAsymptoticallyFlat}. As a set, the Newman--Unti group is given by a subset of \(\SO*(3,1) \times \ck[\infty](\R \times \Sph^2)\), but it has a complicated Lie group structure \cite{prinz2022LieTheoryAsymptoticNU}.

        For completeness, I mention that under the Möbius transformation
        \begin{equation}
            \zeta \to \zeta' = \frac{a \zeta + b}{c \zeta + d}
        \end{equation}
        the metric of the sphere transforms according to (recall \cref{eq: sphere-metric-stereographic})
        \begin{equation}
            2 \gamma_{\zeta' \bar{\zeta}'} \dd{\zeta} \dd{\bar{\zeta}'} = 2 K(\zeta,\bar{\zeta})^2 \gamma_{\zeta \bar{\zeta}} \dd{\zeta} \dd{\bar{\zeta}}
        \end{equation}
        with
        \begin{equation}\label{eq: K-zeta-zetabar}
            K(\zeta,\bar{\zeta}) = \frac{1 + \zeta \bar{\zeta}}{(a \zeta + b)(\bar{a} \bar{\zeta} + \bar{b}) + (c \zeta + d)(\bar{c} \bar{\zeta} + \bar{d})}.
        \end{equation}

        Now we impose that the Carrollian structure is preserved as a whole. This means we also must have that \(\tensor{n}{^a} \to \omega^{-1} \tensor{n}{^a}\). We know already that \(\omega = K(\zeta,\bar{\zeta})\) because we have already restricted the set of possible transformations to the Newman--Unti group. To further restrict the transformations, we notice that the \(u\) parameter is defined by the equation
        \begin{equation}
            \tensor{n}{^a}\tensor{\tilde{\nabla}}{_a}u = 1,
        \end{equation}
        and this equation is gauge-independent. Hence, since the normal vector is known to transform as \(\tensor{n}{^a} \to \omega^{-1} \tensor{n}{^a}\), it follows that we must have 
        \begin{equation}
            \tensor{\tilde{\nabla}}{_a}u \to \omega \tensor{\tilde{\nabla}}{_a}u.
        \end{equation}
        This is better expressed in the notation of differential forms, in which case we find that the transformation for \(u\) must be such that
        \begin{equation}\label{eq: transformation-u-uprime}
            \dd{u} \to \dd{u'} = K(\zeta,\bar{\zeta}) \dd{u},
        \end{equation}
        where we used \(\omega = K(\zeta,\bar{\zeta})\). Integrating \cref{eq: transformation-u-uprime} allows us to conclude that the allowed transformations for \(u\) are those with the form
        \begin{equation}
            u \to u' = K(\zeta,\bar{\zeta})(u + f(\zeta,\bar{\zeta})),
        \end{equation}
        where \(f \in \ck[\infty](\Sph^2)\) is an arbitrary function that occurs as an integration constant. We assume it to be smooth because we are working in a smooth manifold.

        The transformations that preserve the (weak) Carrollian structure are thus given by
        \begin{equation}\label{eq: bms-transformation}
            \left\lbrace
            \begin{aligned}
                \zeta &\to \frac{a \zeta + b}{c \zeta + d}, \\
                u &\to K(\zeta,\bar{\zeta})(u+f(\zeta,\bar{\zeta})),
            \end{aligned}
            \right.
        \end{equation}
        where \(K\) is given by \cref{eq: K-zeta-zetabar}. These are known as \glsxtrfull{BMS} transformations and they form the (restricted) \gls{BMS} group in four dimensions \cite{bondi1962GravitationalWavesGeneral,sachs1962GravitationalWavesGeneral,sachs1962AsymptoticSymmetriesGravitational}, which I will denote as \(\BMS\). The adjective ``restricted'' refers to the fact that we are ignoring time and space reflections.

        Notice that a \gls{BMS} transformation is characterized by a pair \((\Lambda,f) \in \SO*(3,1) \times \ck[\infty](\Sph^2)\), where \(\Lambda\) relates to the coefficients \(a\), \(b\), \(c\), and \(d\) in \cref{eq: bms-transformation} by
        \begin{equation}
            \Pi^{-1}(\Lambda) = \mqty(a & b \\ c & d),
        \end{equation}
        with \(\Pi \colon \SL(2,\Comp)/\Z_2 \to \SO*(3,1)\) being the isomorphism between the two groups. We denote the function \(K\) associated to \(\Lambda\) by \(K_{\Lambda}\). Using this notation, the group product \(\odot\) is given by
        \begin{equation}\label{eq: bms-product}
            (\Lambda',f') \odot (\Lambda,f) = (\Lambda' \Lambda, f + (K_{\Lambda^{-1}} \circ \Lambda) \cdot (f' \circ \Lambda)),
        \end{equation}
        where \(\circ\) denotes composition of mappings and \(\cdot\) denotes the pointwise product of functions. 

        Notice that \(\SO*(3,1)\) is isomorphic to the subgroup 
        \begin{equation}
            \mathcal{L} = \qty{(\Lambda,0); \Lambda \in \SO*(3,1)}
        \end{equation}
        of the \gls{BMS} group. Similarly, \(\ck[\infty](\Sph^2)\) (with the pointwise sum of functions as the group product) is isomorphic to 
        \begin{equation}
            \mathcal{T} = \qty{(\Eins,f); f \in \ck[\infty](\Sph^2)}.
        \end{equation}
        \cref{eq: bms-product} allows us to conclude that \(\mathcal{T}\) is a normal subgroup of the \gls{BMS} group, but \(\mathcal{L}\) is not.
        
        Notice that these two subgroups satisfy 
        \begin{equation}
            \mathcal{L} \cap \mathcal{T} = \qty{(\Eins,0)},
        \end{equation}
        which means their intersection is given by the \gls{BMS} group's identity. Furthermore, using \cref{eq: bms-product} we can see that any element \((\Lambda,f)\) of the \gls{BMS} group can be written in the form
        \begin{equation}
            (\Lambda,f) = (\Lambda,0) \odot (\Eins, f).
        \end{equation}
        Hence, we can write the \gls{BMS} group as \(\mathcal{L}\odot\mathcal{T}\). These results allow us to conclude that the \gls{BMS} group is the semidirect product of \(\mathcal{L}\) and \(\mathcal{T}\). Hence, the \gls{BMS} group is given by \(\mathcal{L} \ltimes \mathcal{T}\). Using the isomorphisms \(\mathcal{L} \cong \SO*(3,1)\) and \(\mathcal{T} \cong \ck[\infty](\Sph^2)\), we can write that the \gls{BMS} group is given by the semidirect product \(\SO*(3,1) \ltimes \ck[\infty](\Sph^2)\).

    \subsection{Alternative Derivation with Conformal Killing Vector Fields}\label{subsec: BMS-derivation-Killing}
        Let us now derive the \gls{BMS} group once again, but this time employing vector field methods. In other words, we want to find which vector fields generate the \gls{BMS} transformations. This will make it easier for us to compare the results with those of the Poincaré group. 

        We know we want to consider confomorphisms. Furthermore, these confomorphisms should also preserve (in a conformal sense) the normal vector \(\tensor{n}{^a}\). We can state these conditions mathematically by stating that we are looking for diffeomorphisms generated by vector fields \(\tensor{\xi}{^a}\) such that
        \begin{equation}\label{eq: bms-lie-derivative-conditions}
            \Lie[\xi]\tensor{\tilde{h}}{_a_b} = \lambda \tensor{\tilde{h}}{_a_b} \qq{and} \Lie[\xi]\tensor{n}{^a} = - \frac{\lambda}{2} \tensor{n}{^a},
        \end{equation}
        where the second condition follows from the fact that \(\tensor{n}{^a}\) transforms as \(\tensor{n}{^a} \to \omega^{-1} \tensor{n}{^a}\) while \(\tensor{\tilde{h}}{_a_b}\) transforms as \(\tensor{\tilde{h}}{_a_b} \to \omega^2 \tensor{\tilde{h}}{_a_b}\). This can be derived by rewriting the transformation in terms of pullbacks and pushforwards and modifying \cref{eq: CKV-lie-derivative-from-pullback} appropriately.

        \cref{eq: bms-lie-derivative-conditions} establishes a system of differential equations for the vector field \(\tensor{\xi}{^a}\). \cref{eq: general-expression-lie-derivative} allows us to express this system of differential equations in terms of \emph{any} covariant derivative on \(\nullfut\). We choose to do so with the covariant derivative \(\tensor{D}{_a}\), which is defined as having the same Christoffel symbols as the Levi-Civita connection on the round sphere, with the remaining Christoffel symbols vanishing. This choice of covariant derivative turns \((\nullfut,\tensor{\tilde{h}}{_a_b},\tensor{n}{^a},\tensor{D}{_a})\) into a strong Carrollian structure \cite{duval2014ConformalCarrollGroups}. 

        Using the condition imposed on the metric, we find that
        \begin{subequations}
            \begin{align}
                \lambda \tensor{\tilde{h}}{_a_b} &= \Lie[\xi] \tensor{\tilde{h}}{_a_b}, \\
                &= \tensor{\xi}{^c}\tensor{D}{_c}\tensor{\tilde{h}}{_a_b} + \tensor{\tilde{h}}{_c_b}\tensor{D}{_a}\tensor{\xi}{^c} + \tensor{\tilde{h}}{_a_c}\tensor{D}{_b}\tensor{\xi}{^c}, \\
                &= \tensor{D}{_a}(\tensor{\tilde{h}}{_c_b}\tensor{\xi}{^c}) + \tensor{D}{_b}(\tensor{\tilde{h}}{_a_c}\tensor{\xi}{^c}).
            \end{align}
        \end{subequations}
        The equation
        \begin{equation}\label{eq: newman-unti-projected-ckv-equation}
            \tensor{D}{_a}(\tensor{\tilde{h}}{_c_b}\tensor{\xi}{^c}) + \tensor{D}{_b}(\tensor{\tilde{h}}{_a_c}\tensor{\xi}{^c}) = \lambda \tensor{\tilde{h}}{_a_b}
        \end{equation}
        is the conformal Killing equation for a one-form \(\tensor{\tilde{h}}{_a_b}\tensor{\xi}{^b}\) defined on the sphere. Therefore, we can conclude that the projection of \(\tensor{\xi}{^a}\) on the sphere is a conformal Killing vector field on the sphere. If we denote this conformal Killing vector field as \(\tensor{Y}{^a}\) we can thus write
        \begin{equation}\label{eq: newman-unti-vector-field}
            \tensor{\xi}{^a} = \tensor{Y}{^a} + F \tensor{n}{^a},
        \end{equation}
        for some smooth function \(F\) which we assume is such that \(\pdv*{F}{u} > 0\). This family of vector fields generate the Newman--Unti transformations. 
    
        At this point, it is useful to notice that \cref{eq: newman-unti-projected-ckv-equation,eq: newman-unti-vector-field} imply
        \begin{equation}
            \tensor{D}{_a}\tensor{Y}{_b} + \tensor{D}{_b}\tensor{Y}{_a} = \lambda \tensor{\tilde{h}}{_a_b}.
        \end{equation}
        This can be understood as an equation on the two-sphere, in which we have an inverse metric available. Contracting the expression with the inverse metric allows us to conclude that
        \begin{equation}\label{eq: newman-unti-lambda-value}
            \tensor{D}{_a}\tensor{Y}{^a} = \lambda.
        \end{equation}
        This last equation can be understood as both on the sphere or on \(\nullfut\), as it is the same statement in both cases.

        As in our previous derivation, we get from the Newman--Unti group to the \gls{BMS} group by considering the behavior of the normal vector \(\tensor{n}{^a}\). We have 
        \begin{subequations}
            \begin{align}
                - \frac{\lambda}{2} \tensor{n}{^a} &= \Lie[\xi] \tensor{n}{^a}, \\
                &= \tensor{\xi}{^b}\tensor{D}{_b}\tensor{n}{^a} - \tensor{n}{^b}\tensor{D}{_b}\tensor{\xi}{^a}, \\
                &= - \tensor{n}{^b}\tensor{D}{_b}\tensor{\xi}{^a},
            \end{align}
        \end{subequations}
        where we used the fact that \(\tensor{D}{_a}\tensor{n}{^b} = 0\). In terms of \(\tensor{D}{_a}\), the differential equation implied by the behavior of \(\tensor{n}{^a}\) is 
        \begin{equation}\label{eq: bms-n-ckv-equation}
            \tensor{n}{^b}\tensor{D}{_b}\tensor{\xi}{^a} = \frac{\lambda}{2} \tensor{n}{^a},
        \end{equation}
        where \(\lambda\) is given by \cref{eq: newman-unti-lambda-value}. We can use \cref{eq: newman-unti-vector-field} to find that
        \begin{subequations}\label{eq: nDxi-BMS}
            \begin{align}
                \tensor{n}{^b}\tensor{D}{_b}\tensor{\xi}{^a} &= \tensor{n}{^b}\tensor{D}{_b}\tensor{Y}{^a} + \tensor{n}{^b}\tensor{D}{_b}(F\tensor{n}{^a}), \\
                &= (\tensor{n}{^b}\tensor{D}{_b}F)\tensor{n}{^a}.
            \end{align}
        \end{subequations}
        This last expression follows from \(\tensor{D}{_a}\tensor{n}{^b} = 0\) and \(\tensor{n}{^b}\tensor{D}{_b}\tensor{Y}{^a} = 0\)---the latter can be obtained by expressing the equation in terms of Christoffel symbols. 

        \cref{eq: bms-n-ckv-equation,eq: nDxi-BMS} can be combined to yield
        \begin{equation}
            (\tensor{n}{^b}\tensor{D}{_b}F)\tensor{n}{^a} = \frac{\lambda}{2} \tensor{n}{^a}.
        \end{equation}
        This implies 
        \begin{equation}
            \tensor{n}{^a}\tensor{D}{_a}F = \frac{\lambda}{2},
        \end{equation}
        which together with \cref{eq: newman-unti-lambda-value} yields
        \begin{equation}
            \tensor{n}{^a}\tensor{D}{_a}F = \frac{\tensor{D}{_a}\tensor{Y}{^a}}{2}.
        \end{equation}
        This is a differential equation that must be respected by the function \(F\) if we want \(\tensor{n}{^a}\) to be preserved up to a gauge transformation. 

        To solve this differential equation, introduce the coordinate \(u\) through \(\tensor{n}{^a}\tensor{D}{_a}u = 1\). Since \(\tensor{D}{_a}\tensor{Y}{^a}\) can be understood as defined on the sphere, it bears no dependence on the parameter \(u\). We can then integrate the differential equation to get 
        \begin{equation}
            F(u,\zeta,\bar{\zeta}) = \frac{\tensor{D}{_a}\tensor{Y}{^a}}{2} u + f(\zeta,\bar{\zeta}),
        \end{equation}
        where \(f \in \ck[\infty](\Sph^2)\) arises as an integration ``constant''. 

        Bringing everything together we find that the generic vector field generating a \gls{BMS} transformation is given by 
        \begin{equation}\label{eq: bms-vector-field}
            \tensor{\xi}{^a} = \tensor{Y}{^a} + \qty(\frac{\tensor{D}{_b}\tensor{Y}{^b}}{2} u + f(\zeta,\bar{\zeta}))\tensor{n}{^a},
        \end{equation}
        with \(\tensor{Y}{^a}\) being some conformal Killing vector field on the two-sphere, \(u\) being defined by \(\tensor{n}{^b}\tensor{D}{_b}u = 1\), and \(f \in \mathcal{C}^{\infty}(\mathbb{S}^2)\). 

    \subsection{Poincaré Group as a \texorpdfstring{\glsfmtshort{BMS}}{BMS} Subgroup}
        Now that we know what the \gls{BMS} group is, it is interesting to understand how the Poincaré group fits within it. 

        We begin by noticing that the \gls{BMS} group is given by \(\BMS = \SO*(3,1) \ltimes \ck[\infty](\Sph^2)\), while the Poincaré group is \(\ISO*(3,1) = \SO*(3,1) \ltimes \R^4\). The semidirect product structure is certainly similar, but we still need to understand how and whether the Poincaré group is a subgroup of the \gls{BMS} group. 

        The physical reason we expect \(\ISO*(3,1)\) to fit inside \(\BMS\) is because \(\BMS\) is the symmetry group at future null infinity, while \(\ISO*(3,1)\) is the symmetry group in the bulk of Minkowski spacetime. It seems reasonable that bulk symmetries should extend to boundary symmetries, so it is reasonable to expect that there is a relation between these two groups. In fact, before the analysis by \textcite{bondi1962GravitationalWavesGeneral,sachs1962GravitationalWavesGeneral,sachs1962AsymptoticSymmetriesGravitational}, it was thought that the symmetry group at infinity should simply be the Poincaré group. Hence, a byproduct of our discussion should be to understand what are the extra symmetries present in the \gls{BMS} group. 

        Our analysis with conformal Killing vector fields is very useful at this point. This is due to the fact that we have clear expressions for the Killing vector fields of the Poincaré group at infinity, which are given on \cref{eq: poincare-generators-infinity}. We can then compare them with \cref{eq: bms-vector-field} to understand how the transformations relate to each other.
        
        Firstly, we notice that Lorentz transformations correspond precisely to the \(\SO*(3,1)\) contributions to the \gls{BMS} group. This is not surprising, since \(\SO*(3,1)\) is indeed the Lorentz group and we already knew the Lorentz group acts on the celestial sphere through conformal transformations (\cref{subsec: night-sky}).
        
        We also see the translations are inside the \gls{BMS} group by means of the \(\ck[\infty](\Sph^2)\) factor. In fact, the \(\ck[\infty](\Sph^2)\) factor generalizes the translations by allowing transformations of the form
        \begin{equation}
            \tensor{\xi}{^a} = f(\zeta,\bar{\zeta})\tensor{\qty(\pdv{u})}{^a},
        \end{equation}
        where \(f \in \ck[\infty](\Sph^2)\), meaning \(f\) is now a linear combination of spherical harmonics with any value for \(l\), as opposed to the values \(l \leq 1\) of \cref{eq: poincare-generators-infinity}. This thus leads us to the concept of generalized translations, or supertranslations\footnote{The term ``supertranslation'' is not related to supersymmetry. In fact, notice that the \gls{BMS} analysis was published in 1962 \cite{bondi1962GravitationalWavesGeneral,sachs1962GravitationalWavesGeneral}, while supersymmetry arose only in the 1970s \cite{weinberg2000Supersymmetry}.}. 

        From this discussion, one may be prompted to conclude that the Poincaré group is a subgroup of the \gls{BMS} group. This is correct, but there is an important caveat: there is not a preferred choice of Poincaré subgroup. More specifically, the Poincaré group is not a normal subgroup.

        To be more clear, let us work with the Lorentz group. This is motivated by the fact that there is actually a preferred choice of translations, but there is no preferred choice of Lorentz subgroup of the \gls{BMS} group. The ``natural'' choice of Lorentz subgroup of \(\BMS\) is given by
        \begin{equation}
            \mathcal{L} = \qty{(\Lambda,0) \in \BMS; \Lambda \in \SO*(3,1)}.
        \end{equation}
        It is correct to say that \(\mathcal{L} \cong \SO*(3,1)\). Hence, the Lorentz group is a subgroup of the \gls{BMS} group. Nevertheless, let \(g \in \ck[\infty](\Sph^2)\). Noticing first that \((\Eins,g)^{-1} = (\Eins,-g)\), I point out that
        \begin{equation}
            (\Eins,-g) \odot (\Lambda,0) \odot (\Eins,g) = (\Lambda, -g + (K_{\Lambda^{-1}} \circ \Lambda) \cdot (g \circ \Lambda)),
        \end{equation}
        which in general is not an element of \(\mathcal{L}\). Hence, \(\mathcal{L}\) is not a normal subgroup. This means that while \(\mathcal{L}\) is a copy of the Lorentz group inside the \gls{BMS} group, so is \(g^{-1} \mathcal{L} g\).

        In some cases, this is not a deep statement. For example, consider Minkowski spacetime. We have a bulk Lorentz group, which induces symmetries at infinity. The induced group can then be chosen to be the correct Lorentz group. However, in the general case of an asymptotically flat spacetime we might not have bulk symmetries to choose which is the correct Lorentz group. Hence, both \(\mathcal{L}\) and \(g^{-1} \mathcal{L} g\) should be considered as possible Lorentz groups. It is impossible to prefer one over the other. This leads, for instance, to difficulties in defining angular momentum at null infinity \cite{winicour1980AngularMomentumGeneral}.

        This does not occur for translations. Sachs \cite{sachs1962AsymptoticSymmetriesGravitational} has shown that the translations are the unique four-dimensional normal subgroup of the \gls{BMS} group. Therefore, although there is not a preferred definition of Lorentz transformations within the \gls{BMS} group, there is a preferred definition of translations. 

    \subsection{Possible Criticisms of the Derivation of the BMS Group}\label{subsec: possible-criticisms-derivation-BMS}
        Since our derivation of the \gls{BMS} group has led us to an infinite-dimensional group, one could challenge some assumptions we made during the derivation. The purpose of this section is to consider two possible critiques of the derivation of the \gls{BMS} group and argue that the results are, in fact, correct. Hence, I will point out possible flaws within the arguments and show that these ``problems'' are in fact necessary to obtain at least the Poincaré group at infinity. 

        An appropriate definition of asymptotic symmetries at null infinity should reproduce \emph{at least} the Poincaré group. In other words, whatever is the symmetry group at infinity, it should have the Poincaré group as a subgroup. This is due to the fact that at null infinity we have ``effectively'' Minkowski spacetime, and the Poincaré group is the symmetry group of Minkowski spacetime. All Poincaré transformations should have an asymptotic version at null infinity, and thus should belong to the group of asymptotic symmetries. Hence, an asymptotic symmetry group that is smaller than the Poincaré group should be considered inadmissible on physical grounds. 

        The first possible criticism one could make to the derivations above is that we considered that \((\nullfut, \tensor{\tilde{h}}{_a_b}, \tensor{n}{^a})\) was preserved only conformally. Hence, instead of using the equivalence relation
        \begin{equation}
            (\nullfut, \tensor{\tilde{h}}{_a_b}, \tensor{n}{^a}) \sim (\nullfut, \omega^2 \tensor{\tilde{h}}{_a_b}, \omega^{-1} \tensor{n}{^a}),
        \end{equation}
        maybe we should have used 
        \begin{equation}
            (\nullfut, \tensor{\tilde{h}}{_a_b}, \tensor{n}{^a}) \sim (\nullfut, \tensor{\tilde{h}}{_a_b}, \tensor{n}{^a}),
        \end{equation}
        which means we should consider only isometries, not confomorphisms. In other words, we should always work with \(\omega = 1\). This is what is done in the bulk of spacetime, and it is how we discussed the Poincaré symmetries in the first place.

        If one goes back to our derivation and considers this new imposition, one will find this does not rule out the supertranslations. Rather, it only affects Lorentz transformations. Boosts, to be more precise. The only transformations with \(\omega \neq 1\) are the Lorentz boosts, which become conformal transformations on the sphere at infinity. Hence, excluding conformal transformations seems to be too strong of a requirement, because it leads to an asymptotic symmetry group that does not have the Poincaré group as a subgroup. We conclude that we must consider the conformal transformations, precisely as we did. 

        The second criticism one could make is that we imposed that the weak Carrollian structure should be preserved, but we could have imposed conservation of the strong Carrollian structure. As shown in Appendix \ref{app: preserving-scs}, it turns out that preserving the strong Carrollian structure rules out supertranslations, but it also rules out spatial translations. Hence, it is too strong of a requirement, and we cannot ask for the strong Carrollian structure to be preserved by the asymptotic symmetries without losing the Poincaré group in the process.

        We conclude, therefore, that supertranslations are a necessity if we want to enjoy the Poincaré group at infinity. They are a feature, not a bug. 

    \subsection{Supertranslations in the Bondi--Sachs Formalism}\label{subsec: BMS-derivation-Bondi-Sachs}
        It is a good idea to rederive the supertranslations in the Bondi--Sachs formalism for completeness and see that the results do not depend on the detailed hypotheses behind a conformal compactification. We will follow the original analysis done by \textcite{sachs1962AsymptoticSymmetriesGravitational}.

        We start with \cref{eq: bondi-gauge-AF}, which is our definition of asymptotically flat spacetime in Bondi gauge. I rewrite it here for convenience.
        \begin{multline}\label{eq: bondi-gauge-CK-supertranslations}
            \dd{s}^2 = - \dd{u}^2 - 2 \dd{u}\dd{r} + r^2 \tensor{\gamma}{_{A}_{B}}\dd{\tensor{x}{^A}}\dd{\tensor{x}{^B}} \\ + \frac{2 m}{r}\dd{u}^2 + r \tensor{C}{_{A}_{B}}\dd{\tensor{x}{^A}}\dd{\tensor{x}{^B}} + \tensor{D}{^{B}}\tensor{C}{_{A}_{B}}\dd{u}\dd{\tensor{x}{^A}} \\ + \frac{1}{r}\qty(\frac{4}{3}\qty(\tensor{N}{_A} + u \tensor{\partial}{_A}m) - \frac{1}{8}\tensor{\partial}{_A}\qty(\tensor{C}{^{B}^{C}}\tensor{C}{_{B}_{C}}))\dd{u}\dd{\tensor{x}{^A}} \\ + \frac{1}{16r^2}\tensor{C}{_A_B}\tensor{C}{^A^B}\dd{u}\dd{r} + \frac{1}{4}\tensor{\gamma}{_A_B}\tensor{C}{_C_D}\tensor{C}{^C^D}\dd{\tensor{x}{^A}}\dd{\tensor{x}{^B}} + \cdots.
        \end{multline}
        Notice that these components have very specific decay properties. Namely, we are assuming they have the form
        \begin{subequations}\label{eq: falloff-bondi-sachs}
            \begin{align}
                \tensor{g}{_u_u} &= -1 + \order{\frac{1}{r}}, \\
                \tensor{g}{_u_r} &= -1 + \order{\frac{1}{r^2}}, \\
                \tensor{g}{_u_A} &= \order{1}, \\
                \tensor{g}{_r_r} &= 0, \\
                \tensor{g}{_r_A} &= 0, \\
                \tensor{g}{_A_B} &= r^2\tensor{\gamma}{_A_B} + \order{r}.
            \end{align}
        \end{subequations}
        In this approach, the \gls{BMS} transformations are defined by demanding that they preserve the Bondi gauge and these falloff conditions. Hence, we will derive them by considering vector fields \(\tensor{\xi}{^a}\) which preserve the asymptotic falloff conditions under the infinitesimal transformation
        \begin{equation}
            \tensor{g}{_\mu_\nu} \to \tensor{g}{_\mu_\nu} + \Lie[\xi]\tensor{g}{_\mu_\nu}.
        \end{equation}

        For a vector field to be well-defined at infinity, it needs to be such that
        \begin{equation}
            \tensor{\xi}{^\mu} = \order{1}\qc \tensor{\xi}{^r} = \order{r}, \qq{and} \tensor{\xi}{^A} = \order{1}.
        \end{equation}
        Hence, we will impose these restrictions from the start. It is also possible to take an alternative route and derive these conditions from the preservation of the Bondi gauge and falloff assumptions, but I will take the easier route to keep calculations less messy. We therefore start with the general Ansatz\footnote{A more general Ansatz than (\ref{eq: first-ansatz-bondi-sachs-supertranslation}) would write \(o(r^{-n})\) (with little o notation) instead of \(\order{r^{-n-1}}\) (with big O notation), since this gives more freedom on the asymptotic behavior of the vector field. I do not think there is much need for this level of precision here, so I will work with big O.}
        \begin{subequations}\label{eq: first-ansatz-bondi-sachs-supertranslation}
            \begin{align}
                \tensor{\xi}{^u} &= c(u,\zeta,\bar{\zeta}) + \frac{d(u,\zeta,\bar{\zeta})}{r} + \order{\frac{1}{r^2}}, \\
                \tensor{\xi}{^r} &= rg(u,\zeta,\bar{\zeta}) + h(u,\zeta,\bar{\zeta}) + \order{\frac{1}{r}}, \\
                \tensor{\xi}{^A} &= \tensor{Y}{^A}(u,\zeta,\bar{\zeta}) + \frac{\tensor{X}{^A}(u,\zeta,\bar{\zeta})}{r} + \order{\frac{1}{r^2}},
            \end{align}
        \end{subequations}
        where the leading and subleading terms are kept. Our analysis will not give information on the subsubleading terms, so we will not pay attention to them.
        
        Let us now compute the necessary Lie derivatives. We can do it component by component. We will work with the covariant derivative on the sphere, and hence we have the general expression
        \begin{equation}
            \Lie[\xi]\tensor{g}{_a_b} = \tensor{\xi}{^c}\tensor{D}{_c}\tensor{g}{_a_b} + \tensor{g}{_c_b}\tensor{D}{_a}\tensor{\xi}{^c} + \tensor{g}{_a_c}\tensor{D}{_b}\tensor{\xi}{^c}.
        \end{equation}
        This choice is convenient because it allows us to perform the calculations without worrying about the complicated spacetime Christoffel symbols, while still simplifying expressions on the sphere. In practice, this means we are choosing to use a covariant derivative with the connection coefficients of the sphere, while all other connection coefficients vanish. In particular, this means \(\tensor{D}{_u} = \tensor{\partial}{_u}\) and \(\tensor{D}{_r} = \tensor{\partial}{_r}\).

        It is convenient to begin by imposing the Bondi gauge is preserved. This implies, first and foremost, that
        \begin{subequations}
            \begin{align}
                0 = \Lie[\xi]\tensor{g}{_r_r} &= \tensor{\xi}{^c}\tensor{D}{_c}\tensor{g}{_r_r} + \tensor{g}{_c_r}\tensor{D}{_r}\tensor{\xi}{^c} + \tensor{g}{_r_c}\tensor{D}{_r}\tensor{\xi}{^c}, \\
                &= 0 + \tensor{g}{_u_r}\tensor{D}{_r}\tensor{\xi}{^u} + 0 + \tensor{g}{_r_u}\tensor{D}{_r}\tensor{\xi}{^u} + 0, \\
                &= 2\tensor{g}{_u_r}\tensor{D}{_r}\tensor{\xi}{^u}, \\
                &= 2\frac{d(u,\zeta,\bar{\zeta})}{r^2}.
            \end{align}
        \end{subequations}
        Other contributions to the Lie derivative vanish due to the choice of gauge. To preserve the Bondi gauge, this expression needs to vanish exactly. Hence, we conclude that \(\tensor{\xi}{^u}\) cannot depend on \(r\). In terms of our original Ansatz, we conclude
        \begin{equation}
            d(u,\zeta,\bar{\zeta}) = 0.
        \end{equation}

        Using this result, we then have
        \begin{subequations}
            \begin{align}
                0 = \Lie[\xi]\tensor{g}{_r_A} &= \tensor{\xi}{^c}\tensor{D}{_c}\tensor{g}{_r_A} + \tensor{g}{_c_A}\tensor{D}{_r}\tensor{\xi}{^c} + \tensor{g}{_r_c}\tensor{D}{_A}\tensor{\xi}{^c}, \\
                &= -\tensor{\gamma}{_A_B}\tensor{X}{^B} - \tensor{D}{_A}c + \order{\frac{1}{r}}.
            \end{align}
        \end{subequations}
        The expression must vanish exactly, but we will only need the leading term. From it, we conclude that
        \begin{equation}
            \tensor{X}{^A}(u,\zeta,\bar{\zeta}) = -\tensor{D}{^A}c(u,\zeta,\bar{\zeta}),
        \end{equation}
        where we are raising uppercase Latin indices with the metric on the sphere.

        We must now impose that \(r\) is still the areal distance after the transformation. This means imposing
        \begin{equation}\label{eq: Lie-det-gAB-BS-BMS}
            \Lie[\xi]\qty(\det(\tensor{g}{_A_B})) = 0,
        \end{equation}
        because we still want to have \(\det(\tensor{g}{_A_B}) = r^4\det(\tensor{\gamma}{_A_B})\) after the transformation. There is, however, a difficulty: \(\det(\tensor{g}{_A_B})\) is a scalar density, not a scalar field. This means \(\det(\tensor{g}{_A_B})\) is a coordinate-dependent object, and its transformation law is actually
        \begin{equation}
            \det(\tensor{g}{_{A'}_{B'}}) = \det(\pdv{\tensor{x}{^C}}{\tensor{x}{^{D'}}})^2 \det(\tensor{g}{_A_B}).
        \end{equation}
        As a consequence, the definition of the Lie derivative gets modified. The Lie derivative of the metric determinant is given by\footnote{Some authors prefer to avoid the Lie derivative of a scalar density and manipulate the expressions to get to expressions involving only Lie derivatives of tensorial quantities. For example, this is done in the solution to Exercise 11 of \cite{strominger2018LecturesInfraredStructure}. To my knowledge, this can lead to a few differences down the calculations. See \vref{footnote: difference-lie-tensor-density-strominger}.} \cite{lovelock1989TensorsDifferentialForms,yano1955LieDerivativesIts}
        \begin{equation}
            \Lie[\xi]\qty(\det(\tensor{g}{_A_B})) = \tensor{\xi}{^c}\tensor{D}{_c}\qty(\det(\tensor{g}{_A_B})) + 2 \qty(\det(\tensor{g}{_A_B})) \tensor{D}{_c}\tensor{\xi}{^c}.
        \end{equation}
        Since \(\det(\tensor{g}{_A_B}) = r^4 \det(\tensor{\gamma}{_A_B})\), we find that
        \begin{subequations}\label{eq: Lie-det-gAB-BS-BMS-up-to-order-1}
            \begin{align}
                0 = \Lie[\xi]\qty(\det(\tensor{g}{_A_B})) &= \tensor{\xi}{^c}\tensor{D}{_c}\qty(r^4\det(\tensor{\gamma}{_A_B})) + 2 r^4\qty(\det(\tensor{\gamma}{_A_B})) \tensor{D}{_c}\tensor{\xi}{^c}, \label{eq: BMS-bondi-sachs-lie-derivative-determinant-zero} \\
                &= 2r^3\det(\tensor{\gamma}{_A_B}) \qty[2\tensor{\xi}{^r} + r \tensor{D}{_c}\tensor{\xi}{^c}], \\
                &= 2r^3\det(\tensor{\gamma}{_A_B}) \qty[r(\tensor{D}{_u}c + 3 g + \tensor{D}{_A}\tensor{Y}{^A}) + \qty(2h-\tensor{D}{_A}\tensor{D}{^A}c) + \order{\frac{1}{r}}].
            \end{align}
        \end{subequations}
        This immediately tells us that
        \begin{equation}\label{eq: g-u-zeta-zetabar-BS-BMS}
            g(u,\zeta,\bar{\zeta}) = - \frac{\tensor{D}{_u}c(u,\zeta,\bar{\zeta})}{3} - \frac{\tensor{D}{_A}\tensor{Y}{^A}(u,\zeta,\bar{\zeta})}{3}
        \end{equation}
        and that
        \begin{equation}
            h(u,\zeta,\bar{\zeta}) = \frac{\tensor{D}{_A}\tensor{D}{^A}c(u,\zeta,\bar{\zeta})}{2}.
        \end{equation}

        Therefore, at this point our Ansatz has become
        \begin{subequations}\label{eq: second-ansatz-bondi-sachs-supertranslation}
            \begin{align}
                \tensor{\xi}{^u} &= c(u,\zeta,\bar{\zeta}), \\
                \tensor{\xi}{^r} &= - \frac{r\tensor{D}{_u}c}{3} - \frac{r\tensor{D}{_A}\tensor{Y}{^A}}{3} + \frac{\tensor{D}{_A}\tensor{D}{^A}c}{2} + \order{\frac{1}{r}}, \\
                \tensor{\xi}{^A} &= \tensor{Y}{^A}(u,\zeta,\bar{\zeta}) - \frac{\tensor{D}{^A}c}{r} + \order{\frac{1}{r^2}},
            \end{align}
        \end{subequations}

        With these results at hand, we can start imposing the preservation of the falloff conditions. We have
        \begin{multline}
            \order{\frac{1}{r}} = \Lie[\xi]\tensor{g}{_u_u} = \frac{2r}{3}\qty(\tensor{D}{_u}\tensor{D}{_u}c + \tensor{D}{_u}\tensor{D}{_A}\tensor{Y}{^A}) - \qty(2\tensor{D}{_u}c + \tensor{D}{_A}\tensor{D}{^A}\tensor{D}{_u}c - \tensor{D}{^B}\tensor{C}{_A_B}\tensor{D}{_u}\tensor{Y}{^A}) + \order{\frac{1}{r}},
        \end{multline}
        where \(\tensor{D}{^B}\tensor{C}{_A_B} = 2\tensor{g}{_u_A}\) (this factor will disappear at the end). At \(\order{r}\), we now learn that 
        \begin{equation}\label{eq: BS-BMS-guu-order-r}
            \tensor{D}{_u}\qty(\tensor{D}{_u}c + \tensor{D}{_A}\tensor{Y}{^A}) = 0,
        \end{equation}
        which amounts to saying \(g\) in \cref{eq: g-u-zeta-zetabar-BS-BMS} is \(u\)-independent. We cannot write this in a more convenient way yet. At order \(\order{1}\) we find the differential equation
        \begin{equation}\label{eq: BS-BMS-guu-order-1}
            \tensor{D}{_u}\qty(2 + \tensor{D}{_A}\tensor{D}{^A})c - \tensor{D}{^B}\tensor{C}{_A_B}\tensor{D}{_u}\tensor{Y}{^A} = 0.
        \end{equation}
        We will keep this equation in this form for now (in fact, we will not need it). 

        The next Lie derivative is
        \begin{equation}
            \order{\frac{1}{r^2}} = \Lie[\xi]\tensor{g}{_u_r} = - g - \tensor{D}{_u}c + \order{\frac{1}{r^2}},
        \end{equation}
        and hence
        \begin{equation}
            g(\zeta,\bar{\zeta}) = -\tensor{D}{_u}c(u,\zeta,\bar{\zeta}).
        \end{equation}
        Since the \gls{LHS} does not depend on \(u\), neither does the \gls{RHS}. We are forced to conclude that \(c\) is at most linear in \(u\). We will therefore decompose \(c\) as
        \begin{equation}
            c(u,\zeta,\bar{\zeta}) = f(\zeta,\bar{\zeta}) + u e(\zeta,\bar{\zeta}).
        \end{equation}
        \cref{eq: BS-BMS-guu-order-r} now implies that
        \begin{equation}
            \tensor{D}{_u}\qty(e(\zeta,\bar{\zeta}) + \tensor{D}{_A}\tensor{Y}{^A}) = 0,
        \end{equation}
        which means \(\tensor{D}{_A}\tensor{Y}{^A}\) cannot depend on \(u\). 

        Next we have
        \begin{equation}
            \order{1} = \Lie[\xi]\tensor{g}{_u_A} = r^2\tensor{\gamma}{_A_B}\tensor{D}{_u}\tensor{Y}{^B} + r\tensor{C}{_A_B}\tensor{D}{_u}\tensor{Y}{^B} + \order{1}.
        \end{equation}
        At order \(\order{r^2}\) we learn that \(\tensor{Y}{^B}\) is \(u\)-independent. Using this fact, the next order becomes trivial. 

        As a summary, so far we have the expressions
        \begin{subequations}\label{eq: third-ansatz-bondi-sachs-supertranslation}
            \begin{align}
                \tensor{\xi}{^u} &= f(\zeta,\bar{\zeta}) + u e(\zeta,\bar{\zeta}), \\
                \tensor{\xi}{^r} &= - \frac{re(\zeta,\bar{\zeta})}{3} - \frac{r\tensor{D}{_A}\tensor{Y}{^A}}{3} + \frac{\tensor{D}{_A}\tensor{D}{^A}f}{2} + \frac{u\tensor{D}{_A}\tensor{D}{^A}e}{2} + \order{\frac{1}{r}}, \\
                \tensor{\xi}{^A} &= \tensor{Y}{^A}(\zeta,\bar{\zeta}) - \frac{\tensor{D}{^A}f}{r} - \frac{u\tensor{D}{^A}e}{r} + \order{\frac{1}{r^2}},
            \end{align}
        \end{subequations}
        where are supplemented by \cref{eq: BS-BMS-guu-order-1}.

        The final equation is
        \begin{equation}
            \order{r} = \Lie[\xi]\tensor{g}{_A_B} = -2r^2 \tensor{\gamma}{_A_B} e + r^2 \tensor{\gamma}{_C_B} \tensor{D}{_A} \tensor{Y}{^C} + r^2 \tensor{\gamma}{_A_C}\tensor{D}{_B}\tensor{Y}{^C} + \order{r}.
        \end{equation}
        We thus find that
        \begin{equation}
            \tensor{D}{_A} \tensor{Y}{_B} + \tensor{D}{_B} \tensor{Y}{_A} = 2 e \tensor{\gamma}{_A_B}.
        \end{equation}
        This is the conformal Killing equation on the sphere, which restricts \(\tensor{Y}{^A}\) to be a conformal Killing vector field on the sphere. Contracting both sides with \(\tensor{\gamma}{^A^B}\) yields
        \begin{equation}
            e = \frac{\tensor{D}{_A} \tensor{Y}{^A}}{2},
        \end{equation}
        which fixes \(e\). 

        At last, our original Ansatz has become
        \begin{subequations}\label{eq: fourth-ansatz-bondi-sachs-supertranslation}
            \begin{align}
                \tensor{\xi}{^u} &= f(\zeta,\bar{\zeta}) + \frac{u\tensor{D}{_A} \tensor{Y}{^A}}{2}, \\
                \tensor{\xi}{^r} &= - \frac{r\tensor{D}{_A} \tensor{Y}{^A}}{2} + \frac{\tensor{D}{_A}\tensor{D}{^A}f}{2} + \frac{u\tensor{D}{_A}\tensor{D}{^A}\tensor{D}{_B} \tensor{Y}{^B}}{4} + \order{\frac{1}{r}}, \\
                \tensor{\xi}{^A} &= \tensor{Y}{^A}(\zeta,\bar{\zeta}) - \frac{\tensor{D}{^A}f}{r} - \frac{u\tensor{D}{^A}\tensor{D}{_B} \tensor{Y}{^B}}{2r} + \order{\frac{1}{r^2}},
            \end{align}
        \end{subequations}
        Notice these equations are complemented by the fact that \(\tensor{Y}{^A}\) is a conformal Killing vector field on the sphere.
        
        We can simplify the \(r\)-term. To see this, it is important to notice that, in the round sphere, \(\tensor{R}{_A_B} = \tensor{\gamma}{_A_B}\) (where \(\tensor{R}{_A_B}\) is the Ricci tensor associated to \(\tensor{\gamma}{_A_B}\)). Furthermore, recall that the commutator of covariant derivatives on an arbitrary manifold is \cite{wald1984GeneralRelativity}
        \begin{equation}
            (\tensor{\nabla}{_a}\tensor{\nabla}{_b} - \tensor{\nabla}{_b}\tensor{\nabla}{_b})\tensor{T}{^{c_1}^{\cdots}^{c_k}_{d_1}_{\cdots}_{d_l}} = - \sum_{i=1}^k \tensor{R}{_a_b_e^{c_i}}\tensor{T}{^{c_1}^{\cdots}^{e}^{\cdots}^{c_k}_{d_1}_{\cdots}_{d_l}} + \sum_{j=1}^l \tensor{R}{_a_b_{d_j}^{e}}\tensor{T}{^{c_1}^{\cdots}^{c_k}_{d_1}_{\cdots}_{e}_{\cdots}_{d_l}}.
        \end{equation}
        Using these results and the conformal Killing equation for \(\tensor{Y}{^A}\), we see that
        \begin{subequations}
            \begin{align}
                \tensor{D}{_A}\tensor{D}{^A}\tensor{D}{_B}\tensor{Y}{^B} &= \tensor{D}{_A}\tensor{D}{_B}\tensor{D}{^A}\tensor{Y}{^B} - \tensor{D}{_A}\qty(\tensor{R}{^A_B_C^B}\tensor{Y}{^C}), \\
                &= \tensor{D}{_A}\tensor{D}{_B}\tensor{D}{^A}\tensor{Y}{^B} - \tensor{D}{_A}\qty(\tensor{R}{^A_C}\tensor{Y}{^C}), \\
                &= \tensor{D}{_A}\tensor{D}{_B}\tensor{D}{^A}\tensor{Y}{^B} - \tensor{D}{_A}\tensor{Y}{^A}, \\
                &= \tensor{D}{_B}\tensor{D}{_A}\tensor{D}{^A}\tensor{Y}{^B} - \tensor{R}{_A_B_C^A}\tensor{D}{^C}\tensor{Y}{^B} - \tensor{R}{_A_B_C^B}\tensor{D}{^A}\tensor{Y}{^C} - \tensor{D}{_A}\tensor{Y}{^A}, \\
                &= \tensor{D}{_B}\tensor{D}{_A}\tensor{D}{^A}\tensor{Y}{^B} - \tensor{D}{_A}\tensor{Y}{^A}, \\
                &= -\tensor{D}{_B}\tensor{D}{_A}\tensor{D}{^B}\tensor{Y}{^A} + \tensor{D}{_B}\tensor{D}{_A}\tensor{D}{^C}\tensor{Y}{^C}\tensor{\gamma}{^A^B} - \tensor{D}{_A}\tensor{Y}{^A}, \\
                &= -\tensor{D}{_B}\tensor{D}{^B}\tensor{D}{_A}\tensor{Y}{^A} + \tensor{D}{_B}\qty(\tensor{R}{_A^B_C^A}\tensor{Y}{^A}) + \tensor{D}{_B}\tensor{D}{^B}\tensor{D}{^C}\tensor{Y}{^C} - \tensor{D}{_A}\tensor{Y}{^A}, \\
                &= - \tensor{D}{_B}\qty(\tensor{R}{^B_C}\tensor{Y}{^A}) - \tensor{D}{_A}\tensor{Y}{^A}, \\
                &= - 2\tensor{D}{_A}\tensor{Y}{^A}.
            \end{align}
        \end{subequations}
        Notice this can be used with the previous results to show that \cref{eq: BS-BMS-guu-order-1} is satisfied, even though we did not impose it explicitly.
        
        At last, we conclude that the vector field
        \begin{multline}\label{eq: bondi-sachs-BMS-transformation}
            \tensor*{\xi}{^a} = \qty[f + \frac{u \tensor{D}{_A}\tensor{Y}{^A}}{2}]\tensor{\qty(\pdv{u})}{^a} \\ + \qty[\frac{\tensor{D}{^A}\tensor{D}{_A}f}{2} - (u+r)\frac{\tensor{D}{_A}\tensor{Y}{^A}}{2} + \order{\frac{1}{r}}]\tensor{\qty(\pdv{r})}{^a} \\ +\qty[-\frac{\tensor{D}{^A}f}{r} + \tensor{Y}{^A} - \frac{u \tensor{D}{^A}\tensor{D}{_B}\tensor{Y}{^B}}{2r} + \order{\frac{1}{r^2}}]\tensor{\qty(\pdv{\tensor{x}{^A}})}{^a}
        \end{multline}
        preserves the asymptotic structure of the asymptotically flat metric in \cref{eq: bondi-gauge-CK-supertranslations}. Above, \(f\) is an arbitrary element of \(\ck[\infty](\Sph^2)\) and \(\tensor{Y}{^A}\) is a conformal Killing vector field on the sphere \(\Sph^2\). If we change the radial coordinate according to \(r=1/l\), as we did elsewhere, we find
        \begin{multline}
            \tensor*{\xi}{^a} = \qty[f + \frac{u \tensor{D}{_A}\tensor{Y}{^A}}{2}]\tensor{\qty(\pdv{u})}{^a} \\ + \qty[\frac{l^2 \tensor{D}{^A}\tensor{D}{_A}f}{2} - (u l^2+l)\frac{\tensor{D}{_A}\tensor{Y}{^A}}{2} + \order{l^3}]\tensor{\qty(\pdv{l})}{^a} \\ +\qty[-l\tensor{D}{^A}f + \tensor{Y}{^A} - \frac{u l \tensor{D}{^A}\tensor{D}{_B}\tensor{Y}{^B}}{2} + \order{l^2}]\tensor{\qty(\pdv{\tensor{x}{^A}})}{^a}
        \end{multline}
        in which case the limit \(l\to0^+\) clearly leads to the vector field
        \begin{equation}
            \qty[f + \frac{u \tensor{D}{_A}\tensor{Y}{^A}}{2}]\tensor{\qty(\pdv{u})}{^a} + \tensor{Y}{^A}\tensor{\qty(\pdv{\tensor{x}{^A}})}{^a}
        \end{equation}
        on \(\nullfut\). This is exactly what we would expect of a general \gls{BMS} transformation based on our previous discussions. Notice that, as before, the conformal Killing vector fields \(\tensor{Y}{^A}\) lead to Lorentz transformations, while the function \(f\) leads to (super)translations.

        An important comment is in order: (\ref{eq: bondi-sachs-BMS-transformation}) is only a \gls{BMS} symmetry (for example, a supertranslation) at infinity. \cref{eq: bondi-sachs-BMS-transformation} is extremely useful in many situations, but it is a gauge-dependent extension to the bulk of an object defined at infinity. Indeed, we obtained it by demanding that the Bondi gauge be preserved. Other choices of gauge could lead to other expressions. However, if worked out correctly, any gauge choice will consistently yield the same limit at \(\nullfut\), which is where the symmetry is actually defined. 

        Now that we know what are the vector fields originating the \gls{BMS} group, we may wonder how exactly they modify the tensor fields \(m\), \(\tensor{C}{_A_B}\), \(\tensor{N}{_A_B}\), and \(\tensor{N}{_A}\). This can be calculated by considering the Lie derivatives of each component of the metric up to the desired order. This requires knowledge of some of the terms we omitted in \cref{eq: bondi-sachs-BMS-transformation}. These terms can be obtained by imposing that the Bondi gauge is preserved at subleading orders. This essentially means improving the precision of the conditions we imposed on \(\tensor{g}{_r_A}\) and on the determinant of \(\tensor{g}{_A_B}\).
        
        With this in mind, consider the new Ansatz
        \begin{subequations}
            \begin{align}
                \tensor{\xi}{^u} &= f(\zeta,\bar{\zeta}) + \frac{u\tensor{D}{_A} \tensor{Y}{^A}}{2}, \\
                \tensor{\xi}{^r} &= -(u+r) \frac{\tensor{D}{_A} \tensor{Y}{^A}}{2} + \frac{\tensor{D}{_A}\tensor{D}{^A}f}{2} + \frac{k(u,\zeta,\bar{\zeta})}{r} + \order{\frac{1}{r^2}}, \\
                \tensor{\xi}{^A} &= \tensor{Y}{^A}(\zeta,\bar{\zeta}) - \frac{\tensor{D}{^A}f}{r} - \frac{u\tensor{D}{^A}\tensor{D}{_B} \tensor{Y}{^B}}{2r} + \frac{\tensor{Z}{^A}(u,\zeta,\bar{\zeta})}{r^2} + \order{\frac{1}{r^3}}.
            \end{align}
        \end{subequations}
        We want to determine \(k\) and \(\tensor{Z}{^A}\). For \(\tensor{g}{_r_A}\) we find
        \begin{equation}
            0 = \Lie[\xi]\tensor{g}{_r_A} = - \frac{2 \tensor{Z}{_A}}{r} + \frac{\tensor{C}{_A_B}\tensor{D}{^B}f}{r} + \frac{u\tensor{C}{_A_B}\tensor{D}{^B}\tensor{D}{_C} \tensor{Y}{^C}}{2r} + \order{\frac{1}{r^2}}.
        \end{equation}
        Hence, imposing \(\Lie[\xi]\tensor{g}{_r_A}\) at order \(\order{\frac{1}{r}}\) fixes \(\tensor{Z}{^A}\) to be given by
        \begin{equation}
            \tensor{Z}{^A} = \frac{\tensor{C}{^A^B}\tensor{D}{_B}f}{2} + \frac{u\tensor{C}{^A^B}\tensor{D}{_B}\tensor{D}{_C} \tensor{Y}{^C}}{4}
        \end{equation}

        To fix the determinant of \(\tensor{g}{_A_B}\) we must consider \cref{eq: Lie-det-gAB-BS-BMS} up to \(\order{1}\). Neglecting the terms that were already considered in \cref{eq: Lie-det-gAB-BS-BMS-up-to-order-1}, we have
        \begin{subequations}\label{eq: Lie-det-gAB-BS-BMS-up-to-order-1/r}
            \begin{align}
                0 = \Lie[\xi]\qty(\det(\tensor{g}{_A_B})) &= 2r^3\det(\tensor{\gamma}{_A_B}) \qty[2\tensor{\xi}{^r} + r \tensor{D}{_c}\tensor{\xi}{^c}], \\
                &= 2r^3\det(\tensor{\gamma}{_A_B}) \qty[\frac{1}{r}\qty(2k + \tensor{D}{_A}\tensor{Z}{^A} - k) + \order{\frac{1}{r^2}}],
            \end{align}
        \end{subequations}
        and hence
        \begin{subequations}
            \begin{align}
                k(u,\zeta,\bar{\zeta}) &= -\tensor{D}{_A}\tensor{Z}{^A}, \\
                &= - \frac{\tensor{D}{_A}(\tensor{C}{^A^B}\tensor{D}{_B}f)}{2} - \frac{u(\tensor{D}{_A}\tensor{C}{^A^B}\tensor{D}{_B}\tensor{D}{_C} \tensor{Y}{^C})}{4}, \\
                &= - \frac{\tensor{D}{_A}\tensor{C}{^A^B}\tensor{D}{_B}f}{2} - \frac{\tensor{C}{^A^B}\tensor{D}{_A}\tensor{D}{_B}f}{2} - \frac{u\tensor{D}{_A}\tensor{C}{^A^B}\tensor{D}{_B}\tensor{D}{_C} \tensor{Y}{^C}}{4} - \frac{u\tensor{C}{^A^B}\tensor{D}{_A}\tensor{D}{_B}\tensor{D}{_C}\tensor{Y}{^C}}{4}.
            \end{align}
        \end{subequations}

        Therefore, at this order, we get the expression
        \begin{equation}
            \tensor*{\xi}{^a} = \tensor*{\xi}{_f^a} + \tensor*{\xi}{_Y^a},
        \end{equation}
        where\footnote{\cref{eq: bondi-sachs-BMS-supertranslation} does not match exactly some similar expressions one can find in the literature---see, \eg,  p. 150 of Ref. \citeonline{strominger2018LecturesInfraredStructure}. This is because we chose to impose the condition on \cref{eq: BMS-bondi-sachs-lie-derivative-determinant-zero}, with a tensor density, instead of working with strictly tensorial quantities as done, for example, in Ref. \citeonline{strominger2018LecturesInfraredStructure}.\label{footnote: difference-lie-tensor-density-strominger}}
        \begin{multline}\label{eq: bondi-sachs-BMS-supertranslation}
            \tensor*{\xi}{_f^a} = f\tensor{\qty(\pdv{u})}{^a} + \left[\frac{\tensor{D}{^A}\tensor{D}{_A}f}{2} - \frac{\tensor{D}{_A}\tensor{C}{^A^B}\tensor{D}{_B}f}{2r} - \frac{\tensor{C}{^A^B}\tensor{D}{_A}\tensor{D}{_B}f}{2r} + \order{\frac{1}{r^2}}\right]\tensor{\qty(\pdv{r})}{^a} \\ +\left[-\frac{\tensor{D}{^A}f}{r} + \frac{\tensor{C}{^A^B}\tensor{D}{_B}f}{2r^2} + \order{\frac{1}{r^3}}\right]\tensor{\qty(\pdv{\tensor{x}{^A}})}{^a}
        \end{multline}
        and
        \begin{multline}\label{eq: bondi-sachs-BMS-superrotation}
            \tensor*{\xi}{_Y^a} = \frac{u \tensor{D}{_A}\tensor{Y}{^A}}{2}\tensor{\qty(\pdv{u})}{^a} + \left[- (u+r)\frac{\tensor{D}{_A}\tensor{Y}{^A}}{2}  - \frac{u\tensor{D}{_A}\tensor{C}{^A^B}\tensor{D}{_B}\tensor{D}{_C} \tensor{Y}{^C}}{4r} \right. \\ \left. - \frac{u\tensor{C}{^A^B}\tensor{D}{_A}\tensor{D}{_B}\tensor{D}{_C}\tensor{Y}{^C}}{4r} + \order{\frac{1}{r^2}}\right]\tensor{\qty(\pdv{r})}{^a} \\ +\left[\tensor{Y}{^A} - \frac{u \tensor{D}{^A}\tensor{D}{_B}\tensor{Y}{^B}}{2r} + \frac{u\tensor{C}{^A^B}\tensor{D}{_B}\tensor{D}{_C} \tensor{Y}{^C}}{4r^2} + \order{\frac{1}{r^3}}\right]\tensor{\qty(\pdv{\tensor{x}{^A}})}{^a}.
        \end{multline}
        \(\tensor*{\xi}{_f^a}\) is a generic generator of supertranslations, while \(\tensor*{\xi}{_Y^a}\) is a generic generator of Lorentz transformations.

        We can now compute the expression
        \begin{equation}
            \Lie[f]\tensor{g}{_u_u} \equiv \Lie[\tensor{\xi}{_f}]\tensor{g}{_u_u} = \frac{2 \tensor{D}{_u}m}{r} + \frac{\tensor{D}{_u}\tensor{D}{_A}\tensor{C}{^A^B}\tensor{D}{_B}f}{r} + \frac{\tensor{D}{_u}\tensor{C}{^A^B}\tensor{D}{_A}\tensor{D}{_B}f}{r} + \order{\frac{1}{r^2}}.
        \end{equation}
        Since
        \begin{equation}
            \tensor{g}{_u_u} = -1 + \frac{2 m}{r} + \order{\frac{1}{r^2}},
        \end{equation}
        we can now see that the variation of the Bondi mass under a transformation generated by \(\tensor*{\xi}{_f^a}\) is
        \begin{equation}\label{eq: Lie-f-mB}
            \delta_{f} m = f\tensor{\partial}{_u}m + \frac{1}{2}\qty[\tensor{D}{_A}\tensor{N}{^A^B}\tensor{D}{_B}f + \tensor{N}{^A^B}\tensor{D}{_A}\tensor{D}{_B}f],
        \end{equation}
        where I used the fact that \(\tensor{D}{_u} = \tensor{\partial}{_u}\) and the definition of the Bondi news tensor as \(\tensor{N}{_A_B} = \tensor{\partial}{_u}\tensor{C}{_A_B}\).

        We can do the same calculation with \(\tensor*{\xi}{_Y^a}\). We have 
        \begin{equation}\label{eq: Lie-Y-mB}
            \delta_{Y}m = \frac{u \tensor{D}{_A}\tensor{Y}{^A} \tensor{\partial}{_u}m}{2} + \tensor{Y}{^A}\tensor{D}{_A}m + \frac{3 m \tensor{D}{_A}\tensor{Y}{^A}}{2} + \frac{\tensor{C}{^A^B}\tensor{D}{_A}\tensor{D}{_B}\tensor{D}{_C}\tensor{Y}{^C}}{4}.
        \end{equation}
        Due to the fact that, for a Lorentz transformation, \(\tensor{Y}{^\zeta}\) is comprised of spherical harmonics with at most \(l=1\), it follows that the last term vanishes. It will be useful to keep it, though, for our future discussions of superrotations on \cref{sec: superrotations}.

        Similar calculations can be carried out for \(\tensor{C}{_A_B}\). By picking the \(\order{r}\) term of \(\Lie[\xi]\tensor{g}{_A_B}\) we find
        \begin{equation}\label{eq: Lie-f-CAB}
            \delta_{f}\tensor{C}{_A_B} = f \tensor{\partial}{_u}\tensor{C}{_A_B} + \qty(\tensor{\gamma}{_A_B}\tensor{D}{_C}\tensor{D}{^C} - 2\tensor{D}{_A}\tensor{D}{_B})f,
        \end{equation}
        and
        \begin{multline}\label{eq: Lie-Y-CAB}
            \delta_{Y}\tensor{C}{_A_B} = \frac{u \tensor{D}{_C}\tensor{Y}{^C}}{2} \tensor{\partial}{_u}\tensor{C}{_A_B} + \tensor{Y}{^C}\tensor{D}{_C}\tensor{C}{_A_B} - \frac{\tensor{D}{_C}\tensor{Y}{^C}}{2}\tensor{C}{_A_B} + 2 \tensor{C}{_C_(_A}\tensor{D}{_B_)}\tensor{Y}{^C} \\ + \frac{u}{2}\qty(\tensor{\gamma}{_A_B}\tensor{D}{_C}\tensor{D}{^C} - 2\tensor{D}{_A}\tensor{D}{_B})\tensor{D}{_D}\tensor{Y}{^D}.
        \end{multline}
        By comparing \cref{eq: Lie-f-CAB,eq: Lie-Y-CAB,eq: shear-decomposition}, we see that the last term in \cref{eq: Lie-f-CAB,eq: Lie-Y-CAB}---which is always the only term that is independent of \(\tensor{C}{_A_B}\)---induces a shift in the memory field \(C\) defined on \cref{eq: shear-decomposition}. Notice also that the term vanishes if \(\tensor{Y}{^A}\) is a Lorentz transformation and if \(f\) is a translation, since spherical harmonics with \(l \leq 1\) are annihilated by the differential operator. 

        Using \cref{eq: Lie-f-CAB,eq: Lie-Y-CAB}, we can obtain the expressions for the Bondi news tensor \(\tensor{N}{_A_B}\). Since \(\tensor{N}{_A_B} = \tensor{\partial}{_u}\tensor{C}{_A_B}\), we have that the variation of \(\tensor{N}{_A_B}\) under the transformation generated by \(\tensor{\xi}{^a}\) is given by
        \begin{equation}
            \tensor{N}{_A_B} + \delta \tensor{N}{_A_B} = \tensor{\partial}{_u}\tensor{C}{_A_B} + \tensor{\partial}{_u}(\delta\tensor{C}{_A_B}).
        \end{equation}
        By differentiating \cref{eq: Lie-f-CAB,eq: Lie-Y-CAB} we thus find
        \begin{equation}\label{eq: Lie-f-NAB}
            \delta_{f} \tensor{N}{_A_B} = f \tensor{\partial}{_u}\tensor{N}{_A_B},
        \end{equation}
        and
        \begin{equation}\label{eq: Lie-Y-NAB}
            \delta_Y \tensor{N}{_A_B} = \frac{u \tensor{D}{_C}\tensor{Y}{^C}}{2} \tensor{\partial}{_u}\tensor{N}{_A_B} + \tensor{Y}{^C}\tensor{D}{_C}\tensor{N}{_A_B} + 2 \tensor{N}{_C_(_A}\tensor{D}{_B_)}\tensor{Y}{^C} + \frac{1}{2}\qty(\tensor{\gamma}{_A_B}\tensor{D}{_C}\tensor{D}{^C} - \tensor{D}{_A}\tensor{D}{_B})\tensor{D}{_D}\tensor{Y}{^D}.
        \end{equation} 

\section{Consequences of Supertranslations}\label{sec: modern-developments}
    Now that we know what supertranslations are, we are ready to discuss some of their physical implications and applications. 

    \subsection{Spontaneous Symmetry Breaking of Supertranslations}\label{subsec: SSB-supertranslations}
        Within the Bondi--Sachs formalism, we can define a classical notion of vacuum at \(\nullfut\) by demanding that the Bondi news tensor vanishes, \(\tensor{N}{_A_B} = 0\). This is a vacuum in the sense that no \glspl{GW} are being propagated to infinity, and hence there are no excitations of the gravitational field. Let us assume as well that no radiation due to matter fields is reaching null infinity, so we can be in a ``true'' vacuum. 

        At leading order, we know from \cref{subsec: bondi-sachs-formalism} that this vacuum is completely determined by the initial data
        \begin{equation}
            \qty{\left.m(\zeta,\bar{\zeta})\right|_{\nullfut_-}, \left.C(\zeta,\bar{\zeta})\right|_{\nullfut_-}},
        \end{equation}
        where we neglected \(\tensor{N}{_A_B}\) because it is assumed to vanish and \(\tensor{N}{_A}\) does not enter the expressions at leading order. 

        Now let us apply a supertranslation to this initial data. Supertranslations are a symmetry of the theory (here considered to be \gls{GR} at null infinity). However, \cref{eq: Lie-f-mB,eq: Lie-f-CAB,eq: Lie-f-NAB} tell us that, under these conditions,
        \begin{subequations}
            \begin{align}
                \delta_f m &= 0, \\
                \delta_f \tensor{C}{_A_B} &= \qty(\tensor{\gamma}{_A_B}\tensor{D}{_C}\tensor{D}{^C} - 2\tensor{D}{_A}\tensor{D}{_B})f, \label{eq: SSB-supertranslation-CAB} \\
                \delta_f \tensor{N}{_A_B} &= 0.
            \end{align}
        \end{subequations}
        From \cref{eq: shear-decomposition}, we see the variation of \(\tensor{C}{_A_B}\) means 
        \begin{equation}\label{eq: SSB-memory-field}
            \delta_f C = f.
        \end{equation}
        Hence, even though the theory is invariant under supertranslations (because they are symmetries of the gravitational field at null infinity), the vacuum is not. 

        In \gls{QFT}, if a theory enjoys a symmetry, but the vacuum does not, we say the symmetry has been spontaneously broken \cite{peskin1995IntroductionQuantumField,weinberg1996ModernApplications,schwartz2014QuantumFieldTheory}. We can thus say that, in \gls{GR}, supertranslation symmetry is spontaneously broken. However, since the operator on the \gls{RHS} of \cref{eq: SSB-supertranslation-CAB} annihilates spherical harmonics with \(l \leq 1\), the translations are left unbroken. 

        Whenever a continuous symmetry is spontaneously broken, there occur massless states in the theory known as \glspl{NGB}. Let me explain. Assume the symmetry group is a Lie group (as is the \gls{BMS} group), so we are considering continuous symmetries. Upon \gls{SSB}, for each broken dimension of the symmetry group one gets one propagating massless mode in the theory---this is the \gls{NGB}. For instance, in the \gls{SM}, the \(\SU(2)_{\text{L}} \times \Ug(1)_{Y}\) symmetry\footnote{The subscript ``L'' stands for ``left'', because only ``left-handed'' particles interact weakly, while the subscript ``\(Y\)'' denotes hypercharge. Later, the subscript ``em'' in \(\Ug(1)_{\text{em}}\) denotes ``electromagnetic'', referring to the electromagnetic interaction.} of electroweak interactions is broken down to a \(\Ug(1)_{\text{em}}\) symmetry. Three dimensions of the group are broken, thus leading to three \glspl{NGB}. These new propagating modes can be interpreted as new polarizations of the \(W^{\pm}\) and \(Z\) bosons, which can then be understood as massive (this is known as the Higgs mechanism \cite{peskin1995IntroductionQuantumField,schwartz2014QuantumFieldTheory,weinberg1996ModernApplications}).

        Within \gls{GR}, we are spontaneously breaking infinitely many supertranslations (one for each spherical harmonic with \(l \geq 2\)). Hence, we get infinitely many \glspl{NGB}. These are the different harmonics comprising the memory field \(C\). Recall from \cref{subsec: bondi-sachs-formalism} that any harmonics of \(C\) with \(l \leq 1\) do not occur in the gravitational field, and hence are unphysical, which is in perfect agreement with the fact that translations are not spontaneously broken.

    \subsection{Physical Realization of Supertranslations}\label{subsec: physical-realization-supertranslations}
        So far, \gls{BMS} transformations, and supertranslations especially, may seem to be purely theoretical. After all, supertranslations are symmetries that are only available at infinity, and infinity is always far away. So let us discuss how a supertranslation can be physically realized.

        To get some intuition, we begin by discussing boosts. Like supertranslations, boosts are diffeomorphisms. As a consequence, two spacetimes that differ by a boost (or by a supertranslation) are to be considered physically the same. There is no experiment that can distinguish between these two possibilities. Nevertheless, it is possible to physically realize a boost. In other words, one can ``induce'' a boost by means of a physical process. 

        Consider a spacetime comprised of a single massive particle of mass \(M\). By Birkhoff's theorem one knows this spacetime is given by Schwarzschild spacetime. This is independent of whether the particle is ``at motion'' or ``at rest'': both solutions are physically equivalent, because they are diffeomorphic. 

        Consider, however, a spacetime as follows. We start with a single massive particle of mass \(M\), but at advanced time \(v_0\) a massless particle (let us say a photon) with energy \(E\) is sent in from infinity \emph{en route} to a collision with the particle of mass \(M\). Upon the collision, the massive particle absorbs the photon. Hence, for advanced time \(v < v_0\) the solution is the Schwarzschild solution with mass \(M\), but for \(v > v_0\) the solution is the Schwarzschild solution with mass \(M+E\). Furthermore, these two solutions are boosted relative to each other, because after the collision the massive particle has energy \(M+E\) and also is moving relative to its initial state. In other words, it was accelerated at the instant of the collision. Notice that in this example the global spacetime is not Schwarzschild spacetime.

        Notice that the time-reversed version of this experiment works as well: if a massive particle emits a photon, the spacetime before the emission is related to the spacetime after the emission by a boost (and a change in the mass parameter). We will focus on this time-reversed situation. Notice as well that the important thing about the massless particle is that it is massless, it does not need to be a photon. For us, a graviton will be just as good.

        Let us translate this idea into symbols. From \cref{eq: BS-EFE-uu} we know that
        \begin{equation}
            \tensor{\partial}{_u}\qty(m - \frac{1}{4}\tensor{D}{^A}\tensor{D}{^B}\tensor{C}{_A_B}) = - \tensor{\mathcal{T}}{_u_u},
        \end{equation}
        with \(\tensor{\mathcal{T}}{_u_u}\) denoting the energy flux reaching infinity in the form of null matter or \glspl{GW}. For us, \(\tensor{\mathcal{T}}{_u_u}\) represents the emission of massless particles from the bulk of spacetime.

        Now integrate this expression from before the emission (\(u = u_0\)) to after the emission (\(u = u_f\)). We find that
        \begin{equation}
             \tensor{D}{^A}\tensor{D}{^B}\qty(\Delta \tensor{C}{_A_B}) = 4 \int_{u_0}^{u_f}  \tensor{\mathcal{T}}{_u_u} \dd{u} + 4 \Delta m.
        \end{equation}
        We can write this expression in a simpler form by employing \cref{eq: DA-DB-CAB}. We find
        \begin{equation}\label{eq: memory-field-from-null-radiation}
             \tensor{D}{_A}\tensor{D}{^A}\qty(\tensor{D}{_B}\tensor{D}{^B} + 2)\Delta C = - 4 \int_{u_0}^{u_f}  \tensor{\mathcal{T}}{_u_u} \dd{u} - 4 \Delta m.
        \end{equation}
        The differential operator on the \gls{LHS} annihilates spherical harmonics with \(l \leq 1\), which are associated with the spacetime's mass and linear momentum. This is expected, since these quantities are determined by \(m\), and its presence in the equation above plays the role of canceling these terms in the stress-energy integral. The remaining harmonics of \(C\) can be determined by inverting the differential operator or by using the fact that spherical harmonics are eigenfunctions of this differential operator.

        The upshot of this discussion is that the emission of massless particles (such as electromagnetic or gravitational radiation) will lead to changes in the Bondi mass and in the memory field \(C\). Since we are in an asymptotically flat spacetime in Bondi gauge both before and after the emission, we know the relative transformation between the spacetime before and after the transformation is a \gls{BMS} transformation. By considering \cref{eq: Lie-f-mB,eq: Lie-f-CAB,eq: Lie-f-NAB,eq: Lie-Y-mB,eq: Lie-Y-CAB,eq: Lie-Y-NAB} we can determine which \gls{BMS} transformation is happening. 

        When considering time-dependent changes in momentum, this transformation has a natural interpretation as a boost, and could be caused by the emission of gravitational radiation, for example. Conservation of momentum requires the bulk to compensate for the momentum emitted to infinity.
        
        For the \(l \geq 2\) harmonics, the shift in the memory field \(C\) can be interpreted as a supertranslation by means of \cref{eq: SSB-memory-field}. Hence, the emission of null radiation to null infinity (be it due to matter or due to gravity itself) is associated with a realization of a supertranslation in the spacetime.

        To my knowledge, this was originally noticed by \textcite{strominger2016GravitationalMemoryBMS}, and it is the first step toward understanding the connections of supertranslations with other predictions in physics to be discussed in \cref{subsec: weinberg-soft-graviton,subsec: gravitational-memory-effect}. While I enjoy the term ``physically induced \gls{BMS} transformation'', I should point out that a clearer terminology is that we just found out about ``states with memory''. In a sense, null infinity remembers about the emission and absorption of null particles, and this memory looks precisely like a supertranslation\footnote{There are caveats in higher dimensions, and sometimes in \(d=4\) as well. For instance, it may be possible to have memory in the \(\Psi\) field defined on \cref{subsec: bondi-sachs-formalism}. See Ref. \citeonline{satishchandran2019AsymptoticBehaviorMassless} for a detailed discussion.}.

    \subsection{The Scattering Problem in Asymptotically Flat Spacetimes}\label{subsec: scattering-problem}
        The \gls{BMS} group is available both at \(\nullfut\) and at \(\nullpas\). The analysis in either case is essentially the same, up to a few sign differences due to the time direction of either case. I will also introduce an extra difference. In \(\nullfut\) we are working with stereographic coordinates defined by \(\zeta = e^{i\phi}\cot(\frac{\theta}{2})\), in accordance with \cref{eq: def-stereographic}. However, in \(\nullpas\) we will work with an antipodal definition of stereographic coordinates, which is given by \(\zeta = -e^{i\phi}\tan(\frac{\theta}{2})\), in accordance with \cref{eq: def-stereographic-antipodal}. This will prove to be convenient later on.
        
        In a spacetime which is asymptotically flat at both past and future null infinity we get a large group \(\BMS^- \times \BMS^+\) (the direct product of \(\BMS^-\) and \(\BMS^+\)). We can always define groups based on some property of spacetime, but the key question is whether they are physically significant. Can we learn something by considering \(\BMS^- \times \BMS^+\)? A possible way of trying to obtain physical information from this group is to consider, as done by \textcite{strominger2014BMSInvarianceGravitational}, whether gravitational scattering is invariant under the symmetries of \(\BMS^- \times \BMS^+\). 
        
        More specifically, suppose we are working in a theory of quantum gravity written in terms of perturbations of an asymptotically flat spacetime. Hence, we decompose the full metric as
        \begin{equation}
            \tensor{g}{_\mu_\nu} = \tensor{\bar{g}}{_\mu_\nu} + \tensor{h}{_\mu_\nu}
        \end{equation}
        and quantize the graviton field \(\tensor{h}{_\mu_\nu}\). The classical background metric \(\tensor{\bar{g}}{_\mu_\nu}\) is assumed to be asymptotically flat. 

        Under this setup, asking whether \(\BMS^- \times \BMS^+\) is a symmetry of gravitational scattering means asking whether, given \(B \in \BMS^- \times \BMS^+\), it holds that
        \begin{equation}
            B S - S B = 0,
        \end{equation}
        where \(S\) is the \(S\)-matrix. In terms of in and out states, we can write
        \begin{equation}
            \mel{\text{out}}{B S - S B}{\text{in}} = 0.
        \end{equation}

        When \(B\) is acting on an incoming state, it must be acting on a state defined on \(\nullpas\). Since \(\BMS^+\) is not defined on \(\nullpas\), we see that we must have \(B\ket{\text{in}} = B^-\ket{\text{in}}\) for some \(B^- \in \BMS^-\) (up to an isomorphism). Similarly, \(\bra{\text{out}}B = \bra{\text{out}}B^+\) for some \(B^+ \in \BMS^+\) (up to an isomorphism). Hence, a symmetry of gravitational scattering is expected to satisfy 
        \begin{equation}
            \mel{\text{out}}{B^+ S - S B^-}{\text{in}} = 0.
        \end{equation}
        There is no reason to expect that this expression will hold for any \(B^+ \in \BMS^+\) and any \(B^- \in \BMS^-\). Rather, it is expected that there must be some relation between \(B^-\) and \(B^+\) if we want this relation to hold. Hence, it does not seem that \(\BMS^- \times \BMS^+\) is a symmetry of gravitational scattering, but a subgroup of it could be. 

        \textcite{strominger2014BMSInvarianceGravitational} noticed that the existence of such a symmetry at the quantum level should also yield, in a suitable limit, a symmetry of classical gravitational scattering. Let us then think about this problem in a classical setting. 

        In the classical theory of asymptotically flat spacetimes, the gravitational scattering problem is the problem of, given the metric near \(\nullpas\), obtaining the metric near\footnote{This discussion neglects the possibility of other sorts of global structures in the spacetime. For instance, the presence of an event horizon would mean \(\nullfut\) cannot be used to obtain information about the gravitational field in the entire spacetime, and hence we would also like to know the gravitational field near the event horizon.} \(\nullfut\). Since we are interested in the asymptotically flat case, we can work with the Bondi gauge under the Ansatz of \cref{eq: bondi-gauge-AF}. Then we know the metric at \(\nullfut\) is completely determined by the initial data on \cref{eq: initial-data-BS-AF}. A similar statement holds for \(\nullpas\). 

        The issue is that, given the initial data at \(\nullpas\), we can get many different data at \(\nullfut\), each related to the others by \gls{BMS} transformations. While the outcome is physically the same, this means the scattering problem is still ill-defined---a choice of Bondi gauge at \(\nullpas\) should define a choice of gauge at \(\nullfut\). 

        \textcite{strominger2014BMSInvarianceGravitational} then noticed that a geodesic that leaves \(\nullpas\), crosses spatial infinity \(\spatinf\), and then reaches \(\nullfut\) arrives at the opposite point in the celestial sphere. This can be illustrated by \cref{fig: minkowski-einstein-cylinder}, which illustrates the antipodal relation between \(\nullfut\) and \(\nullpas\) near \(\spatinf\). Alternatively, one can thing that an (imaginary, and unphysical) signal that leaves the Earth at the North pole, crosses infinity, and comes back arrives at the South pole. 

        With this intuition in mind, \textcite{strominger2014BMSInvarianceGravitational} proposed that one should relate the initial data in \(\nullpas\) and in \(\nullfut\) by imposing the validity of the antipodal map
        \begin{equation}
            \left.m^+(\zeta,\bar{\zeta})\right|_{\nullfut_-} = \left.m^-(\zeta,\bar{\zeta})\right|_{\nullpas_+} \qq{and} \left.C^+(\zeta,\bar{\zeta})\right|_{\nullfut_-} = \left.C^-(\zeta,\bar{\zeta})\right|_{\nullpas_+},
        \end{equation}
        where \(\nullfut_-\) denotes the limit \(u \to - \infty\) in \(\nullfut\), while \(\nullpas_+\) denotes the \(v \to + \infty\) limit in \(\nullpas\). I neglected the role of the angular momentum aspect because it only enters at subleading orders. Notice that our distinct definitions of the stereographic coordinate at \(\nullfut\) and \(\nullpas\) allow for the antipodal condition to be written in a simple manner. Notice also that this identification requires that the initial data appropriately define the gravitational field at past and future null infinities, and while this is ensured for \gls{CK} spacetimes, it may not be ensured for other classes of spacetimes. 

        With the knowledge of the antipodal relation, an identification between elements of \(\BMS^-\) and \(\BMS^+\) becomes clear. Given \((\tensor*{Y}{_+^\zeta},f_+) \in \BMS^+\), the associated \(\BMS^-\) transformation is given by
        \begin{equation}
            f_-(\zeta,\bar{\zeta}) = f_+(\zeta,\bar{\zeta}) \qq{and} \tensor*{Y}{_-^\zeta}(\zeta,\bar{\zeta}) = \tensor*{Y}{_+^\zeta}(\zeta,\bar{\zeta}).
        \end{equation}
        This antipodal relation---which can be explicitly verified for the Poincaré group in Minkowski spacetime---allows us to identify a subgroup of \(\BMS^- \times \BMS^+\), which we will denote by \(\BMS^0\). \(\BMS^0\) is then expected to lead to a symmetry of the quantum scattering problem in asymptotically flat spacetimes. 

        This identification allows us to define an infinite number of conserved supertranslation charges. Namely, define 
        \begin{equation}\label{eq: supertranslation-charge}
            Q^{\pm}_f = \frac{1}{4\pi}\int_{\nullinf^{\pm}_{\mp}} f^{\pm} m^{\pm} \sqrt{\gamma} \dd[2]{x}.
        \end{equation}
        Then the antipodal matching conditions imply that
        \begin{equation}
            Q^{+}_f = Q^{-}_f.
        \end{equation}
        At the quantum level, we then have
        \begin{equation}
            Q^{+}_f S - S Q^{-}_f = 0,
        \end{equation}
        for any \(f \in \ck[\infty](\Sph^2)\). If we insert in and out states, we get the general expression
        \begin{equation}\label{eq: ward-identity-supertranslations}
            \mel{\text{out}}{Q^{+}_f S - S Q^{-}_f}{\text{in}} = 0.
        \end{equation}
        This is known as a Ward identity---a dynamical consequence of a symmetry of the theory. Its explicit form was carried out by Strominger in Ref. \citeonline{strominger2014BMSInvarianceGravitational}. Let us reproduce this calculation. 

        Start from the definition of the supertranslation charges, \cref{eq: supertranslation-charge}. Using the Einstein equation (\ref{eq: BS-EFE-uu}), we notice that 
        \begin{subequations}
            \begin{align}
                Q^{+}_f &= \frac{1}{4\pi}\int_{\nullfut_{-}} f(\zeta,\bar{\zeta}) m(\zeta,\bar{\zeta}) \sqrt{\gamma} \dd[2]{x}, \\
                &= -\frac{1}{4\pi}\int_{\nullfut} f(\zeta,\bar{\zeta}) \tensor{\partial}{_u}m(u,\zeta,\bar{\zeta}) \sqrt{\gamma} \dd{u}\dd[2]{x} + \frac{1}{4\pi}\int_{\nullfut_{+}} f(\zeta,\bar{\zeta}) m(\zeta,\bar{\zeta}) \sqrt{\gamma} \dd[2]{x}, \\
                &= \frac{1}{4\pi}\int_{\nullfut} f(\zeta,\bar{\zeta}) \qty[\tensor{\mathcal{T}}{_u_u} - \frac{1}{4}\tensor{D}{^A}\tensor{D}{^B}\tensor{N}{_A_B}] \sqrt{\gamma} \dd{u}\dd[2]{x} + \frac{1}{4\pi}\int_{\nullfut_{+}} f(\zeta,\bar{\zeta}) m(\zeta,\bar{\zeta}) \sqrt{\gamma} \dd[2]{x}, \\
                &= \frac{1}{4\pi}\int_{\nullfut} f(\zeta,\bar{\zeta}) \qty[\tensor{\mathcal{T}}{_u_u} - \frac{1}{4}\tensor{D}{^A}\tensor{D}{^B}\tensor{N}{_A_B}] \sqrt{\gamma} \dd{u}\dd[2]{x}.
            \end{align}
        \end{subequations}
        In the last step I made the simplifying assumption (also done in Refs. \citeonline{strominger2014BMSInvarianceGravitational,he2015BMSSupertranslationsWeinberg}) that the Bondi mass aspect decays to zero in the far future. \(\nullfut_+\) denotes the \(u \to + \infty\) limit in \(\nullfut\).

        Notice we can write
        \begin{subequations}\label{eq: supertranslation-ward-identity-Fplus}
            \begin{align}
                F^+ &= - \frac{1}{16\pi} \int_{\nullfut} f \tensor{D}{^A}\tensor{D}{^B}\tensor{N}{_A_B} \sqrt{\gamma} \dd{u}\dd[2]{x}, \\ 
                &= - \frac{1}{16\pi} \int_{\nullfut} \tensor{D}{_A}\tensor{D}{_B}f \tensor{N}{^A^B} \sqrt{\gamma} \dd{u}\dd[2]{x}, \\
                &= - \frac{1}{32\pi} \lim_{\omega \to 0^+}\int_{\nullfut} \qty(e^{-i\omega u} + e^{+i\omega u}) \tensor{D}{_A}\tensor{D}{_B}f \tensor{N}{^A^B} \sqrt{\gamma} \dd{u}\dd[2]{x},
            \end{align}
        \end{subequations}
        which in the quantum theory corresponds to an operator creating or annihilating a zero energy graviton with polarization \(\tensor{D}{_A}\tensor{D}{_B}f\). Hence, this term can be interpreted as a soft graviton (a zero-energy graviton) insertion.

        We now notice that, for a given out state, we get
        \begin{equation}
            \bra{\text{out}}Q^+_f = \sum_{\substack{k \\ \text{out}}} E_k^{\text{out}} f(\zeta_k^{\text{out}}) \bra{\text{out}} + \bra{\text{out}} F^+,
        \end{equation}
        where \(\zeta_k^{\text{out}}\) is the angular coordinate in which the \(k\)-th particle punctures\footnote{Notice this implicitly assumes we are dealing with massless particles. It is possible to deal with massive particles by using a suitable hyperbolic slicing of Minkowski spacetime---see Ref. \citeonline{strominger2018LecturesInfraredStructure}.} \(\nullfut\). \(E_k^{\text{out}}\) is the energy of this particle. The first term comes from the effective stress tensor \(\tensor{\mathcal{T}}{_u_u}\), while the second term comes from the soft graviton insertion. 

        A similar calculation yields
        \begin{equation}
            Q^-_f \ket{\text{in}} = \sum_{\substack{l \\ \text{in}}} E_l^{\text{in}} f(\zeta_l^{\text{in}}) \ket{\text{in}} + F^- \ket{\text{in}}.
        \end{equation}

        In total, we see the Ward identity can be written as
        \begin{equation}
            \mel{\text{out}}{F^+ S - S F^-}{\text{in}} = \qty(\sum_{\substack{l \\ \text{in}}} E_l^{\text{in}} f(\zeta_l^{\text{in}}) - \sum_{\substack{k \\ \text{out}}} E_k^{\text{out}} f(\zeta_k^{\text{out}})) \mel{\text{out}}{S}{\text{in}}.
        \end{equation}
        Notice this implicitly uses the antipodal map to relate the supertranslations at past and future null infinity.

        We can further simplify this expression by computing \(F^{\pm}\) explicitly. To do this, we notice that
        \begin{subequations}
            \begin{align}
                F^{+} &= - \frac{1}{16\pi} \int_{\nullfut} f \tensor{D}{^A}\tensor{D}{^B}\tensor{N}{_A_B} \sqrt{\gamma} \dd{u}\dd[2]{x}, \\
                &= - \frac{1}{16\pi} \int_{\nullfut} f \tensor{D}{_A}\tensor{D}{^B}\tensor{N}{_C_B}\tensor{\gamma}{^A^C} \sqrt{\gamma} \dd{u}\dd[2]{x}, \\
                &= - \frac{1}{16\pi} \int_{\nullfut_+} f \tensor{D}{_A}\tensor{D}{^B}\tensor{C}{_C_B}\tensor{\gamma}{^A^C} \sqrt{\gamma} \dd{u}\dd[2]{x} + \frac{1}{16\pi} \int_{\nullfut_-} f \tensor{D}{_A}\tensor{D}{^B}\tensor{C}{_C_B}\tensor{\gamma}{^A^C} \sqrt{\gamma} \dd{u}\dd[2]{x}, \\
                &= - \frac{1}{8\pi} \int_{\nullfut_+} f \tensor{D}{_{\bar{\zeta}}}\tensor{D}{^B}\tensor{C}{_{\zeta}_B}\tensor{\gamma}{^{\bar{\zeta}}^{\zeta}} \sqrt{\gamma} \dd{u}\dd[2]{x} + \frac{1}{8\pi} \int_{\nullfut_-} f \tensor{D}{_{\bar{\zeta}}}\tensor{D}{^B}\tensor{C}{_{\zeta}_B}\tensor{\gamma}{^{\bar{\zeta}}^{\zeta}} \sqrt{\gamma} \dd{u}\dd[2]{x}, \\
                &= - \frac{1}{8\pi} \int_{\nullfut} f \tensor{D}{_{\bar{\zeta}}}\tensor{D}{^B}\tensor{N}{_{\zeta}_B}\tensor{\gamma}{^{\bar{\zeta}}^{\zeta}} \sqrt{\gamma} \dd{u}\dd[2]{x}.
            \end{align}
        \end{subequations}
        In this expression, we used the fact that \(\tensor{D}{_[_A}\tensor{D}{^B}\tensor{C}{_C_]_B} = 0\) at \(\nullfut_{\pm}\) (this follows from the fact shown in \cref{subsec: bondi-sachs-formalism} that, in vacuum, \(\tensor{C}{_A_B}\) is completely determined by the memory field). 

        With these considerations in mind, we can now see that
        \begin{subequations}
            \begin{align}
                F^+ &= - \frac{1}{8\pi} \int_{\nullfut} f \tensor{D}{_{\bar{\zeta}}}\tensor{D}{^B}\tensor{N}{_{\zeta}_B}\tensor{\gamma}{^{\bar{\zeta}}^{\zeta}} \sqrt{\gamma} \dd{u}\dd[2]{x}, \\
                &= \frac{1}{8\pi} \int_{\nullfut} \tensor{D}{_{\bar{\zeta}}} f \tensor{D}{^B}\tensor{N}{_{\zeta}_B}\tensor{\gamma}{^{\bar{\zeta}}^{\zeta}} \sqrt{\gamma} \dd{u}\dd[2]{x}, \\
                &= -\frac{i}{8\pi} \int_{\nullfut} \tensor{\partial}{_{\bar{\zeta}}} f \tensor{D}{^B}\tensor{N}{_{\zeta}_B}\tensor{\gamma}{^{\bar{\zeta}}^{\zeta}} \tensor{\gamma}{_{\bar{\zeta}}_{\zeta}} \dd{u}\dd{\zeta}\dd{\bar{\zeta}}, \\
                &= -\frac{i}{8\pi} \int_{\nullfut} \tensor{\partial}{_{\bar{\zeta}}} f \tensor{D}{^B}\tensor{N}{_{\zeta}_B} \dd{u}\dd{\zeta}\dd{\bar{\zeta}}.
            \end{align}
        \end{subequations}
        Above we used the fact that, in stereographic coordinates\footnote{I am avoiding using the language of differential forms, which is a clearer way of writing these integrals. The convention \(\sqrt{\gamma} = -i \tensor{\gamma}{_{\zeta}_{\bar{\zeta}}}\) will reproduce the results of Appendix \ref{app: complex-analysis}, in which \(\dd{\zeta}\dd{\bar{\zeta}} = \dd{\bar{\zeta}} \wedge \dd{\zeta}\).}, \(\sqrt{\gamma} = -i \tensor{\gamma}{_{\zeta}_{\bar{\zeta}}}\).

        Now pick 
        \begin{equation}
            f(\zeta) = \frac{1}{\zeta - \chi}.
        \end{equation}
        Then \cref{eq: derivative-Dirac-delta-complex} implies
        \begin{subequations}
            \begin{align}
                F^+ &= -\frac{i}{8\pi} \int_{\nullfut} 2 \pi i \delta^{(2)}(\zeta-\chi) \tensor{D}{^B}\tensor{N}{_{\zeta}_B} \dd{u}\dd{\zeta}\dd{\bar{\zeta}}, \\
                &= \frac{1}{4} \int \tensor{D}{^B}\tensor{N}{_{\chi}_B}\qty(u,\chi) \dd{u}.
            \end{align}
        \end{subequations}

        In total, the Ward identity becomes
        \begin{multline}\label{eq: ward-identity-supertranslations-localized}
            \mel**{\text{out}}{\int \tensor{D}{^B}\tensor*{N}{^{\text{out}}_{\zeta}_B}\qty(u,\zeta) \dd{u} S + S \int \tensor{D}{^B}\tensor*{N}{^{\text{in}}_{\zeta}_B}\qty(v,\zeta) \dd{v}}{\text{in}} \\ = 4 \qty(\sum_{\substack{k \\ \text{out}}} \frac{E_k^{\text{out}}}{\zeta - \zeta_k^{\text{out}}} - \sum_{\substack{l \\ \text{in}}} \frac{E_l^{\text{in}}}{\zeta - \zeta_l^{\text{in}}}) \mel{\text{out}}{S}{\text{in}}.
        \end{multline}
        Above, \(\tensor*{N}{^{\text{out}}_{\zeta}_B}\) denotes the Bondi news tensor at \(\nullfut\), while \(\tensor*{N}{^{\text{in}}_{\zeta}_B}\) denotes the Bondi news at \(\nullpas\). The previous version of the Ward identity can be recovered by multiplying by a function \(f(\zeta)\) and then performing a contour integral over \(\zeta\) around all the poles. The result then follows from the Cauchy integral formula (\ref{eq: cauchy-integral-formula}). 

        Further discussion about the scattering problem and supertranslation charges can be found in the original papers, Refs. \citeonline{strominger2014BMSInvarianceGravitational,he2015BMSSupertranslationsWeinberg}, and in the book by \textcite{strominger2018LecturesInfraredStructure}.

    \subsection{Weinberg's Soft Graviton Theorem}\label{subsec: weinberg-soft-graviton}
        This section may seem unconnected to the rest of this text at first. Instead of thinking about supertranslations, we will for now consider a \gls{QFT} including gravitons. More specifically, we will be interested in the role of gravitons with extremely small energies. These are called soft gravitons. The hint for how this relates to the remainder of our discussions is that the \gls{BMS} group is a consequence of \gls{IR} physics, since it considers the behavior of gravity at very large distances. The problem of understanding the role of soft gravitons in a scattering problem also concerns the \gls{IR} behavior of gravity, because it involves gravitons with extremely small energies.
    
        Consider a scattering process in some collider experiment. We collide \(n\) incoming particles and produce \(m\) outgoing particles as a consequence. This is an \(n \to m\) scattering process. A typical task in \gls{QFT} is to compute the cross section for such a scattering. To carry out this calculation we will need, of course, previous knowledge of \gls{QFT}, but I will try to keep the conclusion of this subsection accessible to all readers. In the following, I follow \textcite{weinberg1965InfraredPhotonsGravitons}. See also Refs. \citeonline{weinberg1995Foundations,strominger2018LecturesInfraredStructure}.

        Let us consider such a calculation in a theory involving gravitons. We are interested in computing a scattering amplitude such as
        \begin{equation}
            \vcenter{\hbox{\begin{tikzpicture}
                    \begin{feynman}
                        \node[blob] (c) at (0,0);
                        \vertex (a1) at (180:2);
                        \vertex (a2) at (135:2);
                        \vertex (a3) at (225:2);
                        \vertex (b1) at (30:2);
                        \vertex (b2) at (330:2);
                        \diagram*{
                            (a1) -- [fermion] (c),
                            (a2) -- [fermion] (c),
                            (a3) -- [fermion] (c),
                            (c) -- [fermion] (b1),
                            (c) -- [fermion] (b2),
                        };
                    \end{feynman}
                \end{tikzpicture}}} = \mel{\text{out}}{S^0}{\text{in}},
        \end{equation}
        where \(S^0\) is the \(S\)-matrix. The blob in this diagram includes all loop orders. There is, however, a caveat: we will also assume that none of the gravitons inside the blob has spatial momentum \(\vb{p}\) above a certain cutoff \(\Lambda\), \(\norm{\vb{p}} \leq \Lambda\), which is taken to be very small. This is represented by the \(0\) superscript on \(S^0\). Gravitons with momentum below the cutoff \(\Lambda\) are said to be ``soft''. Gravitons with momentum above \(\Lambda\) are ``hard''. I am also assuming that all the external particles are hard. 

        The next step is then to add in the soft gravitons in the calculation. This means we need to sum over all diagrams including soft graviton exchanges. Schematically, we are now considering the scattering amplitude
        \begin{multline}\label{eq: virtual-soft-corrections}
            \mel{\text{out}}{S}{\text{in}} \, \equiv \,\vcenter{\hbox{\begin{tikzpicture}
            \begin{feynman}
                \node[blob, pattern color=white] (c) at (0,0);
                \vertex (a1) at (180:2);
                \vertex (a2) at (135:2);
                \vertex (a3) at (225:2);
                \vertex (b1) at (30:2);
                \vertex (b2) at (330:2);
                \diagram*{
                    (a1) -- [fermion] (c),
                    (a2) -- [fermion] (c),
                    (a3) -- [fermion] (c),
                    (c) -- [fermion] (b1),
                    (c) -- [fermion] (b2),
                };
            \end{feynman}
        \end{tikzpicture}}} + \vcenter{\hbox{\begin{tikzpicture}
            \begin{feynman}
                \node[blob, pattern color=white] (c) at (0,0);
                \vertex (a1) at (180:2);
                \vertex (a2) at (135:2);
                \vertex (a3) at (225:2);
                \vertex (a4) at (180:1);
                \vertex (a5) at (225:1);
                \vertex (b1) at (30:2);
                \vertex (b2) at (330:2);
                \diagram*{
                    (a4) -- [graviton] (a5),
                    (a1) -- [fermion] (c),
                    (a2) -- [fermion] (c),
                    (a3) -- [fermion] (c),
                    (c) -- [fermion] (b1),
                    (c) -- [fermion] (b2),
                };
            \end{feynman}
        \end{tikzpicture}}} \\ + \vcenter{\hbox{\begin{tikzpicture}
            \begin{feynman}
                \node[blob, pattern color=white] (c) at (0,0);
                \vertex (a1) at (180:2);
                \vertex (a2) at (135:2);
                \vertex (a3) at (225:2);
                \vertex (a4) at (180:1);
                \vertex (a5) at (225:1);
                \vertex (a6) at (135:1);
                \vertex (b1) at (30:2);
                \vertex (b2) at (330:2);
                \vertex (b3) at (30:0.8);
                \diagram*{
                    (a4) -- [graviton] (a5),
                    (a6) -- [graviton] (b3),
                    (a1) -- [fermion] (c),
                    (a2) -- [fermion] (c),
                    (a3) -- [fermion] (c),
                    (c) -- [fermion] (b1),
                    (c) -- [fermion] (b2),
                };
            \end{feynman}
        \end{tikzpicture}}} + \vcenter{\hbox{\begin{tikzpicture}
            \begin{feynman}
                \node[blob, pattern color=white] (c) at (0,0);
                \vertex (a1) at (180:2);
                \vertex (a2) at (135:2);
                \vertex (a3) at (225:2);
                \vertex (a4) at (180:1);
                \vertex (a5) at (225:1);
                \vertex (a6) at (135:1);
                \vertex (b1) at (30:2);
                \vertex (b2) at (330:2);
                \vertex (b3) at (30:0.8);
                \vertex (b4) at (30:1.5);
                \vertex (b5) at (330:1);
                \diagram*{
                    (b4) -- [graviton] (b5),
                    (a4) -- [graviton] (a5),
                    (a6) -- [graviton] (b3),
                    (a1) -- [fermion] (c),
                    (a2) -- [fermion] (c),
                    (a3) -- [fermion] (c),
                    (c) -- [fermion] (b1),
                    (c) -- [fermion] (b2),
                };
            \end{feynman}
        \end{tikzpicture}}} + \cdots,
        \end{multline}
        where each of the diagrams represents all diagrams with that number of exchanged gravitons (which are denoted by coil-like lines). This time, we assume that all soft graviton momenta satisfy \(\norm{\vb{p}} \geq \lambda\). \(\lambda\) (which is different from \(\Lambda\)) is introduced as a regulator and should be taken to zero at the end of the calculation. The absence of the \(0\) superscript in \(S\) indicates we are now considering contributions due to soft gravitons as well.

        How can we do such a calculation? Before getting worried with the infinitely many diagrams at all loop orders, let us consider the simplest case. Let us consider the original diagram and add one single graviton line to it. Hence, we want to consider the diagram
        \begin{equation}
            \vcenter{\hbox{\begin{tikzpicture}
                \begin{feynman}
                    \node[blob, pattern color=white] (c) at (0,0);
                    \vertex (a1) at (180:2);
                    \vertex (a2) at (135:2);
                    \vertex (a3) at (225:2);
                    \vertex (a4) at (225:1);
                    \vertex (a5) at (300:1.5) {\footnotesize\(\mu\nu\)};
                    \vertex (b1) at (30:2);
                    \vertex (b2) at (330:2);
                    \diagram*{
                        (a4) -- [graviton] (a5),
                        (a1) -- [fermion] (c),
                        (a2) -- [fermion] (c),
                        (a3) -- [fermion] (c),
                        (c) -- [fermion] (b1),
                        (c) -- [fermion] (b2),
                    };
                \end{feynman}
            \end{tikzpicture}}}
        \end{equation}
        where the \(\mu\nu\) symbol indicates that the other end of the graviton line must either be attached to another line in the diagram or it must be considered an external particle. \(\mu\) and \(\nu\) are the hanging indices at the end of the line which must be contracted with something else.

        To compute this diagram, we must know what is the vertex for the hard particle--soft graviton interaction. The interactions between gravity and matter arise from the interaction term (\(G = \hbar = c = 1\))
        \begin{equation}
            S_I = S_M,
        \end{equation}
        where \(S_M\) denotes the matter action. If we split the metric as\footnote{The \(\sqrt{32\pi}\) factor is meant to keep the kinetic term of \(\tensor{\hat{h}}{_\mu_\nu}\) canonically normalized. It is usually written as \(\sqrt{32\pi G}\), but I am setting \(G=1\). I am partially writing \(\tensor{\hat{h}}{_\mu_\nu}\) (with a hat) to denote it should be thought of as a quantum field, but also to indicate this difference in normalization relative to the remainder of the text.} \(\tensor{g}{_\mu_\nu} = \tensor{\eta}{_\mu_\nu} + \sqrt{32\pi}\tensor{\hat{h}}{_\mu_\nu}\), then we can make a Taylor expansion of the action to get
        \begin{equation}
            S_I = \sqrt{32\pi} \int \eval{\fdv{S_M}{\tensor{\hat{h}}{^\mu^\nu}(x)}}_{\hat{h}=0}\tensor{\hat{h}}{_\mu_\nu}(x) \dd[4]{x} + \cdots,
        \end{equation}
        where I neglected terms that do not depend on \(\tensor{\hat{h}}{_\mu_\nu}\) or that depend only on higher orders. Using the definition of the stress-energy tensor we get
        \begin{equation}
            S_I = - \sqrt{8\pi}\int \tensor{T}{^\mu^\nu} \tensor{\hat{h}}{_\mu_\nu} \dd[4]{x} + \cdots.
        \end{equation}
        Notice \(\sqrt{-g}=1\) to the order we are working with. For simplicity, we will work with only a minimally coupled scalar field in the matter sector, but the result is general. In this simplified case, we have
        \begin{equation}
            \tensor{T}{_\mu_\nu} = \tensor{\partial}{_\mu}\phi\tensor{\partial}{_\nu}\phi - \frac{1}{2}\tensor{\eta}{_\mu_\nu}\qty[\tensor{\partial}{^\rho}\phi\tensor{\partial}{_\rho}\phi + m^2 \phi^2].
        \end{equation}
        The interaction vertex is then given by \cite{choi1995FactorizationPolarizationLinearized,donoghue2017EPFLLecturesGeneral,jakobsen2020GeneralRelativityQuantum}
        \begin{equation}\label{eq: weinberg-soft-graviton-vertex}
            \vcenter{\hbox{\begin{tikzpicture}
                \begin{feynman}
                    \vertex (c) at (0,0);
                    \vertex (s1) at (210:1);
                    \vertex (s2) at (330:1);
                    \vertex (g) at (90:1)  {\footnotesize\(\mu\nu\)};
                    \diagram*{
                        (c) -- [graviton] (g),
                        (s1) -- [fermion,momentum=\(p_1\)] (c),
                        (c) -- [fermion,momentum=\(p_2\)] (s2),
                    };
                \end{feynman}
            \end{tikzpicture}}} = i\sqrt{8\pi} \qty(\tensor{p}{_1_\mu}\tensor{p}{_2_\nu}+\tensor{p}{_1_\nu}\tensor{p}{_2_\mu} - \tensor{\eta}{_\mu_\nu}(\tensor{p}{_1_\rho}\tensor*{p}{_2^\rho}+m^2)).
        \end{equation}
        Recalling that the scalar propagator is 
        \begin{equation}
            \vcenter{\hbox{\begin{tikzpicture}
                \begin{feynman}
                    \vertex (s1) at (180:0.75);
                    \vertex (s2) at (0:0.75);
                    \diagram*{
                        (s1) -- [fermion,momentum=\(p\)](s2),
                    };
                \end{feynman}
            \end{tikzpicture}}} = \frac{-i}{p^2 + m^2 - i \epsilon}.
        \end{equation}

        Using these results, we see that, if the graviton is attached to an incoming particle,
        \begin{multline}
            \vcenter{\hbox{\begin{tikzpicture}
                \begin{feynman}
                    \node[blob, pattern color=white] (c) at (0,0);
                    \vertex (a1) at (180:2);
                    \vertex (a2) at (135:2);
                    \vertex (a3) at (225:2);
                    \vertex (a4) at (225:1);
                    \vertex (a5) at (300:1.5) {\footnotesize\(\mu\nu\)};
                    \vertex (b1) at (30:2);
                    \vertex (b2) at (330:2);
                    \diagram*{
                        (a4) -- [graviton] (a5),
                        (a4) -- [draw=none,momentum'=\(q\)] (a5),
                        (a1) -- [fermion] (c),
                        (a2) -- [fermion] (c),
                        (a3) -- [fermion,momentum=\(p\)] (a4),
                        (a4) -- [fermion] (c),
                        (c) -- [fermion] (b1),
                        (c) -- [fermion] (b2),
                    };
                \end{feynman}
            \end{tikzpicture}}} \\ = \qty(\frac{\sqrt{8\pi} \qty(\tensor{p}{_\mu}(\tensor{p}{_\nu}-\tensor{q}{_\nu})+\tensor{p}{_\nu}(\tensor{p}{_\mu}-\tensor{q}{_\mu}) - \tensor{\eta}{_\mu_\nu}(\tensor{p}{_\rho}(\tensor{p}{^\rho}-\tensor{q}{^\rho})+m^2))}{(p-q)^2 + m^2 - i \epsilon}) \\ \times \qty(\vcenter{\hbox{\begin{tikzpicture}
                \begin{feynman}
                    \node[blob, pattern color=white] (c) at (0,0);
                    \vertex (a1) at (180:2);
                    \vertex (a2) at (135:2);
                    \vertex (a3) at (225:2);
                    \vertex (b1) at (30:2);
                    \vertex (b2) at (330:2);
                    \diagram*{
                        (a1) -- [fermion] (c),
                        (a2) -- [fermion] (c),
                        (a3) -- [fermion,momentum=\(p\)] (c),
                        (c) -- [fermion] (b1),
                        (c) -- [fermion] (b2),
                    };
                \end{feynman}
            \end{tikzpicture}}}).
        \end{multline}
        This equation is complicated, but we know the graviton is soft! As a consequence, \(\tensor{q}{^\mu}\) is taken to be very small compared to \(\tensor{p}{^\mu}\). Furthermore, since the graviton is connected to an external line we know that \(\tensor{p}{^\mu}\) is nearly on-shell. Hence, we may write
        \begin{equation}\label{eq: soft-graviton-theorem-in}
            \vcenter{\hbox{\begin{tikzpicture}
                \begin{feynman}
                    \node[blob, pattern color=white] (c) at (0,0);
                    \vertex (a1) at (180:2);
                    \vertex (a2) at (135:2);
                    \vertex (a3) at (225:2);
                    \vertex (a4) at (225:1);
                    \vertex (a5) at (300:1.5) {\footnotesize\(\mu\nu\)};
                    \vertex (b1) at (30:2);
                    \vertex (b2) at (330:2);
                    \diagram*{
                        (a4) -- [graviton] (a5),
                        (a4) -- [draw=none,momentum'=\(q\)] (a5),
                        (a1) -- [fermion] (c),
                        (a2) -- [fermion] (c),
                        (a3) -- [fermion,momentum=\(p\)] (a4),
                        (a4) -- [fermion] (c),
                        (c) -- [fermion] (b1),
                        (c) -- [fermion] (b2),
                    };
                \end{feynman}
            \end{tikzpicture}}} = \qty(-\frac{\sqrt{8\pi} \tensor{p}{_\mu}\tensor{p}{_\nu}}{\tensor{p}{^\rho}\tensor{q}{_\rho} + i \epsilon}) \times \qty(\vcenter{\hbox{\begin{tikzpicture}
                \begin{feynman}
                    \node[blob, pattern color=white] (c) at (0,0);
                    \vertex (a1) at (180:2);
                    \vertex (a2) at (135:2);
                    \vertex (a3) at (225:2);
                    \vertex (b1) at (30:2);
                    \vertex (b2) at (330:2);
                    \diagram*{
                        (a1) -- [fermion] (c),
                        (a2) -- [fermion] (c),
                        (a3) -- [fermion,momentum=\(p\)] (c),
                        (c) -- [fermion] (b1),
                        (c) -- [fermion] (b2),
                    };
                \end{feynman}
            \end{tikzpicture}}}) + \order{\tensor{q}{^0}}.
        \end{equation}
        To leading order in \(\tensor{q}{^0}\), the diagram with one extra graviton is proportional to the original diagram! The proportionality factor,
        \begin{equation}
            -\frac{\sqrt{8\pi} \tensor{p}{_\mu}\tensor{p}{_\nu}}{\tensor{p}{^\rho}\tensor{q}{_\rho} + i \epsilon},
        \end{equation}
        was derived for a soft graviton attached to an incoming particle. If we repeated the argument with an outgoing particle, the relation between the momenta would change a bit and we get
        \begin{equation}\label{eq: soft-graviton-theorem-out}
            \vcenter{\hbox{\begin{tikzpicture}
                \begin{feynman}
                    \node[blob, pattern color=white] (c) at (0,0);
                    \vertex (a1) at (180:2);
                    \vertex (a2) at (135:2);
                    \vertex (a3) at (225:2);
                    \vertex (a4) at (225:1);
                    \vertex (a5) at (300:1.5);
                    \vertex (b1) at (30:2);
                    \vertex (b2) at (330:2);
                    \vertex (b3) at (330:1);
                    \vertex (b4) at (0:2) {\footnotesize\(\mu\nu\)};
                    \diagram*{
                        (a1) -- [fermion] (c),
                        (a2) -- [fermion] (c),
                        (a3) -- [fermion] (c),
                        (c) -- [fermion] (b1),
                        (c) -- [fermion] (b3),
                        (b3) -- [fermion,momentum'=\(p\)] (b2),
                        (b3) -- [graviton] (b4),
                        (b3) -- [draw=none,momentum=\(q\)] (b4),
                    };
                \end{feynman}
            \end{tikzpicture}}} = \qty(\frac{\sqrt{8\pi} \tensor{p}{_\mu}\tensor{p}{_\nu}}{\tensor{p}{^\rho}\tensor{q}{_\rho} - i \epsilon}) \times \qty(\vcenter{\hbox{\begin{tikzpicture}
                \begin{feynman}
                    \node[blob, pattern color=white] (c) at (0,0);
                    \vertex (a1) at (180:2);
                    \vertex (a2) at (135:2);
                    \vertex (a3) at (225:2);
                    \vertex (b1) at (30:2);
                    \vertex (b2) at (330:2);
                    \diagram*{
                        (a1) -- [fermion] (c),
                        (a2) -- [fermion] (c),
                        (a3) -- [fermion] (c),
                        (c) -- [fermion] (b1),
                        (c) -- [fermion,momentum'=\(p\)] (b2),
                    };
                \end{feynman}
            \end{tikzpicture}}}) + \order{\tensor{q}{^0}}.
        \end{equation}
        If the graviton was connected to an internal line, there would be no contribution at \(\flatfrac{1}{\tensor{p}{^\rho}\tensor{q}{_\rho}}\) order, because the internal line would not be on-shell, and thus 
        \begin{equation}
            (p-q)^2 + m^2 \not\approx \tensor{p}{^\rho}\tensor{q}{_\rho}.
        \end{equation}
        For our purposes, these contributions can be safely ignored. \cref{eq: soft-graviton-theorem-in,eq: soft-graviton-theorem-out} are known as the Weinberg soft graviton theorem. While I derived it for scalar particles, it also holds for higher spins \cite{weinberg1965InfraredPhotonsGravitons,weinberg1995Foundations}.

        We can now get back to \cref{eq: virtual-soft-corrections}. Using the soft graviton theorem, \textcite{weinberg1965InfraredPhotonsGravitons,weinberg1995Foundations} showed that the scattering amplitudes with and without soft gravitons behave as
        \begin{equation}\label{eq: virtual-IR-divergence}
            \abs{\mel{\text{out}}{S}{\text{in}}}^2 = \abs{\mel{\text{out}}{S^0}{\text{in}}}^2 \qty(\frac{\lambda}{\Lambda})^B,
        \end{equation}
        with \(B>0\) a constant. Recall now that the physical limit is \(\lambda \to 0\), since this is necessary to remove the regulator we inserted when defining which soft gravitons we were considering. Hence, this result means that, once the effects of soft gravitons are considered, all processes have zero probability of happening. All sorts of time evolution or propagation are forbidden. This is known as an \gls{IR} divergence, and it indicates we are missing something.

        Let us add more soft gravitons to our scattering problem. We now consider
        \begin{multline}
            \mel{\text{out}'}{S}{\text{in}} \, = \,\vcenter{\hbox{\begin{tikzpicture}
            \begin{feynman}
                \node[blob] (c) at (0,0) {$\boldsymbol{V}$};
                \vertex (a1) at (180:2);
                \vertex (a2) at (135:2);
                \vertex (a3) at (225:2);
                \vertex (b1) at (30:2);
                \vertex (b2) at (330:2);
                \diagram*{
                    (a1) -- [fermion] (c),
                    (a2) -- [fermion] (c),
                    (a3) -- [fermion] (c),
                    (c) -- [fermion] (b1),
                    (c) -- [fermion] (b2),
                };
            \end{feynman}
        \end{tikzpicture}}} + \vcenter{\hbox{\begin{tikzpicture}
            \begin{feynman}
                \node[blob] (c) at (0,0) {$\boldsymbol{V}$};
                \vertex (a1) at (180:2);
                \vertex (a2) at (135:2);
                \vertex (a3) at (225:2);
                \vertex (a4) at (310:1.5);
                \vertex (a5) at (225:1);
                \vertex (b1) at (30:2);
                \vertex (b2) at (330:2);
                \diagram*{
                    (a4) -- [graviton] (a5),
                    (a1) -- [fermion] (c),
                    (a2) -- [fermion] (c),
                    (a3) -- [fermion] (c),
                    (c) -- [fermion] (b1),
                    (c) -- [fermion] (b2),
                };
            \end{feynman}
        \end{tikzpicture}}} \\ + \vcenter{\hbox{\begin{tikzpicture}
            \begin{feynman}
                \node[blob] (c) at (0,0) {$\boldsymbol{V}$};
                \vertex (a1) at (180:2);
                \vertex (a2) at (135:2);
                \vertex (a3) at (225:2);
                \vertex (a4) at (310:1.5);
                \vertex (a5) at (225:1);
                \vertex (a6) at (5:2.5);
                \vertex (b1) at (30:2);
                \vertex (b2) at (330:2);
                \vertex (b3) at (30:0.8);
                \diagram*{
                    (a4) -- [graviton] (a5),
                    (a6) -- [graviton] (b3),
                    (a1) -- [fermion] (c),
                    (a2) -- [fermion] (c),
                    (a3) -- [fermion] (c),
                    (c) -- [fermion] (b1),
                    (c) -- [fermion] (b2),
                };
            \end{feynman}
        \end{tikzpicture}}} + \vcenter{\hbox{\begin{tikzpicture}
            \begin{feynman}
                \node[blob] (c) at (0,0) {$\boldsymbol{V}$};
                \vertex (a1) at (180:2);
                \vertex (a2) at (135:2);
                \vertex (a3) at (225:2);
                \vertex (a4) at (310:1.5);
                \vertex (a5) at (225:1);
                \vertex (a6) at (5:2.5);
                \vertex (b1) at (30:2);
                \vertex (b2) at (330:2);
                \vertex (b3) at (30:0.8);
                \vertex (b4) at (355:1.8);
                \vertex (b5) at (330:1);
                \diagram*{
                    (b4) -- [graviton] (b5),
                    (a4) -- [graviton] (a5),
                    (a6) -- [graviton] (b3),
                    (a1) -- [fermion] (c),
                    (a2) -- [fermion] (c),
                    (a3) -- [fermion] (c),
                    (c) -- [fermion] (b1),
                    (c) -- [fermion] (b2),
                };
            \end{feynman}
        \end{tikzpicture}}} + \cdots.
        \end{multline}
        As before, each diagram represents all possible diagrams with, say, one outgoing graviton. The ``V'' in the blobs indicate that the blob represents the diagram with all the virtual contributions due to soft gravitons. In other words, the ``V'' blob encapsulates all of the infinitely many diagrams of \cref{eq: virtual-soft-corrections}. The gravitons we are considering this time all have momenta with \(\lambda \leq \norm{\vb{p}}\), but the imposition on the maximum momentum is done over the collection of gravitons, not on each one individually. We require that the total energy carried out by the real gravitons emitted in the process is equal to or less than some value \(E\). This is motivated by the fact that infinitely many gravitons with small energy could amount to a large energy lost in the scattering process, and we want to avoid this scenario. Finally, notice that the outgoing state is no longer \(\ket{\text{out}}\), but \(\ket{\text{out}'}\). The prime indicates the state \(\ket{\text{out}}\) is now accompanied by an arbitrary number of soft gravitons. 

        For completeness, let me write the amplitude for scattering with emission of one extra soft graviton. This is given by
        \begin{equation}\label{eq: one-emitted-graviton}
            \mel{\text{out}+1}{S}{\text{in}} = \sqrt{8\pi}  \qty(\sum_{\substack{i\\\text{out}}} \frac{\tensor*{p}{^i_\mu}\tensor*{p}{^i_\nu}}{\tensor*{p}{_i^\rho}\tensor{q}{_\rho}} - \sum_{\substack{j\\\text{in}}} \frac{\tensor*{p}{^j_\mu}\tensor*{p}{^j_\nu}}{\tensor*{p}{_j^\rho}\tensor{q}{_\rho}}) \tensor{\epsilon}{^\mu^\nu}(q)\mel{\text{out}}{S}{\text{in}}  + \order{\tensor{q}{^0}},
        \end{equation}
        where \(\tensor{\epsilon}{^\mu^\nu}(q)\) is the polarization tensor of the graviton with momentum \(\tensor{q}{^\mu}\). Due to our choice of gauge, the polarization tensor must satisfy \(\tensor{\epsilon}{^\mu^\nu}(q)\tensor{\eta}{_\mu_\nu} = 0\) and \(\tensor{q}{_\mu}\tensor{\epsilon}{^\mu^\nu}(q) = 0\), \ie, it must be transverse and traceless. As a consequence, we can rewrite \cref{eq: one-emitted-graviton} as
        \begin{equation}\label{eq: one-emitted-graviton-TT}
            \mel{\text{out}+1}{S}{\text{in}} = \sqrt{8\pi}  \qty(\sum_{\substack{i\\\text{out}}} \frac{\tensor*{p}{^i_\mu}\tensor*{p}{^i_\nu}}{\tensor*{p}{_i^\rho}\tensor{q}{_\rho}} - \sum_{\substack{j\\\text{in}}} \frac{\tensor*{p}{^j_\mu}\tensor*{p}{^j_\nu}}{\tensor*{p}{_j^\rho}\tensor{q}{_\rho}})^{\text{TT}} \tensor{\epsilon}{^\mu^\nu}(q)\mel{\text{out}}{S}{\text{in}} + \order{\tensor{q}{^0}}.
        \end{equation}

        Using these results, \textcite{weinberg1965InfraredPhotonsGravitons,weinberg1995Foundations} showed that 
        \begin{equation}\label{eq: real-IR-divergence}
            \abs{\mel{\text{out}'}{S}{\text{in}}}^2 = \abs{\mel{\text{out}'}{S}{\text{in}}}^2 b(B) \qty(\frac{E}{\lambda})^B,
        \end{equation}
        for the same positive constant \(B\) of \cref{eq: virtual-IR-divergence} and for some real function \(b(x)\). Using both \cref{eq: virtual-IR-divergence,eq: real-IR-divergence}, we find the final result
        \begin{equation}
            \abs{\mel{\text{out}'}{S}{\text{in}}}^2 = \abs{\mel{\text{out}}{S^0}{\text{in}}}^2 b(B) \qty(\frac{E}{\Lambda})^B,
        \end{equation}
        which is now well-defined and finite in the \(\lambda \to 0\) limit. 

        The lesson to be taken is that in any scattering process an infinite number of soft gravitons will be emitted. This is imposed by the \gls{IR} structure of \gls{GR} and required for the scattering probability to be non-vanishing. 

        The natural question at this point is: what does this have to do with the \gls{BMS} group? 

        Despite both the Weinberg soft theorem and the \gls{BMS} group being discovered in the 1960's, the relation between them would only become clear in the 2010's. \textcite{strominger2014BMSInvarianceGravitational,he2015BMSSupertranslationsWeinberg} noticed that the Weinberg soft graviton theorem is actually a consequence of \gls{BMS} invariance of gravitational scattering. More specifically, the Weinberg soft graviton theorem is equivalent to the Ward identity (\ref{eq: ward-identity-supertranslations}), as we shall show. The upshot is that the Weinberg soft graviton theorem and the supertranslations in the \gls{BMS} group are two related aspects of the \gls{IR} structure of \gls{GR}. 

        To push this analogy further, one can notice that the charges \(Q^{\pm}_f\) act non-trivially on the Minkowski vacuum. This is expected, since we saw in \cref{subsec: SSB-supertranslations} that the \gls{GR} vacuum is not supertranslation-invariant. The action of these supertranslation charges is to create a soft graviton \cite{strominger2014BMSInvarianceGravitational,strominger2018LecturesInfraredStructure}, meaning we can understand the soft gravitons as the \glspl{NGB} associated to the spontaneous symmetry breaking of supertranslation symmetry. 

        This sheds light on the nature of the different vacua in gravity. A choice of vacuum in quantum gravity will determine what is the ``absence of soft gravitons''. Other vacua are then interpreted as having various numbers of soft gravitons. The number of soft gravitons is to be understood only in relational terms, as any of these vacua are equally allowed to be taken as ``the'' vacuum.

        Weinberg predicted that in any scattering experiment there will be the emission of soft gravitons. This is, therefore, a vacuum transition, and it is a quantum point of view of the classical vacuum transitions we found in \cref{subsec: physical-realization-supertranslations}. 

        Let us then prove this connection is real. I follow the discussion of the original paper by \textcite{he2015BMSSupertranslationsWeinberg}. Notice the Weinberg soft graviton theorem was formulated in terms of creation and annihilation operators for the graviton field \(\tensor{h}{_{\mu}_\nu}\), but the Ward identity on \cref{eq: ward-identity-supertranslations-localized} is written in terms of the Bondi news tensor \(\tensor{N}{_A_B}\). Therefore, we need to convert from one to the other. 

        We start by considering the quantization of the gravitational field (see Sec. 2 of Ref. \citeonline{basile2024LecturesQuantumGravity} for an introduction to perturbative quantum gravity). Near null infinity, it can be taken to be approximately free. Therefore, in Cartesian coordinates, we can make an expansion in creation and annihilation operators according to 
        \begin{equation}\label{eq: h-mu-nu-ladder-operators}
            \tensor*{\hat{h}}{^{\text{out}}_\mu_\nu}(x) = \frac{1}{(2\pi)^{3}}\sum_{\lambda = +, -} \int \qty[\tensor*{\epsilon}{^{\lambda}^*_\mu_\nu}(\vb{q}) \hat{a}^{\text{out}}_{\lambda}(\vb{q}) e^{i q \cdot x} + \tensor*{\epsilon}{^{\lambda}_\mu_\nu}(\vb{q}) \hat{a}^{\text{out}}_{\lambda}(\vb{q})^\dagger e^{-i q \cdot x}] \frac{\dd[3]{q}}{2 \omega_{\vb{q}}}.
        \end{equation}
        Notice \(\lambda\) runs over the two polarizations of the graviton. In this convention, the commutation relations for the ladder operators are
        \begin{equation}\label{eq: CCR-weinberg-ward}
            \comm{\hat{a}^{\text{out}}_{\lambda}(\vb{q})}{\hat{a}^{\text{out}}_{\mu}(\vb{p})^\dagger} = (2 \omega_{\vb{q}}) (2\pi)^3\delta_{\lambda,\mu} \delta^{(3)}\qty(\vb{q}-\vb{p}).
        \end{equation}
        This is the same convention we implicitly assumed when deriving the Weinberg soft graviton theorem: an outgoing state with one extra graviton is obtained by acting with \(\hat{a}^{\text{out}}_{\lambda}(\vb{q})\) on the outgoing bra (or with \(\hat{a}^{\text{out}}_{\lambda}(\vb{q})^\dagger\) on the outgoing ket). Furthermore, \cref{eq: h-mu-nu-ladder-operators} implies the polarization of an outgoing graviton is given by \(\tensor{\epsilon}{_\mu_\nu}\), not \(\tensor*{\epsilon}{^*_\mu_\nu}\).

        We will slowly change coordinates to retarded stereographic coordinates. The first step is to rewrite the null vectors \(\tensor{q}{^a}\) in stereographic coordinates. If \(\vb{q}\) points to the point in the sphere with stereographic coordinate \(\chi\), we get the Cartesian components
        \begin{equation}\label{eq: null-vector-from-stereographic}
            \tensor{q}{^\mu} = \frac{\omega_{\vb{q}}}{1 + \chi \bar{\chi}}\qty(1 + \chi \bar{\chi},\chi + \bar{\chi},-i(\chi - \bar{\chi}),-1 + \chi \bar{\chi}).
        \end{equation}
        This expression comes from \cref{eq: Cartesian-from-stereographic}.

        This is useful for us to write the polarization tensors. The polarization tensors must be symmetric, transverse (\(\tensor*{\epsilon}{^{\pm}_{\mu}_{\nu}}(\vb{q})\tensor{q}{^\mu} = 0\)) and traceless (\(\tensor*{\epsilon}{^{\pm}_{\mu}_{\nu}}(\vb{q})\tensor{\eta}{^\mu^\nu} = 0\)). This can be achieved by defining \(\tensor*{\epsilon}{^{\pm}_{\mu}_{\nu}}(\vb{q}) = \tensor*{\epsilon}{^{\pm}_{\mu}}(\vb{q})\tensor*{\epsilon}{^{\pm}_{\mu}}(\vb{q})\), where the polarization vectors \(\tensor*{\epsilon}{^{\pm}_{\mu}}(\vb{q})\) are the same ones used in electrodynamics and can be taken to be 
        \begin{subequations}\label{eq: polarization-stereographic}
            \begin{align}
                \tensor*{\epsilon}{^{+}^{\mu}}(\vb{q}) = \frac{1}{\sqrt{2}}\qty(\bar{\chi},1,- i, \bar{\chi}), \\
                \intertext{and}
                \tensor*{\epsilon}{^{-}^{\mu}}(\vb{q}) = \frac{1}{\sqrt{2}}\qty(\chi,1,+i, \chi).
            \end{align}
        \end{subequations}
        where \(\tensor{q}{^\mu}\) and the stereographic coordinate \(\chi\) are related by \cref{eq: null-vector-from-stereographic}.

        We can now write the shear tensor \(\tensor{C}{_A_B}\) as 
        \begin{equation}
            \tensor{C}{_A_B} = \sqrt{32 \pi} \lim_{r \to + \infty} \pdv{\tensor{x}{^\mu}}{\tensor{x}{^A}} \pdv{\tensor{x}{^\nu}}{\tensor{x}{^B}} \frac{\tensor*{\hat{h}}{^{\text{out}}_\mu_\nu}(u,r,\zeta,\bar{\zeta})}{r}.
        \end{equation}
        In terms of creation and annihilation operators we find
        \begin{equation}\label{eq: CAB-from-hmunu-weinberg-ward}
            \tensor{C}{_A_B}(u,\vu{x}) =  \frac{\sqrt{32 \pi}}{(2\pi)^3} \lim_{r \to + \infty} \pdv{\tensor{x}{^\mu}}{\tensor{x}{^A}} \pdv{\tensor{x}{^\nu}}{\tensor{x}{^B}} \frac{1}{r}\sum_{\lambda = +, -} \int \qty[\tensor*{\epsilon}{^{\lambda}^*_\mu_\nu}(\vb{q}) \hat{a}^{\text{out}}_{\lambda}(\vb{q}) e^{- i \omega_{\vb{q}} u - i \omega_{\vb{q}} r (1 - \vu{q} \vdot \vu{x})} + \Hc] \frac{\dd[3]{q}}{2 \omega_{\vb{q}}},
        \end{equation}
        with \(\Hc\) denoting the Hermitian conjugate.

        We can compute the angular integral in \cref{eq: CAB-from-hmunu-weinberg-ward} to leading order in \(r\) by employing the stationary phase approximation---see Appendix \ref{app: stationary-phase-approximation-sphere} and, in particular, \cref{eq: stationary-phase-sphere}. We find
        \begin{multline}\label{eq: CAB-from-hmunu-weinberg-ward-post-integral}
            \tensor{C}{_A_B}(u,\vu{x}) = - \frac{i \sqrt{32 \pi}}{8\pi^2} \lim_{r \to + \infty} \pdv{\tensor{x}{^\mu}}{\tensor{x}{^A}} \pdv{\tensor{x}{^\nu}}{\tensor{x}{^B}} \frac{1}{r^2}\sum_{\lambda = +, -} \int \left[\tensor*{\epsilon}{^{\lambda}^*_\mu_\nu}(\vb{x}) \hat{a}^{\text{out}}_{\lambda}(\omega\vu{x}) e^{- i \omega u} \right. \\ \left. + \tensor*{\epsilon}{^{\lambda}^*_\mu_\nu}(-\vb{x}) \hat{a}^{\text{out}}_{\lambda}(-\omega\vu{x}) e^{- i \omega u - 2 i \omega r} - \Hc\right] \dd{\omega},
        \end{multline}
        I flipped the sign on \(\Hc\) because we pulled out a factor of \(i\) from inside the integral. The antipodal modes on the second line will not survive up to \(\nullfut\): the exponential behaves as \(e^{-i\omega v}\) and oscillates too much. A stationary phase approximation will ensure the term vanishes at future null infinity (see Theorem 7.7.1 in Ref. \citeonline{hormander2003DistributionTheoryFourier}). We thus find 
        \begin{equation}\label{eq: CAB-from-hmunu-weinberg-ward-post-post-integral}
            \tensor{C}{_A_B}(u,\vu{x}) = - \frac{i \sqrt{32 \pi}}{8\pi^2} \lim_{r \to + \infty} \pdv{\tensor{x}{^\mu}}{\tensor{x}{^A}} \pdv{\tensor{x}{^\nu}}{\tensor{x}{^B}} \frac{1}{r^2}\sum_{\lambda = +, -} \int \left[\tensor*{\epsilon}{^{\lambda}^*_\mu_\nu}(\vb{x}) \hat{a}^{\text{out}}_{\lambda}(\omega\vu{x}) e^{- i \omega u} - \Hc\right] \dd{\omega},
        \end{equation}

        Since \(\tensor{C}{_A_B}\) is real and traceless, the component \(\tensor{C}{_{\zeta}_{\zeta}}\) is sufficient to describe the tensor in stereographic coordinates (\(\tensor{C}{_{\zeta}_{\bar{\zeta}}} = 0\) due to the traceless condition and the other components are obtained by conjugation).

        Using the expressions for the Cartesian coordinates in terms of stereographic coordinates (these were given in \cref{eq: def-stereographic,eq: Cartesian-from-stereographic}), we find with the help of \cref{eq: polarization-stereographic} that
        \begin{subequations}\label{eq: polarization-tensors-stereographic}
            \begin{align}
                \tensor*{\epsilon}{^{+}_\zeta_\zeta}(\chi,\bar{\chi}) &= \pdv{\tensor{x}{^\mu}}{\zeta} \pdv{\tensor{x}{^\nu}}{\zeta} \tensor*{\epsilon}{^{+}_\mu_\nu} = \frac{2 r^2 \bar{\zeta}^2 (\bar{\zeta} - \bar{\chi})^2}{(1 + \zeta \bar{\zeta})^4}, \\
                \intertext{and}
                \tensor*{\epsilon}{^{-}_\zeta_\zeta}(\chi,\bar{\chi}) &= \pdv{\tensor{x}{^\mu}}{\zeta} \pdv{\tensor{x}{^\nu}}{\zeta} \tensor*{\epsilon}{^{-}_\mu_\nu} = \frac{2 r^2 (1 + \bar{\zeta} \chi)^2}{(1 + \zeta \bar{\zeta})^4}.
            \end{align}
            In spite of these expressions, it is important to notice that the definitions of the polarization tensors imply that \(\tensor*{\epsilon}{^{\pm}^*_\mu_\nu} = \tensor*{\epsilon}{^{\mp}_\mu_\nu}\), and hence
            \begin{align}
                \tensor*{\epsilon}{^{+}^*_\zeta_\zeta}(\chi,\bar{\chi}) &= \pdv{\tensor{x}{^\mu}}{\zeta} \pdv{\tensor{x}{^\nu}}{\zeta} \tensor*{\epsilon}{^{-}_\mu_\nu} = \frac{2 r^2 (1 + \bar{\zeta} \chi)^2}{(1 + \zeta \bar{\zeta})^4}, \\
                \intertext{and}
                \tensor*{\epsilon}{^{-}^*_\zeta_\zeta}(\chi,\bar{\chi}) &= \pdv{\tensor{x}{^\mu}}{\zeta} \pdv{\tensor{x}{^\nu}}{\zeta} \tensor*{\epsilon}{^{+}_\mu_\nu} = \frac{2 r^2 \bar{\zeta}^2 (\bar{\zeta} - \bar{\chi})^2}{(1 + \zeta \bar{\zeta})^4}.
            \end{align}
        \end{subequations}
        
        At \(\vu{q} = \vu{x}\) we get \(\zeta = \chi\) (and \(\bar{\zeta} = \bar{\chi}\)). Hence, \cref{eq: CAB-from-hmunu-weinberg-ward-post-post-integral,eq: polarization-tensors-stereographic} now yield
        \begin{equation}\label{eq: Czz-from-hmunu-weinberg-ward-final}
            \tensor{C}{_{\zeta}_{\zeta}}(u,{\zeta},\bar{\zeta}) = \frac{-i\sqrt{32 \pi}}{4\pi^2 (1 + \zeta\bar{\zeta})^2} \int \left[\hat{a}^{\text{out}}_{+}(\omega,\zeta,\bar{\zeta}) e^{- i \omega u} - \hat{a}^{\text{out}}_{-}(\omega,\zeta,\bar{\zeta})^{\dagger} e^{+ i \omega u}\right] \dd{\omega}.
        \end{equation}

        We now have the shear tensor. The next step is to get to the Bondi news tensor. A time derivative yields
        \begin{equation}\label{eq: Nzz-from-hmunu-weinberg-ward-1}
            \tensor*{N}{^{\text{out}}_{\zeta}_{\zeta}}(u,{\zeta},\bar{\zeta}) = \frac{-\sqrt{32 \pi}}{4\pi^2 (1 + \zeta\bar{\zeta})^2} \int_0^{+\infty} \omega \left[\hat{a}^{\text{out}}_{+}(\omega,\zeta,\bar{\zeta}) e^{- i \omega u} + \hat{a}^{\text{out}}_{-}(\omega,\zeta,\bar{\zeta})^{\dagger} e^{+ i \omega u}\right] \dd{\omega},
        \end{equation}
        where the ``out'' superscript indicates this is the Bondi news tensor at \(\nullfut\) (as opposed to \(\nullpas\)). 

        Notice that
        \begin{equation}\label{eq: Nzz-from-hmunu-weinberg-ward-2}
            \int \tensor*{N}{^{\text{out}}_{\zeta}_{\zeta}}(u,{\zeta},\bar{\zeta}) \dd{u} = \frac{-\sqrt{32 \pi}}{4\pi (1 + \zeta\bar{\zeta})^2} \lim_{\omega \to 0^+}\qty[\omega \hat{a}^{\text{out}}_{+}(\omega,\zeta,\bar{\zeta}) + \omega\hat{a}^{\text{out}}_{-}(\omega,\zeta,\bar{\zeta})^{\dagger}],
        \end{equation}
        where we used that the \(\omega = 0\) Fourier mode is the average of the limits \(\omega \to 0^+\) and \(\omega \to 0^-\). This is the definition used in Ref. \citeonline{he2015BMSSupertranslationsWeinberg} and it is necessary because the integral on \cref{eq: Nzz-from-hmunu-weinberg-ward-1} only runs over half of the real line.

        To meet with \cref{eq: ward-identity-supertranslations-localized}, we now consider the object
        \begin{equation}
            \mel**{\text{out}}{\int \tensor*{N}{^{\text{out}}_{\zeta}_\zeta}\qty(u,\zeta,\bar{\zeta}) \dd{u} S}{\text{in}} = \frac{-\sqrt{32 \pi}}{4\pi (1 + \zeta\bar{\zeta})^2} \lim_{\omega \to 0^+} \omega \mel**{\text{out}}{\hat{a}^{\text{out}}_{+}(\omega,\zeta,\bar{\zeta}) S}{\text{in}}. 
        \end{equation}
        In the previous expressions, we used the fact that \(\hat{a}^{\text{out}}_{-}(\omega,\zeta)^{\dagger}\) annihilates the outgoing state for \(\omega \to 0\). If \(\ket{\text{out}}\) contains soft gravitons, this still holds, because our definition of annihilation operators multiplies the state by a factor of \(\omega^{\frac{1}{2}}\) (see \cref{eq: CCR-weinberg-ward}), which will lead to annihilation in the \(\omega \to 0\) limit.
        
        Similarly we find that
        \begin{equation}
            \mel**{\text{out}}{S \int \tensor*{N}{^{\text{in}}_{\zeta}_\zeta}\qty(v,\zeta,\bar{\zeta}) \dd{v}}{\text{in}} = \frac{-\sqrt{32 \pi}}{4\pi (1 + \zeta\bar{\zeta})^2} \lim_{\omega \to 0^+} \omega \mel**{\text{out}}{S \hat{a}^{\text{in}}_{-}(\omega,\zeta,\bar{\zeta})}{\text{in}}. 
        \end{equation}
        Diagramatically we can see that an incoming graviton with negative helicity with energy \(\omega\) coming in from the direction \(\zeta\) (as defined on \(\nullpas\)) contributes to the overall amplitude in precisely the same way as an outgoing graviton with positive helicity, energy \(\omega\), and puncturing \(\nullfut\) at \(\zeta\). Therefore, we may write
        \begin{multline}
            \mel**{\text{out}}{\int \tensor*{N}{^{\text{out}}_{\zeta}_\zeta}\qty(u,\zeta,\bar{\zeta}) \dd{u} S + S \int \tensor*{N}{^{\text{in}}_{\zeta}_\zeta}\qty(v,\zeta,\bar{\zeta}) \dd{v}}{\text{in}} \\ = \frac{-\sqrt{32 \pi}}{2\pi (1 + \zeta\bar{\zeta})^2} \lim_{\omega \to 0^+} \omega \mel**{\text{out}}{\hat{a}^{\text{out}}_{+}(\omega,\zeta,\bar{\zeta}) S}{\text{in}}. 
        \end{multline}
        The \gls{LHS} of this expression is one covariant derivative away from the \gls{LHS} of \cref{eq: ward-identity-supertranslations-localized}
        
        To simplify the \gls{RHS}, we apply Weinberg's soft graviton theorem as stated on \cref{eq: one-emitted-graviton}. We get
        \begin{multline}
            \mel**{\text{out}}{\int \tensor*{N}{^{\text{out}}_{\zeta}_\zeta}\qty(u,\zeta,\bar{\zeta}) \dd{u} S + S \int \tensor*{N}{^{\text{in}}_{\zeta}_\zeta}\qty(v,\zeta,\bar{\zeta}) \dd{v}}{\text{in}} \\ = \frac{-8}{(1 + \zeta\bar{\zeta})^2} \qty(\sum_{\substack{i\\\text{out}}} \frac{\tensor*{p}{^i_\mu}\tensor*{p}{^i_\nu}}{\tensor*{p}{_i^\rho}\tensor{k}{_\rho}} - \sum_{\substack{j\\\text{in}}} \frac{\tensor*{p}{^j_\mu}\tensor*{p}{^j_\nu}}{\tensor*{p}{_j^\rho}\tensor{k}{_\rho}}) \tensor{\epsilon}{^+^\mu^\nu}(\zeta,\bar{\zeta}) \mel**{\text{out}}{S}{\text{in}}, 
        \end{multline}
        where \(\tensor{k}{^\mu} = (1, \vu{x})\) and \(\vu{x}\) is the unit vector pointing to \(\zeta\) in the sphere. Since we know that \(\tensor{\epsilon}{^+^\mu^\nu}(\zeta) = \tensor{\epsilon}{^+^\mu}(\zeta) \tensor{\epsilon}{^+^\nu}(\zeta)\), we can write
        \begin{multline}
            \mel**{\text{out}}{\int \tensor*{N}{^{\text{out}}_{\zeta}_\zeta}\qty(u,\zeta,\bar{\zeta}) \dd{u} S + S \int \tensor*{N}{^{\text{in}}_{\zeta}_\zeta}\qty(v,\zeta,\bar{\zeta}) \dd{v}}{\text{in}} \\ = \frac{-8}{(1 + \zeta\bar{\zeta})^2} \qty(\sum_{\substack{i\\\text{out}}} \frac{[\tensor*{p}{^i} \cdot\tensor{\epsilon}{^+}(\zeta,\bar{\zeta})]^2}{\tensor*{p}{^i}\cdot k} - \sum_{\substack{j\\\text{in}}} \frac{[\tensor*{p}{^j} \cdot\tensor{\epsilon}{^+}(\zeta,\bar{\zeta})]^2}{\tensor*{p}{^j}\cdot k}) \mel**{\text{out}}{S}{\text{in}}.
        \end{multline}

        To compute this expression, we write the Cartesian components of the vectors in terms of the steoreographic angles according to\footnote{\cref{eq: outgoing-cartesian-stereographic-weinberg-ward} actually considers the outgoing case, but the incoming case is similar.}
        \begin{subequations}\label{eq: outgoing-cartesian-stereographic-weinberg-ward}
            \begin{align}
                \tensor{k}{^\mu} &= \frac{1}{1 + \zeta \bar{\zeta}}\qty(1 + \zeta \bar{\zeta},\zeta + \bar{\zeta},-i(\zeta - \bar{\zeta}),-1 + \zeta \bar{\zeta}), \\
                \tensor*{p}{_i^\mu} &= \frac{E_i}{1 + \zeta_i \bar{\zeta}_i}\qty(1 + \zeta_i \bar{\zeta}_i,\zeta_i + \bar{\zeta}_i,-i(\zeta_i - \bar{\zeta}_i),-1 + \zeta_i \bar{\zeta}_i), \\
                \tensor*{\epsilon}{^{+}^{\mu}}(\zeta,\bar{\zeta}) &= \frac{1}{\sqrt{2}}\qty(\bar{\zeta},1,-i,\bar{\zeta}).
            \end{align}
        \end{subequations}
        These expressions all follow from \cref{eq: polarization-stereographic,eq: null-vector-from-stereographic}. \(\zeta_i\) denote the points in the celestial sphere punctured by the particles with momenta \(\tensor*{p}{_i^\mu}\). \(E_i\) denote the energy carried by each of these particles. Using these expressions we find
        \begin{multline}
            \mel**{\text{out}}{\int \tensor*{N}{^{\text{out}}_{\zeta}_\zeta}\qty(u,\zeta,\bar{\zeta}) \dd{u} S + S \int \tensor*{N}{^{\text{in}}_{\zeta}_\zeta}\qty(v,\zeta,\bar{\zeta}) \dd{v}}{\text{in}} \\ = \frac{8}{(1 + \zeta\bar{\zeta})} \qty(\sum_{\substack{i\\\text{out}}} \frac{E_i^{\text{out}} (\bar{\zeta} - \bar{\zeta}_i^{\text{out}})}{(\zeta - \zeta_i^{\text{out}})(1 + \zeta_i^{\text{out}} \bar{\zeta}_i^{\text{out}})} - \sum_{\substack{j\\\text{in}}} \frac{E_j^{\text{in}} (\bar{\zeta} - \bar{\zeta}_j^{\text{in}})}{(\zeta - \zeta_j^{\text{in}})(1 + \zeta_j^{\text{in}} \bar{\zeta}_j^{\text{in}})}) \mel**{\text{out}}{S}{\text{in}}.
        \end{multline}

        One can check that
        \begin{equation}
            \tensor{D}{^A}\tensor{N}{_\zeta_A} = \tensor{\gamma}{^{\zeta}^{\bar{\zeta}}}\tensor{\partial}{_{\bar{\zeta}}}\tensor{N}{_\zeta_\zeta}.
        \end{equation}
        Using this expression, we conclude that
        \begin{multline}
            \mel**{\text{out}}{\int \tensor{D}{^A}\tensor*{N}{^{\text{out}}_{\zeta}_A}\qty(u,\zeta,\bar{\zeta}) \dd{u} S + S \int \tensor{D}{^A}\tensor*{N}{^{\text{in}}_{\zeta}_A}\qty(v,\zeta,\bar{\zeta}) \dd{v}}{\text{in}} \\ = 4 \qty(\sum_{\substack{i\\\text{out}}} \frac{E_i^{\text{out}}}{\zeta - \zeta_i^{\text{out}}} - \sum_{\substack{j\\\text{in}}} \frac{E_j^{\text{in}}}{\zeta - \zeta_j^{\text{in}}}) \mel**{\text{out}}{S}{\text{in}} \\ + 4 \qty(\sum_{\substack{i\\\text{out}}} \frac{E_i^{\text{out}} \bar{\zeta}_i^{\text{out}}}{1 + \zeta_i^{\text{out}} \bar{\zeta}_i^{\text{out}}} - \sum_{\substack{j\\\text{in}}} \frac{E_j^{\text{in}} \bar{\zeta}_j^{\text{in}}}{1 + \zeta_j^{\text{in}} \bar{\zeta}_j^{\text{in}}}) \mel**{\text{out}}{S}{\text{in}}.
        \end{multline}
        Momentum conservation forces the last line to vanish. We thus find that the Weinberg soft graviton theorem can be written as 
        \begin{multline}
            \mel**{\text{out}}{\int \tensor{D}{^A}\tensor*{N}{^{\text{out}}_{\zeta}_A}\qty(u,\zeta,\bar{\zeta}) \dd{u} S + S \int \tensor{D}{^A}\tensor*{N}{^{\text{in}}_{\zeta}_A}\qty(v,\zeta,\bar{\zeta}) \dd{v}}{\text{in}} \\ = 4 \qty(\sum_{\substack{i\\\text{out}}} \frac{E_i^{\text{out}}}{\zeta - \zeta_i^{\text{out}}} - \sum_{\substack{j\\\text{in}}} \frac{E_j^{\text{in}}}{\zeta - \zeta_j^{\text{in}}}) \mel**{\text{out}}{S}{\text{in}}.
        \end{multline}
        Reverting the calculations allows us to reobtain the soft theorem from this expression. Furthermore, this expression is identical to \cref{eq: ward-identity-supertranslations-localized}, which in turn is equivalent to the Ward identity for supertranslations. We conclude that the Weinberg soft graviton theorem is indeed the Ward identity for supertranslations.

    \subsection{Gravitational Memory Effect}\label{subsec: gravitational-memory-effect}
        The \gls{GW} displacement memory effect \cite{zeldovich1974RadiationGravitationalWaves,braginsky1985KinematicResonanceMemory,braginsky1987GravitationalwaveBurstsMemory,christodoulou1991NonlinearNatureGravitation,thorne1992GravitationalwaveBurstsMemory,blanchet1992HereditaryEffectsGravitational} was originally discovered in linearized gravity by \textcite{zeldovich1974RadiationGravitationalWaves}. It consists on the prediction that the passage of a \gls{GW} permanently displaces the relative positions of two nearby inertial detectors stationed near infinity. It turns out that this effect is too a reflection of \gls{BMS} transformations. Let us then derive the memory effect and see how this correspondence occurs. I will work in the framework of linearized gravity, but there are more powerful results in full \gls{GR}. I follow \textcite{braginsky1987GravitationalwaveBurstsMemory,bieri2024GravitationalWaveDisplacement}.

        We consider two nearby inertial detectors, initially at rest relative to each other, stationed ``near infinity''. This means that when we perform a large-\(r\) (or small-\(l\)) expansion we can neglect the subleading terms. Write the separation between these two detectors as \(\tensor{d}{^i} + \tensor{\Delta x}{^i}(t)\), where \(\tensor{\Delta x}{^i}(0) = 0\) and we assume \(\tensor{\Delta x}{^i}(t)\) is small compared to \(\tensor{d}{^i}\). In this case, the acceleration of the relative position between the two detectors is described by the geodesic deviation equation, given by \cite{wald1984GeneralRelativity}
        \begin{equation}
            \dv[2]{\tensor{\Delta x}{^j}}{t} = - \tensor{R}{_t_i_t^j}\tensor{d}{^i}.
        \end{equation}
        Our task is then to compute \(\tensor{R}{_t_i_t^j}\). To linear order in the perturbation \(\tensor{h}{_\mu_\nu} = \tensor{g}{_\mu_\nu} - \tensor{\eta}{_\mu_\nu}\) and in the \gls{TT} gauge, the expression is given by \cite{carroll2019SpacetimeGeometryIntroduction}
        \begin{equation}
            \tensor{R}{_t_i_t_j} = - \frac{1}{2}\tensor{\partial}{_t}\tensor{\partial}{_t}\tensor*{h}{^{\text{TT}}_i_j},
        \end{equation}
        where the \glsxtrshort{TT} superscript indicates the \gls{TT} projection---see Appendix \ref{app: linearized-gravity}. We thus conclude the geodesic deviation due to the passage of a \gls{GW} is, in the linearized theory,
        \begin{equation}
            \dv[2]{\tensor{\Delta x}{^j}}{t} = \frac{\tensor{d}{^i}}{2} \tensor{\partial}{_t}\tensor{\partial}{_t}\tensor*{h}{^{\text{TT}}_i_j}.
        \end{equation}

        Consider now the following physical scenario: in a Minkowski background, a number of massive bodies scatter off each other. All the bodies are asymptotically inertial and we count only bound states as ``bodies'' (\eg, two black holes orbiting each other count as a single body). This was originally considered in this context by \textcite{braginsky1987GravitationalwaveBurstsMemory}, and the key question is: what is the total deviation \(\tensor{\Delta x}{_j}\) after all interactions have settled down? In other words, if \(\tensor{\Delta x}{_j} = 0\) long before the massive bodies interact, what will be \(\tensor{\Delta x}{_j}\) long after their interaction? 

        We can then integrate the geodesic deviation equation twice to get\footnote{There is no occurrence of an additional velocity (which would occur due to the integration constant in the first integration) because we assumed the detectors are initially at rest relative to each other.}
        \begin{subequations}\label{eq: memory-from-perturbation}
            \begin{align}
                \tensor{\Delta x}{_j} &= \frac{\tensor{d}{^i}}{2} \eval{\tensor*{h}{^{\text{TT}}_i_j}}^{+\infty}_{-\infty}, \\
                &= \frac{\tensor{d}{^i}}{2} \tensor*{\Delta h}{^{\text{TT}}_i_j},
            \end{align}
        \end{subequations}
        where \(\tensor*{\Delta h}{^{\text{TT}}_i_j}\) is the overall change in the gravitational field. Let us compute it.

        In this scenario, the stress tensor at early times is given by\footnote{Completely analogous expressions hold at late times. I will focus on a single case for simplicity, and consider both cases at the end.} \cite{poisson2011MotionPointParticles}
        \begin{equation}
            \tensor*{T}{^{\text{in}}_a_b}(x) = \sum_{\substack{p \\ \text{in}}} \int \tensor*{p}{^{(p)}_a} \tensor*{u}{^{(p)}_b} \delta^{(4)}\qty(x - \tensor{z}{^{(p)}}(\tau)) \dd{\tau},
        \end{equation}
        where \(\tensor*{p}{^{(p)}_a}\) is the four-momentum of the \(p\)-th massive body, \(\tensor*{u}{^{(p)}_a}\) its four-velocity, and \(\tensor{z}{^{(p)}}(\tau)\) its worldline as a function of its proper time. Since the bodies are moving nearly inertially at early times, we can write
        \begin{equation}
            \tensor*{z}{_{(p)}^0}(\tau) = \gamma_p \tau \qq{and} \vb{z}_{(p)} = \gamma_p \vb{v}_p \tau,
        \end{equation}
        where \(\vb{v}_p\) is the three-velocity of the \(p\)-th body and \(\gamma_p^{-2} = 1 - \norm{\vb{v}_p}^2\).

        Using these expressions for the worldline, we may write
        \begin{subequations}
            \begin{align}
                \tensor*{T}{^{\text{in}}_a_b}(t,\vb{x}) &= \sum_{\substack{p \\ \text{in}}} \int \tensor*{p}{^{(p)}_a} \tensor*{u}{^{(p)}_b} \delta\qty(t - \gamma_p \tau) \delta^{(3)}\qty(\vb{x} - \gamma_p \vb{v}_p \tau) \dd{\tau}, \\
                &= \sum_{\substack{p \\ \text{in}}} \frac{\tensor*{p}{^{(p)}_a} \tensor*{u}{^{(p)}_b}}{\gamma_p} \delta^{(3)}\qty(\vb{x} - t\vb{v}_p).
            \end{align}
        \end{subequations}
        
        Consider now \cref{eq: hmunu-linear-taylor-multipole}, which yields the gravitational field far away from the source in terms of the stress tensor. We find that, at early times, in de Donder gauge, and in Cartesian components, 
        \begin{subequations}
            \begin{align}
                \tensor*{\bar{h}}{^{\text{in}}_\mu_\nu}(t,\vb{x}) &= \frac{4}{r} \sum_{\substack{p \\ \text{in}}} \frac{\tensor*{p}{^{(p)}_\mu} \tensor*{u}{^{(p)}_\nu}}{\gamma_p} \sum_{n=0}^{+\infty}  \frac{1}{n!} \left\lfloor\dv[n]{t} \int \delta^{(3)}\qty(\vb{y} - t\vb{v}_p) \qty(\vu{x} \vdot \vb{y})^n \dd[3]{y}\right\rfloor + \order{\frac{1}{r^2}}, \\
                &= \frac{4}{r} \sum_{\substack{p \\ \text{in}}} \frac{\tensor*{p}{^{(p)}_\mu} \tensor*{u}{^{(p)}_\nu}}{\gamma_p} \sum_{n=0}^{+\infty}  \frac{1}{n!} \left\lfloor\dv[n]{t} t^n \qty(\vu{x} \vdot \vb{v}_p)^n\right\rfloor + \order{\frac{1}{r^2}}, \\
                &= \frac{4}{r} \sum_{\substack{p \\ \text{in}}} \frac{\tensor*{p}{^{(p)}_\mu} \tensor*{u}{^{(p)}_\nu}}{\gamma_p} \sum_{n=0}^{+\infty} \qty(\vu{x} \vdot \vb{v}_p)^n + \order{\frac{1}{r^2}}, \\
                &= \frac{4}{r} \sum_{\substack{p \\ \text{in}}} \frac{\tensor*{p}{^{(p)}_\mu} \tensor*{u}{^{(p)}_\nu}}{\gamma_p\qty(1 - \vu{x} \vdot \vb{v}_p)} + \order{\frac{1}{r^2}}.
            \end{align}
        \end{subequations}
        Above, as in Appendix \ref{app: linearized-gravity}, \(\lfloor\cdot\rfloor\) denotes evaluation at retarded time \(u=t-r\), but this turned out to be irrelevant. 

        Introduce the null vector \(\tensor{k}{^\mu} = (1, \vu{x})\), which points from the source to the observation point. In terms of this null vector we find that
        \begin{equation}
            \tensor*{\bar{h}}{^{\text{in}}_\mu_\nu}(t,\vb{x}) = -\frac{4}{r} \sum_{\substack{p \\ \text{in}}} \frac{\tensor*{p}{^{(p)}_\mu} \tensor*{u}{^{(p)}_\nu}}{\tensor{k}{_\rho} \tensor*{u}{^{(p)}^\rho}} + \order{\frac{1}{r^2}}.
        \end{equation}

        At last, we find that the change in the gravitational field between early and late times is given, in \gls{TT} gauge, by
        \begin{equation}\label{eq: braginsky-thorne}
            \tensor*{\Delta h}{^{\text{TT}}_i_j}(\vb{x}) = -\frac{4}{r} \qty[\sum_{\substack{p \\ \text{out}}} \frac{\tensor*{p}{^{(p)}_i} \tensor*{p}{^{(p)}_j}}{\tensor{k}{_\rho} \tensor*{p}{^{(p)}^\rho}} - \sum_{\substack{q \\ \text{in}}} \frac{\tensor*{p}{^{(q)}_i} \tensor*{p}{^{(q)}_j}}{\tensor{k}{_\rho} \tensor*{p}{^{(q)}^\rho}}]^{\text{TT}} + \order{\frac{1}{r^2}},
        \end{equation}
        where we used \(\tensor*{p}{^{(q)}^a} = m_q \tensor*{u}{^{(q)}^a}\) to get an expression in terms of momenta only. We could get rid of the negative sign by changing the definition of \(\tensor{k}{^\mu}\) so that it is past-directed and points from observation point to the source (as originally done by \textcite{braginsky1987GravitationalwaveBurstsMemory}), but our definition finds a better physical interpretation in terms of the momentum of the gravitons emitted by the source. 

        \cref{eq: braginsky-thorne} is known as the Braginsky--Thorne formula for the linear memory. Applying it to \cref{eq: memory-from-perturbation} allows us to obtain the physical deviation in the two inertial detectors near infinity. We learn that, in a general scattering problem, there will be a permanent displacement between the inertial detectors stationed at infinity due to the passage of a \gls{GW}.
        
        To see an example of how the memory contribution may look like, let us choose a specific model for the source. A common model for \gls{GW} generation is a binary in a circular orbit, but this would be a single bound state and there would be no change in the second derivative of the quadrupole moment, and hence it will not present (linear, ordinary) memory. Hence, let us choose instead an unbounded problem: two massive bodies in a hyperbolic orbit. An advantage of this model is that we can also compute the metric perturbation during the interaction, and hence see how the metric undergoes a permanent shift. We assume their velocities to be non-relativistic, so the Braginsky--Thorne formula may be approximated by
        \begin{equation}\label{eq: braginsky-thorne-non-relativistic}
            \tensor*{\Delta h}{^{\text{TT}}_i_j}(t,\vb{x}) = \frac{4}{r} \qty[\sum_{\substack{p \\ \text{out}}} m_p \tensor*{v}{^{(p)}_i} \tensor*{v}{^{(p)}_j} - \sum_{\substack{q \\ \text{in}}} m_q \tensor*{v}{^{(q)}_i} \tensor*{v}{^{(q)}_j}]^{\text{TT}} + \order{\frac{1}{r^2}}.
        \end{equation}
        Notice that, in fact, each of these terms corresponds to the second derivative of the quadrupole moment of the source, since 
        \begin{equation}
            \tensor{\ddot{Q}}{_i_j} = 2\sum_{p} m_p \tensor*{v}{^{(p)}_i} \tensor*{v}{^{(p)}_j}.
        \end{equation}
        In fact, \cref{eq: braginsky-thorne-non-relativistic} can be obtained directly by working in the quadrupole approximation \cite{bieri2024GravitationalWaveDisplacement}. Notice that, in terms of the quadrupole moment, we find
        \begin{equation}
            \tensor*{\Delta h}{^{\text{TT}}_i_j}(t,\vb{x}) = \frac{2}{r} \tensor*{\Delta \ddot{Q}}{^{\text{TT}}_i_j}(t-r) + \order{\frac{1}{r^2}}.
        \end{equation}

        Consider then two bodies with masses \(m_1\) and \(m_2\) and position vectors \(\vb{r}_1\) and \(\vb{r}_2\). We assume the center of mass to be at rest at the origin. It is convenient to introduce the notation
        \begin{equation}
            \vb{r} = \vb{r}_1 - \vb{r}_2\qc \mu = \frac{m_1 m_2}{m_1 + m_2}, \qq{and} \alpha = G m_1 m_2,
        \end{equation}
        which is inverted by
        \begin{equation}
            \vb{r}_1 = \frac{\mu \vb{r}}{m_1}, \qq{and} \vb{r}_2 = - \frac{\mu \vb{r}}{m_2}.
        \end{equation}
        Then, following \cite{landau1976Mechanics}, we know that the Newtonian solution for \(\vb{r}(t)\) is given parametrically as
        \begin{subequations}
            \begin{align}
                t &= \frac{\alpha}{2 E}\sqrt{\frac{\mu}{2 E}} \qty(e \sinh\xi - \xi), \\
                x &= \frac{\alpha}{2 E} \qty(e - \cosh\xi), \\
                y &= \frac{\alpha}{2 E} \sqrt{e^2 - 1}\sinh\xi,
            \end{align}
        \end{subequations}
        where \(E\) is the energy and \(e > 1\) is the orbit's eccentricity. Using these results, one can find the quadrupole moment. It is given by 
        \begin{equation}
            \tensor{Q}{_i_j} = \frac{\alpha^2 \mu}{4 E^2}\mqty((e - \cosh\xi)^2 & \sqrt{e^2-1}\sinh\xi (e - \cosh\xi) & 0 \\ \sqrt{e^2-1}\sinh\xi (e - \cosh\xi) & (e^2 - 1)\sinh^2\xi & 0 \\ 0 & 0 & 0).
        \end{equation}
        We still need to project to the \gls{TT} components. Once we do this by using \(\tensor{\Lambda}{_i_j^k^l}(\vu{z})\), we get to 
        \begin{equation}
            \tensor*{Q}{^{\text{TT}}_i_j} = \frac{\alpha^2 \mu}{8 E^2}\mqty((e - \cosh\xi)^2 - (e^2 - 1)\sinh^2\xi & 2\sqrt{e^2-1}\sinh\xi (e - \cosh\xi) & 0 \\ 2\sqrt{e^2-1}\sinh\xi (e - \cosh\xi) & -(e - \cosh\xi)^2 + (e^2 - 1)\sinh^2\xi & 0 \\ 0 & 0 & 0).
        \end{equation}

        The next step would be differentiating \(\tensor*{Q}{^{\text{TT}}_i_j}\) with respect to time. Since we only have \(\tensor*{Q}{^{\text{TT}}_i_j}\) as a function of \(\xi\), and \(t\) as a function of \(\xi\), we need to write 
        \begin{subequations}
            \begin{align}
                \dv[2]{\tensor*{Q}{^{\text{TT}}_i_j}}{t} &= \dv{\xi}{t}\dv{\xi}\qty(\dv{\tensor*{Q}{^{\text{TT}}_i_j}}{t}), \\
                &= \dv{\xi}{t}\dv{\xi}\qty(\dv{\xi}{t}\dv{t}\dv{\tensor*{Q}{^{\text{TT}}_i_j}}{\xi}), \\
                &= \qty(\dv{t}{\xi})^{-1}\dv{\xi}\qty(\qty(\dv{t}{\xi})^{-1}\dv{t}\dv{\tensor*{Q}{^{\text{TT}}_i_j}}{\xi}).
            \end{align}
        \end{subequations}

        The resulting components are 
        \begin{subequations}
            \begin{align}
                \tensor*{\ddot{Q}}{^{\text{TT}}_x_x}(t) &= - \tensor*{\ddot{Q}}{^{\text{TT}}_y_y}(t) = \frac{2 E \qty[2(e^2 - 1) \sinh^2\xi - \cosh\xi (e \cosh\xi -1)(e + (e^2 -2)\cosh\xi)]}{(e \cosh\xi - 1)^3}, \\
                \intertext{and}
                \tensor*{\ddot{Q}}{^{\text{TT}}_x_y}(t) &=  \tensor*{\ddot{Q}}{^{\text{TT}}_y_x}(t) = \frac{2 E \sqrt{e^2-1}\sinh\xi \qty(4 \cosh\xi - e (3 + \cosh(2\xi)))}{(e \cosh\xi - 1)^3},
            \end{align}
        \end{subequations}
        where \(\xi = \xi(t)\).

\begin{figure}
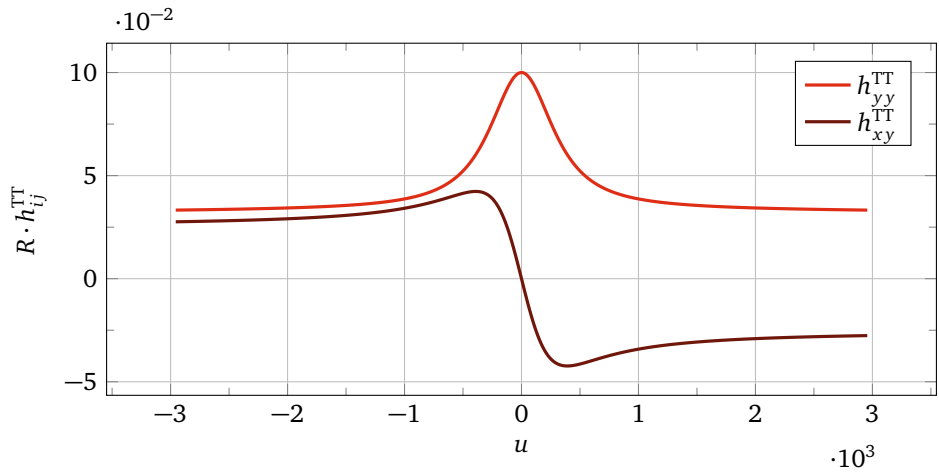

            \centering
            \includestandalone{memory-gw}
            \caption{Time-dependent gravitational field in the \gls{TT} gauge for a pair of masses with the same mass \(m_1=m_2=\num{1}\), Newtonian energy \(E = \num{0.01}\) and impact parameter \(b=\num{100}\). \(R\) is the distance from the sources to the observation point (assumed to be perpendicular to the plane of motion of the sources). \(u\) is the retarded time. Since we are using \(G=c=1\) units, all values are to be compared to the scales set by the masses \(m_1=m_2\). In the plot, notice how after the masses interact in the neighborhood of \(u=0\) there is a lasting change in the gravitational field (specifically in the \(\tensor*{h}{^{\text{TT}}_x_y}\) component).}
            \label{fig: memory-kepler}
        \end{figure}

        With these results at hand, we can make a parametric plot of each component of \(\tensor*{h}{^{\text{TT}}_i_j}\) against time \(t(\xi)\) (or, more precisely, retarded time). This is shown in \cref{fig: memory-kepler}, which clearly exhibits that the values of \(\tensor*{h}{^{\text{TT}}_i_j}\) at early and late times differ by a finite amount.

        Notice that a finite difference between the gravitational field at late and early times is exactly what we found in \cref{subsec: physical-realization-supertranslations} when considering the emission of null radiation to null infinity. Incidentally, the scattering process we considered this time emits null radiation to null infinity in the form of \glspl{GW}. Therefore, the memory effect we are observing in this derivation is a manifestation of a physical realization of a supertranslation. We had foreshadowed this connection by referring to the physical realization of a supertranslation as a ``state with memory''. This discussion now justifies the name ``memory field'' for the \(C\) field defined in \cref{eq: shear-decomposition}.

        Above, I described the linear memory effect, originally discussed by \textcite{zeldovich1974RadiationGravitationalWaves,braginsky1985KinematicResonanceMemory,braginsky1987GravitationalwaveBurstsMemory}. Nevertheless, the memory effect is a much deeper effect. \textcite{christodoulou1991NonlinearNatureGravitation}, and independently \textcite{blanchet1992HereditaryEffectsGravitational}, noticed that the memory effect also has a nonlinear counterpart. \textcite{thorne1992GravitationalwaveBurstsMemory} soon interpreted this counterpart as being the linear memory due to the emission of gravitons by the matter sources. Indeed, consider \cref{eq: memory-field-from-null-radiation,eq: BS-EFE-uu}. We see that the emission of \glspl{GW} to infinity also induces a memory shift. In a black hole binary merger, for example, we get a bound state, and hence the quadrupole will not change significantly at linear order. However, the \glspl{GW} emitted by the binary during the merger can be considered as bodies moving asymptotically inertially. One may then compute the linear memory related to these emitted \glspl{GW} and find the nonlinear contribution. This can be understood as the meaning behind the Bondi news contribution to \(\tensor{\mathcal{T}}{_u_u}\) occurring in \cref{eq: memory-field-from-null-radiation}.

        A great interest in the memory effect is due to the fact it can be measured in principle, and hence provides an experimental test of the \gls{IR} structure of \gls{GR}. Furthermore, since black hole binaries involve nonlinear memory, it is also a test of the nonlinear nature of Einstein gravity. Current \gls{GW} detectors, such as \glsxtrshort{LIGO}, Virgo, and \glsxtrshort{KAGRA}, cannot detect memory in individual events, the reason being they are not sensitive to the very small frequencies associated to the effect. However, it is possible to combine the measurements of many different events and obtain a coherent result, but a large number of events is necessary. Future experiments, such as \glsxtrshort{LISA}, may be able to detect memory in individual events. For details on measuring memory, see, for example, \cite{favata2010GravitationalwaveMemoryEffect,grant2023OutlookDetectingGravitationalwave}.

        For further discussion about the memory effect, see the excellent discussion by \textcite{bieri2024GravitationalWaveDisplacement}, which also discusses the relation between nonlinear memory and \gls{CK} spacetimes.

    \subsection{The \texorpdfstring{\Glsfmtlong{IR}}{Infrared} Triangle}\label{subsec: infrared-triangle}
        In \cref{subsec: weinberg-soft-graviton} we saw that supertranslation symmetry and the Weinberg soft graviton theorem are two aspects of the same physical phenomenon. We also reached a similar conclusion concerning supertranslations and the memory effect on \cref{subsec: gravitational-memory-effect}. This naturally raises a question: are the soft graviton theorem and the memory effect related? If so, how? 

        They are indeed related, and the connection was given by \textcite{strominger2016GravitationalMemoryBMS}. It is merely a Fourier transform and a change of notation and conventions: we need to write the expressions in position space (even though the Weinberg soft theorem is written in momentum space) and account for the different normalizations \(\tensor{g}{_\mu_\nu} = \tensor{\eta}{_\mu_\nu} + \tensor{h}{_\mu_\nu}\) (which we used in the memory effect) and \(\tensor{g}{_\mu_\nu} = \tensor{\eta}{_\mu_\nu} + \sqrt{32\pi}\tensor{\hat{h}}{_\mu_\nu}\) (which we used in the Weinberg soft graviton theorem). 

        The expectation value for the graviton field fluctuation sourced by the \(\ket{\text{in}}\) to \(\ket{\text{out}}\) transition is given by
        \begin{subequations}
            \begin{align}
                \tensor*{h}{^{\text{TT}}_\mu_\nu}(t,\vb{x}) &= \frac{\sqrt{32\pi}}{\mel{\text{out}}{S}{\text{in}}}\qty[\mel**{\text{out}}{\tensor*{\hat{h}}{^{\text{out}}_\mu_\nu}(t,\vb{x}) S}{\text{in}} + \mel**{\text{out}}{S \tensor*{\hat{h}}{^{\text{in}}_\mu_\nu}(t,\vb{x})}{\text{in}}], \\
                &= \frac{\sqrt{32\pi}}{(2\pi)^3\mel{\text{out}}{S}{\text{in}}} \sum_{\lambda} \int \mel**{\text{out}}{\tensor*{\epsilon}{^{\lambda}^*_\mu_\nu}(\vb{q}) \hat{a}^{\text{out}}_\lambda(\vb{q}) e^{i q \cdot x} S + \tensor*{\epsilon}{^{\lambda}_\mu_\nu}(\vb{q}) S \hat{a}^{\text{in}}_\lambda(\vb{q})^\dagger e^{-i q \cdot x}}{\text{in}} \frac{\dd[3]{q}}{2 \omega_{\vb{q}}}.
            \end{align}
        \end{subequations}
        I dropped the Hermitian conjugate terms that would annihilate the states they are being applied to at the end of the calculation. 
        
        We are interested in the large \(r\) limit, and we know from \cref{subsec: weinberg-soft-graviton} that a stationary phase approximation yields
        \begin{multline}\label{eq: memory-from-soft-theorem-omega-integral}
            \tensor*{h}{^{\text{TT}}_\mu_\nu}(u,r\vu{x}) = \frac{-i\sqrt{32\pi}}{2(2\pi)^2 r \mel{\text{out}}{S}{\text{in}}} \\ \times \sum_{\lambda} \int_0^{+\infty} \mel**{\text{out}}{\tensor*{\epsilon}{^{\lambda}^*_\mu_\nu}(\vu{x}) \hat{a}^{\text{out}}_\lambda(\omega \vu{x}) e^{- i \omega u} S - \tensor*{\epsilon}{^{\lambda}_\mu_\nu}(-\vu{x}) S \hat{a}^{\text{in}}_\lambda(-\omega \vu{x})^\dagger e^{+ i \omega u}}{\text{in}} \dd{\omega},
        \end{multline}
        with \(u = t - r\) and \(r = \norm{\vb{x}}\).

        Next we can write 
        \begin{subequations}
            \begin{align}
                \tensor*{\Delta h}{^{\text{TT}}_\mu_\nu}(r\vu{x}) &= \eval{\tensor{h}{_\mu_\nu}(u,r\vu{x})}_{-\infty}^{+\infty}, \\
                &= \int_{-\infty}^{+\infty} \tensor{\partial}{_u}\tensor{h}{_\mu_\nu}(u,r\vu{x}) \dd{u}, \\
                &= \frac{-\sqrt{32\pi}}{8\pi r \mel{\text{out}}{S}{\text{in}}} \notag \\ &\qquad\qquad \times \sum_{\lambda} \lim_{\omega\to0^+}\qty[\mel**{\text{out}}{\omega\tensor*{\epsilon}{^{\lambda}^*_\mu_\nu}(\vu{x}) \hat{a}^{\text{out}}_\lambda(\omega \vu{x}) S + \omega \tensor*{\epsilon}{^{\lambda}_\mu_\nu}(-\vu{x}) S \hat{a}^{\text{in}}_\lambda(-\omega \vu{x})^\dagger}{\text{in}}], \\
                &= \frac{-\sqrt{32\pi}}{4\pi r \mel{\text{out}}{S}{\text{in}}} \sum_{\lambda} \lim_{\omega\to0^+}\qty[\omega\tensor*{\epsilon}{^{\lambda}^*_\mu_\nu}(\vu{x}) \mel**{\text{out}}{\hat{a}^{\text{out}}_\lambda(\omega \vu{x}) S}{\text{in}}],
            \end{align}
        \end{subequations}
        where I once again accounted for the factor of \(2\) that we get for the \(\omega = 0\) Fourier mode. This is related to the fact that the integral over \(\omega\) on \cref{eq: memory-from-soft-theorem-omega-integral} only runs over half the real line.

        Using the Weinberg soft graviton theorem as stated on \cref{eq: one-emitted-graviton-TT} we now get that
        \begin{subequations}\label{eq: braginsky-thorne-from-weinberg}
            \begin{align}
                \tensor*{\Delta h}{^{\text{TT}}_\mu_\nu}(r\vu{x}) &= \frac{-4}{r} \sum_{\lambda} \tensor*{\epsilon}{^{\lambda}^*_\mu_\nu}(\vu{x}) \qty(\sum_{\substack{i\\\text{out}}} \frac{\tensor*{p}{^i_\alpha}\tensor*{p}{^i_\beta}}{\tensor*{p}{_i^\rho}\tensor{k}{_\rho}} - \sum_{\substack{j\\\text{in}}} \frac{\tensor*{p}{^j_\alpha}\tensor*{p}{^j_\beta}}{\tensor*{p}{_j^\rho}\tensor{k}{_\rho}})^{\text{TT}} \tensor*{\epsilon}{^{\lambda}^\alpha^\beta}(\vu{x}), \\
                &= \frac{-4}{r} \qty(\sum_{\substack{i\\\text{out}}} \frac{\tensor*{p}{^i_\mu}\tensor*{p}{^i_\nu}}{\tensor*{p}{_i^\rho}\tensor{k}{_\rho}} - \sum_{\substack{j\\\text{in}}} \frac{\tensor*{p}{^j_\mu}\tensor*{p}{^j_\nu}}{\tensor*{p}{_j^\rho}\tensor{k}{_\rho}})^{\text{TT}},
            \end{align}
        \end{subequations}
        where we wrote \(\tensor{k}{^\mu} = (1,\vu{x})\) and we used the fact that \(\sum_{\lambda} \tensor*{\epsilon}{^{\lambda}^*_\mu_\nu}(\vu{x}) \tensor*{\epsilon}{^{\lambda}^\alpha^\beta}(\vu{x})\) is an identity on the space of \gls{TT} symmetric tensors. Notice \cref{eq: braginsky-thorne-from-weinberg} is precisely the Braginsky--Thorne formula given on \cref{eq: braginsky-thorne}.

        The first point to notice is that the memory effect is associated with very small frequencies. This is intuitive, since it consists of a build-up along a long period of time. For this reason, it is common to borrow the term ``\gls{DC}'' from electromagnetism and say that the memory effect is a ``\gls{DC} effect'', \ie, a zero-frequency effect.

        It may come as a surprise that these two processes are related. After all, Braginsky and Thorne were scattering stars and black holes, while Weinberg was scattering elementary particles. Yet, they recover the same result. This happens because, in the deep \gls{IR} limit, elementary particles and black holes all look the same. They are treated as pointlike massive particles, and end up having the very same \gls{IR} structure.

\begin{figure}
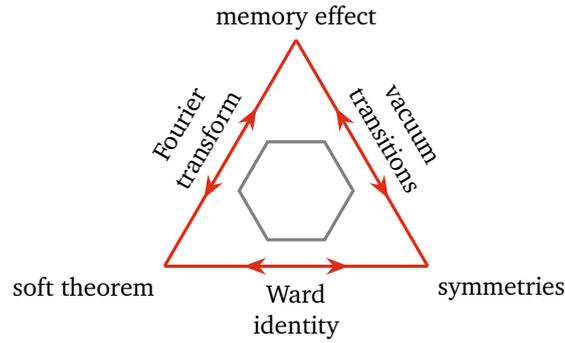

            \centering
            \includestandalone{ir-triangle}
            \caption{There is a triangular relation between three topics in \gls{IR} physics: asymptotic symmetries (such as supertranslations), memory effects (such as the \gls{GW} displacement memory effect), and soft theorems (such as Weinberg's soft graviton theorem). Understanding one corner of the triangle helps understanding the remaining ones.}
            \label{fig: IR-triangle}
        \end{figure}

        With this result at hand, we have found a triangular relation between supertranslations, the Weinberg soft graviton theorem, and the memory effect, with the links between these topics being given by Ward identities, a Fourier transform, and vacuum transitions. This is known as the \gls{IR} triangle or \gls{PSZ} triangle \citeonline{pasterski2016NewGravitationalMemories} and it is illustrated in \cref{fig: IR-triangle}. 

        Perhaps the most interesting feature of the \gls{IR} triangle is the fact that it is not restricted to the discussions we made. As put by Strominger \cite{strominger2018LecturesInfraredStructure}, the triangle echoes throughout gravity and gauge theory and presents itself in many different forms. For instance, in quantum electrodynamics there are analogs of supertranslations known as large gauge symmetries \cite{he2014NewSymmetriesMassless}, there is a soft photon theorem \cite{weinberg1965InfraredPhotonsGravitons}, and there is an electromagnetic memory effect \cite{bieri2013ElectromagneticAnalogGravitational}---and these three phenomena enjoy a triangular relation identical to that of \cref{fig: IR-triangle} \cite{pasterski2017AsymptoticSymmetriesElectromagnetic,strominger2018LecturesInfraredStructure}. We will discuss this in more depth on \cref{sec: electrodynamics}. It is also possible to discuss similar results in non-Abelian gauge theory and at subleading orders. The subleading \gls{IR} triangle in gravity latter leads us to the notion of superrotations, discussed on \cref{sec: superrotations}.

        This relation is extremely useful. It teaches us that we can learn about one corner of the triangle by studying the other ones. In particular, one can discuss \gls{IR} physics by starting from any particular corner of the \gls{IR} triangle. While I chose to start with \gls{BMS} symmetries, it would be equally possible to start with the Weinberg soft graviton theorem or with the gravitational memory effect.

    \subsection{Construction of Hadamard States}\label{subsec: hadamard-states-asymptotically-flat}
        There are four basic forces in our universe: gravity, electromagnetism, and the strong and weak nuclear forces. While gravity is well-described by \gls{GR}, the \gls{SM} provides a good description of the remaining three \cite{schwartz2014QuantumFieldTheory,langacker2017StandardModel,donoghue2023DynamicsStandardModel}. However, while \gls{GR} is a classical field theory, the \gls{SM} is a \gls{QFT} constructed on flat spacetime. Thus, in principle, \gls{GR} cannot explain how the fields of the \gls{SM} fall because classical \gls{GR} knows nothing about how to deal with quantum fields. Similarly, the \gls{SM} knows nothing about gravity, because it is typically constructed on flat spacetime. 

        While a full description of gravity at the quantum level would require a knowledge of a full theory of quantum gravity, which at the present is only a dream, we can make do with an approximation. \Gls{QFTCS} is the description of how quantum fields evolve upon a curved background spacetime. In this approximation, the background spacetime is assumed to be fixed, so that the quantum fields do not affect the background geometry. Nevertheless, the curved structure of the spacetime can affect the evolution of the quantum fields.

        An interesting approach to \gls{QFT} that is particularly useful in the curved spacetime formulation is known as the algebraic approach---see, \eg, Refs. \citeonline{aguiaralves2023NonperturbativeAspectsQuantum,dappiaggi2017HadamardStatesLightlike,fewster2020AlgebraicQuantumField,hollands2015QuantumFieldsCurved,khavkine2015AlgebraicQFTCurved,wald1994QuantumFieldTheory}. In this approach, one takes a dual view of the theory. One part of the theory is described by the algebra of observables, while the other is the space of states. Let us start with the algebra of observables.
        
        The algebra of observables is the collection of all observables to be considered in the theory and, as the name suggests, it has the structure of an algebra, which is a vector space with a product between vectors satisfying some conditions. On a generic globally hyperbolic spacetime\footnote{Global hyperbolicity is a causality condition. It essentially means the spacetime admits the formulation of initial value problems for fields propagating on it. This notion is well-discussed in Refs. \citeonline{hawking1973LargeScaleStructure,wald1984GeneralRelativity,wald1994QuantumFieldTheory}.} the basic elements of the algebra of observables for a free scalar field are the smeared field operators
        \begin{equation}
            \phi(f) = \int \phi(x) f(x) \sqrt{-g} \dd[4]{x},
        \end{equation}
        where \(f\) is some smooth and compactly supported function on the spacetime. We get the full algebra of observables by taking products of these operators---such as \(\phi(f)\phi(g)\), or \(\phi(f)\phi(g)\phi(h)\)---and by considering the identity as well. This is known as the free algebra generated by \(\qty{\phi(f) \mid f \in \ck[\infty]_0(M)}\). Finally, we impose the algebra respects what we would expect of a (free) quantum field. Namely, 
        \begin{enumerate}
            \item \(\phi(f + \alpha g) = \phi(f) + \alpha \phi(g)\) for \(f, g \in \ck[\infty]_0(M)\) and \(\alpha \in \Comp\), 
            \item \(\phi(f)^\dagger = \phi(\bar{f})\),
            \item \(\phi(P f) = 0\) (where \(P\) is the Klein--Gordon operator,
            \item \(\comm{\phi(f)}{\phi(g)} = i E(f,g) \Eins\), where \(E\) is the advanced Green function minus the retarded Green function. 
        \end{enumerate}
        These conditions are meant to ensure we are dealing with smeared fields, that the field satisfies the equation of motion (recall this is a free theory), and that it obeys the \gls{CCR}. See Refs.  \citeonline{aguiaralves2023NonperturbativeAspectsQuantum,dappiaggi2017HadamardStatesLightlike,fewster2020AlgebraicQuantumField,hollands2015QuantumFieldsCurved,khavkine2015AlgebraicQFTCurved,wald1994QuantumFieldTheory} for further details. The algebra thus described, which I will denote by \(\mathcal{A}_{\text{bulk}}\), is the algebra in the bulk of the asymptotically flat spacetime.
        
        Next we need to discuss states on this algebra. States are defined as normalized and positive linear functionals on the algebra, \ie, they are linear maps \(\omega \colon \mathcal{A}_{\text{bulk}} \to \Comp\) with the properties that
        \begin{enumerate}
            \item \(\omega(A^* A) \geq 0\) for any \(A \in \mathcal{A}_{\text{bulk}}\), 
            \item \(\omega(\Eins) = 1\).
        \end{enumerate}
        These conditions are meant so that the state represents an expectation value. Namely, \(\omega(A)\) will be interpreted as the expectation value of the observable \(A\). An \(n\)-point function in the state \(\omega\) is then written as
        \begin{equation}
            \omega(\phi(f_1)\cdots\phi(f_n)) = \int \omega(\phi(x_1)\cdots\phi(x_n)) f_1(x_1) \cdots f_n(x_n) \sqrt{-g(x_1)}\dd[4]{x_1} \cdots \sqrt{-g(x_n)}\dd[4]{x_n}.
        \end{equation}

        Within \gls{AQFT}, the construction of the algebra of observables is usually the simplest step. For example, the construction of the algebra of observables for a Klein--Gordon field is essentially what we discussed above, up to nomenclature. However, even in the case of a non-interacting theory it is very hard to construct physically meaningful states. 

        By ``physically meaningful'' I mean that the state should allow for the expectation value of the stress-energy-momentum tensor to be well-defined. This is a non-trivial condition known as the Hadamard condition and is discussed in depth in Ref. \citeonline{khavkine2015AlgebraicQFTCurved}, for example. See also Sec. 2.B of Ref. \citeonline{aguiaralves2023NonperturbativeAspectsQuantum} for a pedagogical introduction.

        The reason for this difficulty can be traced back to our expression for the algebra. Notice we made sense of the object 
        \begin{equation}
            \phi(f)\phi(g) = \int \phi(x)\phi(y)f(x) g(y) \sqrt{-g(x)} \dd[4]{x} \sqrt{-g(y)} \dd[4]{y},
        \end{equation}
        and hence we can take products of fields. However, since these two fields are smeared separately, we cannot make sense of the object
        \begin{equation}
            \phi(x)^2,
        \end{equation}
        which involves the product of two fields at the same point in spacetime. Notice that the stress-energy tensor has the same difficulty, since it is quadratic in the fields. In particular, it is well-known that the expected value of the (\emph{not} normal-ordered) Hamiltonian in flat spacetime is infinite.
        
        The Hadamard condition allows us to perform a notion of normal-ordering with the stress tensor, even in curved spacetimes. However, not all possible states satisfy this condition. If a given state is not Hadamard, it cannot be used to understand the stress tensor of the quantum field, and in this sense it is not ``physically meaningful''.
        
        \textcite{dappiaggi2017HadamardStatesLightlike,dappiaggi2006RigorousStepsHolography,dappiaggi2011RigorousConstructionHadamard,moretti2006UniquenessTheoremBMSInvariant,moretti2008QuantumOutstatesHolographically} describe a method to construct physically meaningful states by exploiting the \gls{BMS} group. Namely, one does as follows. Assume a spacetime which is asymptotically flat at future null and timelike infinities. Notice this requires the existence of the timelike infinity, and they work on a spacetime admitting a conformal compactification. They construct an algebra of observables for a \gls{QFT} living on the boundary \(\nullfut \cup \qty{\timefut}\) of spacetime, \(\mathcal{A}_{\text{bound.}}\). Interestingly, one can prove (as they do) the existence of an injective map 
        \begin{equation}
            \imath \colon \mathcal{A}_{\text{bulk}} \to \mathcal{A}_{\text{bound.}}.
        \end{equation}
        Hence, all observables in the bulk can be understood by looking at the behavior on the boundary. This is quite similar to the popular holographic ideas in quantum gravity \cite{bousso2002HolographicPrinciple}.

        Suppose now one is given a state \(\omega_{\text{bound.}}\) on the boundary algebra \(\mathcal{A}_{\text{bound.}}\). We may then define a state \(\omega_{\text{bulk}} \colon \mathcal{A}_{\text{bulk}} \to \Comp\) by writing
        \begin{equation}
            \omega_{\text{bulk}}(A) = \imath_*\omega_{\text{bound.}}(A) \equiv  \omega_{\text{bound.}}(\imath(A)),
        \end{equation}
        where the notation \(\imath_*\) is reminiscent of the notion of pullback introduced in \cref{sec: symmetries-curved-spacetime}. Diagrammatically, we can write
        \begin{center}
            \includestandalone{pullback-state}
        \end{center}

        This is a very useful construction. In a general asymptotically flat spacetime, the bulk may not have enough symmetries for us to define a particularly interesting state. In Minkowski spacetime, for example we define the Minkowski vacuum (or Poincaré vacuum) as the unique Poincaré-invariant state. If the Poincaré group is not available, we may have an increased difficulty in identifying physically meaningful states. Nevertheless, in an asymptotically flat spacetimes we have an infinite-dimensional group of symmetries at the conformal boundary: the \gls{BMS} group. It is then possible to construct a state in the boundary algebra by demanding it to be the unique \gls{BMS}-invariant state. One then pulls this state back to the bulk algebra and the resulting bulk state turns out to be a Hadamard state. 

        The elementary idea is that one exploits the infinite-dimensional symmetry group at the boundary to define a state, which can then be pulled-back to the bulk. This construction and hypotheses are reviewed in detail in Ref. \citeonline{dappiaggi2017HadamardStatesLightlike}. It is also possible to generalize these ideas to de Sitter-like spacetimes, and we will revisit this on \cref{sec: infrared-de-sitter}. 

    \subsection{Soft Hair on Black Holes}
        Finally, we discuss some results concerning black holes obtained by \textcite{hawking2016SoftHairBlack,hawking2017SuperrotationChargeSupertranslation}.

        A known prediction in \gls{BHT} is the information loss puzzle. \textcite{hawking1974BlackHoleExplosions,hawking1975ParticleCreationBlack} originally noticed that the behavior of quantum fields on a spacetime containing a black hole formed by gravitational collapse leads the black hole to emit radiation. This radiation, now known as Hawking radiation, is approximately thermal with temperature inversely proportional to the black hole's mass (in the case of a Schwarzschild black hole). This means that the black hole loses mass to the radiation bath, and the temperature rises as the black hole becomes smaller. The increase in temperature implies the black hole loses mass even faster, until it eventually completely evaporates. 

        The complete evaporation argument assumes the calculations---which are performed using \gls{QFTCS}---to hold up to the last instants of evaporation. Hence, the argument cannot be fully trusted, since it is expected that full quantum gravity effects will eventually kick in and contribute significantly. This means the assumptions that go in \gls{QFTCS} are no longer valid---for example, the background can no longer be assumed to be classical or unaffected by the quantum fields---and thus the prediction is no longer trustworthy. Yet, it is interesting to consider the consequences of the evaporation process with the goal of obtaining more information about what may or not change in a full theory of quantum gravity. 

        One of the main puzzles is known as the information loss puzzle, or information loss paradox. It was noticed by Hawking \cite{hawking1976BreakdownPredictabilityGravitational} that if the quantum field starts in a pure state at very early times, it will be in a mixed state after the black hole evaporates. The basic idea is that the mixed state for the fields outside the black hole is a partial trace of the full pure state. However, once the black hole evaporates, the outer part is all that is left, and this is a mixed state. This means the time evolution of the quantum field state is not unitary. 

        This setup can be cast in a pictorial version. Consider two black holes formed by gravitational collapse. One of them is formed by the spherically symmetric collapse of a star made of an equal number of protons and electrons, possibly with a number of neutrons as well. This black hole has a given mass \(M\) (determined by the number of protons, neutrons, and electrons), electric charge \(Q = 0\) (because the electrons cancel out the protons), and spin \(J = 0\) (because the collapse is spherically symmetric). Similarly, consider the situation in which all particles are replaced by their antiparticles. In this case, the original stars are different in their compositions, but the resulting black holes have the same mass, electric charge, and angular momentum. 

        A famous result in black hole physics is the no-hair conjecture (see Ref. \citeonline{chrusciel2012StationaryBlackHoles} for a review). This conjecture states that all stationary black holes are diffeomorphic to the Kerr--Newman spacetime with the same mass, spin, and electric charge. Hence, the geometries of both black holes considered above are equal. It follows that the black holes must evaporate in the same way, and after their evaporation is completed there is no longer any information on the spacetime about which black hole was originally composed of antimatter.

        This conclusion is seen by many physicists as troublesome, but there is nothing problematic with it in principle \cite{unruh2017InformationLoss}. It does involve problematic arguments, though, due to the subtleties of the operator algebras describing the quantum fields and how entanglement works in \gls{QFT}. I will not discuss these details in here, but more information can be found in Refs. \citeonline{witten2018APSMedalExceptional,raju2022LessonsInformationParadox}.

        What is mainly of interest for us is noticing which sort of information can be preserved in black hole evaporation. The no-hair theorem states that stationary solutions are always diffeomorphic to the Kerr--Newman spacetime. Nevertheless, we can still consider the vacuum transitions of \cref{subsec: physical-realization-supertranslations}. Different vacua of \gls{GR} may be distinguished by the action of symmetries. In this point of view, we should distinguish two black holes which are boosted relative to each other. Indeed, thinking in this manner allows us to notice that the momentum in the spacetime is conserved, and hence allows us to obtain some information about the original matter that formed the black hole (its original net momentum). In this sense, we get eleven ``hairs'' on a black hole: one for each Poincaré symmetry and another for electric charge.

        We have seen, however, that supertranslations are also asymptotic symmetries of asymptotically flat spacetimes. \textcite{hawking2016SoftHairBlack,hawking2017SuperrotationChargeSupertranslation} noticed this (and its electromagnetic analog) and proposed that black holes should have infinitely many soft hairs, which are then associated to supertranslations and large gauge transformations (the electromagnetic analog of supertranslations). 
        
        In principle, it appears this could lead to more information conservation in black hole spacetimes. Nevertheless, it was then argued by \textcite{bousso2017SoftHairSoft} that the soft and hard modes actually decouple, meaning the soft hair on black holes cannot provide extra information on the ``hard'' aspects of black hole information loss.

\section{Electrodynamics}\label{sec: electrodynamics}
    While the main topic of these notes is gravitation, electrodynamics also enjoys a similar \gls{IR} triangle (\cref{subsec: infrared-triangle}): it has \gls{IR} symmetries \cite{he2014NewSymmetriesMassless}, a soft photon theorem \cite{weinberg1965InfraredPhotonsGravitons}, and a memory effect \cite{bieri2013ElectromagneticAnalogGravitational}. This section briefly discusses these ideas. The equivalences between these results are well-known and the proofs are similar to the gravitational case. Hence, I refer to Refs. \citeonline{he2014NewSymmetriesMassless,pasterski2017AsymptoticSymmetriesElectromagnetic,strominger2018LecturesInfraredStructure} for the discussions on the electromagnetic \gls{IR} triangle.

    \subsection{Large Gauge Transformations}
        We begin by describing the analysis due to \textcite{he2014NewSymmetriesMassless} that clarified the notion of asymptotic symmetries in electrodynamics. Following the same conventions as \textcite{strominger2018LecturesInfraredStructure}, we write the action for Maxwell's electromagnetism in Gaussian units as
        \begin{equation}
            S = - \frac{1}{16\pi} \int \tensor{F}{_a_b}\tensor{F}{_c_d}\tensor{g}{^a^c}\tensor{g}{^b^d} \sqrt{-g} \dd[4]{x} + S_M,
        \end{equation}
        where \(\tensor{F}{_a_b}\) is the Faraday tensor, \(\tensor{g}{_a_b}\) is the spacetime metric, and \(S_M\) stands for the action of the charged matter fields.

        At least locally, it is always possible to write the Faraday tensor in the form
        \begin{equation}
            \tensor{F}{_a_b} = 2\tensor{\nabla}{_[_a}\tensor{A}{_b_]}.
        \end{equation}
        In Minkowski spacetime and in the absence of magnetic monopoles, this is true globally. The ``electromagnetic field'' is the four-potential \(\tensor{A}{_a}\) up to gauge equivalence \cite{wald2022AdvancedClassicalElectromagnetism}. The equations of motion are 
        \begin{equation}
            \tensor{\nabla}{^a}\tensor{F}{_a_b} = 4\pi \tensor{j}{_b},
        \end{equation}
        where
        \begin{equation}
            \tensor{j}{_a} = - \fdv{S_M}{\tensor{A}{^a}}
        \end{equation}
        is the charge current for the coupled fields.

        The charge contained in spacetime is then given by 
        \begin{subequations}
            \begin{align}
                Q &= \frac{1}{4\pi} \oint_{\Sph^2} \tensor{\epsilon}{_a_b_c_d}\tensor{F}{^c^d}, \\
                &= \frac{1}{4\pi} \oint_{\Sph^2} \tensor{F}{_r_u} r^2 \sin\theta \dd{\theta}\dd{\phi}, \\
                &= \frac{1}{4\pi} \oint_{\Sph^2} r^2 \vb{E} \vdot \dd{\vb{S}},
            \end{align}
        \end{subequations}
        where the integral is understood over a sphere in the limit as its radius becomes infinite. This is just an application of Gauss's law in our system of units. \(\tensor{\epsilon}{_a_b_c_d}\) is known as the volume form on spacetime and it will be only relevant when we are writing integrals. See App. B of Ref. \citeonline{wald1984GeneralRelativity} for an account.

        When studying supertranslations, we could choose between a geometric analysis as in \cref{subsec: BMS-derivation-global-geometric,subsec: BMS-derivation-Killing}, or fix a gauge and work in the Bondi--Sachs formalism as in \cref{subsec: BMS-derivation-Bondi-Sachs}. The geometric formulation of electrodynamics and gauge theory in general relies on the theory of principal fiber bundles \cite{hamilton2017MathematicalGaugeTheory,tu2017DifferentialGeometryConnections}, which is out of our scope. Hence we will fix a gauge and work in similarity to the Bondi--Sachs formalism. This is the approach originally taken by \textcite{he2014NewSymmetriesMassless}.

        The gauge transformations in electrodynamics are given by 
        \begin{equation}\label{eq: gauge-transformation-electrodynamics}
            \tensor{A}{_a} \to \tensor{A}{_a} + \tensor{\nabla}{_a}\varepsilon
        \end{equation}
        for some choice of function \(\varepsilon\). Since we want to consider the hypersurface \(\nullfut\) of Minkowski spacetime, which is parameterized by retarded time and by angular variables, we will choose a gauge in which the radial component does not occur. This is known as radial gauge, and is characterized by
        \begin{equation}
            \tensor{A}{_r} = 0,
        \end{equation}
        where \(r\) stands for the radial coordinate, not for an abstract index. Given any electromagnetic potential \(\tensor{A}{_a}\), we can always go to radial gauge by choosing a gauge parameter \(\varepsilon\) according to
        \begin{equation}\label{eq: gauge-parameter-for-radial-gauge}
            \tensor{\partial}{_r} \varepsilon = - \tensor{A}{_r},
        \end{equation}
        which can always be solved. Recall that the covariant and partial derivatives act on scalar functions in the same way.

        This choice of gauge leaves a residual gauge symmetry. Namely, once we are in radial gauge, any further gauge transformation (\ref{eq: gauge-transformation-electrodynamics}) with \(\tensor{\partial}{_r}\varepsilon = 0\) will preserve radial gauge, although it can still change the electromagnetic potential. With this in mind, we will make one further gauge choice. We will choose the electromagnetic potential at \(\nullfut\) to be in retarded gauge,
        \begin{equation}
            \tensor{A}{_u}|_{\nullfut} = 0.
        \end{equation}
        This equation fixes a gauge for the electromagnetic field at \(\nullfut\). Given any electromagnetic potential \(\tensor{A}{_a}\), we can go to retarded radial gauge by choosing \(\varepsilon\) satisfying \cref{eq: gauge-parameter-for-radial-gauge} with the boundary condition
        \begin{equation}
            \tensor{\partial}{_u}\varepsilon|_{\nullfut} = - \tensor{A}{_u}|_{\nullfut}.
        \end{equation}

        Retarded radial gauge will play the same role for electromagnetism that the Bondi gauge played for gravity. Notice, in particular, that any field configuration admits retarded radial gauge. Our next goal is to impose adequate falloff conditions to describe the electromagnetic field of an isolated system, similarly to how we imposed falloff conditions on \cref{subsec: bondi-sachs-formalism} to ensure an asymptotically flat metric. 

        The most basic condition we may impose on the electromagnetic field is that it is still defined at infinity. This means the field strength tensor \(\tensor{F}{_a_b}\) should be defined at \(\nullfut\). By using the inverse radial coordinate \(l = 1/r\), this means imposing 
        \begin{equation}
            \tensor{F}{_u_l}, \tensor{F}{_u_{\zeta}}, \tensor{F}{_l_{\zeta}}, \tensor{F}{_{\zeta}_{\bar{\zeta}}} = \order{l^0}.
        \end{equation}
        Once we express this in terms of \(r=1/l\), this is translated to 
        \begin{equation}
            \tensor{F}{_u_r}, \tensor{F}{_r_{\zeta}} = \order{\frac{1}{r^2}}, \qq{and} \tensor{F}{_u_{\zeta}}, \tensor{F}{_{\zeta}_{\bar{\zeta}}} = \order{r^0}.
        \end{equation}
        In retarded radial gauge, this implies 
        \begin{equation}\label{eq: retarded-radial-gauge-falloff}
            \tensor{A}{_u} = \order{\frac{1}{r}}\qc \tensor{A}{_r} = 0, \qq{and} \tensor{A}{_{\zeta}} = \order{1}.
        \end{equation}

        Even with all of these conditions, we still have residual gauge freedom. This residual freedom is to be interpreted as due to asymptotic symmetries in elecctrodynamics. This is completely analogous to how supertranslations were a residual gauge freedom after Bondi gauge was fixed and appropriate falloff conditions were imposed. In order to preserve \cref{eq: retarded-radial-gauge-falloff}, we can only choose gauge parameters \(\varepsilon\) with the form 
        \begin{equation}\label{eq: large-gauge-transformations}
            \varepsilon(\zeta,\bar{\zeta}) \in \ck[\infty](\mathbb{S}^2).
        \end{equation}
        These are then gauge transformations that ``survive'' at infinity. As such, they are called large gauge transformations.
        
    \subsection{Antipodal Matching Conditions}
        Next, let us consider the Liénard--Wiechert solutions, which describe the fields for a system of finitely many charged particles moving in Minkowski spacetime. See, for example, Refs. \citeonline{wald2022AdvancedClassicalElectromagnetism,zangwill2013ModernElectrodynamics} and especially Ref. \citeonline{lechner2018ClassicalElectrodynamicsModern} for an explicitly covariant treatment. For a system of \(n\) particles with charges \(Q_k\) and three-velocities \(\vb{v}_k\) that pass through the origin at time zero, one finds that the radial electric field is given by \cite{strominger2018LecturesInfraredStructure}
        \begin{equation}
            \tensor{F}{_r_t}(t,\vb{x}) = \sum_{k=1}^n \frac{Q_k \gamma_k (r - t \vu{x} \vdot \vu{v}_k)}{\abs{\gamma_k^2 (t - r \vu{x} \vdot \vu{v}_k)^2 - t^2 + r^2}^{\frac{3}{2}}}.
        \end{equation}
        \(\gamma_k^{-2} = 1 - \norm{\vb{v}_k}^2\) is the Lorentz factor for each charge and we wrote \(r = \norm{\vb{x}}^2\), \(\vb{x} = r \vu{x}\), as usual. 

        What is particularly interesting about this field is its behavior near spatial infinity (\(\spatinf\) in \cref{fig: minkowskipenrose}). Defining \(r = 1/l\) we find 
        \begin{equation}
            \tensor{F}{_l_t}(t,\vb{x}) = - \sum_{k=1}^n \frac{Q_k \gamma_k (1 - l t \vu{x} \vdot \vu{v}_k)}{\abs{\gamma_k^2 (lt - \vu{x} \vdot \vu{v}_k)^2 - l^2 t^2 + 1}^{\frac{3}{2}}}.
        \end{equation}
        We can now take the limit to \(\spatinf\) (which means \(l \to 0\) at constant \(t\)) to find
        \begin{equation}
            \lim_{l \to 0} \tensor{F}{_l_t} = - \sum_{k=1}^n \frac{Q_k \gamma_k}{\abs{\gamma_k^2 (\vu{x} \vdot \vu{v}_k)^2 + 1}^{\frac{3}{2}}}.
        \end{equation}
        Notably, the result is direction-dependent (\ie, it depends on \(\vu{x}\)), even though \(\spatinf\) is a single point. This means \(\tensor{F}{_l_t}\), and thus \(\tensor{F}{_a_b}\) itself, is not a continuous tensor across \(\spatinf\). 

        What happens when we take the limit \(l \to 0\) along lines of constant \(u = t-r\) or \(v = t+r\), and then take the limit to \(\spatinf\) by taking \(u \to - \infty\) or \(v \to +\infty\)? In these cases, we find
        \begin{equation}
            \tensor{F}{_l_t} = \tensor{F}{_l_u} = \tensor{F}{_l_v}
        \end{equation}
        and thus
        \begin{equation}
            \lim_{u \to -\infty}\lim_{l \to 0} \tensor{F}{_l_u} = - \sum_{k=1}^n \frac{Q_k}{\gamma_k^2 (1 - \vu{x} \vdot \vu{v}_k)^2}.
        \end{equation}
        and
        \begin{equation}
            \lim_{v \to +\infty}\lim_{l \to 0} \tensor{F}{_l_v} = - \sum_{k=1}^n \frac{Q_k}{\gamma_k^2 (1 + \vu{x} \vdot \vu{v}_k)^2}.
        \end{equation}

        We thus see that taking the limit of the electromagnetic field to \(\nullfut\) and then taking \(u \to -\infty\) to reach \(\spatinf\) does not yield the same result as taking first a limit to \(\nullpas\) and then \(v \to +\infty\) to reach \(\spatinf\). In fact, the two solutions are antipodally related:
        \begin{equation}\label{eq: antipodal-matching-EM}
            \lim_{v \to +\infty}\lim_{l \to 0} \tensor{F}{_l_v}(\vu{x}) = \lim_{u \to -\infty}\lim_{l \to 0} \tensor{F}{_l_u}(-\vu{x}).
        \end{equation}

        This antipodal discontinuity seems natural once one considers Minkowski spacetime wrapped around the Einstein cylinder, as in \cref{fig: minkowski-einstein-cylinder}, for in this case \(\nullfut\) and \(\nullpas\) are antipodal continuations of each other. In this case, the ``field lines'' of the electromagnetic field are continuous across \(\spatinf\), but they are interpreted as an antipodal discontinuity from a finite perspective. 

        The condition (\ref{eq: antipodal-matching-EM}) is expected to always hold. Even in the presence of more complicated fields, a finite-energy configuration cannot have electromagnetic waves at infinitely early (\(u \to -\infty\)) or late (\(v \to + \infty\)) times, and hence we expect the Coulombic behavior to be general. This is verified \emph{a posteriori}. For now, if we assume this to be true, we see that any smooth function \(\varepsilon\) on Minkowski spacetime with 
        \begin{equation}\label{eq: antipodal-matching-condition}
            \lim_{v \to +\infty}\lim_{\substack{l \to 0 \\v = \text{const.}}} \varepsilon(\vu{x}) = \lim_{u \to -\infty}\lim_{\substack{l \to 0 \\u = \text{const.}}} \varepsilon(-\vu{x})
        \end{equation}
        gives rise to a conserved charge by defining 
        \begin{equation}
            Q_{\varepsilon}^+ = \frac{1}{4\pi} \int_{\nullfut_-} \varepsilon \tensor{\epsilon}{_a_b_c_d}\tensor{F}{^c^d} = \frac{1}{e^2} \int_{\nullpas_+} \varepsilon \tensor{\epsilon}{_a_b_c_d}\tensor{F}{^c^d} = Q_{\varepsilon}^-,
        \end{equation}
        where \(\nullfut_-\) is the sphere at the limit \(u \to -\infty\) of \(\nullfut\), and similarly \(\nullpas_+\) is the sphere at the limit \(v \to +\infty\) of \(\nullpas\). Notice \(Q_{\varepsilon}^+ = Q_{\varepsilon}^-\) trivially. Since there are infinitely many functions \(\varepsilon\) satisfying the antipodal matching condition, there are infinitely many conserved charges. 

        For \(\varepsilon = 1\), the conserved charge is total electric charge. Incidentally, this is the case of \cref{eq: large-gauge-transformations} in which the gauge transformation is global. The global \(\Ug(1)\) gauge symmetry of electromagnetism is exactly the origin of the conservation of electric charge, according to Noether's theorem. Our results then suggest that the other conserved charges are the results of the large gauge transformations considered in the previous section. In fact, the ``diagonal subgroup'' of large gauge transformations at \(\nullfut\) and \(\nullpas\) that obey the antipodal matching condition (\ref{eq: antipodal-matching-condition}) gives rise to the conserved charges \(Q_{\varepsilon}^\pm\) \cite{he2014NewSymmetriesMassless,strominger2018LecturesInfraredStructure}. Finally, the Ward identities
        \begin{equation}
            \mel{\text{out}}{Q_{\varepsilon}^{+}S - S Q_{\varepsilon}^-}{\text{in}} = 0
        \end{equation}
        can be seen to be equivalent to Weinberg's soft photon theorem \cite{he2014NewSymmetriesMassless}. 

    \subsection{Weinberg's Soft Photon Theorem}
        The Ward identity for large gauge transformations is the Weinberg soft photon theorem. While we will not prove this equivalence (for that, see Refs. \citeonline{he2014NewSymmetriesMassless,strominger2018LecturesInfraredStructure}), we will now derive the theorem as originally done by \textcite{weinberg1965InfraredPhotonsGravitons}. The approach is very similar to how we derived the soft graviton theorem in \cref{subsec: weinberg-soft-graviton}, and therefore I will only sketch the calculations.

        The main difference between the soft photon and graviton theorems is that \cref{eq: weinberg-soft-graviton-vertex} is replaced by
        \begin{equation}\label{eq: weinberg-soft-photon-vertex}
            \vcenter{\hbox{\begin{tikzpicture}
                \begin{feynman}
                    \vertex (c) at (0,0);
                    \vertex (s1) at (210:1);
                    \vertex (s2) at (330:1);
                    \vertex (g) at (90:1)  {\footnotesize\(\mu\)};
                    \diagram*{
                        (c) -- [photon] (g),
                        (s1) -- [fermion,momentum=\(p_1\)] (c),
                        (c) -- [fermion,momentum=\(p_2\)] (s2),
                    };
                \end{feynman}
            \end{tikzpicture}}} = -i eQ (\tensor*{p}{_1_\mu} + \tensor*{p}{_2_\mu}).
        \end{equation}
        This follows from the fact that a charged scalar field with electric charge \(eQ\) (\(Q \in \Z\)) has a current
        \begin{equation}
            \tensor{j}{_\mu} = i Q \qty(\phi \tensor{\partial}{_\mu}\phi^* - \phi^*\tensor{\partial}{_\mu}\phi).
        \end{equation}

        It follows that \cref{eq: soft-graviton-theorem-in} becomes
        \begin{equation}\label{eq: soft-photon-theorem-in}
            \vcenter{\hbox{\begin{tikzpicture}
                \begin{feynman}
                    \node[blob, pattern color=white] (c) at (0,0);
                    \vertex (a1) at (180:2);
                    \vertex (a2) at (135:2);
                    \vertex (a3) at (225:2);
                    \vertex (a4) at (225:1);
                    \vertex (a5) at (300:1.5) {\footnotesize\(\mu\)};
                    \vertex (b1) at (30:2);
                    \vertex (b2) at (330:2);
                    \diagram*{
                        (a4) -- [photon] (a5),
                        (a4) -- [draw=none,momentum'=\(q\)] (a5),
                        (a1) -- [fermion] (c),
                        (a2) -- [fermion] (c),
                        (a3) -- [fermion,momentum=\(p\)] (a4),
                        (a4) -- [fermion] (c),
                        (c) -- [fermion] (b1),
                        (c) -- [fermion] (b2),
                    };
                \end{feynman}
            \end{tikzpicture}}} = \qty(-\frac{e Q \tensor{p}{_\mu}}{\tensor{p}{^\rho}\tensor{q}{_\rho} + i \epsilon}) \times \qty(\vcenter{\hbox{\begin{tikzpicture}
                \begin{feynman}
                    \node[blob, pattern color=white] (c) at (0,0);
                    \vertex (a1) at (180:2);
                    \vertex (a2) at (135:2);
                    \vertex (a3) at (225:2);
                    \vertex (b1) at (30:2);
                    \vertex (b2) at (330:2);
                    \diagram*{
                        (a1) -- [fermion] (c),
                        (a2) -- [fermion] (c),
                        (a3) -- [fermion,momentum=\(p\)] (c),
                        (c) -- [fermion] (b1),
                        (c) -- [fermion] (b2),
                    };
                \end{feynman}
            \end{tikzpicture}}}) + \order{\tensor{q}{^0}}.
        \end{equation}
        Similarly, \cref{eq: soft-graviton-theorem-out} becomes
        \begin{equation}\label{eq: soft-photon-theorem-out}
            \vcenter{\hbox{\begin{tikzpicture}
                \begin{feynman}
                    \node[blob, pattern color=white] (c) at (0,0);
                    \vertex (a1) at (180:2);
                    \vertex (a2) at (135:2);
                    \vertex (a3) at (225:2);
                    \vertex (a4) at (225:1);
                    \vertex (a5) at (300:1.5);
                    \vertex (b1) at (30:2);
                    \vertex (b2) at (330:2);
                    \vertex (b3) at (330:1);
                    \vertex (b4) at (0:2) {\footnotesize\(\mu\)};
                    \diagram*{
                        (a1) -- [fermion] (c),
                        (a2) -- [fermion] (c),
                        (a3) -- [fermion] (c),
                        (c) -- [fermion] (b1),
                        (c) -- [fermion] (b3),
                        (b3) -- [fermion,momentum'=\(p\)] (b2),
                        (b3) -- [photon] (b4),
                        (b3) -- [draw=none,momentum=\(q\)] (b4),
                    };
                \end{feynman}
            \end{tikzpicture}}} = \qty(\frac{eQ \tensor{p}{_\mu}}{\tensor{p}{^\rho}\tensor{q}{_\rho} - i \epsilon}) \times \qty(\vcenter{\hbox{\begin{tikzpicture}
                \begin{feynman}
                    \node[blob, pattern color=white] (c) at (0,0);
                    \vertex (a1) at (180:2);
                    \vertex (a2) at (135:2);
                    \vertex (a3) at (225:2);
                    \vertex (b1) at (30:2);
                    \vertex (b2) at (330:2);
                    \diagram*{
                        (a1) -- [fermion] (c),
                        (a2) -- [fermion] (c),
                        (a3) -- [fermion] (c),
                        (c) -- [fermion] (b1),
                        (c) -- [fermion,momentum'=\(p\)] (b2),
                    };
                \end{feynman}
            \end{tikzpicture}}}) + \order{\tensor{q}{^0}}.
        \end{equation}

        Using these results, we see \cref{eq: one-emitted-graviton} becomes
        \begin{equation}\label{eq: one-emitted-photon}
            \mel{\text{out}+1}{S}{\text{in}} =   e \qty(\sum_{\substack{i\\\text{out}}} \frac{Q_i \tensor*{p}{^i_\mu}}{\tensor*{p}{_i^\rho}\tensor{q}{_\rho}} - \sum_{\substack{j\\\text{in}}} \frac{Q_j\tensor*{p}{^j_\mu}}{\tensor*{p}{_j^\rho}\tensor{q}{_\rho}}) \tensor{\epsilon}{^\mu}(q)\mel{\text{out}}{S}{\text{in}}  + \order{\tensor{q}{^0}},
        \end{equation}
        which is the Weinberg soft photon theorem. 

    \subsection{Electromagnetic Memory Effect}
        Finally, let us briefly discuss the electromagnetic memory effect. This effect was discovered by \textcite{bieri2013ElectromagneticAnalogGravitational} and can be considered in full detail within electromagnetism, or approximated in the slow-velocity regime. Ref. \citeonline{bieri2013ElectromagneticAnalogGravitational} discusses both cases, and Ref. \citeonline{pasterski2017AsymptoticSymmetriesElectromagnetic} connects the result to the other vertices of the electromagnetic \gls{IR} triangle. I will consider only the simplest case of the slow motion of a charge upon the passage of an electromagnetic wave. 

        For a slowly moving charges, the Lorentz force can be written as 
        \begin{equation}
            m \dv[2]{\vb{x}}{t} = q \vb{E},
        \end{equation}
        where we are neglecting the magnetic field due to the assumption that the velocity is small (it is possible to consider a more detailed approach). If an electromagnetic wave passes, then the charge will receive a velocity kick given by 
        \begin{equation}
            \Delta \vb{v} = \frac{q}{m} \int_{-\infty}^{+\infty} \vb{E} \dd{t}.
        \end{equation}
        For a slowly moving source far from the charge, we can approximate the electric field by its dipole contribution \cite{wald2022AdvancedClassicalElectromagnetism,lechner2018ClassicalElectrodynamicsModern,zangwill2013ModernElectrodynamics}
        \begin{equation}
            \vb{E} = \frac{1}{r}\qty[\dv[2]{\vb{p}}{t}]^{\text{T}} + \order{\frac{1}{r^2}},
        \end{equation}
        where \(\vb{p}\) is the dipole moment and the superscript ``T'' indicates a transversal projection removing the radial component. One thus concludes that
        \begin{equation}
            \Delta \vb{v} = \frac{q}{m r}\qty[\eval{\dv{\vb{p}}{t}}_{+\infty} - \eval{\dv{\vb{p}}{t}}_{-\infty}]^{\text{T}}.
        \end{equation}

        To make an analogy with the Braginsky--Thorne formula (\ref{eq: braginsky-thorne}), we can assume that at early and late times the field sources are freely moving charges. In this case, 
        \begin{equation}
            \eval{\dv{\vb{p}}{t}}_{\pm\infty} = \sum_{\substack{p \\ \text{in/out}}} q_p \vb{v}_p,
        \end{equation}
        and we conclude that
        \begin{equation}
            \Delta \vb{v} = \frac{q}{m r}\qty[\sum_{\substack{p \\ \text{out}}} q_p \vb{v}_p - \sum_{\substack{p \\ \text{in}}} q_p \vb{v}_p]^{\text{T}}.
        \end{equation}

        As previously mentioned, it is possible to obtain more precise calculations. Physically, the most important aspect is the observation that the electromagnetic memory effect leaves a ``kick'' on the charge, while the gravitational memory effect yields merely a displacement\footnote{There is a \gls{GW} velocity memory effect, discussed for example by \textcite{bieri2024GravitationalWaveDisplacement}, but this is not of interest for us.}. This is due to the fact that freely moving charges can still have a nonvanishing change in the electric dipole, but freely moving masses cannot have a nonvanishing third-derivative of the quadrupole moment.

\section{Superrotations}\label{sec: superrotations}
    In \cref{subsec: CK-spacetimes}, we noticed that the conformal compactification definition of asymptotic flatness is not as general as one could wish. For example, it has limitations when the spacetime involves emission of \glspl{GW}, which surely is a phenomenon of interest. Even when considering the more general class of \gls{CK} spacetimes, I mentioned one can consider other asymptotic properties and get a different definition.

    This subtlety raises an important difficulty in theoretical physics: figuring out which definitions better suit the theory. As an example, we saw that there were many possible choices of structures at infinity one could choose to preserve when defining an asymptotic symmetry group, notably the weak and strong Carrollian structures. We chose to work with the weak Carrollian structure due to the fact we expected to obtain at least the Poincaré group at infinity (see \cref{subsec: possible-criticisms-derivation-BMS}).

    With this in mind, there is an interesting question we can raise. When studying the conformal group on the sphere on \cref{subsec: conformal-isometries-sphere}, I chose to discard many ``candidates to conformal Killing vector fields'' based on the fact they did not lead to global transformations. In other words, their flows were not complete. The exceptions were the conformal Killing vector fields generating the Möbius transformations, which I kept and we later saw they yielded the Lorentz transformations in the \gls{BMS} group. 

    What if we decided to keep those other vector fields? Banks \cite{banks2003CritiquePureString} (see footnote 17 on p. 29), and Barnich and Troessaert \cite{barnich2010AspectsBMSCFT,barnich2010SymmetriesAsymptoticallyFlat} suggested this is precisely what should be done, with motivation coming from the study of two-dimensional \glspl{CFT} \cite{belavin1984InfiniteConformalSymmetry}. This extends the symmetries of the \gls{BMS} group to include the so-called superrotations, but at the price of the collection of symmetries no longer being a Lie group. Instead, we only get a Lie algebra, which is a collection of infinitesimal symmetries.
    
    The goal of this section is then to explore this approach and understand what superrotations are. This will require a few more results on Lie groups and algebras in order to understand the new structures that we will now find.

    \subsection{Lie Algebras}
        Vector fields can be understood as infinitesimal diffeomorphisms of a curved spacetime. Given a vector field \(\tensor{v}{^a}\), its flow \(\qty{\phi_t}\) is given by solving
        \begin{equation}
            \left\lbrace
            \begin{aligned}
                \tensor*{\dot{\gamma}}{_p^a}(t) &= \tensor{v}{^a}, \\
                \tensor{\gamma}{_p}(0) &= p, \\
                \phi_t(p) &= \tensor{\gamma}{_p}(t),
            \end{aligned}
            \right.
        \end{equation}
        where in the first equation the vector field is being evaluated at \(\phi_t(p)\). Hence, integrating the vector field leads us to a diffeomorphism. I am abusing terminology in saying ``diffeomorphism'', since we are actually headed toward vector fields whose flow is not globally defined (because not even the vector fields themselves are).

        The idea of an infinitesimal symmetry is natural in this context, given that we want to investigate the remaining vector fields that respect the conformal Killing equation in the sphere. With this in mind, the first question we should address is: what is the mathematical structure of infinitesimal symmetries? 

        If we want to discuss infinitesimal transformations, we certainly must be working with a Lie group. Otherwise, the word ``infinitesimal'' would be meaningless. Let \(G\) then be some Lie group with identity \(e\). An infinitesimal transformation should be a transformation that is infinitesimally close to the identity. Thinking in geometrical terms, it makes sense that we should then look at the tangent space \(\T[e]G\), which is composed of all tangent vectors to \(G\) at \(e\). This is precisely the space we get when we pay attention to group elements that are infinitesimally close to the identity. As a consequence, it seems that a good notion of ``infinitesimal symmetry'' should require that the collection of these infinitesimal symmetries has a vector space structure.

        Notice, though, that we have still not exploited the fact that \(G\) is a group. To do so, we define the map 
        \begin{equation}
            \begin{aligned}
                l_g \colon G &\to G \\
                h &\mapsto g h.
            \end{aligned}
        \end{equation}
        \(l_g\) is known as left translation. Due to \(G\) being a Lie group, \(l_g\) is a diffeomorphism, which allows us to consider the pushforward \(l_g^*\). In particular, we find there is a map \(l_g^* \colon \T[e]G \to \T[g]G\). This means if we know \(\T[e]G\), we actually know \(\T[g]G = l_g^* \T[e]G\) for any \(g\)! 

        It is particularly interesting to look at the so-called left-invariant vector fields. These are vector fields \(v\) (I am omitting the indices again) such that \(l_g^* v = v\) for all \(g \in G\). If we denote the value of the vector field \(v\) at the point \(h\) by \(v|_h\), this condition states that
        \begin{equation}
            l_g^* v|_h = v|_{gh}.
        \end{equation}
        In particular, notice that a left-invariant vector field is uniquely determined by its value at the identity, because
        \begin{equation}
            v|_g = l_g^* v|_e.
        \end{equation}
        This actually establishes an isomorphism: \(\T[e]G\) is isomorphic to the real vector space of left-invariant vector fields.

        Why is this important? Because we just learned that the space \(\T[e]G\) of infinitesimal symmetries is equipped with some structure it borrowed from the space of vector fields. Notably, it is equipped with a bilinear map \(\comm{\cdot}{\cdot} \colon \T[e]G \times \T[e]G \to \T[e]G\). This is true because the property of being a left-invariant vector field is preserved by the commutator of vector fields. Indeed,
        \begin{subequations}
            \begin{align}
                l_g^* \comm{v}{w}(f) &= \comm{v}{w}(f \circ l_g), \\
                &= v(w(f \circ l_g)) - w(v(f \circ l_g)), \\
                &= v(l_g^* w(f)) - w(l_g^* v(f)), \\
                &= v(w(f)) - w(v(f)), \\
                &= \comm{v}{w}(f).
            \end{align}
        \end{subequations}

        Hence, the space of infinitesimal symmetries of a Lie group \(G\) is given by the pair \((\T[e]G, \comm{\cdot}{\cdot})\). This will be our basic model for a collection of infinitesimal symmetries. Now we just need to figure out what are the basic properties of \(\comm{\cdot}{\cdot}\). From its definition, one can show that
        \begin{enumerate}
            \item it is bilinear,
            \item it is antisymmetric, meaning \(\comm{v}{w} = - \comm{w}{v}\),
            \item it satisfies the Jacobi identity, \(\comm{v}{\comm{w}{x}} + \comm{w}{\comm{x}{v}} + \comm{x}{\comm{v}{w}} = 0\).
        \end{enumerate}
        These will be the basic objects for our definition of a Lie algebra. 

        \begin{definition}[Lie Algebra]
            A \emph{Lie algebra} over \(\K\) (assumed to be \(\R\) or \(\Comp\)) is a \(\K\)-vector space \(\mathfrak{g}\) equipped with a map \(\comm{\cdot}{\cdot} \colon \mathfrak{g} \times \mathfrak{g} \to \mathfrak{g}\) which 
            \begin{enumerate}
                \item is bilinear,
                \item is antisymmetric,
                \item and satisfies the Jacobi identity.
            \end{enumerate}
            \(\comm{\cdot}{\cdot}\) is known as the Lie bracket, or commutator. If \(G\) is a Lie group, it is common to denote its Lie algebra in fraktur font: \(\mathfrak{g}\).
        \end{definition}

        Let us then carry out some examples. One option would be to understand the manifold structure of the Lie groups we are interested in and carry out the calculations in that way. Sometimes this will be simple. Consider \(\SO(2)\), for example. All elements of \(\SO(2)\) can be written in the form
        \begin{equation}
            R(\theta) = \mqty(\cos\theta & \sin\theta \\ - \sin\theta & \cos\theta)
        \end{equation}
        for \(\theta \in [0,2\pi)\). We then see that, as a manifold, \(\SO(2)\) is the circle \(\Sph^1\), and it follows that \(\so(2)\) (the Lie algebra of \(\SO(2)\)) is \(\R\) with the trivial Lie bracket \(\comm{J}{J} = 0\).

        Nevertheless, other groups can get considerably more complicated. \(\SO(3)\), as a manifold, is the three-dimensional real projective space \(\mathbb{P}\R^3\), which is the space of lines passing through the origin of \(\R^4\). Equivalently, \(\mathbb{P}\R^3\) is a sphere with antipodal points identified. Working directly with \(\SO(3)\) as a manifold and computing the Lie algebra in this manner would be laborious at least, and one hardly would be able to generalize this calculation for more complicated groups. 

        This difficulty can be bypassed. In the case of matrix groups, one can trade the commutator of vector fields by the matrix commutator \cite{tu2011IntroductionManifolds,lee2012IntroductionSmoothManifolds,hamilton2017MathematicalGaugeTheory} and arrive at many results much more easily. For example, consider \(\SO(n)\). If we formally write an infinitesimal rotation as
        \begin{equation}
            R(\epsilon) = \Eins + \sum_{i=1}^{\frac{n(n-1)}{2}} \epsilon_i T_i + \order{\epsilon^2}.
        \end{equation}
        Then the orthogonality condition \(R^\intercal R = \Eins\) becomes, at leading order in \(\epsilon\),
        \begin{equation}
            \Eins + \sum_{i=1}^{\frac{n(n-1)}{2}} \epsilon_i T_i^\intercal + \sum_{i=1}^{\frac{n(n-1)}{2}} \epsilon_i T_i + \order{\epsilon^2} = \Eins,
        \end{equation}
        from which we conclude that the generators \(T_i\) must satisfy
        \begin{equation}
            T_i^\intercal = -T_i,
        \end{equation}
        \ie, they are antisymmetric. Fix \(n=3\), for concreteness. Then a basis for the space of three-dimensional antisymmetric real matrices is provided by
        \begin{equation}
            J_x = \mqty(0 & 1 & 0 \\ -1 & 0 & 0 \\ 0 & 0 & 0)\qc J_y = \mqty(0 & 0 & -1 \\ 0 & 0 & 0 \\ 1 & 0 & 0), \qq{and} J_z = \mqty(0 & 0 & 0 \\ 0 & 0 & 1 \\ 0 & -1 & 0).
        \end{equation}
        These three matrices lead to the commutation relations
        \begin{equation}\label{eq: lie-algebra-so3}
            \comm{J_i}{J_j} = \sum_{k=1}^3 \epsilon_{ijk} J_k,
        \end{equation}
        which establishes the Lie algebra of \(\SO(3)\). \(\epsilon_{ijk}\) is the Levi-Civita symbol.

        In curved spacetime, however, we were not computing Lie groups directly. We were first computing conformal Killing vector fields. Furthermore, our discussion of Lie algebras started with vector fields and the elements of a Lie algebra are vector fields on the Lie group. A key question is then: can we obtain information about a Lie algebra directly from the vector fields on a spacetime \(M\) whose flow yields the Lie group? 

        The answer is positive. It is also quite general: one can formulate it for an arbitrary Lie group acting on an arbitrary manifold, even if it does not have a metric. A full proof is given in Sec. III.D.5 of Ref. \citeonline{choquet-bruhat2004AnalysisManifoldsPhysics}. The basic idea is the following. Consider a Lie group \(G\), a manifold \(M\), and a smooth injective homomorphism
        \begin{equation}
            \begin{aligned}
                G &\to \Diff(M)\\
                g &\mapsto \phi_g.
            \end{aligned}
        \end{equation}
        The pushforward of this homomorphism can be used to obtain a vector field on \(M\) for each element in \(\T[g]G\). This determines a vector space isomorphism, and computing the vector fields on both manifolds (\(M\) and \(G\)) explicitly leads to the result \cite[pp. 164--165]{choquet-bruhat2004AnalysisManifoldsPhysics}. 

        Notice that, in particular, this result means that commutators of Killing vector fields are Killing vector fields, and similarly for conformal Killing vector fields. Furthermore, this result implies we can compute the isometry Lie algebra of a pseudo-Riemannian manifold by simply computing its Killing vector fields. The conformal Lie algebra is similarly obtained by computing the conformal Killing vector fields. 

    \subsection{Case Study: Poincaré Algebra}\label{subsec: poincare-lie-algebra-bulk}
        As our first case study, let us work out the Poincaré algebra in \(n\) dimensions, \(\iso*(n-1,1)\).

        \cref{eq: poincare-generators} is the main equation in this calculation. We simply need to pick the Killing vector fields for the Poincaré group and compute their commutators. It will be less cumbersome if we omit the abstract indices. 

        The procedure is the following. We pick two arbitrary Killing vector fields from \cref{eq: poincare-generators} and compute their commutator by throwing in a placeholder function \(f\) just to keep track of how the derivatives are acting. Let me show an example. If we consider the vector fields \(\tensor{P}{_\mu}\) and \(\tensor{J}{_\rho_\sigma}\), we get the commutator
        \begin{subequations}
            \begin{align}
                \comm*{\tensor{P}{_\mu}}{\tensor{J}{_\rho_\sigma}}(f) &= \pdv{\tensor{x}{^\mu}}\qty(\tensor{x}{_\rho}\pdv{f}{\tensor{x}{^\sigma}} - \tensor{x}{_\sigma}\pdv{f}{\tensor{x}{^\rho}}) - \qty(\tensor{x}{_\rho}\pdv{\tensor{x}{^\sigma}} - \tensor{x}{_\sigma}\pdv{\tensor{x}{^\rho}})\pdv{f}{\tensor{x}{^\mu}}, \\
                &= \tensor{\eta}{_\mu_\rho}\pdv{f}{\tensor{x}{^\sigma}} - \tensor{\eta}{_\mu_\sigma}\pdv{f}{\tensor{x}{^\rho}}, \\
                &= \tensor{\eta}{_\mu_\rho}\tensor{P}{_\sigma}f - \tensor{\eta}{_\mu_\sigma}\tensor{P}{_\rho}f.
            \end{align}
        \end{subequations}

        The other commutators can be computed in a similar fashion. The final result is that
        \begin{subequations}\label{eq: poincare-algebra}
            \begin{align}
                \comm*{\tensor{P}{_\mu}}{\tensor{P}{_\nu}} &= 0, \\
                \comm*{\tensor{P}{_\mu}}{\tensor{J}{_\rho_\sigma}} &= \tensor{\eta}{_\mu_\rho}\tensor{P}{_\sigma} - \tensor{\eta}{_\mu_\sigma}\tensor{P}{_\rho}, \\
                \comm*{\tensor{J}{_\mu_\nu}}{\tensor{J}{_\rho_\sigma}} &= \tensor{\eta}{_\nu_\rho}\tensor{J}{_\mu_\sigma} - \tensor{\eta}{_\mu_\rho}\tensor{J}{_\nu_\sigma} - \tensor{\eta}{_\sigma_\mu}\tensor{J}{_\rho_\nu} + \tensor{\eta}{_\sigma_\nu}\tensor{J}{_\rho_\mu},
            \end{align}
        \end{subequations}
        where I removed the placeholder function \(f\) and the equalities should be understood as equalities between vector fields on spacetime.

        We know the Poincaré group is the semidirect product \(\ISO*(n-1,1) = \SO*(n-1,1) \ltimes \R^n\). This is reflected in the structure of the Poincaré algebra, which is a semidirect sum of Lie algebras\footnote{The term ``semidirect product of Lie algebras'' is also common.}. We begin with some auxiliary definitions.

        \begin{definition}[Lie Subalgebra]
            Let \(\mathfrak{g}\) be a Lie algebra with Lie bracket \(\comm{\cdot}{\cdot}\). A vector subspace \(\mathfrak{h} \subeq \mathfrak{g}\) is said to be a \emph{Lie subalgebra} of \(\mathfrak{g}\) if 
            \begin{equation}
                \comm{h_1}{h_2} \in \mathfrak{h}, \Forall h_1, h_2 \in \mathfrak{h}.
            \end{equation}
            In other words, a Lie subalgebra is a vector subspace closed under the Lie bracket. Notice \(\mathfrak{h}\) inherits the Lie bracket from \(\mathfrak{g}\).
        \end{definition}

        Just as normal subgroups are a special class of subgroups, ideals are a special class of Lie subalgebras.

        \begin{definition}[Ideal]
            Consider a Lie algebra \(\mathfrak{g}\) with Lie bracket \(\comm{\cdot}{\cdot}\). A Lie subalgebra \(\mathfrak{a} \subeq \mathfrak{g}\) is said to be an \emph{ideal} if 
            \begin{equation}
                \comm{a}{g} \in \mathfrak{a}, \Forall a \in \mathfrak{a}, g \in \mathfrak{g}.
            \end{equation}
            Notice how this is similar to the definition of a normal subgroup.
        \end{definition}

        Earlier, I said that a group \(G\) is the semidirect product of the subgroups \(H, N \subeq G\) if
        \begin{enumerate}
            \item \(N\) is a normal subgroup of \(G\), \(GN = NG\),
            \item \(G\) can be written as \(HN\) (or, equivalently, \(NH\)), 
            \item \(N \cap H = \qty{e}\).
        \end{enumerate}
        Our definition of the semidirect sum of Lie algebras will be similar. The notion of a normal subgroup will be traded for that of an ideal, while the remaining conditions will be combined into the notion of a direct sum of vector spaces\footnote{Given a vector space \(U\) with subspaces \(V\) and \(W\), we say \(U\) is the direct sum of \(V\) and \(W\) if \(V\) and \(W\) span \(U\) and their intersection is only the zero vector. See, for example, Ref. \citeonline{axler2015LinearAlgebraDone} for details.}.

        \begin{definition}[Semidirect Sum of Lie Algebras]
            Let \(\mathfrak{g}\) be a Lie algebra, \(\mathfrak{h}\) be a Lie subalgebra, and \(\mathfrak{a}\) be an ideal. If, as vector spaces, \(\mathfrak{g}\) is the direct sum of \(\mathfrak{h}\) and \(\mathfrak{a}\), then we say that, as Lie algebras, \(\mathfrak{g}\) is the semidirect sum of \(\mathfrak{h}\) and \(\mathfrak{a}\). We then write \(\mathfrak{g} = \mathfrak{h} \loplus \mathfrak{a}\).
        \end{definition}
        As with the semidirect product of groups, there is an extrinsic definition of the semidirect sum. See, for example, Ref. \citeonline{onishchik1993FoundationsLieTheory}.

        From \cref{eq: poincare-algebra} we can immediately see that 
        \begin{equation}
            \iso*(n-1,1) = \so*(n-1,1) \loplus \mathcal{Y}{_1}
        \end{equation}
        where \(\mathcal{Y}{_1}\) is the trivial Lie algebra of translations in \(n\) dimensions. The notation is not standard, but makes reference to the fact that translations are associated to spherical harmonics with \(l \leq 1\).

        There is a special case of the semidirect sum when both the components are ideals. In this case, we call it a direct sum.
        
        \begin{definition}[Direct Sum of Lie Algebras]
            Let \(\mathfrak{g}\) be a Lie algebra with \(\mathfrak{g} = \mathfrak{a}_1 \loplus \mathfrak{a}_2\). If \(\mathfrak{a}_1\) is an ideal of \(\mathfrak{g}\), we say \(\mathfrak{g}\) is a \emph{direct sum} of \(\mathfrak{a}_1\) and \(\mathfrak{a}_2\) and write \(\mathfrak{g} = \mathfrak{a}_1 \oplus \mathfrak{a}_2\).
        \end{definition}

    \subsection{Case Study: Poincaré Algebra at Null Infinity}\label{subsec: poincare-lie-algebra-boundary}
        Let us work out the Poincaré algebra in a slightly different approach, which will be a warm up to the \gls{BMS} group and to superrotations. Specifically, we will now compute the Poincaré algebra by working with a different basis and starting from the expression for the Poincaré Killing vector fields at null infinity, \cref{eq: poincare-generators-infinity}. 

        Following \textcite{barnich2010AspectsBMSCFT}, we define the complex vector fields 
        \begin{equation}\label{eq: def-ln}
            \tensor*{l}{_n^a} = - \zeta^{n+1} \tensor{\qty(\pdv{\zeta})}{^a} \qq{and} \tensor*{\bar{l}}{_n^a} = - \bar{\zeta}^{n+1} \tensor{\qty(\pdv{\bar{\zeta}})}{^a}.
        \end{equation}
        We also define the complex functions
        \begin{equation}
            t_{mn} = \frac{2 \zeta^m \bar{\zeta}^n}{1 + \zeta \bar{\zeta}}.
        \end{equation}
        We could have chosen to work with spherical harmonics instead of \(t_{mn}\), which would amount to a different choice of basis in the Lie algebra.

        Using these auxiliary definitions, we get to the main expressions we are interested in. Define now the complex vector fields
        \begin{gather}
            \tensor*{L}{_n^a} = \tensor*{l}{_n^a} + \frac{u \tensor{D}{_b}\tensor*{l}{_n^b}}{2}\tensor{\qty(\pdv{u})}{^a}, \\
            \tensor*{\bar{L}}{_n^a} = \tensor*{\bar{l}}{_n^a} + \frac{u \tensor{D}{_b}\tensor*{\bar{l}}{_n^b}}{2}\tensor{\qty(\pdv{u})}{^a}, \\
            \intertext{and}
            \tensor*{T}{_m_n^a} = t_{mn}\tensor{\qty(\pdv{u})}{^a}.
        \end{gather}
        For future reference, I mention that
        \begin{equation}
             \tensor{D}{_a}\tensor*{l}{^a_n} = - \frac{(1+n)}{2}t_{n0} + \frac{(1-n)}{2}t_{n+1,1}.
        \end{equation}
        
        These vector fields can be used to obtain a basis for the Poincaré algebra. This can be seen by writing the generators of \cref{eq: poincare-generators-infinity} in terms of these new vectors. By inspection one can tell that
        \begin{subequations}\label{eq: poincare-generators-LT}
            \begin{align}
                \tensor*{\mathcal{P}}{^a_t} &= \frac{1}{2}\qty(\tensor*{T}{_0_0^a} + \tensor*{T}{_1_1^a}), \\
                \tensor*{\mathcal{P}}{^a_x} &= -\frac{1}{2}\qty(\tensor*{T}{_1_0^a} + \tensor*{T}{_0_1^a}), \\
                \tensor*{\mathcal{P}}{^a_y} &= \frac{i}{2}\qty(\tensor*{T}{_1_0^a} - \tensor*{T}{_0_1^a}), \\
                \tensor*{\mathcal{P}}{^a_z} &= \frac{1}{2}\qty(\tensor*{T}{_0_0^a} - \tensor*{T}{_1_1^a}), \\
                \tensor*{\mathcal{J}}{^a_y_z} &= -\frac{i}{2}\qty(\tensor*{L}{_{-1}^a} - \tensor*{L}{_1^a}) + \frac{i}{2}\qty(\tensor*{\bar{L}}{_{-1}^a} - \tensor*{\bar{L}}{_1^a}), \\
                \tensor*{\mathcal{J}}{^a_z_x} &= \frac{1}{2}\qty(\tensor*{L}{_{-1}^a} + \tensor*{L}{_1^a}) + \frac{1}{2}\qty(\tensor*{\bar{L}}{_{-1}^a} + \tensor*{\bar{L}}{_1^a}), \\
                \tensor*{\mathcal{J}}{^a_x_y} &= -i\tensor*{L}{_0^a} + i\tensor*{\bar{L}}{_0^a}, \\
                \tensor*{\mathcal{J}}{^a_x_t} &= -\frac{1}{2}\qty(\tensor*{L}{_{-1}^a} - \tensor*{L}{_1^a}) - \frac{1}{2}\qty(\tensor*{\bar{L}}{_{-1}^a} - \tensor*{\bar{L}}{_1^a}), \\
                \tensor*{\mathcal{J}}{^a_y_t} &= -\frac{i}{2}\qty(\tensor*{L}{_{-1}^a} + \tensor*{L}{_1^a}) + \frac{i}{2}\qty(\tensor*{\bar{L}}{_{-1}^a} + \tensor*{\bar{L}}{_1^a}), \\
                \tensor*{\mathcal{J}}{^a_z_t} &= -\tensor*{L}{_0^a} - \tensor*{\bar{L}}{_0^a}.
            \end{align}
        \end{subequations}
        This means that the real linear combinations\footnote{Notice I do not mean linear combinations with real coefficients. I mean linear combinations whose imaginary part is zero. For this to happen, imaginary coefficients may be used. \(\tensor*{\mathcal{J}}{^a_x_y}\) in \cref{eq: poincare-generators-LT} is an example.} of the vector fields span the Poincaré algebra. 

        It is convenient to work with the complexified algebra\footnote{For some vector space \(V\), we can think of its complexification \(V^\Comp\) as the vector space \(V^\Comp = V \oplus iV\), meaning we allow all vectors to now have an imaginary part. \(\oplus\) denotes the direct sum of vector spaces. Both the real and imaginary parts of the complexified vectors are elements of \(V\), and \(V^{\Comp}\) has a natural vector space structure given by the direct sum. The same construction can be used for Lie algebras. In terms of a tensor product of vector spaces one could also write \(V^\Comp = V \otimes \Comp\).}, since it allows us to work with \(\tensor*{L}{_n^a}\), \(\tensor*{\bar{L}}{_n^a}\), and \(\tensor*{T}{_m_n^a}\) directly instead of restricting us to real linear combinations. We can later recover the actual Poincaré algebra by considering only the real elements of the complexified algebra.

        One can now compute the algebra in the exact same fashion as we did on \cref{subsec: poincare-lie-algebra-bulk}. I will just quote the results. I will also omit the abstract indices to avoid cluttering the notation. The results are
        \begin{subequations}\label{eq: extended-bms4-algebra}
            \begin{align}
                \comm*{\tensor{T}{_m_n}}{\tensor{T}{_o_p}} &= 0, \\
                \comm*{\tensor{L}{_l}}{\tensor{T}{_m_n}} &= \qty(\frac{1+l}{2} - m) \tensor{T}{_{l+m,n}}, \\
                \comm*{\tensor{\bar{L}}{_l}}{\tensor{T}{_m_n}} &= \qty(\frac{1+l}{2} - n) \tensor{T}{_{m,l+n}}, \\
                \comm*{\tensor*{L}{_m}}{\tensor{\bar{L}}{_n}} &= 0, \\
                \comm*{\tensor*{\bar{L}}{_m}}{\tensor{\bar{L}}{_n}} &= (m-n)\tensor*{\bar{L}}{_{m+n}}, \\
                \comm*{\tensor*{L}{_m}}{\tensor{L}{_n}} &= (m-n)\tensor*{L}{_{m+n}}.
            \end{align}
        \end{subequations}

        Define the Lie subalgebras 
        \begin{align}
            \witt_{1} &= \vspan\qty{\tensor*{L}{_{-1}}, \tensor*{L}{_0}, \tensor*{L}{_1}}, \\
            \overline{\witt}_{1} &= \vspan\qty{\tensor{\bar{L}}{_{-1}}, \tensor{\bar{L}}{_0}, \tensor{\bar{L}}{_1}}, \\
            \mathcal{Y}_1 &= \vspan\qty{\tensor{T}{_0_0}, \tensor{T}{_1_0}, \tensor{T}{_0_1}, \tensor{T}{_1_1}}.
        \end{align}
        Notice that these are indeed Lie subalgebras. Furthermore, \(\witt_{1}\) and \(\overline{\witt}_{1}\) are isomorphic as Lie algebras. The (complexified) Poincaré algebra is then given by\footnote{I am leaving implicit in the notation that we are considering the complexified algebra. In any case, one can get the real algebra by taking the real part.}
        \begin{equation}
            \iso*(3,1) = \qty(\witt_{1} \oplus \overline{\witt}_{1}) \loplus \mathcal{Y}_1.
        \end{equation}
        We see then that \(\witt_{1} \oplus \overline{\witt}_{1}\) is the Lorentz algebra, while \(\mathcal{Y}_1\) are the translations. This could also be verified explicitly by using \cref{eq: poincare-generators-LT} in \cref{eq: extended-bms4-algebra} to obtain \cref{eq: poincare-algebra}.

    \subsection{Case Study: \texorpdfstring{\glsfmtshort{BMS}}{BMS} Algebra}
        As a final stop before we get to superrotations, let me briefly mention the \gls{BMS} algebra in four dimensions, \(\bms\). 

        The \gls{BMS} group is obtained by adding supertranslations to the Poincaré group. Instead of considering translations that can be written as linear combinations of spherical harmonics with \(l \leq 1\) on the celestial sphere, we allow for any smooth function on \(\Sph^2\). 

        In the language of the Lie algebra given in \cref{eq: extended-bms4-algebra}, this means we now consider all possible generators of the form \(\tensor{T}{_m_n}\). Hence, we now consider
        \begin{equation}
            \mathcal{Y} = \vspan\left\lbrace\tensor{T}{_m_n} \middle| m,n \in \N_0\right\rbrace,
        \end{equation}
        and the complexified \gls{BMS} algebra is given by 
        \begin{equation}
            \bms = \qty(\witt_{1} \oplus \overline{\witt}_{1}) \loplus \mathcal{Y},
        \end{equation}
        where \(\mathcal{Y}\) is the algebra of supertranslations.

    \subsection{The Extended \texorpdfstring{\glsfmtshort{BMS}}{BMS} Algebra}
        At this point, the notation we are employing makes it natural to ask what happens when we trade \(\witt_1\) and \(\overline{\witt}_1\) for the infinite-dimensional Lie algebras
        \begin{equation}
            \witt = \vspan\left\lbrace\tensor{L}{_m} \middle| m \in \Z\right\rbrace \qq{and} \overline{\witt} = \vspan\left\lbrace\tensor{\bar{L}}{_m} \middle| m \in \Z\right\rbrace.
        \end{equation}
        These algebras (which are isomorphic) are known as the Witt algebra. The direct sum \(\witt \oplus \overline{\witt}\) is the local conformal algebra of the two-dimensional sphere (and of the plane), and it plays a prominent role in two-dimensional \gls{CFT} \cite{schottenloher2008MathematicalIntroductionConformal,blumenhagen2009IntroductionConformalField,difrancesco1997ConformalFieldTheory,polchinski1998IntroductionBosonicString}. Before I argue why it is interesting to enlarge the Lie algebra in this way, let me first argue against it. Why did we neglect these extra generators in the first place? 

        The vectors \(\tensor*{l}{^a_m}\) (resp. \(\tensor*{\bar{l}}{^a_m}\)) do not lead to globally defined vector fields on the sphere if \(\abs{m} > 1\). More specifically, for \(m<-1\) we can tell from \cref{eq: def-ln} that \(\tensor*{l}{^a_m}\) (\(\tensor*{\bar{l}}{^a_m}\)) is ill-defined at \(\zeta = 0\). For \(m>1\), we can change coordinates to \(\xi = - 1/\bar{\zeta}\) and we find the vectors with \(m>1\) are ill-defined for \(\xi=0\) (\(\zeta=\infty\)). This was argued in \cref{subsec: conformal-isometries-sphere}. Hence, these new vector fields are not properly defined on all of the sphere. This is quite different from our original definition of conformal Killing vector field, which even assumed the flow of the vector field was complete. 

        A second reason to oppose the inclusion of these new vector fields is that they fail to satisfy the conformal Killing equation. The conformal Killing equation on the sphere was written on \cref{eq: ckv-sphere}. In particular, it implies that
        \begin{equation}
            \pdv{\bar{\zeta}} \tensor*{l}{_m^\zeta} = 0,
        \end{equation}
        for example. It may seem this equation is always satisfied, since \(\tensor*{l}{_m^\zeta} = - \zeta^{m+1}\) is a function of \(\zeta\) only, but this is inaccurate. As mentioned in Appendix \ref{app: complex-analysis}, the condition \(\pdv{\bar{\zeta}} \tensor*{l}{_m^\zeta} = 0\) is equivalent to the Cauchy--Riemann equations, and therefore it can only be satisfied by holomorphic \(\tensor*{l}{_m^\zeta}\), which implies \(m\geq -1\) (and \(m \leq 1\) if one changes coordinates to \(\xi = -1/\bar{\zeta}\) and performs the same analysis). This means the conformal Killing equation will be violated at some points. This happens exactly at the poles of \(\tensor*{l}{_m^\zeta}\), which are precisely the points in which \(\tensor*{l}{_m^\zeta}\) fails to be holomorphic. 

        As an example, for \(m=-2\), we have
        \begin{equation}
            \tensor*{l}{_{-2}^\zeta} = -\frac{1}{\zeta}.
        \end{equation}
        Using \cref{eq: derivative-Dirac-delta-complex} we learn that
        \begin{equation}
            \tensor{\partial}{_{\bar{\zeta}}} \tensor*{l}{_{-2}^\zeta} = - 2\pi i \delta^{(2)}(\zeta),
        \end{equation}
        and hence the conformal Killing equation is violated at the South pole. This is precisely the point in which the vector field is ill-defined. Notice this also means the commutators in \cref{eq: extended-bms4-algebra} can be problematic at the poles.

        For these sorts of reasons, we left the extra vector fields \(\tensor*{L}{^a_m}\) and \(\tensor*{\bar{L}}{^a_m}\) out of the \gls{BMS} group. However, there are also reasons in favor of introducing them. For instance, the arguments we just discussed mean only that \(\tensor*{L}{^a_m}\) and \(\tensor*{\bar{L}}{^a_m}\) are local objects when \(\abs{m} > 1\): they do not lead to a group, but they still carry information about the infinitesimal symmetries of null infinity. 

        As mentioned at the beginning of this section, figuring out the appropriate definitions for a physical problem is a difficult task. Including these new generators is a mathematical possibility, but the true question is whether this leads to any new physical insights. After all, the purpose of studying symmetries in physics is always to obtain a better understanding of the problem at hand. What we should really ask is: can we learn something from these new generators? The answer turns out to be yes.

        Firstly, notice the root of this problem is the question ``what are the conformal symmetries of a two-dimensional sphere?'' This question is central to the study of \glspl{CFT} in two-dimensions, and enlarging the conformal algebra to include the two copies of the Witt algebra has been shown to be useful \cite{difrancesco1997ConformalFieldTheory,schottenloher2008MathematicalIntroductionConformal,blumenhagen2009IntroductionConformalField,polchinski1998IntroductionBosonicString,belavin1984InfiniteConformalSymmetry}. Considering a \gls{QFT} with an infinite-dimensional symmetry group considerably constrains its behavior, and allows to a much more detailed study and even to complete solutions of some models. Furthermore, the generators of the Witt algebra (in fact, of an extension of it known as the Virasoro algebra) naturally appear when studying the behavior of the stress tensor in a two-dimensional \gls{CFT}.

        Secondly, we will see that this perspective yields interesting new insights into physics. Namely, pursuing these sorts of symmetries leads to a new \gls{IR} triangle, and thus the notion of superrotation echoes in other effects in physics.

        With this optimist point of view in mind, we define the extended (and complexified) \gls{BMS} algebra as\footnote{The notation \(\ebms\) for the extended \gls{BMS} algebra is not standard. In Ref. \citeonline{barnich2010AspectsBMSCFT}, for example, the extended \gls{BMS} algebra is written simply \(\bms\), but I prefer to make a clear distinction between the \gls{BMS} algebra (here understood as the Lie algebra for the \gls{BMS} group without superrotations) and the extended \gls{BMS} algebra.}
        \begin{equation}
            \ebms = (\witt \oplus \overline{\witt}) \loplus \mathcal{Y}.
        \end{equation}

        The new generators of the Lie algebra, \ie, the vector fields \(\tensor*{L}{^a_m}\) and \(\tensor*{\bar{L}}{^a_m}\) with \(\abs{m} > 1\), are called superrotations. Just as supertranslations are a natural generalization of translations, superrotations are a natural generalization of Lorentz transformations. Notice that considering \(\tensor*{L}{^a_n}\) for \(n \in \Z\) also naturally invites us to extend supertranslations to consider \(\tensor*{T}{^a_m_n}\) with \(m, n \in \Z\).

        The term ``superrotation'' appears with different meanings in the literature, because there are many different enhancements of the \gls{BMS} group. These enhancements ultimately depend on the precise definition of an asymptotically flat spacetime and, in the Bondi--Sachs formalism, what are the falloff conditions imposed on the gravitational field. Above, we considered the ``extended \gls{BMS} algebra'', attributed to \textcite{banks2003CritiquePureString,barnich2010AspectsBMSCFT,barnich2010SymmetriesAsymptoticallyFlat} (see also Refs. \citeonline{flanagan2017ConservedChargesExtended,barnich2016FiniteBMSTransformations}). These superrotations are ``Virasoro-like'', in the sense they correspond to local conformal transformations in the celestial sphere. Another commonly studied extension is the generalized \gls{BMS} group \cite{campiglia2014AsymptoticSymmetriesSubleading,campiglia2015NewSymmetriesGravitational,compere2018SuperboostTransitionsRefraction}, which trades the group \(\SO*(3,1)\) for the diffeomorphism group\footnote{In this case, some authors prefer the term ``super-Lorentz transformation'' and reserve the terms ``superrotation'' and ``superboost'' for specific transformations within this class \cite{aneesh2022CelestialHolographyLectures}.} \(\mathrm{Diff}(\Sph^2)\). Both the generalized and extended \gls{BMS} transformations are further extended by the Weyl \gls{BMS} group \cite{freidel2021WeylBMSGroup,geiller2022PartialBondiGauge}. Yet another generalization is the \gls{ACKH} group \cite{aguiaralves2025NullInfinityKilling}, which we will discuss in \cref{sec: sky-killing-horizon}.

    \subsection{The Subleading \texorpdfstring{\Glsfmtlong{IR}}{Infrared} Triangle}
        The key question about superrotations is whether they lead to interesting new insights. If so, then they have an important significance in the \gls{IR} structure of gravity. 

        Arguably, the interest in superrotations was hinted already in the 1980s, although in a very different scenario. In a seminal paper, \textcite{brown1986CentralChargesCanonical} considered the asymptotic symmetries of \(d=2+1\) \gls{AdS} spacetime. Three-dimensional gravity has the technical advantage of being much more simple than its four-dimensional counterpart (for instance, there are no gravitational waves, which makes it easier to handle), and the isometry group of \gls{AdS} spacetime is always the conformal group in one less dimension. \textcite{brown1986CentralChargesCanonical} showed that the asymptotic symmetries not only also yield the local conformal symmetries, but they also include a central charge. This means the asymptotic symmetries do not form two copies of the Witt algebra, but actually two copies of the Virasoro algebra, which is the symmetry algebra of \emph{quantum} \glspl{CFT}! Later, \textcite{strominger1998BlackHoleEntropy} built on top of these results and uses methods from \gls{CFT} to compute the entropy of a three-dimensional black hole. Excitingly, the value matches what is obtained using standard \gls{QFTCS}. This suggests these sorts of local conformal symmetries should be taken seriously. See Sec. 2.2.3 in Ref. \citeonline{compere2019AdvancedLecturesGeneral} for a pedagogical introduction.
    
        Three-dimensional gravity is not the only context in which we can pursue a physical motivation behind superrotations. In \cref{subsec: infrared-triangle} we discussed the leading-order gravitational \gls{IR} triangle: there is a triangular relation between supertranslations, the Weinberg soft graviton theorem, and the \gls{GW} displacement memory effect. In \cref{sec: electrodynamics} we discussed the electromagnetic analog. Now that we are dealing with a new collection of asymptotic symmetries (the superrotations), we have a new \gls{IR} triangle, which one can call the subleading-order gravitational \gls{IR} triangle. Or simply subleading \gls{IR} triangle. One of the corners of this new triangle is already known to us: the superrotation symmetries. We still expect to find a new soft theorem and a new memory effect.

        The subleading soft graviton theorem was given by Cachazo and Strominger in Ref. \citeonline{cachazo2014EvidenceNewSoft}, and it provides a universal soft factor at \(\order{\omega^0}\) (where \(\omega\) is the soft graviton energy) for amplitudes involving the emission of one soft graviton. \textcite{kapec2014SemiclassicalVirasoroSymmetry} have later shown that this soft theorem is the Ward identity for a diagonal subalgebra of the extended \gls{BMS} algebra. The Cachazo--Strominger soft theorem and the extended \gls{BMS} algebra were then related to a new memory effect---known as the spin memory effect---by \textcite{pasterski2016NewGravitationalMemories}. The existence of the Cachazo--Strominger soft theorem and the prediction of the spin memory effect in classical \gls{GR} suggest that superrotations should be taken as a relevant symmetry of asymptotically flat spacetimes. See Ref. \citeonline{pasterski2019ImplicationsSuperrotations} for further discussions about the subleading \gls{IR} triangle.

        While I will not discuss the Cachazo--Strominger theorem in detail, the spin memory effect is easy to derive. Following \textcite{pasterski2016NewGravitationalMemories,pasterski2019ImplicationsSuperrotations}, we imagine some objects describing a closed orbit. These could be particles at the \glsxtrshort{LHC}, signals exchanged between the \glsxtrshort{LISA} satellites, and so on. These orbits are assumed to describe a circle \(\mathcal{C}\) (for simplicity) with radius \(L\) near null infinity. By ``near null infinity'' I mean \(L \ll r_0\), where \(r_0\) is the radius of a sphere on which the circle is circumscribed as measured from the source. The center of the circle is (in stereographic coordinates) \(\zeta_0\).

        We can parameterize \(\mathcal{C}\) by an angle \(\phi \in [0,2\pi)\). Since \(L \ll r_0\), the circle on the sphere is approximately mapped to a circle on the complex plane. Since the circle is centered at \(\zeta_0\), it has the parameterization
        \begin{equation}
            Z(\phi) = \zeta_0 + \ell e^{i \phi} + \order{\frac{L^2}{r_0^2}},
        \end{equation}
        where \(\ell\) needs to be fixed so that the physical circle on the sphere has radius \(L\). We can do this by computing the circumference of the circle on the complex plane (using the correct metric on the sphere) and imposing it is given by \(2\pi L\). Upon doing this, we find that (up to an innocuous shift on the definition of \(\phi\))
        \begin{equation}
            Z(\phi) = \zeta_0\qty[1 + \frac{L e^{i \phi} (1 + \zeta_0 \bar{\zeta_0})}{2 r_0 \sqrt{\zeta_0 \bar{\zeta_0}}}] + \order{\frac{L^2}{r_0^2}}.
        \end{equation}

        Consider now a \gls{CK} spacetime with metric given in the Bondi gauge by \cref{eq: bondi-gauge-AF}. For a light-ray moving along the circle \(\mathcal{C}\) (meaning \(\zeta = Z(\phi)\) and \(r = r_0\)), we find
        \begin{subequations}
            \begin{align}
                0 &= \frac{\dd{s}^2}{\dd{u}^2}, \\
                &= -1 + r_0^2 \tensor{\gamma}{_{A}_{B}} \pdv{\tensor{Z}{^A}}{u} \pdv{\tensor{Z}{^B}}{u} + 2 \frac{m}{r_0} + r_0 \tensor{C}{_{A}_{B}} \pdv{\tensor{Z}{^A}}{u}\pdv{\tensor{Z}{^B}}{u} + \tensor{D}{^{B}}\tensor{C}{_{A}_{B}}\pdv{\tensor{Z}{^A}}{u} + \cdots,
            \end{align}
        \end{subequations}
        where the dots stand for subleading terms. 

        Notice only the last term is odd under \(\pdv*{\tensor{Z}{^A}}{u} \to - \pdv*{\tensor{Z}{^A}}{u}\). If two light rays are set to orbit on \(\mathcal{C}\) on different directions, then, on the large-\(r_0\) limit, there will be a delay in the time it takes for each them to conclude one orbit---\ie, one of them will run a full orbit before the other. This delay will be proportional to the integral of odd term in square brackets, \ie, to
        \begin{equation}
            \Delta P = \oint_{\mathcal{C}} \tensor{D}{^{B}}\tensor{C}{_{A}_{B}}\dd{\tensor{x}{^A}}.
        \end{equation}
        While our original calculation assumed a circular motion, notice this conclusion holds for any small closed contour \(\mathcal{C}\).

        It is important to notice this effect does not happen in vacuum. Indeed, consider the decomposition (\ref{eq: shear-decomposition}) for \(\Psi = 0\), which is the vacuum case. Then 
        \begin{subequations}
            \begin{align}
                \tensor{D}{^B}\tensor{C}{_A_B} &= \qty(\tensor{D}{_{A}}\tensor{D}{_C}\tensor{D}{^C} - 2 \tensor{D}{^{B}}\tensor{D}{_B}\tensor{D}{_A})C, \\
                &= -\tensor{D}{_{A}}\qty(\tensor{D}{_C}\tensor{D}{^C} + 2)C,
            \end{align}
        \end{subequations}
        where the omitted manipulations are mere commutations of covariant derivatives and the remark that the Ricci tensor on the unit sphere matches the metric. Using this, we find that, in vacuum, 
        \begin{equation}
            \Delta P = - \oint_{\mathcal{C}} \tensor{D}{_{A}}\qty(\tensor{D}{_C}\tensor{D}{^C} + 2)C \dd{\tensor{x}{^A}} = 0,
        \end{equation}
        which vanishes because it is the integral of a total derivative over a closed contour. We see that there is only desynchronization when there is passage of gravitational radiation. 

        Finally, one may integrate over many different cycles to obtain the full delay effect. The result is 
        \begin{equation}
            \Delta^+ u = \frac{1}{2\pi L} \int  \oint_{\mathcal{C}} \tensor{D}{^{B}}\tensor{C}{_{A}_{B}}\dd{\tensor{x}{^A}} \dd{u}.
        \end{equation}
        This is the spin memory effect---the passage of a gravitational wave can introduce a relative delay on the orbital periods of light-rays spinning in opposite directions. 

        As with displacement memory, there is hope of measuring the spin memory effect. See Ref. \citeonline{grant2023OutlookDetectingGravitationalwave} for details, for example.

\section{\texorpdfstring{\Glsfmtlong{IR}}{Infrared} Symmetries in Expanding Universes}\label{sec: infrared-de-sitter}
    So far, we have been working with asymptotically flat spacetimes, which are adequate to describe isolated bodies in \gls{GR}. Nevertheless, when we consider the Universe as a whole we learn that the asymptotic behavior is different than that. More specifically, the \(\Lambda\)CDM model of cosmology tells us that the Universe is more closely related to de Sitter spacetime, not Minkowski spacetime. This invites to consider the behavior of a different class of spacetimes: spacetimes with a cosmological particle horizon. Understanding the \gls{IR} structure of these sorts of spacetimes is our present goal. While our particular investigations will follow an analysis originally due to \textcite{dappiaggi2009CosmologicalHorizonsReconstruction,dappiaggi2009DistinguishedQuantumStates,dappiaggi2017HadamardStatesLightlike} concerning the cosmological horizon itself, other authors have also studied \gls{IR} aspects of cosmological spacetimes \cite{bonga2020BMSlikeSymmetriesCosmology,enriquez-rojo2023AsymptoticDynamicsCharges}. A particularly noteworthy example are the \(\Lambda\)-\gls{BMS} symmetries considered by \textcite{compere2019LBMS4GroupdS4}, which have recently been related to memory effects \cite{compere2024QuadrupolarRadiationSitter}. Instead of considering the cosmological horizon, these investigations focus on the future infinity of de Sitter spacetime, which is a spacelike hypersurface.

    \subsection{Concordance Cosmology and de Sitter Spacetime}
        As a motivating step, let us begin by discussing a few standard results in cosmology. This will allow us to understand in more detail why spacetimes with past cosmological horizons are of interest for us. Our discussions mimic those that can be found in any recent book in \gls{GR} or cosmology---see, for example, Refs. \citeonline{baumann2022Cosmology,weinberg2008Cosmology}.

        The large scale geometry of the Universe can be described by the concordance model (or \(\Lambda\)CDM model) of cosmology \cite{planckcollaboration2020AI}. This model can be approximated by a spatially flat \gls{FLRW} spacetime, \ie, by the metric
        \begin{equation}\label{eq: flat-FLRW}
            \dd{s}^2 = -\dd{\tau}^2 + a(\tau)^2 \dd{\ell}^2,
        \end{equation}
        where \(\dd{\ell}^2\) denotes the flat Euclidean metric. The function \(a(\tau)\) is known as the scale factor and it is obtained by solving the Einstein equations. \(\tau\) is known as cosmic time and corresponds to the proper time of the inertial observers that observe the Universe as being spatially homogeneous and isotropic. These observers are known as cosmic observers. Their four-velocities commute with the Killing fields of the spatial sections with constant \(\tau\).

        One assumes the Universe is filled with a perfect fluid of energy density \(\rho\) and pressure \(P\), the stress tensor of which corresponds to
        \begin{equation}\label{eq: stress-tensor-perfect-fluid}
            \tensor{T}{_a_b} = \rho(\tau) \tensor{u}{_a}\tensor{u}{_b} + P(\tau) \qty(\tensor{g}{_a_b} + \tensor{u}{_a}\tensor{u}{_b}),
        \end{equation}
        where
        \begin{equation}
            \tensor{u}{^a} = \tensor{\qty(\pdv{\tau})}{^a}
        \end{equation}
        is the fluid's four-velocity. It corresponds to the four-velocity of the cosmic observers. This stress tensor is mandated by the symmetries imposed on the metric. The Einstein equations are then written as 
        \begin{equation}\label{eq: EFE-cosmological-constant}
            \tensor{G}{_a_b} + \Lambda \tensor{g}{_a_b} = 8 \pi \tensor{T}{_a_b},
        \end{equation}
        with a cosmological constant \(\Lambda\).

        \cref{eq: flat-FLRW,eq: stress-tensor-perfect-fluid,eq: EFE-cosmological-constant} lead to the Friedmann equations for the scale factor,
        \begin{align}
            \frac{\dot{a}^2}{a^2} &= \frac{\Lambda + 8 \pi \rho}{3}, \label{eq: friedmann-equation}\\
            \intertext{and}
            \frac{\ddot{a}}{a} &= \frac{\Lambda - 4 \pi (\rho + 3 P)}{3}. \label{eq: acceleration-equation}
        \end{align}
        \cref{eq: friedmann-equation} is known as the Friedmann equation and \cref{eq: acceleration-equation} is known as the Raychaudhuri, or acceleration, equation. The dots denote differentiation with respect to cosmic time \(\tau\). Differentiating the Friedmann equation and plugging in the Raychaudhuri equation yields
        \begin{equation}\label{eq: cosmological-energy-conservation}
            \dot{\rho} = - \frac{3 \dot{a}}{a}(\rho + P),
        \end{equation}
        which expresses energy conservation. 

        It is standard to consider matter models based on a simple equation of state given by
        \begin{equation}
            P = w \rho,
        \end{equation}
        for some constant parameter \(w\). For \(w=-1\) we recover the same results given by a cosmological constant. For \(w = 0\) we have a model for dust-like particles with vanishing pressures, which is a useful way of modelling non-relativistic (``cold'') matter. \(w = 1/3\) models radiation and ultrarelativistic particles. 

        By using \(w=0\) on \cref{eq: cosmological-energy-conservation} we find that
        \begin{equation}
            a\frac{\dot{\rho}}{\dot{a}} = -3\rho,
        \end{equation}
        and therefore
        \begin{equation}
            \rho_{m}(\tau) \propto \frac{1}{a(\tau)^3}.
        \end{equation}
        The subscript \(m\) stands for ``matter''. Similarly, plugging in \(w=1/3\) leads to
        \begin{equation}
            \rho_{r}(\tau) \propto \frac{1}{a(\tau)^4},
        \end{equation}
        where \(r\) stands for ``radiation''.

        Using these facts, a model for the Universe can be provided by using the Ansatz
        \begin{equation}\label{eq: LCDM-ansatz}
            \frac{\Lambda + 8 \pi \rho}{3} = H_0^2 \qty[\Omega_{\Lambda,0} + \Omega_{m,0} \qty(\frac{a_0}{a})^3 + \Omega_{r,0} \qty(\frac{a_0}{a})^4].
        \end{equation}
        Above, \(a_0\) is the present-day value of \(a(\tau)\) and \(H_0\) is the present-day value of the Hubble parameter
        \begin{equation}
            H(\tau) = \frac{\dot{a}}{a}.
        \end{equation}
        The constants \(\Omega_{\Lambda,0}\), \(\Omega_{m,0}\), and \(\Omega_{r,0}\) are known as density parameters and should be fixed experimentally. Notice that the Friedmann equation now becomes
        \begin{equation}\label{eq: friedmann-LCDM}
            \frac{\dot{a}^2}{a^2} = H_0^2 \qty[\Omega_{\Lambda,0} + \Omega_{m,0} \qty(\frac{a_0}{a})^3 + \Omega_{r,0} \qty(\frac{a_0}{a})^4].
        \end{equation}
        If we evaluate the expression at present-time we find the constraint
        \begin{equation}
            1 = \Omega_{\Lambda,0} + \Omega_{m,0} + \Omega_{r,0}.
        \end{equation}

\begin{figure}
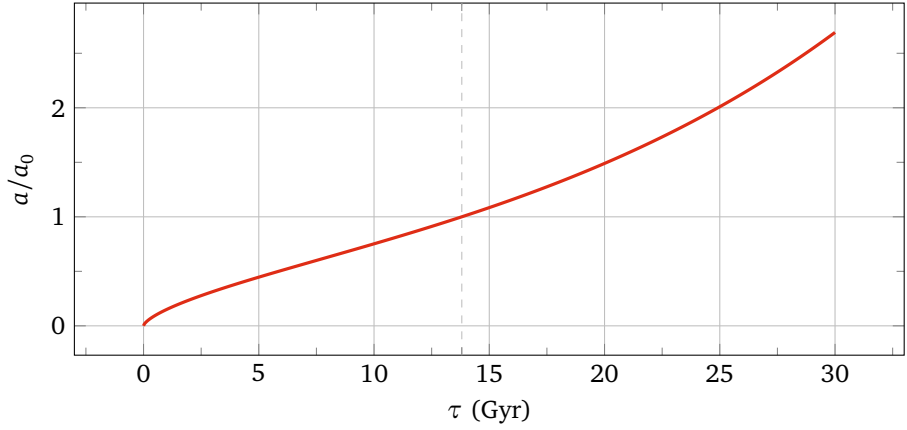

            \centering
            \includestandalone{scale-factor-lcdm}
            \caption{Scale factor (in units of \(a_0\)) as a function of cosmic time (in billions of years) obtained by numerically solving \cref{eq: friedmann-LCDM} with the data on \cref{eq: experimental-data-LCDM}. One has \(a(0) = 0\) (Big Bang) and present-day has \(\tau_0 \approx \SI{13.8}{\giga\year}\) (indicated with a dashed line).}
            \label{fig: LCDM-scale-factor}
        \end{figure}

        Experimentally, one has the values \cite{planckcollaboration2020AVI,particledatagroupcollaboration2024ReviewParticlePhysics,baumann2022Cosmology}
        \begin{subequations}\label{eq: experimental-data-LCDM}
            \begin{align}
                H_0 &= \SI{67.7(4)}{\kilo\meter\per\second\per\mega\parsec} = \SI{6.91(4)e-2}{\per\giga\year}, \\
                \Omega_{\Lambda,0} &= \num{0.689(6)}, \\
                \Omega_{m,0} &= \num{0.311(6)}, \\
                \Omega_{r,0} &= \num{9.14(25)e-5}.
            \end{align}
        \end{subequations}
        Using these data, we can solve \cref{eq: friedmann-LCDM} numerically. The resulting scale factor is plotted on \cref{fig: LCDM-scale-factor}. It is found that there is a point in the past in which the scale factor vanishes---this is the Big Bang. The Ricci scalar diverges at the Big Bang, which is therefore a curvature singularity. If we choose to measure cosmic time starting from the Big Bang, the values above lead to \(\tau_0 \approx \SI{13.8}{\giga\year}\) for the age of the Universe. 

\begin{figure}
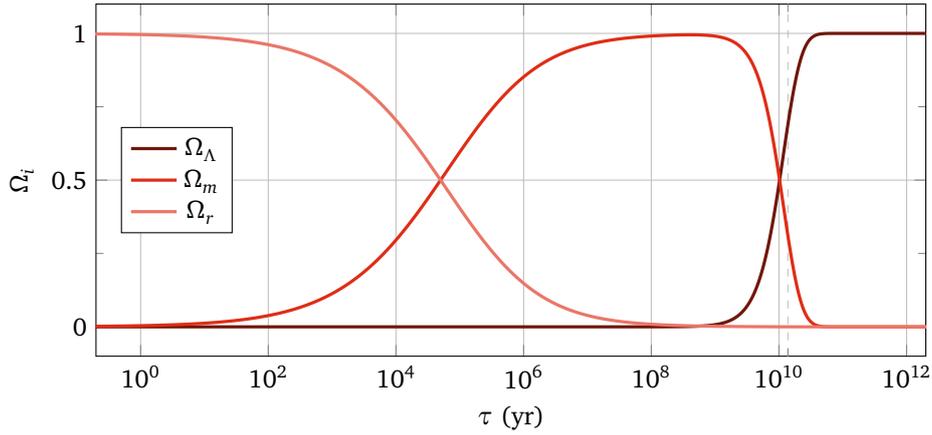

            \centering
            \includestandalone{density-parameters}
            \caption{Values of the density parameters \(\Omega_i\) defined on \cref{eq: LCDM-omega-tilde} as functions of cosmic time. A dashed line indicates present-day.}
            \label{fig: LCDM-domination}
        \end{figure}

        To understand the role that the cosmological constant, matter, and radiation play at each point in the evolution of the Universe we can make plots of the quantities
        \begin{subequations}\label{eq: LCDM-omega-tilde}
            \begin{align}
                \Omega_{\Lambda}(\tau) &= \frac{\Omega_{\Lambda,0}}{\Omega_{\Lambda,0} + \Omega_{m,0} \qty(\frac{a_0}{a(\tau)})^3 + \Omega_{r,0} \qty(\frac{a_0}{a(\tau)})^4}, \\
                \Omega_{m}(\tau) &= \frac{\Omega_{m,0}a_0^3 a(\tau)^{-3}}{\Omega_{\Lambda,0} + \Omega_{m,0} \qty(\frac{a_0}{a(\tau)})^3 + \Omega_{r,0} \qty(\frac{a_0}{a(\tau)})^4}, \\
                \Omega_{r}(\tau) &= \frac{\Omega_{r,0} a_0^4 a(\tau)^{-4}}{\Omega_{\Lambda,0} + \Omega_{m,0} \qty(\frac{a_0}{a(\tau)})^3 + \Omega_{r,0} \qty(\frac{a_0}{a(\tau)})^4}.
            \end{align}
        \end{subequations}
        These three quantities add to one. Hence, they can be used to understand which fraction of the matter of the Universe is comprised by a cosmological constant, by cold matter, or by radiation. Indeed, as attested by \cref{eq: friedmann-LCDM}, the largest of them at each time is responsible for dictating the approximate evolution of the Universe. They are plotted on \cref{fig: LCDM-domination}

        As one can read from \cref{fig: LCDM-domination}, for \(\tau \lesssim \SI{10000}{\year}\) radiation dominates and dictates the evolution of the early Universe. Later, for \(\SI{10000}{\year} \lesssim \tau \lesssim \SI{10}{\giga\year}\) matter dominates. At late times (\(\tau \gtrsim \SI{10}{\giga\year}\)), the cosmological constant dominates and the evolution of the Universe is essentially equal to that of a vacuum spacetime with a cosmological constant. In this region, the Friedmann equation becomes
        \begin{equation}\label{eq: Friedmann-deSitter}
            \frac{\dot{a}^2}{a^2} \approx H_0^2 \Omega_{\Lambda,0} = \frac{\Lambda}{3} > 0,
        \end{equation}
        where we used \cref{eq: LCDM-ansatz} with \(\rho = 0\). The inequality follows from the experimental values of \(H_0\) and \(\Omega_{\Lambda,0}\).

        \cref{eq: Friedmann-deSitter} can be solved exactly and yields
        \begin{equation}\label{eq: scale-factor-de-sitter}
            a(\tau) = a_0 \exp(\sqrt{\frac{\Lambda}{3}} (\tau - \tau_0)),
        \end{equation}
        where \(a(\tau_0) = a_0\) (present-day). In this case, the metric is given by
        \begin{equation}
            \dd{s}^2 = - \dd{\tau}^2 + a_0^2 \exp(2 \sqrt{\frac{\Lambda}{3}} (\tau - \tau_0)) \dd{\ell}^2.
        \end{equation}
        This is the metric for de Sitter spacetime in planar coordinates \cite{spradlin2001HouchesLecturesSitter}. 

\begin{figure}
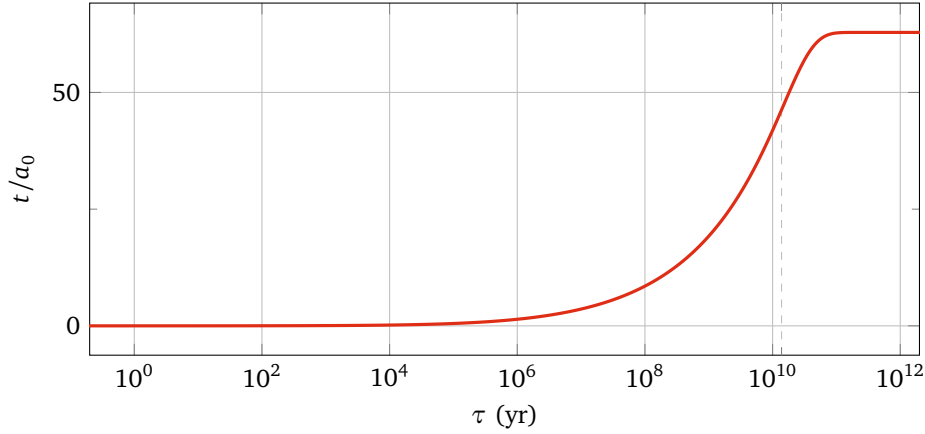

            \centering
            \includestandalone{conformal-time}
            \caption{Conformal time as a function of cosmic time for the \(\Lambda\)CDM model. Present-day is indicated by a dashed line. For large values of cosmic time, we get the asymptotic value of \(t \approx \num{63.06} a_0\).}
            \label{fig: LCDM-conformal-time}
        \end{figure}

        To get a global picture of the geometry of the Universe, let us compute the Carter--Penrose diagram. To do this, it is useful to begin by changing the time coordinate from cosmic time \(\tau\) to conformal time \(t\), which is defined by
        \begin{equation}\label{eq: conformal-time}
            t(\tau) = a_0 \int_0^{\tau} \frac{\dd{\tau'}}{a(\tau')}.
        \end{equation}
        It is possible to choose a different initial condition for \(t\), but we will work with the assumption that \(t(0) = 0\). This means conformal time vanishes at the Big Bang. \cref{fig: LCDM-conformal-time} shows the dependence of conformal time with respect to cosmic time for the experimental data given in \cref{eq: experimental-data-LCDM}. It is interesting to notice that \(t\) tends to a finite value as \(\tau \to +\infty\).

\begin{figure}[tbp]
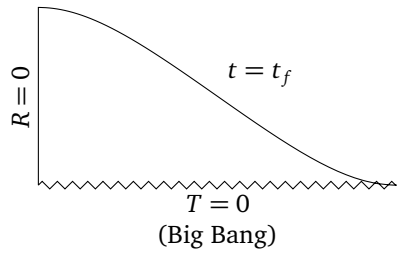

            \centering
            \includestandalone{lcdm-penrose}
            \caption{Carter--Penrose diagram for the \(\Lambda\)CDM spacetime. The zigzag line (representing the Big Bang) is a singularity. The \(t=t_f\) line denotes timelike and null infinity.}
            \label{fig: LCDM-penrose}
        \end{figure}

        The usefulness of conformal time is that the spatially flat \gls{FLRW} geometry can now be written as
        \begin{equation}
            \dd{s}^2 = a(t)^2 \qty[-\dd{t}^2 + \dd{\ell}^2],
        \end{equation}
        and it is now clear that the spacetime metric is related to that of Minkowski spacetime by a conformal transformation. This holds for any spatially flat \gls{FLRW} geometry. By repeating the analysis of \cref{subsec: infinity-minkowski} we can find the expression
        \begin{equation}\label{eq: conformal-einstein-FLRW}
            \dd{s}^2 = 4 a(t)^2 \sec[2](\frac{T-R}{2})\sec[2](\frac{T+R}{2}) \qty[-\dd{T}^2 + \dd{R}^2 + \sin^2{R} \dd{\Sph}^2],
        \end{equation}
        where conformal time is given by 
        \begin{equation}
            t = \frac{1}{2}\qty[\tan(\frac{T-R}{2}) + \tan(\frac{T+R}{2})]
        \end{equation}
        and the compactified coordinates \(T\) and \(R\) obey 
        \begin{equation}
            0 < T < \pi \qq{and} 0 \leq R < \pi - T.
        \end{equation}
        This holds for any scale factor \(a(t)\). However, if for a specific geometry we have that conformal time is bounded in the interval \(t \in (t_0,t_f)\), then we also have the constraints 
        \begin{equation}\label{eq: constraints-T-R-conformal-time}
            t_0 < t = \frac{1}{2}\qty[\tan(\frac{T-R}{2}) + \tan(\frac{T+R}{2})] < t_f.
        \end{equation}
        As attested by \cref{fig: LCDM-conformal-time}, our model of the Universe has \(t_0 = 0\) and \(t_f \approx \num{63.06}a_0\). The Carter--Penrose diagram can then be obtained by plotting the ranges of \(T\) and \(R\). The result is shown on \cref{fig: LCDM-penrose}.

\begin{figure}[tbp]
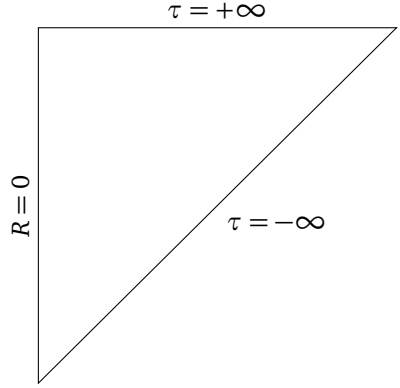

            \centering
            \includestandalone{cosmological-patch}
            \caption{Carter--Penrose diagram for the cosmological patch of de Sitter spacetime.}
            \label{fig: cosmological-desitter-penrose}
        \end{figure}

        We can also compute the Carter--Penrose diagram for de Sitter spacetime. In this case, conformal time is given by 
        \begin{subequations}\label{eq: conformal-time-de-sitter}
            \begin{align}
                t &= a_0 \int \frac{\dd{\tau}}{a(\tau)}, \\
                &= \int \exp(-\sqrt{\frac{\Lambda}{3}}(\tau - \tau_0))\dd{\tau}, \\
                &= - \sqrt{\frac{3}{\Lambda}}\qty[\exp(-\sqrt{\frac{\Lambda}{3}}(\tau - \tau_0))-1] + t_0,
            \end{align}
        \end{subequations}
        with \(t(\tau_0) = t_0\). Notice then that, since \(\tau \in \R\) (de Sitter spacetime does not present a Big Bang), we have
        \begin{equation}
            - \infty < t < t_0 + \sqrt{\frac{3}{\Lambda}}.
        \end{equation}
        It is harmless and convenient to pick \(t_0 = - \sqrt{{3}/{\Lambda}}\). In this case, we find that the coordinates \(T\) and \(R\) obey
        \begin{equation}\label{eq: cosmological-patch-de-sitter}
            -\pi < T < 0 \qq{and} 0 \leq R < \pi + T.
        \end{equation}
        Plotting the ranges of \(T\) and \(R\) gives the Carter--Penrose diagram for the cosmological patch of de Sitter spacetime, \ie, for the metric we originally considered as de Sitter spacetime without any extensions. It is shown on \cref{fig: cosmological-desitter-penrose}. It is important to notice that the \(t = 0\) boundary in de Sitter spacetime corresponds to the \(t= t_f\) boundary in the \(\Lambda\)CDM spacetime. They look different in \cref{fig: LCDM-penrose,fig: cosmological-desitter-penrose} because in de Sitter spacetime we had the freedom of using a conformal transformation to ``flatten'' the boundary, but in the \(\Lambda\)CDM spacetime we had already used this freedom when dealing with the Big Bang. 

        We can extend the original de Sitter spacetime. When dealing with the Schwarzschild spacetime, one can extend the exterior Schwarzschild geometry to obtain the full Kruskal extension. This extension includes not only the asymptotically flat region, but also the black hole region and two other regions corresponding to a white hole and to a ``parallel universe'' \cite{wald1984GeneralRelativity,choquet-bruhat2015IntroductionGeneralRelativity,chrusciel2019ElementsGeneralRelativity,carroll2019SpacetimeGeometryIntroduction}. We can do a similar procedure with de Sitter spacetime. 

\begin{figure}[tbp]
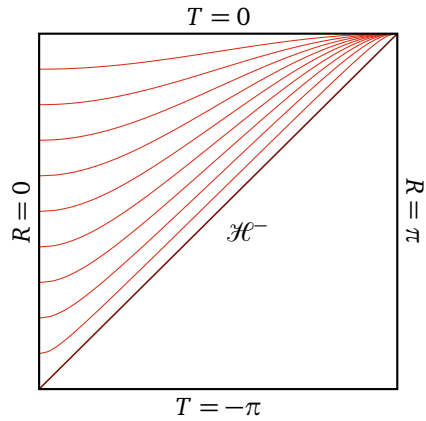

            \centering
            \includestandalone{ds-penrose}
            \caption{Global Carter--Penrose diagram for de Sitter spacetime. The thin lines depict spatially flat planes foliating the cosmological patch. These planes asymptote to the hypersurface \(\horpas\), which is a cosmological horizon. Figure adapted from Ref. \citeonline{aguiaralves2025NullInfinityKilling}.}
            \label{fig: global-desitter-penrose}
        \end{figure}

        The first step to seeing this is computing the scale factor as a function of conformal time. Using \cref{eq: scale-factor-de-sitter,eq: conformal-time-de-sitter} we see that
        \begin{equation}\label{eq: conformal-time-scale-factor-de-sitter}
            a(t) = - \sqrt{\frac{3}{\Lambda}} \frac{1}{t}.
        \end{equation}
        Using this result on \cref{eq: conformal-einstein-FLRW} and rewriting \(t\) in terms of \(T\) and \(R\) we find that the de Sitter metric can be written as
        \begin{equation}\label{eq: global-coordinates-de-sitter}
            \dd{s}^2 = \frac{3 \csc^2 T}{\Lambda}\qty[-\dd{T}^2 + \dd{R}^2 + \sin^2{R} \dd{\Sph}^2].
        \end{equation}
        From this expression it is clear that we can extend the original coordinate ranges on \cref{eq: cosmological-patch-de-sitter} to allow \(0 \leq R \leq \pi\). In this case, \(R\) is now an angular variable on a three-dimensional sphere \(\Sph^3\). The Carter--Penrose diagram for this extended spacetime (which is the full de Sitter spacetime) is shown in \cref{fig: global-desitter-penrose}.

        We notice that the boundary of the original cosmological patch is now a hypersurface on the full de Sitter spacetime. This hypersurface, indicated as \(\horpas\) on \cref{fig: global-desitter-penrose}, is known as a (past) particle horizon \cite{rindler2006RelativitySpecialGeneral,rindler1956VisualHorizonsWorld} for the cosmic observer sitting at \(R = 0\)---this observer cannot send any information to the events beyond the horizon. In fact, for de Sitter spacetime this statement is even stronger: no cosmic observer can send any information beyond the horizon and the notion of cosmic observer as we originally defined does not make sense beyond the horizon. 

        For this reason, \(\horpas\) will be referred to as a cosmological horizon: it is a distinguished null hypersurface in the spacetime that can be interpreted as the boundary of the cosmological patch of de Sitter spacetime. It seems fair then that we consider the original cosmological patch as the ``bulk'' and the global de Sitter spacetime as an extension. \(\horpas\) is then understood as a boundary similar to \(\nullfut\) (or, even better, \(\nullpas\)) in asymptotically flat spacetimes. Our expectation is then that studying this cosmological horizon we may find results similar to those we found when working with null infinity. 

        There is, however, an important point to be addressed: is this horizon physically meaningful? As \cref{fig: LCDM-penrose} clearly shows, the global structure of the \(\Lambda\)CDM universe is very different from that of de Sitter spacetime. In particular, we cannot make a past extension to get a past cosmological horizon, because the Big Bang prevents us from doing so. 

        In spite of this, at late times we know that the cosmological constant gives the most relevant contribution to the Universe's evolution. Hence, at late times, we can say the Universe is asymptotically de Sitter. It seems fair then that we can consider the past cosmological horizon of de Sitter spacetime as an auxiliary structure to understand the behavior of the Universe at late times. In fact, this does not seem so different from what we do with asymptotically flat spacetimes: we know the Universe is not asymptotically flat in large scales, but we can still use the concept of null infinity to obtain information about the behavior of regions of spacetime. In fact, the Weinberg soft graviton theorem shows that we can even use the notion of null infinity to obtain insights into the behavior of elementary particles in a locally flat patch. 

    \subsection{Inflation}
        There is still a second reason to consider ``asymptotically de Sitter'' spacetimes: it is believed that, near its beginning, the Universe underwent a stage of very fast expansion known as inflation. This period of fast expansion is very similar to a de Sitter solution. In this sense, the de Sitter solution is interesting at very early times in the history of the Universe, just as at very late times. In this section, I will very briefly review some ideas behind inflation as a motivation for considering asymptotically de Sitter spacetimes. Further information can be found in any modern book on cosmology---such as Refs. \citeonline{baumann2022Cosmology,weinberg2008Cosmology}---or on specialized reviews such as Refs. \citeonline{baumann2012TASILecturesInflation,kinney2009TASILecturesInflation}.

        While the \(\Lambda\)CDM model is very successful, it has a few drawbacks. For instance, one may inquire why we see no magnetic monopoles, even though they are a standard prediction of \glspl{GUT} \cite{thooft1974MagneticMonopolesUnified,polyakov1974ParticleSpectrumQuantum,weinberg1996ModernApplications,weinberg2008Cosmology}. Alternatively, one may inquire how the universe is spatially flat, given that spatial flatness can be an extremely unstable feature in \gls{FLRW} cosmologies \cite{baumann2022Cosmology,weinberg2008Cosmology}. 

        I will focus on a different aspect, known as the horizon problem. We observe the Universe as being extremely homogeneous and isotropic \cite{planckcollaboration2020AI}. For instance, the relative variations on the temperature of the \gls{CMB} are of the order of \(\num{e-5}\) over the whole sky \cite{planckcollaboration2020AI}. This suggests regions of the sky that are very far away from each other were once in contact. 

\begin{figure}
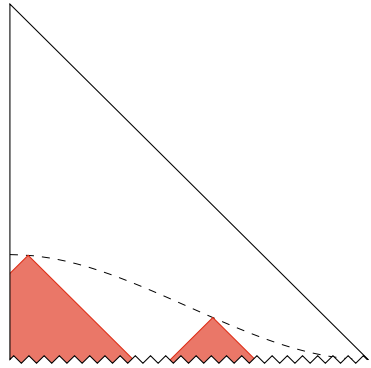

            \centering
            \includestandalone{pre-inflation-penrose}
            \caption{Carter--Penrose diagram for a radiation-filled, spatially flat universe. The zigzag at the bottom of the diagram indicates the Big Bang. The dashed line indicates the cosmic time at photon decoupling. Two light cones on different regions of the universe are indicated. Notice they are not in causal contact. Since the Carter--Penrose diagram omits the angular variables, this does not illustrate the causal relations between different regions in the night sky, but rather between regions closer and farther from Earth.}
            \label{fig: penrose-radiation-filled}
        \end{figure}

        Let us then draw a Carter--Penrose diagram of the universe and, in particular, draw the lightcones of two different points in the sky at the time of emission of the \gls{CMB}. For them to be in thermal equilibrium, they should have been in causal contact before. This is shown in \cref{fig: penrose-radiation-filled} for a radiation-filled universe---\ie, ignoring the late-time behavior of the \(\Lambda\)CDM model, which is irrelevant for this discussion anyway. One sees in the diagram that regions farther and closer to the Earth were not in causal contact, despite observational data suggesting otherwise. 

        \textcite{guth1981InflationaryUniversePossible} noticed this behavior (and the other problems I briefly mentioned) could be remedied if the universe had undergone a stage of rapid expansion before entering the radiation era. The radiation era involves most of the observable predictions of the Big Bang theory (formation of nuclei, formation of the first few elements, and so on), but it can be modified at very early times without affecting the predictions that match with experiment. 

\begin{figure}
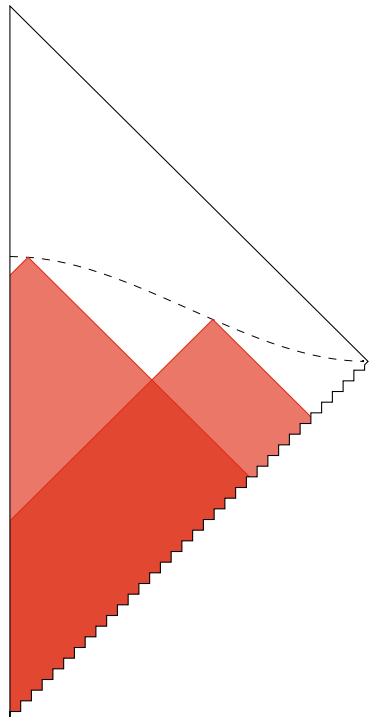

            \centering
            \includestandalone{inflation-penrose}
            \caption{Carter--Penrose diagram for a spatially flat universe dominated by a single component violating the \gls{SEC} (but different from curvature or a cosmological constant). The zigzag at the bottom of the diagram indicates the Big Bang, which has been pushed to conformal infinity. The dashed line indicates the cosmic time at photon decoupling. Two light cones on different regions of the universe are indicated. Notice they now are in causal contact. Since the Carter--Penrose diagram omits the angular variables, this does not illustrate the causal relations between different regions in the night sky, but rather between regions closer and farther from Earth.}
            \label{fig: penrose-SEC-violating}
        \end{figure}

        If one constructs a \gls{FLRW} geometry in which the universe starts with an approximately constant Hubble parameter---thus mimicking a cosmological constant---then the diagram on \cref{fig: penrose-radiation-filled} gets considerably altered. This can happen if the universe is dominated by some matter type with \(\rho + 3 P < 0\) (it is said that this matter type violates the \gls{SEC}), such as a scalar field with a suitably chosen potential. If this is the case, the diagram may become as in \cref{fig: penrose-SEC-violating}. Notice how the diagrams on \cref{fig: cosmological-desitter-penrose,fig: penrose-SEC-violating} are similar apart from the Big Bang on the latter.

        These sorts of theoretical motivations inspire the consideration of a de Sitter-like era at the beginning of the universe. Quantum effects are particularly relevant during this stage. In fact, many of the modern cosmological observations are based on the perturbations of the \gls{FLRW} metric, which can be explained by quantum fluctuations of a scalar field during inflation (this is often called an inflaton). See Refs. \citeonline{baumann2022Cosmology,weinberg2008Cosmology,lesgourgues2013TASILecturesCosmological}, for example for discussions on this. Hence, it would be extremely interesting to understand the behavior of gravity and of \gls{QFT} in regions of the universe similar to de Sitter spacetime.
        
        With these motivations, we may then proceed to the study of spacetimes with cosmological horizons with the goal of understanding their \gls{IR} symmetries.

    \subsection{\texorpdfstring{\glsfmtshort{FLRW}}{FLRW} Spacetimes with Cosmological Horizons}
        Our present goal is to get to an appropriate definition of an ``expanding universe with cosmological horizon''. This definition was provided by Dappiaggi, Moretti, and Pinamonti \cite{dappiaggi2009CosmologicalHorizonsReconstruction} (see also their review, Ref. \citeonline{dappiaggi2017HadamardStatesLightlike}). To gain some intuition, we will start by considering \gls{FLRW} spacetimes resembling de Sitter spacetime. After studying this particular case we will be able to give a more general definition---analogous to the definition of asymptotic flatness---which will no longer depend on the symmetry assumptions that lie behind \gls{FLRW}. Refs. \citeonline{dappiaggi2009CosmologicalHorizonsReconstruction,dappiaggi2017HadamardStatesLightlike} are our main references.

        We start by choosing the spacetime manifold. We want a universe with flat spatial sections, since this is the most relevant case from a cosmological perspective. We may thus prescribe the manifold
        \begin{equation}
            M = (t_0,t_f) \times \R^3,
        \end{equation}
        with \(-\infty \leq t_0 < t_f \leq +\infty\). \(M\) is diffeomorphic to \(\R^4\), but we write \((t_0,t_f)\) instead of \(\R\) to make explicit the range of conformal time. We will require that either \((t_0,t_f) = (0,+\infty)\) or \((t_0,t_f) = (-\infty,0)\). The latter case is the one we had with expanding de Sitter spacetime, while the former would be given by a different coordinate patch of de Sitter spacetime corresponding to a contracting universe. This is the patch below \(\horpas\) on \cref{fig: global-desitter-penrose}.

        The metric is given by a spatially flat \gls{FLRW} metric. In terms of conformal time \(t\), we write
        \begin{equation}
            \dd{s}^2 = a(t)^2\qty[-\dd{t}^2 + \dd{\ell}^2],
        \end{equation}
        with \(\dd{\ell}^2\) standing for the flat Euclidean metric as before. To get a spacetime resembling de Sitter spacetime we ask that there is a constant \(H\) such that
        \begin{equation}\label{eq: asymptotic-scale-factor-DMP}
            a(t) = -\frac{1}{H t} + \order{\frac{1}{t^2}}
        \end{equation}
        as \(t \to \pm \infty\) (according to whether \((t_0,t_f) = (0,+\infty)\) or \((t_0,t_f) = (-\infty,0)\)). If conformal time ranges over \((t_0,t_f) = (0,+\infty)\) we demand \(H < 0\), while for \((t_0,t_f) = (-\infty,0)\) we ask that \(H > 0\). This is intended to keep the scale factor positive. Notice that \cref{eq: asymptotic-scale-factor-DMP} is intended to resemble \cref{eq: conformal-time-scale-factor-de-sitter}. 

        This class of \gls{FLRW} geometries is a very general family that still is asymptotically de Sitter at late (or early) conformal time, which is enforced by the condition (\ref{eq: asymptotic-scale-factor-DMP}) on the scale factor. This decay property will be sufficient to ensure the presence of a cosmological horizon.

        From this point onward, I shall focus on the \((t_0,t_f) = (-\infty,0)\) case for concreteness. The other case can be dealt with in a completely analogous manner, and hence we are not losing any generality.

        Let us show we have a past cosmological horizon. To do this, we work as before and write the metric in null coordinates \((U,V)\) as defined on \cref{subsec: infinity-minkowski}. The metric is then, as in the previous section,
        \begin{equation}
            \dd{s}^2 = 4 a(t)^2 \sec[2](\frac{T-R}{2})\sec[2](\frac{T+R}{2}) \qty[-\dd{T}^2 + \dd{R}^2 + \sin^2{R} \dd{\Sph}^2],
        \end{equation}
        with
        \begin{equation}
            t = \frac{1}{2}\qty[\tan(\frac{T-R}{2}) + \tan(\frac{T+R}{2})],
        \end{equation}
        and the coordinates \(T\) and \(R\) respect
        \begin{equation}
            -\pi < T < 0 \qq{and} 0 \leq R < \pi + T.
        \end{equation}
        This is identical to what we had with de Sitter space, except for the expression of the scale factor itself. This time, it is only asymptotically de Sitter, rather than exactly de Sitter. This means we get
        \begin{equation}
            \dd{s}^2 = \qty(\frac{\csc^2 T}{H^2} + \order{\frac{1}{t^3}})\qty[-\dd{T}^2 + \dd{R}^2 + \sin^2{R} \dd{\Sph}^2].
        \end{equation}
        We thus see that the metric is well-behaved in the \(t \to - \infty\) limit and we can extend it beyond the original patch. This means we can take \(R \geq \pi + T\), but we cannot know how far \(R\) can be extended without further knowledge of the metric: it may happen that the behavior of the extension of the spacetime across the boundary \(R-T=\pi\) (or, less rigorously, \(t = -\infty\)) is very different from what we had in de Sitter spacetime (although it is similar to de Sitter in a neighborhood of \(R-T=\pi\)). 

        This ambiguity in the extension is expected: the physical spacetime is to be taken only as the original manifold \(M\). What we have now found is an extended spacetime \(\tilde{M}\), with 
        \begin{equation}
            \horpas = \qty{(T,R,\zeta,\bar{\zeta}) \in \tilde{M} \mid R-T=\pi}
        \end{equation}
        being the boundary of the original spacetime inside this extension. We see this extension as unphysical for two main reasons:
        \begin{enumerate}
            \item the cosmic observers on the spacetime can only probe the original spacetime \(M\) and are unable to reach beyond \(\horpas\), which is a particle horizon for all cosmic observers;
            \item since the horizon \(\horpas\) is unreachable (and in fact an idealization arising from the assumption that \(t_0 = - \infty\)), it is possible to obtain multiple different extensions \((\tilde{M},\tensor{\tilde{g}}{_a_b})\) and none of them should be preferred over the others, for they are experimentally undistinguishable. 
        \end{enumerate}

        Notice then that the extended spacetime \(\tilde{M}\) plays a role similar to that of the unphysical spacetime considered in asymptotically flat spacetimes---it is experimentally inaccessible, and therefore should be understood as a mere mathematical abstraction. 

        At this point, it is interesting to notice that, in fact, \((\tilde{M},\tensor{\tilde{g}}{_a_b})\) is a conformal completion of the asymptotically flat spacetime \((M,a^{-2}\tensor{g}{_a_b})\), which is just a submanifold of Minkowski spacetime. In this point of view, we notice that the past cosmological horizon \(\horpas\) of \((M,\tensor{g}{_a_b})\) coincides with the past null infinity of \((M,a^{-2}\tensor{g}{_a_b})\).

        This establishes the existence of cosmological horizons on this class of \gls{FLRW} spacetimes. It will still be useful, however, to show a few extra properties these spacetimes enjoy. 

        First we define the vector \(\tensor{X}{^a}\) as
        \begin{equation}
            \tensor{X}{^a} = H \tensor{\qty(\pdv{t})}{^a}.
        \end{equation}
        An explicit computation shows that
        \begin{equation}
            \Lie[X]\tensor{g}{_a_b} = - 2 H \pdv{t}\qty(\log a) \tensor{g}{_a_b} = - 2 \tensor{X}{^c}\tensor{\nabla}{_c}(\log a) \tensor{g}{_a_b}.
        \end{equation}
        Therefore, \(\tensor{X}{^a}\) is a conformal Killing vector field for the metric \(\tensor{g}{_a_b}\). Notice also that, as \(t \to -\infty\),
        \begin{subequations}
            \begin{align}
                \Lie[X]\tensor{g}{_a_b} &= - 2 H \frac{a'}{a} \tensor{g}{_a_b}, \\
                &= - 2 H \frac{\frac{1}{H t^2} + \order{\frac{1}{t^3}}}{-\frac{1}{H t} + \order{\frac{1}{t^2}}} \tensor{g}{_a_b}, \\
                &= \frac{2 H}{t} \tensor{g}{_a_b} + \order{\frac{1}{t^2}},
            \end{align}
        \end{subequations}
        and hence \(\Lie[X]\tensor{g}{_a_b} = 0\) at \(\horpas\). Above, we used a prime to denote differentiation with respect to conformal time. 

        Since \(\horpas\) is the surface of constant \(t = - \infty\), \(\tensor{X}{^a}\) is tangent to \(\horpas\) (more rigorously, one can change coordinates from \(t\) to \(a\) in the extended spacetime \(\tilde{M}\) and then make this argument). Furthermore, notice that
        \begin{subequations}
            \begin{align}
                \tensor{\nabla}{^b}a &= \tensor{g}{^b^c} a' \tensor{\qty(\dd{t})}{_c}, \\
                &= - \frac{a'}{a^2} \tensor{\qty(\pdv{t})}{^b}.
            \end{align}
        \end{subequations}
        For large \(t\), we see then that
        \begin{subequations}
            \begin{align}
                \tensor{\nabla}{^b}a &= - H \tensor{\qty(\pdv{t})}{^b} + \order{\frac{1}{t}}, \\
                &= - \tensor{X}{^b} + \order{\frac{1}{t}}.
            \end{align}
        \end{subequations}

        At last, we will check what is the form of the metric on \(\horpas\). We will write the metric near \(\horpas\) as
        \begin{subequations}
            \begin{align}
                \dd{s}^2 &= a(t)^2 \qty[- \dd{t}^2 + \dd{r}^2 + r^2 \dd{\Sph}^2], \\
                &= a(t)^2 \qty[- 2\dd{t}\dd{v} + \dd{v}^2 + (t-v)^2 \dd{\Sph}^2], \\
                &= \frac{1}{H^2 t^2} \qty[- 2\dd{t}\dd{v} + \dd{v}^2 + (t-v)^2 \dd{\Sph}^2] + \cdots, \\
                &= \frac{1}{H^2} \qty[- \frac{2}{t^2}\dd{t}\dd{v} + \frac{1}{t^2}\dd{v}^2 + \qty(1-\frac{v}{t})^2 \dd{\Sph}^2] + \cdots, \\
                &= \frac{1}{H^2} \qty[- 2H a'\dd{t}\dd{v} + H^2 a^2 \dd{v}^2 + \qty(1+H v a)^2 \dd{\Sph}^2] + \cdots, \\
                &= \frac{1}{H^2} \qty[- 2H \dd{a}\dd{v} + H^2 a^2 \dd{v}^2 + \qty(1+H v a)^2 \dd{\Sph}^2] + \cdots, \\
                &= \frac{1}{H^2} \qty[- 2H \dd{a}\dd{v} + \dd{\Sph}^2] + \cdots, \\
                &= \frac{1}{H^2} \qty[- 2 \dd{a}\dd{\tilde{v}} + \dd{\Sph}^2] + \cdots.
            \end{align}
        \end{subequations}
        Above we introduced advanced time \(v = t + r\), neglected the subleading terms in \(a\) (indicated by the trailing dots), and defined \(\tilde{v} = H v\) at the end. The conclusion is that the extended metric \(\tensor{\tilde{g}}{_a_b}\) restricted to \(\horpas\) can be written as 
        \begin{equation}
            \dd{\tilde{s}}^2|_{\horpas} = \frac{1}{H^2} \qty[- 2 \dd{a}\dd{\tilde{v}} + \dd{\Sph}^2].
        \end{equation}
        Up to the factor of \(H\), this is the Bondi form of the metric. Notice that the induced metric is given by 
        \begin{equation}
            \dd{\tilde{\sigma}}^2|_{\horpas} = \frac{1}{H^2}  \dd{\Sph}^2.
        \end{equation}
        Finally, one should notice that 
        \begin{equation}\label{eq: X-tildev-FLRW}
            \tensor{X}{^a}\tensor{\nabla}{_a}\tilde{v} = H^2,
        \end{equation}
        and hence \(\tilde{v}\) can be understood as the parameter along the integral lines of \(X\) on \(\horpas\).

    \subsection{Expanding Universes with Cosmological Horizons}\label{subsec: expanding-universes-cosmological-horizons}
        We are now ready to provide a definition of an expanding universe with geodesically complete cosmological particle horizon, originally given by \textcite{dappiaggi2009CosmologicalHorizonsReconstruction}. We follow Refs. \citeonline{dappiaggi2009CosmologicalHorizonsReconstruction,dappiaggi2017HadamardStatesLightlike}.

        \begin{definition}[Expanding Universe with Geodesically Complete Cosmological Particle Horizon]\label{def: expanding-universe-cosmological-horizon}
            Consider a spacetime \((M,\tensor{g}{_a_b})\). We assume \(M\) has the topology \(\R \times \Sigma\), with \(\Sigma\) a Riemannian manifold (interpreted as the spatial section of \(M\)). Suppose there are 
            \begin{enumerate}
                \item an unphysical spacetime \((\tilde{M},\tensor{g}{_a_b})\),
                \item a smooth isometric embedding\footnote{\(\psi\) is isometric if, and only if, \(\psi^*\tensor{g}{_a_b} = \tensor{g}{_a_b}|_{\psi(M)}\).} \(\psi \colon M \to \tilde{M}\) with \(\psi(M)\) open in \(\tilde{M}\),
                \item a function \(a\colon M \to \R\) with \(a > 0\),
                \item a future-directed timelike vector field \(\tensor{X}{^a}\), 
                \item a positive constant \(H > 0\).
            \end{enumerate}
            Suppose these objects satisfy the following conditions.
            \begin{enumerate}
                \item \(\psi(M)\) is the interior of a manifold with boundary, where the boundary \(\horpas = \bound\psi(M)\) is an embedded three-manifold in \(\tilde{M}\) and such that \(\horpas \cap \causalf(\psi(M);\tilde{M}) = \varnothing\).
                \item The function \(a\) can be extended to a smooth function on \(\tilde{M}\) such that \(a|_{\horpas} = 0\) and \(\tensor{\tilde{\nabla}}{_b}a|_{\horpas} \neq 0\).
                \item The vector field \(X\) is a conformal Killing vector field for the metric \(\tensor{\tilde{g}}{_a_b}\) in a neighborhood of \(\horpas\), with
                \begin{equation}\label{eq: Lie-X-gab-cosmological-horizon}
                    \Lie[X]\tensor{\tilde{g}}{_a_b} = - 2 \tensor{X}{^c}\tensor{\nabla}{_c}(\log a) \tensor{\tilde{g}}{_a_b}.
                \end{equation}
                Furthermore, \(\tensor{X}{^c}\tensor{\nabla}{_c}(\log a) \to 0\) approaching \(\horpas\) and \(\tensor{X}{^a}\) does not tend identically to zero approaching \(\horpas\).
                \item As a manifold, the diffeomorphism \(\horpas \cong \R \times \Sph^2\) holds. Furthermore, the metric \(\tensor{\tilde{g}}{_a_b}\) has the Bondi form up to a constant at \(\horpas\),
                \begin{equation}\label{eq: extended-metric-horpas}
                    \dd{\tilde{s}}^2|_{\horpas} = H^{-2}[-2\dd{a}\dd{v} + \dd{\Sph}^2].
                \end{equation}
                In particular, the geodesics given by \(v \mapsto (v,\theta,\phi)\) on \(\horpas\) are complete.
            \end{enumerate}
            If all of these conditions are met, we say that \((M,\tensor{g}{_a_b},a,\tensor{X}{^a},H)\) is an \emph{expanding universe with geodesically complete cosmological particle horizon}---or simply an \emph{expanding universe with cosmological horizon}. \(\horpas\) is said to be the \emph{cosmological (particle) horizon} of \((M,\tensor{g}{_a_b})\) and the integral parameter of \(X\) is the \emph{conformal (cosmological) time}.
        \end{definition}
        
        Notice this definition is intended to mimic the behavior of \gls{FLRW} spacetimes with cosmological horizons, but lifts the requirements of homogeneity and isotropy that are present in the \gls{FLRW} case. Hence, this definition is much more general. 

        At \(\horpas\) we can consider the object \(\tensor{\tilde{\nabla}}{^b}a = \tensor{\tilde{g}}{^b^c}\tensor{\tilde{\nabla}}{_c}a\). It is given by 
        \begin{equation}\label{eq: nabla-a-cosmological-horizon}
            \tensor{\tilde{\nabla}}{^b}a = - H^2 \tensor{\qty(\pdv{v})}{^b}.
        \end{equation}

        One has the following result concerning the behavior of these structures \cite{dappiaggi2009CosmologicalHorizonsReconstruction}. 
        \begin{proposition}
            Consider an expanding universe with cosmological horizon \((M,\tensor{g}{_a_b},a,\tensor{X}{^a},H)\) with cosmological particle horizon \(\horpas\). Let \(v\) be the affine parameter along the geodesics of \(\horpas\) such that \cref{eq: extended-metric-horpas} holds. Then the following statements hold true.
            \begin{enumerate}
                \item The vector field \(\tensor{X}{^a}\) defined on \(M\) admits a unique smooth extension \(\tensor{\tilde{X}}{^a}\) to \(M \cup \horpas\). On \(\horpas\), \(\tensor{\tilde{X}}{^a}\) vanishes at most on a set with empty interior. Furthermore, \(\tensor{\tilde{X}}{^a}\) satisfies the Killing equation for the extended metric \(\tensor{\tilde{g}}{_a_b}\) on \(\horpas\).
                \item In terms of the coordinate \(v\) and the angular coordinates \((\theta,\phi) \in \Sph^2\) parameterizing \(\horpas\), it holds that
                \begin{equation}\label{eq: X-fsphere-cosmological-horizon}
                    \tensor{\tilde{X}}{^a} = f(\theta,\phi)\tensor{\qty(\pdv{v})}{^a},
                \end{equation}
                for some function \(f \in \ck[\infty](\Sph^2)\).
            \end{enumerate}
        \end{proposition}

        Notice that in the case of an \gls{FLRW} spacetime the function \(f\) is a constant, as shown in \cref{eq: X-tildev-FLRW}. In this sense, one can interpret the angular dependency of the function \(f\) as a measure of how the spacetime fails to be isotropic at the past cosmological horizon.

    \subsection{Case Study: de Sitter Spacetime}
        Before moving on to studying the \gls{DMP} group of symmetries on expanding universes with cosmological horizons, we should make a case study to gain some intuition about the symmetries we expect to find. Since our model universe is de Sitter spacetime (just as Minkowski spacetime was the model for asymptotically flat geometries), we should first study the isometries of de Sitter spacetime. 

        Since de Sitter spacetime is conformally Minkowskian, they share the same conformal Killing vector fields. Furthermore, we know all Killing vector fields for de Sitter spacetime must be conformal Killing vector fields as well---which follows from the definition of (conformal) Killing vector fields. Hence, we can compute the conformal Killing vector fields of Minkowski spacetime and then check by explicit computation which of them yield Killing vector fields for de Sitter spacetime. 

        In any spacetime dimension \(n > 2\) the conformal Killing vector fields for Minkowski spacetime are spanned by 
        \begin{subequations}
            \begin{align}
                \tensor*{P}{^a_\mu} &= \tensor{\qty(\pdv{\tensor{x}{^\mu}})}{^a}, \\
                \tensor*{J}{^a_\mu_\nu} &= \tensor{x}{_\mu}\tensor{\qty(\pdv{\tensor{x}{^\nu}})}{^a} - \tensor{x}{_\nu}\tensor{\qty(\pdv{\tensor{x}{^\mu}})}{^a}, \\
                \tensor*{D}{^a} &= \tensor{x}{^\mu}\tensor{\qty(\pdv{\tensor{x}{^\mu}})}{^a}, \\
                \tensor*{K}{^a_\mu} &= 2\tensor{x}{_\mu}\tensor{x}{^\nu}\tensor{\qty(\pdv{\tensor{x}{^\nu}})}{^a} - \tensor{x}{^\nu}\tensor{x}{_\nu}\tensor{\qty(\pdv{\tensor{x}{^\mu}})}{^a},
            \end{align}
        \end{subequations}
        where the generators of translations (\(\tensor{P}{_\mu}\)) and Lorentz transformations (\(\tensor{J}{_\mu_\nu}\)) are taken from \cref{eq: poincare-generators}. All Greek index contractions are performed with the Minkowski metric. \(\tensor*{D}{^a}\) generates transformations known as dilations, which simply scale all coordinates by an equal amount. \(\tensor*{K}{^a_\mu}\) generate transformations known as \glspl{SCT}. The assumption \(n > 2\) is necessary, as in \(n=2\) the algebra of conformal Killing vector fields is infinite-dimensional and corresponds to two copies of the Witt algebra, as discussed on \cref{sec: superrotations}. These topics are often discussed in references on \gls{CFT}, such as \citeonline{difrancesco1997ConformalFieldTheory,schottenloher2008MathematicalIntroductionConformal,blumenhagen2009IntroductionConformalField}.

        One can now perform explicit computations of each of the commutators of these vector fields to find that the Lie algebra they span is given by 
        \begin{subequations}\label{eq: decomposed-conformal-algebra-for-de-sitter}
            \begin{align}
                \comm*{\tensor{P}{_\mu}}{\tensor{P}{_\nu}} &= 0, \\
                \comm*{\tensor{P}{_\mu}}{\tensor{J}{_\rho_\sigma}} &= \tensor{\eta}{_\mu_\rho}\tensor{P}{_\sigma} - \tensor{\eta}{_\mu_\sigma}\tensor{P}{_\rho}, \\
                \comm*{\tensor{J}{_\mu_\nu}}{\tensor{J}{_\rho_\sigma}} &= \tensor{\eta}{_\nu_\rho}\tensor{J}{_\mu_\sigma} - \tensor{\eta}{_\mu_\rho}\tensor{J}{_\nu_\sigma} - \tensor{\eta}{_\sigma_\mu}\tensor{J}{_\rho_\nu} + \tensor{\eta}{_\sigma_\nu}\tensor{J}{_\rho_\mu}, \\
                \comm*{D}{D} &= 0, \\
                \comm*{D}{\tensor{P}{_\mu}} &= -\tensor{P}{_\mu}, \\
                \comm*{D}{\tensor{J}{_\rho_\sigma}} &= 0, \\
                \comm*{D}{\tensor{K}{_\mu}} &= \tensor{K}{_\mu}, \\
                \comm*{\tensor{K}{_\mu}}{\tensor{K}{_\nu}} &= 0, \\
                \comm*{\tensor{K}{_\mu}}{\tensor{P}{_\nu}} &= -2(\tensor{\eta}{_\mu_\nu}D + \tensor{J}{_\mu_\nu}), \\
                \comm*{\tensor{K}{_\mu}}{\tensor{J}{_\rho_\sigma}} &= \tensor{\eta}{_\mu_\rho}\tensor{K}{_\sigma} - \tensor{\eta}{_\mu_\sigma}\tensor{K}{_\rho}.
            \end{align}
        \end{subequations}

        To better understand this Lie algebra it is convenient to define \cite{blumenhagen2009IntroductionConformalField}
        \begin{subequations}
            \begin{align}
                \tensor*{L}{^a_\mu_\nu} &= \tensor*{J}{^a_\mu_\nu}, \\
                \tensor*{L}{^a_{-2}_,_{-1}} &= \tensor{D}{^a}, \\
                \tensor*{L}{^a_{-2}_,_{\mu}} &= \frac{1}{2}\qty(\tensor*{P}{^a_\mu}-\tensor*{K}{^a_\mu}), \\
                \tensor*{L}{^a_{-1}_,_{\mu}} &= \frac{1}{2}\qty(\tensor*{P}{^a_\mu}+\tensor*{K}{^a_\mu}).
            \end{align}
        \end{subequations}
        Then we have the algebra
        \begin{equation}\label{eq: conformal-algebra-pre-de-sitter}
            \comm{\tensor{L}{_a_b}}{\tensor{L}{_c_d}} = \tensor{\tilde{\eta}}{_b_c}\tensor{L}{_a_d} - \tensor{\tilde{\eta}}{_a_c}\tensor{L}{_b_d} - \tensor{\tilde{\eta}}{_d_a}\tensor{L}{_c_b} + \tensor{\tilde{\eta}}{_d_b}\tensor{L}{_c_a},
        \end{equation}
        with \(a,b = -2, \ldots, n\) and \(\tensor{\tilde{\eta}}{_a_b}\) being the Minkowski metric with signature \((p+1,q+1)\) (where \((p,q)\) is the signature of the original space). One may notice from \cref{eq: poincare-algebra} that, for a four-dimensional Lorentzian spacetime, this is precisely the Lorentz algebra \(\so*(4,2)\).

        To get to the de Sitter isometries, we need to figure out which of these conformal transformations are isometries of de Sitter spacetime. Through explicit computation, one finds the answer is \(\tensor*{P}{^a_i}\), \(\tensor*{J}{^a_i_j}\), \(\tensor*{D}{^a}\), and \(\tensor*{K}{^a_i}\) for \(i,j = 1,2,3\). Hence, we find that the isometry algebra for de Sitter spacetime is given by (\ref{eq: conformal-algebra-pre-de-sitter}) for \(a,b,c,d \neq 0\). This means the algebra is given by \(\so*(4,1)\). One can then integrate the Killing vector fields to find that the isometry group is \(\SO*(4,1)\), known as the de Sitter group. This is a ten-dimensional group, corresponding to the fact that de Sitter spacetime is maximally symmetric. 

        There is still an unaddressed subtlety: while \(\SO*(4,1)\) is indeed the (or, in fact, a) Lie group with Lie algebra \(\so*(4,1)\), it may happen that the Lie algebra of Killing vector fields does not integrate to a \(\SO*(4,1)\) isometry group. In fact, integrating the transformation generated by \(\tensor*{P}{^a_i}\), \(\tensor*{J}{^a_i_j}\), and \(\tensor*{D}{^a}\) yields
        \begin{equation}
            \tensor{x}{^\mu} \to e^{\lambda} \tensor{R}{^\mu_\nu}\tensor{x}{^\nu} + \tensor{b}{^\mu},
        \end{equation}
        with \(\lambda \in \R\), \(R \in \SO(3)\), and \(b \in \R^3\). \(\lambda\) yields the dilation, \(R\) yields a rotation and \(b\) yields a spatial translation. Meanwhile, the transformation generated by \(\tensor*{K}{^a_i}\) is given by
        \begin{equation}
            \tensor{x}{^\mu} \to \frac{\tensor{x}{^\mu} - \tensor{x}{^\nu}\tensor{x}{_\nu}\tensor{c}{^\mu}}{1 - 2 \tensor{x}{^\nu}\tensor{c}{_\nu} + \tensor{x}{^\nu}\tensor{x}{_\nu}\tensor{c}{^\rho}\tensor{c}{_\rho}},
        \end{equation}
        for \(c \in \R^3\). Greek indices are contracted with the Minkowski metric. \(c\) yields a \gls{SCT}. Notice, however, that this transformation takes some points to infinity. This is not a problem, and is the reason one usually considers a conformal compactification when discussing conformal symmetry. 

        The true issue we have is that \glspl{SCT} map points in the cosmological patch to other regions of de Sitter spacetime. In other words, they take points from the physical region of de Sitter spacetime---which is the relevant region in cosmology---and take them to the unphysical region. For our purposes, we are only focusing in symmetries of the cosmological patch, and hence transformations that do not preserve the cosmological patch are not of interest.

        To see this it will be useful to employ coordinates appropriate to the almost-Bondi gauge we are using at \(\horpas\). This is due to two reasons: Bondi-like coordinates extend from the cosmological patch to a neighborhood of \(\horpas\) (and hence we can see which transformations preserve the cosmological patch), and later this coordinate system is the one we will employ when discussing asymptotic symmetries. In Bondi gauge, with \(a = -(Ht)^{-1}\) and \(v = H(t+r)\), the de Sitter metric is written as
        \begin{equation}
            \dd{s}^2 = H^{-2}\qty[a^2 \dd{v}^2 - 2 \dd{a} \dd{v} + (1 + a v)^2 \dd{\Sph}^2].
        \end{equation}
        In the \(a \to 0\) limit (\ie, at \(\horpas\)) the Killing vector fields are then given by 
        \begin{subequations}\label{eq: de-sitter-killing-vector-fields}
            \begin{align}
                \tensor*{P}{^a_x} &= H \cos\phi \sin\theta \tensor{\qty(\pdv{v})}{^a}, \\
                \tensor*{P}{^a_y} &= H \sin\phi \sin\theta \tensor{\qty(\pdv{v})}{^a}, \\
                \tensor*{P}{^a_z} &= H \cos\theta \tensor{\qty(\pdv{v})}{^a}, \\
                \tensor*{D}{^a} &= v \tensor{\qty(\pdv{v})}{^a}, \\
                \tensor*{J}{^a_x_y} &= \tensor{\qty(\pdv{\phi})}{^a}, \\
                \tensor*{J}{^a_y_z} &= -\sin\phi \tensor{\qty(\pdv{\theta})}{^a} - \cot\theta \cos\phi \tensor{\qty(\pdv{\phi})}{^a}, \\
                \tensor*{J}{^a_z_x} &= \cos\phi \tensor{\qty(\pdv{\theta})}{^a} - \cot\theta \sin\phi \tensor{\qty(\pdv{\phi})}{^a}, \\
                \tensor*{K}{^a_x} &= - \frac{\sin\theta\cos\phi}{H}\qty[2\tensor{\qty(\pdv{a})}{^a} - v^2 \tensor{\qty(\pdv{v})}{^a} + 2 v \cot\theta \tensor{\qty(\pdv{\theta})}{^a} - 2 v \csc^2\theta \tan\phi \tensor{\qty(\pdv{\phi})}{^a}], \\
                \tensor*{K}{^a_y} &= - \frac{\sin\theta\sin\phi}{H}\qty[2\tensor{\qty(\pdv{a})}{^a} - v^2 \tensor{\qty(\pdv{v})}{^a} + 2 v \cot\theta \tensor{\qty(\pdv{\theta})}{^a} + 2 v \csc^2\theta \cot\phi \tensor{\qty(\pdv{\phi})}{^a}], \\
                \tensor*{K}{^a_z} &= - \frac{\cos\theta}{H}\qty[2\tensor{\qty(\pdv{a})}{^a} - v^2 \tensor{\qty(\pdv{v})}{^a} - 2 v \tan\theta \tensor{\qty(\pdv{\theta})}{^a}].
            \end{align}
        \end{subequations}
        Notice that the vectors \(\tensor*{K}{^a_i}\) have components along the \(\tensor{\qty(\pdv*{a})}{^a}\) direction. This means they have a component transversal to the cosmological horizon, meaning their flow can take points in the cosmological patch to the remaining of de Sitter spacetime, and vice-versa. 

        With these considerations in mind, we see that the isometry group of the cosmological patch of de Sitter spacetime only requires us to consider the subalgebra of (\ref{eq: decomposed-conformal-algebra-for-de-sitter}) spanned by \(\tensor*{P}{^a_i}\) (spatial translations), \(\tensor{D}{^a}\) (dilations), and \(\tensor*{J}{^a_i_j}\) (spatial rotations). The resulting algebra is then
        \begin{equation}
            \cds(3,1) = \so(3) \loplus \qty(\R \loplus \R^3),
        \end{equation}
        where I wrote\footnote{The nomenclature and notations for the cosmological de Sitter algebra and group are not standard in the literature.} \(\cds(3,1)\) for the ``cosmological de Sitter algebra in \(3+1\) dimensions''. The group of isometries is then the ``cosmological de Sitter group in \(3+1\) dimensions'' 
        \begin{equation}\label{eq: CdS31}
            \CdS(3,1) = \SO(3) \ltimes \qty(\R \ltimes \R^3).
        \end{equation}
        \(\SO(3)\) corresponds to spatial rotations, \(\R^3\) corresponds to spatial translations, and \(\R\) corresponds to dilations. 

        It is interesting to notice that even though de Sitter spacetime is a maximally symmetric spacetime, the cosmological patch of de Sitter spacetime is not maximally symmetric. This is meant in the sense that not all of the de Sitter group \(\SO*(4,1)\) preserves the cosmological patch. Instead, only the subgroup \(\CdS(3,1)\) does. 
 
    \subsection{The \texorpdfstring{\glsfmtlong{DMP}}{Dappiaggi--Moretti--Pinamonti} Group}
        The \gls{BMS} group is constructed by considering the behavior at infinity of asymptotically flat spacetimes. It is based on studying the Carrollian manifold obtained at future (or past) null infinity. However, what happens if we now consider an expanding universe with cosmological horizon?

        \textcite{dappiaggi2009CosmologicalHorizonsReconstruction,dappiaggi2009DistinguishedQuantumStates,dappiaggi2017HadamardStatesLightlike} considered this problem and proved the existence of an infinite-dimensional symmetry group at the cosmological horizon analogous to the \gls{BMS} group. I will refer to this group as the \glsxtrfull{DMP} group and denote it by \(\DMP\). Our present goal is to derive this group by working with the framework of Carrollian manifolds we previously employed. 

        We first need to identify what is the Carrollian structure of \(\horpas\). By considering \cref{def: expanding-universe-cosmological-horizon} we see that the Carrollian manifold is \(\horpas = \R \times \Sph^2\) (up to diffeomorphism) and the induced Carrollian metric is always the round metric on the sphere with radius \(H^{-1}\). One natural candidate for the vector \(\tensor{n}{^a}\) giving the kernel of the Carrollian metric is the vector \(\tensor{\tilde{\nabla}}{^b}a\).

        What are the allowed symmetries of the Carrollian structure? The basic transformation we are allowed to make that will affect the Carrollian structure is a ``gauge transformation'' \(a \to \omega a\), where \(\omega \in \ck[\infty](\tilde{M})\) does not vanish at \(\horpas\). Such a transformation will act on the standard Carrollian triple according to 
        \begin{equation}
            (\R \times \Sph^2, \dd{\Sph}^2, \tensor{\tilde{\nabla}}{^b}a) \to (\R \times \Sph^2, \dd{\Sph}^2, \omega \tensor{\tilde{\nabla}}{^b}a).
        \end{equation}
        If we want to preserve the manifold structure, we must have diffeomorphisms. If we want to preserve the metric, we must have a sphere isometry. Hence, so far the most general transformation is of the form
        \begin{equation}
            v \to h(v,\theta,\phi)\qc (\theta,\phi) \to R(\theta,\phi),
        \end{equation}
        where \(R \in \SO(3)\) and \(h \in \ck[\infty](\R \times \Sph^2)\). We still need to impose that \(\tensor{\tilde{\nabla}}{^b}a\) is preserved up to a conformal transformation.
        
        This transformation is expected to be accompanied by a suitable transformation of the \(v\) coordinate, which is effectively defined on \(\horpas\) by the condition
        \begin{equation}
            \tensor{\tilde{\nabla}}{^b}a \tensor{\tilde{\nabla}}{_b}v|_{\horpas} = - H^2.
        \end{equation}
        Using this equation and the fact that \(a\) vanishes at \(\horpas\) we find that \(a \to \omega a\) must be accompanied by the transformation \(\dd{v} \to \omega^{-1} \dd{v}\). This expression implies that the \(a \to \omega a\), \(\dd{v} \to \omega^{-1} \dd{v}\) transformation is isometric for \(\horpas\), as (\ref{eq: extended-metric-horpas}) is kept unchanged. 

        To further restrict the possible transformations we employ \cref{eq: Lie-X-gab-cosmological-horizon}. We know that the transformation of the ``scale factor'' \(a\) must still respect the condition
        \begin{subequations}
            \begin{align}
                \Lie[X]\tensor{\tilde{g}}{_a_b} &= - 2 \tensor{X}{^c}\tensor{\nabla}{_c}(\log \omega a) \tensor{\tilde{g}}{_a_b}, \\
                &= - 2 \tensor{X}{^c}\tensor{\nabla}{_c}(\log a) \tensor{\tilde{g}}{_a_b} - 2 \tensor{X}{^c}\tensor{\nabla}{_c}(\log \omega) \tensor{\tilde{g}}{_a_b}.
            \end{align}
        \end{subequations}
        This then implies that \(\tensor{X}{^c}\tensor{\nabla}{_c}(\log \omega) = 0\). It then follows from \cref{eq: X-fsphere-cosmological-horizon} that \(\omega\) cannot depend on \(v\). We now see that \(\dd{v} \to \omega^{-1} \dd{v}\) implies
        \begin{equation}
            v \to \exp(f(\theta,\phi)) v + g(\theta,\phi),
        \end{equation}
        with \(f, g \in \ck[\infty](\Sph^2)\) (I defined \(f(\theta,\phi) = -\log \omega(\theta,\phi)\)).

        We thus conclude that the most general transformation that preserves the cosmological horizon is given by
        \begin{equation}\label{eq: general-DMP-transformation}
            v \to \exp(f(\theta,\phi)) v + g(\theta,\phi)\qc (\theta,\phi) \to R(\theta,\phi),
        \end{equation}
        with \(R \in \SO(3)\) and \(f, g \in \ck[\infty](\Sph^2)\). 

        The transformation (\ref{eq: general-DMP-transformation}) can be written in a more compact notation as the triple \((R,f,g)\). It can be seen then that the composition law for these transformations is 
        \begin{equation}
            (R,f,g) \odot (R',f',g') = (R \circ R',f + f' \circ R,\exp(f' \circ R) \cdot g + g'),
        \end{equation}
        where \(\circ\) denotes composition and \(\cdot\) denotes the pointwise product of functions. This product defines a Lie group structure on the manifold \(\SO(3) \times \ck[\infty](\Sph^2) \times \ck[\infty](\Sph^2)\). The identity is given by \((\Eins, 0, 0)\). The inverse is given by
        \begin{equation}
            (R,f,g)^{-1} = (R^{-1},-f\circ R^{-1},- \exp(-f)\cdot g).
        \end{equation}
        This group is the \gls{DMP} group, henceforth denoted by \(\DMP\). 

        Next we notice that 
        \begin{subequations}
            \begin{align}
                (\Eins, f, g) \odot (\Eins, 0, g') \odot (\Eins, f, g)^{-1} &= (\Eins, f, g + g') \odot (\Eins, f, g)^{-1}, \\
                &= (\Eins, f, g + g') \odot (\Eins, -f, -e^{-f} \cdot g), \\
                &= (\Eins, 0, e^{-f} \cdot g + e^{-f} \cdot g' -e^{-f} \cdot g), \\
                &= (\Eins, 0, e^{-f} \cdot g').
            \end{align}
        \end{subequations}
        This shows \(\qty{(\Eins, 0, g) | g \in \ck[\infty](\Sph^2)}\) is a normal subgroup of \(\qty{(\Eins, f, g) | f, g \in \ck[\infty](\Sph^2)}\). We also notice that 
        \begin{equation}
            (\Eins, f, 0) \odot (\Eins, 0, g) = (\Eins, f, g).
        \end{equation}
        By adding in the fact that 
        \begin{equation}
            \qty{(\Eins, f, 0) | f \in \ck[\infty](\Sph^2)} \cap \qty{(\Eins, 0, g) | g \in \ck[\infty](\Sph^2)} = \qty{(\Eins,0,0)}
        \end{equation}
        we conclude that 
        \begin{subequations}
            \begin{align}
                \qty{(\Eins, f, g) | f, g \in \ck[\infty](\Sph^2)} &= \qty{(\Eins, f, 0) | f \in \ck[\infty](\Sph^2)} \ltimes \qty{(\Eins, 0, g) | g \in \ck[\infty](\Sph^2)}, \\
                &= \ck[\infty](\Sph^2) \ltimes \ck[\infty](\Sph^2).
            \end{align}
        \end{subequations}

        This still does not yield the full structure of \(\DMP\), which has the manifold structure \(\SO(3) \times \ck[\infty](\Sph^2) \times \ck[\infty](\Sph^2)\). To get to this, we must now consider the behavior of the rotations. We find that
        \begin{subequations}
            \begin{align}
                (R,f,g) \odot (\Eins,f',g') &\odot (R,f,g)^{-1} \notag, \\ &= (R,f,g) \odot (\Eins,f',g') \odot (R^{-1},-f\circ R^{-1},- \exp(-f)\cdot g), \\
                &= (R,f + f' \circ R,\exp(f' \circ R) \cdot g + g') \odot (R^{-1},-f\circ R^{-1},- \exp(-f)\cdot g), \\
                &= (\Eins,f' \circ R,\exp(f' \circ R - f) \cdot g + \exp(-f)g' - \exp(-f)\cdot g),
            \end{align}
        \end{subequations}
        which means \(\ck[\infty](\Sph^2) \ltimes \ck[\infty](\Sph^2)\) is a normal subgroup of \(\DMP\). Furthermore, we notice that
        \begin{equation}
            (R,0,0) \odot (\Eins,f,g) = (R,f,g).
        \end{equation}
        Furthermore, 
        \begin{equation}
            \qty{(R,0,0) \mid R \in \SO(3)} \cap (\ck[\infty](\Sph^2) \ltimes \ck[\infty](\Sph^2)) = \qty{(\Eins,0,0)}.
        \end{equation}
        Therefore, we conclude that the \gls{DMP} group has the structure of an iterated semidirect product, 
        \begin{equation}
            \DMP = \SO(3) \ltimes (\ck[\infty](\Sph^2) \ltimes \ck[\infty](\Sph^2)).
        \end{equation}
        This generalizes the cosmological de Sitter group \(\CdS(3,1)\) presented in \cref{eq: CdS31}. It is interesting to notice that it would not be possible to find a similar decomposition for the de Sitter group \(\SO*(4,1)\), because it does not admit nontrivial normal subgroups (it is said to be a simple Lie group).

        We may also perform the symmetry analysis using conformal Killing vector fields, which will yield us immediately the \gls{DMP} algebra \(\dmp\). The calculation is completely analogous to the calculation we performed when studying the \gls{BMS} algebra in \cref{sec: bms-group}. 

        We are looking for vector fields \(\tensor{\xi}{^a}\) on \(\horpas\) such that
        \begin{subequations}
            \begin{align}
                \Lie[\xi]\tensor{\gamma}{_a_b} &= 0, \label{eq: Lie-xi-gamma-DMP-algebra} \\
                \intertext{and} 
                \Lie[\xi]\tensor{n}{^a} &= \lambda \tensor{n}{^a}, \label{eq: Lie-xi-n-DMP-algebra}
            \end{align}
        \end{subequations}
        where \(\tensor{\gamma}{_a_b}\) is the round metric on the sphere with radius \(H^{-1}\), \(\tensor{n}{^b} = \tensor{\tilde{\nabla}}{^b}a\), and \(\lambda \in \ck[\infty](\Sph^2)\).

        As we did in the asymptotically flat scenario, we decompose \(\tensor{\xi}{^a}\) according to
        \begin{equation}
            \tensor{\xi}{^a} = \tensor{Y}{^a} + F \tensor{n}{^a},
        \end{equation}
        where \(\tensor{Y}{^a}\) is a vector field on the sphere. \cref{eq: Lie-xi-gamma-DMP-algebra} implies this is a Killing vector field on the sphere. There is no room for extra fields playing the role of superrotations because only genuine Killing vector fields satisfy the Killing equation locally on the sphere. 

        The remaining condition, \cref{eq: Lie-xi-n-DMP-algebra}, implies that
        \begin{subequations}
            \begin{align}
                \lambda \tensor{n}{^a} &= \Lie[\xi] \tensor{n}{^a}, \\
                &= \tensor{\xi}{^b}\tensor{D}{_b}\tensor{n}{^a} - \tensor{n}{^b}\tensor{D}{_b}\tensor{\xi}{^a}, \\
                &= - \tensor{n}{^b}\tensor{D}{_b}\tensor{\xi}{^a}, \\
                &= - (\tensor{n}{^b}\tensor{D}{_b}F)\tensor{n}{^a}.
            \end{align}
        \end{subequations}
        Using \cref{eq: nabla-a-cosmological-horizon} we conclude that
        \begin{equation}
            \lambda(\theta,\phi) = H^2 \pdv{F}{v},
        \end{equation}
        which we may now integrate to find \(F\) is given by
        \begin{equation}
            F(v,\theta,\phi) = \lambda(\theta,\phi)v + \mu(\theta,\phi),
        \end{equation}
        where the factor of \(H^2\) was absorbed in the definition of \(\lambda\).

        In conclusion, we find that the most general vector field generating the \gls{DMP} symmetries has the form
        \begin{equation}
            \tensor{\xi}{^a} = \qty(\lambda(\theta,\phi)v + \mu(\theta,\phi))\tensor{n}{^a} + \tensor{Y}{^a},
        \end{equation}
        with \(\tensor{Y}{^a}\) a Killing vector field on the sphere. Notice this generalizes the cosmological de Sitter Killing vector fields presented in \cref{eq: de-sitter-killing-vector-fields}.

    \subsection{Hadamard States in Expanding Universes with Cosmological Horizons}
        The analysis by \textcite{dappiaggi2009CosmologicalHorizonsReconstruction,dappiaggi2009DistinguishedQuantumStates,dappiaggi2017HadamardStatesLightlike} was directed at the construction of Hadamard states in expanding universes with cosmological horizons. The core ideas are analogous to the asymptotically flat case, discussed in \cref{subsec: hadamard-states-asymptotically-flat}. 
        
        One considers a boundary algebra of observables \(\mathcal{A}_{\text{bound.}}\) defined on the cosmological horizon \(\horpas\). Simultaneously, one considers a bulk algebra of observables \(\mathcal{A}_{\text{bulk}}\) defined on the cosmological patch of the spacetime. It is then possible to construct a \gls{DMP}-invariant state on the boundary and pull-it-back to the bulk. Within \gls{FLRW} spacetimes, this state can be shown to be Hadamard. In the particular case of de Sitter spacetime this construction yields the so-called Bunch--Davies vacuum \cite{bunch1978QuantumFieldTheory}. In particular, this provides a way of constructing physically meaningful quantum states in inflationary scenarios. 

\section{The Sky as a Killing Horizon}\label{sec: sky-killing-horizon}
    As a final topic, I would like to discuss the work I am currently developing in collaboration with Landulfo \cite{aguiaralves2025NullInfinityKilling} and try to unify the perspectives we had so far toward null infinity in asymptotically flat spacetimes and cosmological horizons in expanding universes. 

    When dealing with null infinity in asymptotically flat spacetimes, the natural globally-defined group we encountered was the \gls{BMS} group,
    \begin{equation}
        \BMS = \SO*(3,1) \ltimes \ck[\infty](\Sph^2),
    \end{equation}
    which we discussed at length in \cref{sec: bms-group}. Nevertheless, in \cref{sec: infrared-de-sitter} we considered the case of expanding universes with cosmological horizons and found the \gls{DMP} group, 
    \begin{equation}
        \DMP = \SO(3) \ltimes (\ck[\infty](\Sph^2) \ltimes \ck[\infty](\Sph^2)).
    \end{equation}

    While these groups look similar, they are also still quite different. The \gls{BMS} group considers the conformal group on the sphere, \(\SO*(3,1)\), which on \cref{sec: superrotations} led us to the notion of superrotations. \gls{DMP}, however, only allows isometries on the sphere, meaning we are stuck with \(\SO(3)\). There is no such thing as a ``local isometry on the sphere'', so we cannot promote \(\SO(3)\) to a larger group in a similar way\footnote{A possible alternative would be to consider the group \(\mathrm{Diff}(\Sph^2)\) of sphere diffeomorphisms, as some authors do \cite{campiglia2014AsymptoticSymmetriesSubleading,campiglia2015NewSymmetriesGravitational,compere2018SuperboostTransitionsRefraction}, but this makes no use of the metric or conformal structure on the sphere.}. On the other hand, \(\DMP\) has two factors of \(\ck[\infty](\Sph^2)\), while \(\BMS\) has only one. It is at least curious that the two groups are so similar, yet so different. 

    One reason for this to be an interesting problem is precisely how the presence of \(\SO*(3,1)\) in \(\BMS\) calls for superrotations. The fact we were able to move from \(\BMS\) to \(\ebms\) on \cref{sec: superrotations} meant we suddenly were in the realm of two-dimensional \gls{CFT}, which opens up many new possible developments and techniques. It would be interesting to try to replicate this in expanding universes, but \(\SO(3)\) is not large enough for us to do that. 

    A natural question is then how to unify the definitions of the \gls{BMS} and \gls{DMP} groups in a manner that clarifies their similarities and differences. This is, in a sense, a quest to extend the definitions of asymptotic symmetries to a wider class of null hypersurfaces with the goal of keeping the original setups (null infinity and cosmological horizons) as particular cases of interest. 

    There have been many efforts along these lines (although not necessarily with the same motivation). The works of \textcite{duval2014CarrollNewtonGalilei,duval2014ConformalCarrollGroups,duval2014ConformalCarrollGroupsa}---which based our discussions of Carrollian manifolds---discuss precisely how to consider these sorts of ``conformal Carroll groups'' on null hypersurfaces. Other notable efforts are the works by \textcite{chandrasekaran2018SymmetriesChargesGeneral}---who extend the \gls{DMP} group by replacing \(\SO(3)\) with \(\mathrm{Diff}(\Sph^2)\)---and \textcite{compere2019LBMS4GroupdS4,ruzziconi2020VariousExtensionsBMS}---who use a Lie algebroid \cite{meinrenken2025LieAlgebroids} to obtain a structure in (anti) de Sitter spacetime that recovers the \gls{BMS} results in the limit in which the cosmological constant tends to zero. 

    The approach I took with Landulfo \cite{aguiaralves2025NullInfinityKilling} is heavily inspired by structures and results found in \gls{GR} and \gls{QFTCS}. The existence of Killing vector fields often leads to interesting null hypersurfaces known as Killing horizons. This notion was originally introduced by \textcite{carter1966CompleteAnalyticExtension,carter1969KillingHorizonsOrthogonally} and is defined as a null hypersurface tangent to a Killing vector field. These structures will be at the core of our discussion for the rest of this section, and hence we will now take a moment to discuss them in more depth. 

    \subsection{Killing Horizons}
        Let us begin by defining what we mean by a Killing horizon. 

        \begin{definition}[Killing Horizon]
            Let \((M,\tensor{g}{_a_b})\) be a Lorentzian spacetime. Suppose there is a Killing vector field \(\tensor{X}{^a}\) on \((M,\tensor{g}{_a_b})\). A null hypersurface \(\mathcal{N}\) normal to \(\tensor{X}{^a}\) is said to be a \emph{Killing horizon}.
        \end{definition}

\begin{figure}[tpbh]
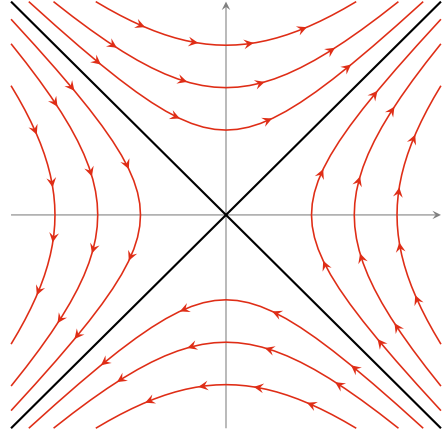

            \centering
            \includestandalone{minkowski-boost}
            \caption{Minkowski diagram exhibiting the flow of the Killing vector field \(\tensor*{J}{^a_x_t}\), which represents a boost. Notice the vector field is null precisely at the surfaces shown in black, which are the hypersurfaces with \(x^2 = t^2\). These are Killing horizons.}
            \label{fig: killing-horizon-minkowski}
        \end{figure}

        A few examples are in place. Let us begin with Minkowski spacetime. We will consider the boost-generating vector field (see \cref{eq: poincare-generators})
        \begin{equation}
            \tensor{K}{^a} = \tensor*{J}{^a_x_t} = x\tensor{\qty(\pdv{t})}{^a} +t\tensor{\qty(\pdv{x})}{^a}.
        \end{equation}
        This is a Killing vector field, as we derived on \cref{sec: symmetries-curved-spacetime}. Notice, though, that
        \begin{equation}
            \tensor{K}{^a}\tensor{K}{_a} = -x^2 + t^2,
        \end{equation}
        which means the vector field is null on the hyperplanes with \(x^2 = t^2\). These are Killing horizons. This behavior is illustrated on \cref{fig: killing-horizon-minkowski}.

        One may notice from \cref{fig: killing-horizon-minkowski} that the boost vector field is timelike on the right wedge of the spacetime (known as the right Rindler wedge). This means we can consider a family of observers whose worldlines are precisely the orbits of the Killing field. These turn out to be uniformly accelerated observers. An important prediction of \gls{QFTCS} is that these observers perceive the standard Minkowski vacuum in Minkowski spacetime as a thermal state at temperature
        \begin{equation}
            T = \frac{a}{2\pi},
        \end{equation}
        where \(a\) is their acceleration. This is known as the (Fulling--Davies--)Unruh effect \cite{fulling1973NonuniquenessCanonicalField,davies1975ScalarProductionSchwarzschild,unruh1976NotesBlackholeEvaporation}, reviewed, for example, in Refs. \citeonline{aguiaralves2023NonperturbativeAspectsQuantum,birrell1982QuantumFieldTheory,crispino2008UnruhEffectIts,mukhanov2007IntroductionQuantumEffects,parker2009QuantumFieldTheory,wald1994QuantumFieldTheory}.

        Notice that there are two Killing horizons on \cref{fig: killing-horizon-minkowski}---one is given by the hypersurface \(x = t\), and the other by the hypersurface \(x = - t\). These two horizons cross at a bifurcation surface in which \(x = t = 0\). Notice this is indeed a surface, since it is a two-dimensional plane in the \((y,z)\)-directions. We say the full structure is a bifurcate Killing horizon. This notion was originally introduced by \textcite{boyer1969GeodesicKillingOrbits} and often occurs when there is a fixed point of an isometry group in spacetime. As the next examples will show, this is an important structure in \gls{QFTCS}. 

\begin{figure}[tpbh]
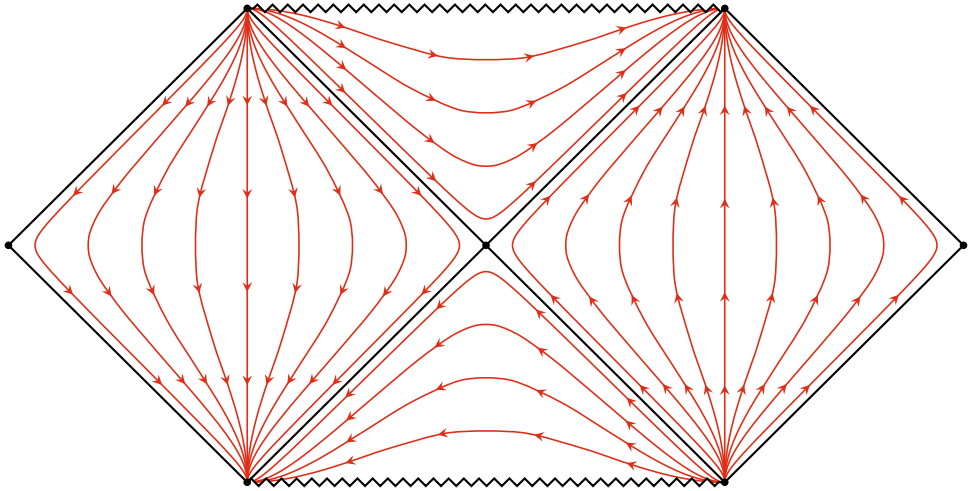

            \centering
            \includestandalone{killing-schwarzschild}
            \caption{Orbits of the Killing vector field \(\tensor{\qty(\pdv*{t})}{^a}\) in the maximally extended Schwarzschild spacetime. Notice how the Killing field is similar to the boost field we considered in \cref{fig: killing-horizon-minkowski}. This time, the (bifurcate) Killing horizon is the event horizon for the white and black hole regions in the spacetime. }
            \label{fig: killing-schwarzschild}
        \end{figure}

        Let us consider Schwarzschild spacetime. We will work in the full Kruskal extension and consider the action of the vector field locally defined in Schwarzschild coordinates as
        \begin{equation}
            \tensor{K}{^a} = \tensor{\qty(\pdv{t})}{^a}.
        \end{equation}
        The resulting orbits are shown in \cref{fig: killing-schwarzschild}. Notice, in particular, that the Killing vector field is timelike on the right wedge (where it is parallel to the worldlines of stationary observers), but near the horizon the field becomes null. Hence, the event horizon of the black (and white) hole(s) is a bifurcate Killing horizon.

        As in Minkowski spacetime, we have a thermal behavior for observers in this spacetime. Namely, stationary observers on the right wedge perceive the natural vacuum on this spacetime---known as the Hartle--Hawking vacuum \cite{hartle1976PathintegralDerivationBlackhole}---as being thermal at temperature
        \begin{equation}\label{eq: hawking-temperature}
            T = \frac{1}{8 \pi M}.
        \end{equation}
        In this sense, the black hole is ``hot''. This particular quantum state is similar to a black hole in a radiation box at the temperature \eqref{eq: hawking-temperature}. This phenomenon is a curved spacetime version of the Unruh effect, and it is similar (but conceptually different from) the Hawking effect \cite{hawking1974BlackHoleExplosions,hawking1975ParticleCreationBlack}, which concerns a black hole formed from gravitational collapse as opposed to an eternal black hole. 

\begin{figure}[tpb]
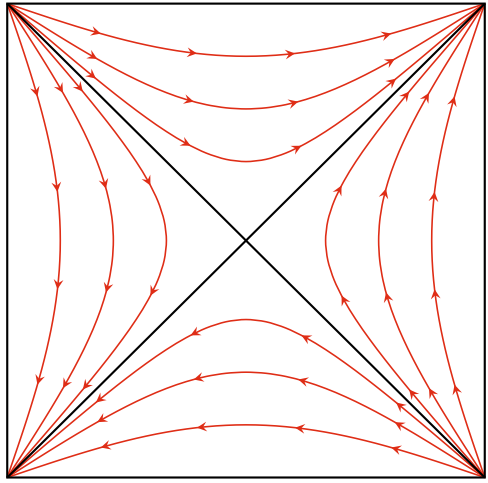

            \centering
            \includestandalone{killing-desitter}
            \caption{Orbits of the Killing vector field \eqref{eq: killing-vector-field-de-sitter} in de Sitter spacetime. Notice how the Killing field is similar to ones we considered in \cref{fig: killing-horizon-minkowski,fig: killing-schwarzschild}. The (bifurcate) Killing horizon is comprised by the cosmological horizons of each cosmological patch.}
            \label{fig: killing-desitter}
        \end{figure}

        We also get a similar behavior in de Sitter spacetime. Consider the de Sitter metric in global coordinates, \cref{eq: global-coordinates-de-sitter}. Consider the vector field
        \begin{equation}\label{eq: killing-vector-field-de-sitter}
            \tensor{K}{^a} = \cos R \sin T \tensor{\qty(\pdv{T})}{^a} + \cos T \sin R \tensor{\qty(\pdv{R})}{^a},
        \end{equation}
        which is a Killing vector field. This vector field is null when
        \begin{equation}
            0 = \tensor{K}{^a}\tensor{K}{_a} = \frac{3}{\Lambda}\qty(-1 + \csc^2 T \sin^2 R),
        \end{equation}
        which is satisfied at the cosmological horizons \(T = R - \pi\) and \(T = - R\). This is illustrated on \cref{fig: killing-desitter}. In this context, we also get a thermal spectrum for the so-called Bunch--Davies vacuum \cite{bunch1978QuantumFieldTheory}, a fact originally noticed by \textcite{gibbons1977CosmologicalEventHorizons}. 

        The pattern between bifurcate Killing horizons and thermal effects is general. It has been shown by \textcite{kay1991TheoremsUniquenessThermal} that, if there is a stationary Hadamard state, then this state is unique and observers following the Killing orbits perceive it as a thermal state. Here, ``stationary'' refers to the state being invariant under the action of the Killing vector field. 

        The temperature of this quantum state is naturally defined by the geometric structure of the Killing horizon. Namely, at the horizon the Killing vector field \(\tensor{K}{^a}\) satisfies
        \begin{equation}
            \tensor{K}{^a}\tensor{\nabla}{_a}\tensor{K}{^b} = \kappa \tensor{K}{^b}
        \end{equation}
        for some function \(\kappa\). When properly normalized (typically by imposing that the Killing vector field is a unit timelike vector field at spatial infinity), we say this function \(\kappa\) is the ``surface gravity'' of the Killing horizon. For black holes, this is similar to a ``renormalized'' measurement of the gravitational acceleration at the horizon. In \gls{BHT}, the temperature of the horizon is given by
        \begin{equation}
            T = \frac{\kappa}{2\pi}.
        \end{equation}
        See Refs. \citeonline{wald1994QuantumFieldTheory,witten2024IntroductionBlackHole,wald2001ThermodynamicsBlackHoles,wall2018SurveyBlackHole} for further information on \gls{BHT} and related topics.

        Hence, Killing horizons are intimately related to the thermal phenomena we encounter when studying \gls{QFTCS}. The interest in them does not stop at that. Recently, it has been shown by \textcite{danielson2022BlackHolesDecohere,danielson2022KillingHorizonsDecohere} that Killing horizons decohere quantum superpositions. In a very short description, once a quantum particle with nonvanishing charge and/or mass is put into spatial superposition, the Coulombic electromagnetic and/or gravitational field induced by the particle is also put in superposition. The quantum state of the background quantum field is thus put into superposition, and this can be shown to lead to a flow of soft radiation into a black hole (or other Killing horizon). These soft particles carry information about the superposition across the Killing horizon, causing the decoherence of the original system. See Refs. \citeonline{danielson2022BlackHolesDecohere,danielson2022KillingHorizonsDecohere} for further details, and Refs. \citeonline{danielson2022BlackHolesDecohere,danielson2022KillingHorizonsDecohere,gralla2024DecoherenceHorizonsGeneral,danielson2025LocalDescriptionDecoherence,danielson2025HowMinimizeDecoherence} for further developments on this topic.

        With these ideas in mind, it becomes apparent that Killing horizons play a key role in gravitational physics. They encode the symmetries of spacetime in distinguished null hypersurfaces, and thus provide natural places for us to understand asymptotic symmetries. In fact, the works of \textcite{danielson2022BlackHolesDecohere,danielson2022KillingHorizonsDecohere} already exhibit interesting physics associated with very low energy particles occurring on Killing horizons. This suggests we should look at Killing horizons to learn more about infrared physics. 

        This has been pursued before. For instance, \textcite{koga2001AsymptoticSymmetriesKilling} had considered \gls{BMS}-like symmetries on Killing horizons and found a group similar to the Newman--Unti group. Later, these ideas were employed by \textcite{hawking2016SoftHairBlack,hawking2017SuperrotationChargeSupertranslation}, for example, in understanding soft hair in black holes. Nevertheless, a very interesting observation is that the notion of Killing horizon can be suitably modified to encompass null infinity as well. This will lead us to the notion of \glsxtrfull{ACKH} introduced in Ref. \citeonline{aguiaralves2025NullInfinityKilling}.

    \subsection{Asymptotic (Conformal) Killing Horizons}
        Let us begin by recalling the definition of expanding universe with cosmological horizon. As defined on \cref{subsec: expanding-universes-cosmological-horizons}, a key property of an expanding universe with cosmological horizon is the existence of a conformal Killing vector field \(\tensor{X}{^a}\) for the metric \(\tensor{g}{_a_b}\) such that (\cref{eq: Lie-X-gab-cosmological-horizon})
        \begin{equation}
            \Lie[X]\tensor{g}{_a_b} = - 2 \tensor{X}{^c}\tensor{\nabla}{_c}\qty(\log a) \tensor{g}{_a_b},
        \end{equation}
        where I omitted the tildes on the extended versions of the metric and \(a\) is some function such that \(\tensor{X}{^c}\tensor{\nabla}{_c}\qty(\log a) \to 0\) as one approaches the horizon. 

        Under this construction, we see that \(\tensor{X}{^a}\) is not necessarily a genuine Killing vector field away from the horizon, but becomes a Killing vector field asymptotically. We thus get a notion of Killing horizon which holds at the horizon itself, but it is not too constraining elsewhere. Hence, while the horizon itself still has a rich symmetry structure (as we will soon see), we do not need to assume the spacetime itself to enjoy a symmetry on an open set. 

        This idea can be generalized to a domain far beyond that of expanding universes with cosmological horizons. Following Ref. \citeonline{aguiaralves2025NullInfinityKilling}, we define an \gls{AKH} in the following manner.

        \begin{definition}[Asymptotic Killing Horizon]
            Let \((M,\tensor{g}{_a_b})\) be a Lorentzian spacetime with an isometric extension \((\tilde{M},\tensor{g}{_a_b})\) (we use the same symbol for the extended metric and the original metric for simplicity). Consider a null surface \(N\) contained in the closure of \(M\) in \(\tilde{M}\)---\ie, a null surface in \(\tilde{M}\) ``touching'' \(M\). We say \(N\) is an \emph{asymptotic Killing horizon} if there is a vector field \(\tensor{X}{^a}\) in \(M\) such that
            \begin{enumerate}
                \item the vector field \(\tensor{X}{^a}\) can be extended to \(N\) without vanishing identically;
                \item \(\tensor{X}{^a}\tensor{X}{_a}\) can be extended to \(N\) and vanishes identically thereon;
                \item \(\Lie[X]\tensor{g}{_a_b}\) can be extended to \(N\) and vanishes identically thereon.
            \end{enumerate}
        \end{definition}

        Hence, we ask that \(\tensor{X}{^a}\) is a Killing vector field at the \gls{AKH}, but not necessarily near it. A similar idea was employed by \textcite{koga2001AsymptoticSymmetriesKilling}, although in a less technical version.

        What is really interesting about thinking in terms of Killing horizons, though, is that it also encompasses null infinity. In this way, we will be able to construct the notion of asymptotic symmetry in a manner that generalizes both the \gls{BMS} and the \gls{DMP} groups in a natural way. To do so, let us work in Bondi gauge and consider the metric
        \begin{equation}
            \dd{\tilde{s}}^2 = - \Omega^2 \dd{u}^2 - 2 \dd{u} \dd{\Omega} + \tensor{\gamma}{_A_B}\dd{\tensor{x}{^A}}\dd{\tensor{x}{^B}} + \cdots,
        \end{equation}
        which is conformally related to some asymptotically flat metric \(\tensor{g}{_a_b}\). We know one possible \gls{BMS} transformation is 
        \begin{equation}
            \tensor{X}{^a} = \tensor{\qty(\pdv{u})}{^a} + \order{\Omega}.
        \end{equation}
        Interestingly, we can see that
        \begin{equation}
            \Lie[X]\tensor{\tilde{g}}{_a_b} = \order{\Omega},
        \end{equation}
        meaning \(\tensor{X}{^a}\) is an asymptotic Killing vector field for \(\tensor{\tilde{g}}{_a_b}\) as \(\Omega \to 0\). Notice, however, that
        \begin{equation}
            \Lie[X](\Omega^2 \tensor{g}{_a_b}) = \Omega^2 \Lie[X]\tensor{g}{_a_b} + \Lie[X](\Omega^2) \tensor{g}{_a_b},
        \end{equation}
        which means \(\Lie[X](\Omega^2 \tensor{g}{_a_b}) = 0\) is equivalent to
        \begin{equation}
            \Lie[X]\tensor{g}{_a_b} = - 2 \Omega^{-1} \Lie[X]\Omega \tensor{g}{_a_b}
        \end{equation}
        as long as both sides of the equation are well-defined. Hence, \(\tensor{X}{^a}\) is, in this sense, an ``asymptotic conformal Killing vector field'' for \(\tensor{g}{_a_b}\).

        With this in mind, we can introduce the following definition. 
        \begin{definition}[Asymptotic Conformal Killing Horizon]
            Let \((M,\tensor{g}{_a_b})\) be a Lorentzian spacetime with a conformal extension \((\tilde{M},\tensor{\tilde{g}}{_a_b})\). Consider a null surface \(N\) contained in the closure of \(M\) in \(\tilde{M}\)---\ie, a null surface in \(\tilde{M}\) ``touching'' \(M\). We say \(N\) is an \emph{asymptotic conformal Killing horizon} if there is a vector field \(\tensor{X}{^a}\) in \(M\) such that
            \begin{enumerate}
                \item the vector field \(\tensor{X}{^a}\) can be extended to \(N\) without vanishing identically;
                \item \(\tensor{X}{^a}\tensor{X}{_a}\) can be extended to \(N\) and vanishes identically thereon;
                \item we can define a positive real function \(\Omega\) in \(M\) satisfying
                \begin{equation}\label{eq: Lie-X-Omega-2}
                    \Lie[X](\Omega^2) = - \frac{1}{2}\Omega^2 \tensor{\nabla}{_a}\tensor{X}{^a}
                \end{equation}
                and such that \(\Lie[X](\Omega^2 \tensor{g}{_a_b})\) can be extended to \(N\) and vanishes identically thereon.
            \end{enumerate}
        \end{definition}

        Notice \cref{eq: Lie-X-Omega-2} is intended to make \(\tensor{X}{^a}\) be an asymptotic Killing vector field for \(\Omega^2 \tensor{g}{_a_b}\). Hence, \(\tensor{X}{^a}\) is an asymptotic Killing vector field for a metric conformal to \(\tensor{g}{_a_b}\), meaning in particular that it preserves the conformal structure in the way we would like. 

        When referring to either an \gls{AKH} or an \gls{ACKH}, I like to write ``\gls{A(C)KH}''. \Glspl{A(C)KH} pose natural structures on which we can consider asymptotic symmetries, and as we shall see they lead to very natural extensions of the \gls{BMS} and \gls{DMP} groups.

    \subsection{Symmetries on Asymptotic Horizons}
        We are now ready to discuss asymptotic symmetry groups on \glspl{A(C)KH}. The procedure is completely analogous to how we discussed the \gls{BMS} and \gls{DMP} groups in terms of Carrollian structures. 

        We start with \glspl{AKH}. Given an \gls{AKH}, we want to obtain a Carrollian structure associated to it. This means a manifold \(N\), a Carrollian metric \(\tensor{h}{_a_b}\), and a normal vector \(\tensor{n}{^a}\). Since an \gls{AKH} is a submanifold of an isometric extension of the original spacetime, we already have one natural choice for the manifold structure (this will be the horizon itself) and for the metric (this will be the induced metric of the extended spacetime on the horizon). Notice these choices are unique up to isometries. 

        It remains to choose a normal vector. While in principle the asymptotic Killing vector \(\tensor{X}{^a}\) may seem like an interesting choice, picking it would not give any information about the ambient spacetime metric along the null direction---and this information is also not provided by the induced metric. For this reason, Ref. \citeonline{aguiaralves2025NullInfinityKilling} prefers to choose the normal null vector
        \begin{equation}
            \tensor{n}{^a} = \sigma \tensor{X}{^a},
        \end{equation}
        where \(\sigma\) is a solution to the differential equation
        \begin{equation}\label{eq: Lie-X-sigma-kappa}
            \Lie[X] \sigma = -\kappa\sigma,
        \end{equation}
        and \(\kappa\) is the ``inaffinity'' for \(\tensor{X}{^a}\): \(\tensor{X}{^b}\tensor{\nabla}{_b}\tensor{X}{^a} = \kappa \tensor{X}{^a}\). This construction will ensure that 
        \begin{equation}
            \tensor{n}{^b}\tensor{\nabla}{_b}\tensor{n}{^a} = 0.
        \end{equation}
        Thus, \(\tensor{n}{^a}\) ``remembers'' information about the ambient metric in the sense it contains information about what are the geodesic vectors in the ambient spacetime. 

        Notice there is freedom in selecting \(\tensor{n}{^a}\). When solving \cref{eq: Lie-X-sigma-kappa}, we are free to pick any initial value for \(\sigma\), which effectively amounts to a choice of function on the cross section of \(N\). In more detail, let \(\Sigma\) be the space of geodesic generators of \(N\). Then, at least locally, \(N\) resembles\footnote{In technical terms, we say \(N\) is a fiber bundle with base space \(\Sigma\). See Refs. \citeonline{hamilton2017MathematicalGaugeTheory,kobayashi1963FoundationsDifferentialGeometry,kolar1993NaturalOperationsDifferential,tu2017DifferentialGeometryConnections} for detailed discussions of fiber bundles and Ref. \citeonline{ciambelli2019CarrollStructuresNull} for a characterization of Carrollian structures in these terms.} \(\R \times \Sigma\). In fact, by simply following the generators of \(N\) we will find that it has the structure \(\R \times \Sigma\). Hence, this generalizes \(\nullfut\) and \(\horpas\) from \(\R \times \Sph^2\) to \(\R \times \Sigma\), where \(\Sigma\) can be a more general choice of cross section. In particular, it can be higher-dimensional if we choose to work in \(d > 4\) spacetimes. The freedom in selecting \(\tensor{n}{^a}\) is a freedom of choosing an initial value for \cref{eq: Lie-X-sigma-kappa}, which amounts to choosing a function on \(\Sigma\). 

        At last, we have found a Carrollian structure for the \gls{AKH}. The freedom we get in choosing this structure corresponds to us having to consider as equivalent any two choices such that
        \begin{equation}
            (N, \tensor{h}{_a_b}, \tensor{n}{^a}) \sim (N, \tensor{h}{_a_b}, \sigma \tensor{n}{^a})
        \end{equation}
        for \(\sigma \in \ck[\infty](\Sigma)\). At the level of Killing vectors, this means we have 
        \begin{subequations}
            \begin{align}
                \Lie[\xi] \tensor{h}{_a_b} &= 0, \\
                \Lie[\xi] \tensor{n}{^a} &= \lambda \tensor{n}{^a},
            \end{align}
        \end{subequations}
        for \(\lambda \in \ck[\infty](\Sigma)\). This is completely analogous to what we found when studying the \gls{DMP} group, and therefore we get to the general vector field
        \begin{equation}\label{eq: AKH-vector-field}
            \tensor{\xi}{^a} = \tensor{Y}{^a} + \qty(g(\tensor{x}{^A})u + f(\tensor{x}{^A}))\tensor{n}{^a},
        \end{equation}
        where \(u\) is an affine parameter along the generators of \(N\), \(\tensor{x}{^A} \in \Sigma\), \(\tensor{Y}{^a}\) is a Killing vector field for \(\tensor{h}{_a_b}\) and \(\tensor{n}{^a}\) is the normal vector we constructed earlier.

        One may be worried about what it means for \(\tensor{Y}{^a}\) to be a Killing vector field for \(\tensor{h}{_a_b}\), because in principle \(\tensor{h}{_a_b}\) could be different at each different value of \(u\). Thanks to the \gls{AKH} structure, this does not happen. We know \(N\) has the structure \(\R \times \Sigma\), but let us now choose a way of slicing \(N\) in many copies of \(\Sigma\) and see how the metric looks like in each of them. For that, we choose a continuous map \(s \colon \Sigma \to N\) with the additional property that for each \(x \in \Sigma\), \(s(x)\) is a point in the generator \(x\) (as a one-dimensional submanifold of \(N\)). Such a map \(s\) is said to be a section. Notice \(s(\Sigma)\) is a submanifold of \(N\) and we can consider the induced metric on \(s(\Sigma)\). It turns out the induced metric does not depend on the choice of section \(s\). On the ambient spacetime \(M\), we have the equation
        \begin{equation}
            \Lie[\lambda X] \tensor{g}{_a_b} = \tensor{\nabla}{_(_a}\lambda \tensor{X}{_b_)} + \lambda \tensor{\nabla}{_(_a}\tensor{X}{_b_)},
        \end{equation}
        with \(\lambda\) an arbitrary function on \(M\). We can now restrict this equation to \(N\)---more precisely, we pull-it-back under the inclusion map \(\imath \colon N \to M\)---and find that 
        \begin{equation}\label{eq: Lie-lambda-X-h}
            \Lie[\lambda X] \tensor{h}{_a_b} = 0
        \end{equation}
        because the pullback of \(\tensor{X}{_a}\) to \(N\) vanishes and \(\tensor{\nabla}{_(_a}\tensor{X}{_b_)} = 0\) on \(N\) because \(N\) is an \gls{AKH}! Hence, if we choose different functions \(s\), we are just sliding \(s(\Sigma)\) up and down along the generators, and this cannot change the induced metric on \(s(\Sigma)\) thanks to \cref{eq: Lie-lambda-X-h}. Notice that, on \cref{eq: Lie-lambda-X-h}, \(\lambda\) is an arbitrary function on \(N\).

        Due to this, we can talk about \(\Sigma\) as a Riemannian manifold with the metric induced on \(s(\Sigma)\) for any choice of \(s\). Let us denote it by \((\Sigma, \tensor{h}{_a_b})\), for instance. Then the transformations generated by the vector fields on \cref{eq: AKH-vector-field} will lead to the group
        \begin{equation}
            G_{\text{AKH}} = \mathrm{Isom}(\Sigma) \ltimes (\ck[\infty](\Sigma) \ltimes \ck[\infty](\Sigma)),
        \end{equation}
        where \(\mathrm{Isom}(\Sigma)\) is the isometry group of \((\Sigma, \tensor{h}{_a_b})\). Notice that \(\mathrm{Isom}(\Sigma)\) is only defined in the first place because we are working with an \gls{AKH}, and a more general null hypersurface could have cross sections that are not isometric. 

        For \glspl{ACKH} we can perform a very similar procedure. A null geodesic is a null geodesic in any spacetime related by a confomorphism. Hence, the key difference is that now the induced metric is only defined up to a confomorphism. Hence, we get the equivalence relation
        \begin{equation}
            (N,\tensor{\tilde{h}}{_a_b},\tensor{\tilde{n}}{^a}) \sim (N,\omega^2 \tensor{\tilde{h}}{_a_b},\sigma \tensor{\tilde{n}}{^a}),
        \end{equation}
        which leads to the equations
        \begin{subequations}
            \begin{align}
                \Lie[\xi] \tensor{\tilde{h}}{_a_b} &= \mu \tensor{\tilde{h}}{_a_b}, \\
                \Lie[\xi] \tensor{\tilde{n}}{^a} &= \lambda \tensor{\tilde{n}}{^a},
            \end{align}
        \end{subequations}
        for \(\mu, \lambda \in \ck[\infty](\Sigma)\). This time we get to the vector field
        \begin{equation}\label{eq: ACKH-vector-field}
            \tensor{\xi}{^a} = \tensor{Y}{^a} + \qty(g(\tensor{x}{^A})u + f(\tensor{x}{^A}))\tensor{n}{^a},
        \end{equation}
        with \(\tensor{Y}{^a}\) a \emph{conformal} Killing vector field as opposed to a genuine Killing vector field. We are thus led to the \gls{ACKH} group
        \begin{equation}
            G_{\text{ACKH}} = \mathrm{Conf}(\Sigma) \ltimes (\ck[\infty](\Sigma) \ltimes \ck[\infty](\Sigma)),
        \end{equation}
        where \(\mathrm{Conf}(\Sigma)\) is the confomorphism group of \((\Sigma, \tensor{\tilde{h}}{_a_b})\). As with \glspl{AKH}, \(\mathrm{Conf}(\Sigma)\) is only defined because we are working with an \gls{ACKH}---and a more general null hypersurface could have cross sections that are not confomorphic.

        For \(\Sigma = \Sph^2\) we find
        \begin{equation}
            G_{\text{AKH}} = \SO(3) \ltimes (\ck[\infty](\Sph^2) \ltimes \ck[\infty](\Sph^2)) = \DMP
        \end{equation}
        and
        \begin{equation}
            G_{\text{ACKH}} = \SO*(3,1) \ltimes (\ck[\infty](\Sph^2) \ltimes \ck[\infty](\Sph^2)).
        \end{equation}
        \(G_{\text{ACKH}}\) is larger than the \gls{BMS} group, but smaller than the Newman--Unti group. In Ref. \citeonline{aguiaralves2025NullInfinityKilling}, the additional transformations are called ``superdilations'', for they are direction-dependent dilations. 

    \subsection{Superdilations}
        The original motivation for considering \glspl{A(C)KH} was to find a bridge between the \gls{BMS} and \gls{DMP} groups. In particular, this was motivated by a search for a notion of superrotation in the de Sitter cosmological horizon. By providing a unified definition of the \gls{BMS} and \gls{DMP} groups, it would be possible to attempt at an extension of the \gls{DMP} group.

        Curiously, the \gls{AKH} group is exactly the same as the \gls{DMP} group, and leaves no clear room for extensions. There is no such thing as a ``local isometry'', so we cannot trade the isometry group of the cross section for a local isometry algebra. In this sense, we cannot get a notion of superrotation in the cosmological horizon. Meanwhile, the \gls{ACKH} group provides a new extension of the \gls{BMS} group in the form of superdilations. These transformations were already present in the \gls{DMP} group (due to its double semidirect product structure), but had not been seriously considered as a symmetry at infinity. 

        Other authors had stumbled upon transformations similar to superdilations. For instance, as previously mentioned, the \gls{ACKH} group is smaller than the Newman--Unti group \cite{newman1962BehaviorAsymptoticallyFlat}, and the original analysis of superrotations by \textcite{barnich2010AspectsBMSCFT,barnich2010SymmetriesAsymptoticallyFlat} also addressed the possibility of considering the extra Newman--Unti transformations. However, these approaches consider transformations much more general than superdilations, because one does not restrict the conformal scaling of the normal vector \(\tensor{\tilde{n}}{^a}\) in any way.

        If on the one hand one may not constrain the conformal scaling of \(\tensor{\tilde{n}}{^a}\), on the other hand it is possible to overconstrain it and remove the superdilations altogether. This is what is typically done in the derivation of the \gls{BMS} group, in which the conformal scaling of \(\tensor{\tilde{n}}{^a}\) is tied to that of the induced metric \(\tensor{\tilde{h}}{_a_b}\). In asymptotically flat spacetimes, this occurs due to the role played by the conformal factor \(\Omega\). In the \gls{ACKH} approach, however, we never imposed any conditions about asymptotic flatness, but rather focused on the more general geometric structure of an \gls{ACKH}. This invites us to keep this freedom and consider superdilations more seriously. Other authors, such as \textcite{ciambelli2019CarrollStructuresNull}, also worked in more general setups and decided to tie the conformal scalings of \(\tensor{\tilde{n}}{^a}\) and \(\tensor{\tilde{h}}{_a_b}\) together to preserve a residual form of conformal invariance and exclude the superdilations. 
        
\begin{figure}
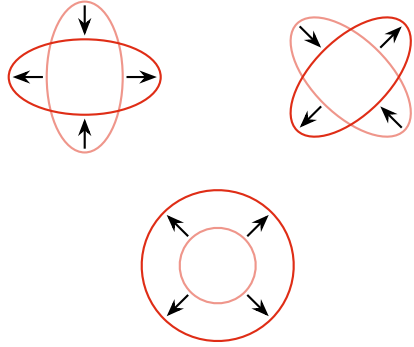

            \centering
            \includestandalone{breathing}
            \caption{Gravitons are transverse and traceless, and thus it follows that \glspl{GW} in \gls{GR} passing through an array of inertial detectors (orange lines) can only cause it to shear in the \(+\) and \(\times\) polarizations (top row). However, modified theories of gravity may involve other polarizations, such as a transverse breathing mode. Such a polarization would affect the array of inertial detectors by making it expand and contract (bottom row). Figure taken from Ref. \citeonline{aguiaralves2025NullInfinityKilling}.}
            \label{fig: GW-shear-or-not-shear}
        \end{figure}

        At the end of the day, the usefulness of a definition comes through its implications. In the case of superdilations, it seems natural to pursue further connections by means of an infrared triangle. For example, \glspl{GW} in \gls{GR} are always transverse and traceless, because so are the two graviton degrees of freedom. Nevertheless, modified theories of gravity may include extra degrees of freedom \cite{eardley1973GravitationalWaveObservationsPRD,eardley1973GravitationalWaveObservationsPRL,will2014ConfrontationGeneralRelativity}. In particular, some theories may include a ``transverse breathing mode'', which corresponds to a transverse, but ``tracefull'', polarization. Instead of causing shears on an array of inertial detectors, this polarization causes the array to expand and contract. This is illustrated on \cref{fig: GW-shear-or-not-shear}.

        Since the \gls{ACKH} approach is purely geometrical and makes no assumptions about the underlying theory of gravity, superdilations could be a hint at a modified theory of gravity, in the sense it is possible they could be mapped to a memory effect in the transverse breathing mode. A guess would be that such a theory can be described by including an extra scalar field, which is what happens in Brans--Dicke gravity \cite{brans1961MachsPrincipleRelativistic}. Some aspects of the infrared behavior of Brans--Dicke gravity have been investigated \cite{hou2021GravitationalMemoryEffects,seraj2021GravitationalBreathingMemory,du2016GravitationalWaveMemory}, and it is known that massless scalar fields also enjoy an interesting infrared structure \cite{campiglia2018CanScalarsHave,campiglia2019ScalarAsymptoticCharges}. Hence, by modifying \gls{GR} in a suitable manner, superdilations could be manifested in memory effects and \glspl{GW} in general, which could lead to observable effects \cite{eardley1973GravitationalWaveObservationsPRD,eardley1973GravitationalWaveObservationsPRL,du2016GravitationalWaveMemory}.

        The interest in this approach is that we use geometry and symmetry considerations and infrared physics to guess (or ``bootstrap'') what should be the degrees of freedom of gravity. The existence of a new scalar mode is suggested by considerations that arise from pure Lorentzian geometry, and may later be tested through \gls{GW} observations. 

    \subsection{Applications}
        In addition to the exploration of modified gravity theories, the \gls{A(C)KH} approach has a few other prospective applications.

        First and foremost, it disfavors the idea of generalizing superrotations to de Sitter spacetime in a \gls{CFT}-friendly manner. This means it may be difficult (or incorrect) to pursue a conformally-invariant holographic duality in the cosmological horizon. 

        Secondly, we may now come back to the notion of bifurcate Killing horizons and wonder what could happen with a bifurcate \gls{ACKH}. With an appropriate definition, it could be that we may understand the relation between \(\nullfut\) and \(\nullpas\) is terms of a bifurcate horizon, and thus shed more light into Strominger's antipodal matching condition \cite{strominger2014BMSInvarianceGravitational}, which was essential for us to identify \gls{BMS} symmetries and the Weinberg soft graviton theorem in \cref{subsec: weinberg-soft-graviton}.

        \Glspl{A(C)KH} are still too young to be fully understood, but it is exciting to see how purely geometric insights may bring new inspiration and results into high-energy physics.

\section{Final Remarks}\label{sec: conclusion}
    The \glsxtrlong{IR} structure of gravity is fascinating. While a first glance could make it seem that very large distances have a simple or almost boring behavior, this guess could not be further from the truth. As we have seem, there is a very rich structure hiding away at the infinitely distant sky.

    In these notes, we have only managed to scratch the surface of these topics. While I wanted to cover many more topics and make this a much more comprehensive reference, there is an old saying---attributed to far too many people to have a single author---that texts are never finished, only abandoned. Since this particular text starts with basic group theory, it seemed an overreach to aim at even more cutting edge topics. 

    There are many directions one could pursue after finishing these notes. Some of the ones that draw my attention the most are the following, in no particular order.
    \begin{itemize}
        \item The covariant phase space formalism \cite{lee1990LocalSymmetriesConstraints,iyer1994PropertiesNoetherCharge,iyer1995ComparisonNoetherCharge}, which allows for a deeper understanding of what are asymptotic symmetries and their associated conserved charges \cite{wald2000GeneralDefinitionConserved,barnich2002CovariantTheoryAsymptotic,barnich2011BMSChargeAlgebra}. See, for example, Refs. \citeonline{gieres2023CovariantCanonicalFormulations,compere2019AdvancedLecturesGeneral,fiorucci2021LeakyCovariantPhase} for introductions. In particular, one sees that a supertranslation is more than a diffeomorphism due to its nontrivial action on the boundary of the spacetime, where it does not behave as a mere gauge transformation.
        \item The celestial holography program \cite{pasterski2021CelestialHolography,aneesh2022CelestialHolographyLectures,raclariu2021LecturesCelestialHolography,pasterski2025ChapterCelestialHolography,pasterski2021LecturesCelestialAmplitudes,donnay2024CelestialHolographyAsymptotic}, which pursues a dual \gls{CFT} to quantum gravity in asymptotically flat spacetimes by exploiting the infrared structure of gravity and gauge theory. It investigates how to formulate quantum gravity in asymptotically flat spacetimes in terms of a two-dimensional Euclidean \gls{CFT} living on the celestial sphere.
        \item The Carrollian holography program \cite{donnay2022CarrollianPerspectiveCelestial,donnay2023BridgingCarrollianCelestial,bagchi2022ScatteringAmplitudesCelestial}, which also pursues a holographic formulation of quantum gravity, but through a three-dimensional Carrollian \gls{CFT} at null infinity.
        \item The study of infrared finite scattering theory in the algebraic approach \cite{prabhu2022InfraredFiniteScattering,prabhu2024InfraredFiniteScatteringAmplitudesSoftTheorems,prabhu2024InfraredFiniteScatteringStatesRepresentations}, which attempts at understanding how to avoid harmful infrared divergences that keep one from defining an \(S\)-matrix.
        \item The deeper relations between scattering amplitudes and classical physics \cite{holstein2004ClassicalPhysicsQuantum,kosower2022SAGEXReviewScattering,mohanty2023GravitationalWavesQuantum}, which is the fundamental reason for why soft theorems and memory effects must be related.
        \item The behavior of asymptotic flatness and the memory effect in higher-dimensional spacetimes \cite{hollands2005AsymptoticFlatnessBondi,hollands2017BMSSupertranslationsMemory,satishchandran2018MemoryEffectParticle,satishchandran2019AsymptoticBehaviorMassless}, which exhibits, for example, how the notion of \gls{BMS} transformations appears to be special in four-dimensional spacetimes.
        \item Improving on the construction of Carrollian manifolds by viewing them as fiber bundles \cite{ciambelli2019CarrollStructuresNull}, thus mimicking what one has in Newton--Cartan theory (see, \eg, Ref. \citeonline{malament2012TopicsFoundationsGeneral}). 
        \item Studying in more depth the symmetries and degrees of freedom of null hypersurfaces in the bulk of spacetime \cite{ciambelli2023AsymptoticSymmetriesCorner,adami2021NullBoundaryPhase,chandrasekaran2018SymmetriesChargesGeneral,grumiller2020Horizons2020,grumiller2020SpacetimeStructureGeneric,donnay2016SupertranslationsSuperrotationsBlack}. In particular, our definition of expanding universe with cosmological horizon imposes a time-independent metric on the horizon, which forbids the passage of gravitational radiation. This can be quite different from the behavior at null infinity---\cref{eq: BS-EFE-uu}, for example, shows how the Bondi mass aspect changes as radiation ``leaks'' through \(\nullfut\)---and taking these effects into account can be relevant in general scenarios.
        \item The relevance of soft gravitons in the quantum decoherence of spatial superpositions by Killing horizons \cite{danielson2022BlackHolesDecohere,danielson2022KillingHorizonsDecohere,gralla2024DecoherenceHorizonsGeneral,danielson2024LocalDescriptionDecoherence}.
    \end{itemize}

    Of course, there are still many topics I could barely scratch along the notes, such as \gls{CK} spacetimes, the construction of Hadamard states in spacetimes with distinguished null hypersurfaces, and much more. What I hope to have shown is that infinity has never been closer, and the sky has never looked so promising.

\section*{Acknowledgments}\addcontentsline{toc}{section}{Acknowledgments}
    These lecture notes and the course to which they originally corresponded would not be possible without the aid of the organizers and contributors of the \href{https://graspschool.github.io/2024/}{I São Paulo School on Gravitational Physics}. I thus thank them, and the institutional support provided by the University of São Paulo. I find it important to thank by name Bruno Arderucio Costa, Caio César Rodrigues Evangelista, Rafael Grossi e Fonseca, and Victor Hugo dos Santos Lins.
    
    The considerable extension of the notes to the present form was motivated by Peter Horvathy's kind encouragement to make them available on the arXiv. Some of the latest additions benefited a lot from discussions with participants of the \href{https://www.ictp-saifr.org/wqag2024/}{Witnessing Quantum Aspects of Gravity in a Lab} conference, both during and after the event itself. Hence, I thank both the organizers and the helpful conversations with Gautam Satishchandran, Daine L. Danielson, Maitá C. Micol, and Gabriel S. Menezes. My recurrent conversations with André G. S. Landulfo, George E. A. Matsas, and João C. A. Barata have, as they often do, helped shape my understanding of all of these topics and of gravitation as a whole. I finally thank two anonymous referees for helpful comments on the manuscript.
    
    Part of the calculations in this work were carried out with the aid of \texttt{Mathematica 14.0} \cite{wolframresearch2024Mathematica140} (and in particular the \texttt{OGRe} package \cite{shoshany2021OGReObjectOrientedGeneral}) in a license to the University of São Paulo (my \emph{alma mater}). Feynman diagrams were drawn with \texttt{tikz-feynman} \cite{ellis2017TikZFeynmanFeynmanDiagrams}. My work has been supported by the São Paulo Research Foundation (FAPESP) under grant \href{bv.fapesp.br/en/bolsas/229389/the-sky-as-a-killing-horizon-and-other-topics-on-the-infrared-structure-of-gravity/}{2025/05161-0} and by the Coordenação de Aperfeiçoamento de Pessoal de Nível Superior---Brasil (CAPES)---Finance Code 001. 

\begin{appendix}
\section{Complex Analysis}\label{app: complex-analysis}
    In a few points throughout the text we need to employ some results from complex analysis, which involves Wirtinger derivatives and surface integrals over the complex plane. In this appendix I briefly summarize these ideas. The main reference is the book by Hörmander \cite{hormander1990IntroductionComplexAnalysis}. I assume the reader is familiar with more elementary facts about complex analysis, such as the general notions of what is a a holomorphic function, the Cauchy--Riemann equations, and so on. These topics are well discussed in Ref. \citeonline{brown2014ComplexVariablesApplications}.

    We will often identify the complex plane \(\Comp\) with the real plane \(\R^2\). Given some function \(f\) on \(\Comp\) (which is not necessarily holomorphic), we can write its differential as 
    \begin{equation}
        \dd{f} = \pdv{f}{x}\dd{x} + \pdv{f}{y}\dd{y}.
    \end{equation}
    This is real analysis. To get to the complex domain, we introduce the complex coordinates
    \begin{equation}
        \zeta = x + iy \qq{and} \bar{\zeta} = x - iy.
    \end{equation}
    If we define the Wirtinger derivatives by
    \begin{equation}
        \pdv{\zeta} = \frac{1}{2}\qty(\pdv{x} - i \pdv{y}) \qq{and} \pdv{\bar{\zeta}} = \frac{1}{2}\qty(\pdv{x} + i \pdv{y}),
    \end{equation}
    then we can also write the differential of a function of a complex variable as 
    \begin{equation}
        \dd{f} = \pdv{f}{\zeta}\dd{\zeta} + \pdv{f}{\bar{\zeta}}\dd{\bar{\zeta}}.
    \end{equation}
    Hence, we see \(\zeta\) and its conjugate \(\bar{\zeta}\) as coordinates on the plane. It is convenient to treat these coordinates as independent, which intrinsically means extending \(x, y \in \R\) to \(x, y \in \Comp\), so that \((\zeta,\bar{\zeta}) \in \Comp^2\). At the end of the day, we remember the physical results fix \(\bar{\zeta}\) as the complex conjugate of \(\zeta\).

    The advantage of taking this point of view is that we can now discuss functions which are differentiable in the real plane, but still not holomorphic. For a function to be holomorphic, it must satisfy the Cauchy--Riemann equations, and in this notation this means that
    \begin{equation}\label{eq: cauchy-riemann-wirtinger}
        \pdv{f}{\bar{\zeta}} = 0.
    \end{equation}
    Hence, if this condition holds, \(f\) is holomorphic and we are dealing with the usual notions of complex analysis. For this reason, it is common---for example in the \gls{CFT} literature---to write \(f(\zeta)\) for holomorphic \(f\), but \(f(\zeta,\bar{\zeta})\) for more general functions. Notice that if \cref{eq: cauchy-riemann-wirtinger} does not hold, we can still talk about derivatives and calculus by borrowing the concepts from the real plane, but we are no longer discussing holomorphic functions. Therefore, this approach allows us to discuss and do calculus with more functions than the more usual notions of complex analysis would allow.  

    Now let us discuss integrals. Consider some bounded open set \(\Omega \subeq \Comp\). Notice that Green's theorem allows us to write
    \begin{subequations}\label{eq: complex-line-integral-surface-integral}
        \begin{align}
            \oint_{\bound\Omega} f \dd{\zeta} &= \oint_{\bound\Omega} f \dd{x} + i f \dd{y}, \\
            &= \int_{\Omega} \qty(i \pdv{f}{x} - \pdv{f}{y}) \dd{x}\wedge\dd{y}, \\
            &= i \int_{\Omega} \qty(\pdv{f}{x} + i\pdv{f}{y}) \dd{x}\wedge\dd{y}, \\
            &= 2i \int_{\Omega} \pdv{f}{\bar{\zeta}} \dd{x}\wedge\dd{y}, \\
            &= 2i \int_{\Omega} \pdv{f}{\bar{\zeta}} \frac{\dd{\bar{\zeta}}\wedge\dd{\zeta}}{2i}, \\
            &= \int_{\Omega} \pdv{f}{\bar{\zeta}} \dd{\bar{\zeta}}\wedge\dd{\zeta}, \\
            &\equiv \int_{\Omega} \pdv{f}{\bar{\zeta}} \dd{\zeta}\dd{\bar{\zeta}}.
        \end{align}
    \end{subequations}
    In particular, the integral vanishes if \(f\) is holomorphic (a standard result). The correct way of writing these integrals would be in the language of differential forms (making the wedge products explicit, as in the previous calculation). This, however, can overload the notation. I will typically write \(\dd{\zeta}\dd{\bar{\zeta}}\) and it should be implicitly understood that this means \(\dd{\bar{\zeta}}\wedge\dd{\zeta}\) (as opposed to \(\dd{\zeta}\wedge\dd{\bar{\zeta}}\), which differs by a sign). This is common in the \gls{CFT} literature.

    For holomorphic \(f\), we know the Cauchy integral formula holds. It states that, for \(\chi \in \Omega\), 
    \begin{equation}\label{eq: cauchy-integral-formula}
        f(\chi) = \frac{1}{2\pi i} \oint_{\bound\Omega} \frac{f(\zeta)}{\zeta - \chi} \dd{\zeta}.
    \end{equation}
    What about non-holomorphic functions? 
    
    Suppose \(f\) is real-differentiable in \(\Omega\), but not necessarily holomorphic. We pick \(\chi \in \Omega\) and let \(d_{\chi}\) be the distance from \(\chi\) to \(\Comp \setminus \Omega\). We may now define the sets
    \begin{equation}
        \Omega_{\epsilon} = \qty{\zeta \in \Omega \mid \abs{\zeta - \chi} > \epsilon}
    \end{equation}
    for all \(0 < \epsilon < d_{\chi}\). Notice that 
    \begin{equation}
        \frac{1}{\zeta - \chi}
    \end{equation}
    is holomorphic away from \(\chi\). Hence, its \(\bar{\zeta}\) derivative vanishes in \(\Omega_{\epsilon}\). We can thus use \cref{eq: complex-line-integral-surface-integral} to write
    \begin{subequations}
        \begin{align}
            \int_{\Omega_{\epsilon}} \pdv{f}{\bar{\zeta}} \frac{1}{\zeta - \chi} \dd{\zeta}\dd{\bar{\zeta}} &= \int_{\Omega_{\epsilon}} \pdv{\bar{\zeta}}(\frac{f(\zeta,\bar{\zeta})}{\zeta - \chi}) \dd{\zeta}\dd{\bar{\zeta}}, \\
            &= \oint_{\bound\Omega_{\epsilon}} \frac{f(\zeta,\bar{\zeta})}{\zeta - \chi} \dd{\zeta}, \\
            &= \oint_{\bound\Omega} \frac{f(\zeta,\bar{\zeta})}{\zeta - \chi} \dd{\zeta} - \oint_{\abs{\zeta-\chi}=\epsilon} \frac{f(\zeta,\bar{\zeta})}{\zeta - \chi} \dd{\zeta}, \\
            &= \oint_{\bound\Omega} \frac{f(\zeta,\bar{\zeta})}{\zeta - \chi} \dd{\zeta} - \int_{0}^{2\pi} \frac{f(\chi + \epsilon e^{i\phi}, \bar{\chi} + \epsilon e^{-i\phi})}{\epsilon e^{i\phi}} \epsilon i e^{i \phi} \dd{\phi}, \\
            &= \oint_{\bound\Omega} \frac{f(\zeta,\bar{\zeta})}{\zeta - \chi} \dd{\zeta} - i \int_{0}^{2\pi} f(\chi + \epsilon e^{i\phi}, \bar{\chi} + \epsilon e^{-i\phi})\dd{\phi}.
        \end{align}
    \end{subequations}
    \(1/(\zeta-\chi)\) is integrable over \(\Omega\) and \(\pdv*{f}{\bar{\zeta}}\) is continuous on the same set. We can then take the \(\epsilon \to 0\) limit and find
    \begin{equation}\label{eq: cauchy-pompeiu}
        f(\chi,\bar{\chi}) = \frac{1}{2\pi i}\oint_{\bound\Omega} \frac{f(\zeta,\bar{\zeta})}{\zeta - \chi} \dd{\zeta} - \frac{1}{2\pi i} \int_{\Omega} \pdv{f}{\bar{\zeta}} \frac{1}{\zeta - \chi} \dd{\zeta}\dd{\bar{\zeta}}.
    \end{equation}
    This is known as the Cauchy--Pompeiu formula. 

    We can still manipulate the Cauchy--Pompeiu formula to an expression that will serve us better. Using \cref{eq: complex-line-integral-surface-integral,eq: cauchy-pompeiu}, we see that 
    \begin{subequations}
        \begin{align}
            f(\chi,\bar{\chi}) &= \frac{1}{2\pi i}\oint_{\bound\Omega} \frac{f(\zeta,\bar{\zeta})}{\zeta - \chi} \dd{\zeta} - \frac{1}{2\pi i} \int_{\Omega} \pdv{f}{\bar{\zeta}} \frac{1}{\zeta - \chi} \dd{\zeta}\dd{\bar{\zeta}}, \\
            &= \frac{1}{2\pi i}\int_{\Omega} \pdv{\bar{\zeta}}\qty(\frac{f(\zeta,\bar{\zeta})}{\zeta - \chi}) \dd{\zeta}\dd{\bar{\zeta}} - \frac{1}{2\pi i} \int_{\Omega} \pdv{f}{\bar{\zeta}} \frac{1}{\zeta - \chi} \dd{\zeta}\dd{\bar{\zeta}}, \\
            &= \frac{1}{2\pi i}\int_{\Omega} f(\zeta,\bar{\zeta})\pdv{\bar{\zeta}}\qty(\frac{1}{\zeta - \chi}) \dd{\zeta}\dd{\bar{\zeta}}.
        \end{align}
    \end{subequations}
    The derivative is meant in a distributional sense, since it vanishes everywhere but at \((\chi,\bar{\chi})\). We may then write
    \begin{equation}\label{eq: derivative-Dirac-delta-complex}
        \pdv{\bar{\zeta}}\qty(\frac{1}{\zeta - \chi}) = 2 \pi i \delta^{(2)}(\zeta - \chi),
    \end{equation}
    where the Dirac delta is defined by the condition that, for any real-differentiable\footnote{To discuss distributions we actually need smooth functions, but if one refrains from differentiating the Dirac delta we can still consider this wider class of differentiable functions.} (but not necessarily holomorphic) function 
    \begin{equation}
        \int_{\Omega} f(\zeta,\bar{\zeta})\delta^{(2)}\qty(\zeta - \chi) \dd{\zeta}\dd{\bar{\zeta}} = f(\chi,\bar{\chi}).
    \end{equation}

    I mention some authors---especially in the \gls{CFT} literature---may omit the \(i\) factor in \cref{eq: derivative-Dirac-delta-complex}, but they then omit other factors of \(i\) that would occur in the integrals. I choose to keep both factors of \(i\).

\section{The Double Cover of the Lorentz Group}\label{app: double-cover-lorentz}
    In this appendix I prove \cref{eq: lorentz-double-cover},
    \begin{equation}
        \SL(2,\Comp)/\Z_2 \cong \SO*(3,1),
    \end{equation}
    in order to establish that the conformal group of the two-sphere indeed is the Lorentz group. \(\Z_2\) is the two-element group \(\qty{-1,1}\) with the product given by multiplication. Since this group only has two elements, \(\SL(2,\Comp)/\Z_2 \cong \SO*(3,1)\) has the meaning that there are two elements of \(\SL(2,\Comp)\) for each element of \(\SO*(3,1)\). \(\SL(2,\Comp)\) is said to be the double cover of \(\SO*(3,1)\). This topic has many aspects studied in algebraic topology \cite{hatcher2002AlgebraicTopology,lima2003FundamentalGroupsCovering} and spin geometry \cite{hamilton2017MathematicalGaugeTheory,lawson1989SpinGeometry}, but I will keep the discussion to the bare essentials and focus only on the group theoretic aspects. Our discussion mimics those of Refs. \citeonline{zee2016GroupTheoryNutshell,weinberg1995Foundations,hamilton2017MathematicalGaugeTheory}.

    Let \(x \in \R^4\) be a point in Minkowski spacetime and let \(L \in \SL(2,\Comp)\). This means, for our purposes, that \(L\) is a \(2 \times 2\) complex matrix with unit determinant. Using the Cartesian coordinates of \(x\), consider the matrix 
    \begin{equation}\label{eq: xsigma}
        \tensor{x}{^\mu} \sigma_\mu,
    \end{equation}
    where \(\sigma_0 = \Eins\) and \(\sigma_i\) are the Pauli matrices,
    \begin{equation}
        \sigma_1 = \mqty(0 & 1 \\ 1 & 0)\qc \sigma_2 = \mqty(0 & -i \\ i & 0), \qq{and} \sigma_3 = \mqty(1 & 0 \\ 0 & -1).
    \end{equation}
    \cref{eq: xsigma}, with \(\tensor{x}{^\mu}\) understood to be real, is the most general expression for a \(2 \times 2\) Hermitian matrix with complex entries. 
    
    Consider now
    \begin{equation}
        X' = L \tensor{x}{^\mu} \sigma_\mu L^\dagger,
    \end{equation}
    where \(\dagger\) denotes Hermitian conjugation. \(X'\) is surely a \(2 \times 2\) complex matrix, and it is still Hermitian as one can show by direct computation. We can thus write 
    \begin{equation}
        \tensor{x}{^\prime^\mu} \sigma_\mu = L \tensor{x}{^\mu} \sigma_\mu L^\dagger.
    \end{equation}
    Since \(L\) has unit determinant, we find that
    \begin{equation}
        \tensor{x}{^\prime^\mu}\tensor*{x}{^\prime_\mu} = - \det(\tensor{x}{^\prime^\mu} \sigma_\mu) = - \det L \det(\tensor{x}{^\mu} \sigma_\mu) \det L^\dagger = - \det(\tensor{x}{^\mu} \sigma_\mu) = \tensor{x}{^\mu}\tensor{x}{_\mu}.
    \end{equation}
    Hence, the map \(x \mapsto x'\) is a Lorentz transformation, because it preserves the invariant interval, and thus the Minkowski metric. We can write 
    \begin{equation}
        L \tensor{x}{^\mu} \sigma_\mu L^\dagger = \tensor{\Lambda(L)}{^\mu_\nu}\tensor{x}{^\nu} \sigma_\mu.
    \end{equation}
    Hence, we found a map \(\Lambda\) from \(\SL(2,\Comp)\) to \(\Og(3,1)\). We now need to show that
    \begin{enumerate}
        \item it is a homomorphism,
        \item its range is \(\SO*(3,1)\), 
        \item it is two-to-one.
    \end{enumerate}

    It is direct to show it is a homomorphism. Let \(M \in \SL(2,\Comp)\). Then 
    \begin{subequations}
        \begin{align}
            \tensor{\Lambda(ML)}{^\mu_\nu}\tensor{x}{^\nu} \sigma_\mu &= (ML) \tensor{x}{^\mu} \sigma_\mu (ML)^\dagger, \\ 
            &= ML \tensor{x}{^\mu} \sigma_\mu L^\dagger M^\dagger, \\
            &= M \tensor{\Lambda(L)}{^\mu_\nu}\tensor{x}{^\nu} \sigma_\mu M^\dagger, \\
            &= \tensor{\Lambda(M)}{^\mu_\nu}\tensor{\Lambda(L)}{^\nu_\rho}\tensor{x}{^\rho} \sigma_\mu, \\
            &= \tensor{(\Lambda(M)\Lambda(L))}{^\mu_\rho}\tensor{x}{^\rho} \sigma_\mu.
        \end{align}
    \end{subequations}
    Since \(\Lambda(ML) = \Lambda(M)\Lambda(L)\), we conclude \(\Lambda\) is indeed a homomorphism.

    Next we want to show that \(\Ran\Lambda = \SO*(3,1)\). An inner product in the space of \(2 \times 2\) complex Hermitian matrices can be introduced by defining the Frobenius inner product
    \begin{equation}
        \braket{X}{Y} = \frac{1}{2}\Tr[X^\dagger Y] = \frac{1}{2}\Tr[XY].
    \end{equation}
    The identity matrix and the Pauli matrices then establish an orthonormal basis with respect to this inner product. Exploiting this fact we see that
    \begin{equation}\label{eq: x'-L-tranformation}
        \tensor{x}{^\prime^\mu} = \frac{\tensor{x}{^\nu}}{2}\Tr[L \sigma_{\nu} L^\dagger \bar{\sigma}^{\mu}],
    \end{equation}
    where \(\bar{\sigma}^0 = - \sigma^0 = \Eins\) and \(\bar{\sigma}^i = \sigma^i\) (we raise the indices of \(\sigma_\mu\) with the Minkowski metric). Therefore, the Lorentz matrix \(\Lambda(L)\) associated to \(L\) is given by
    \begin{equation}
        \tensor{\Lambda(L)}{^\mu_\nu} = \frac{1}{2}\Tr[L \sigma_{\nu} L^\dagger \bar{\sigma}^{\mu}].
    \end{equation}
    Notice
    \begin{equation}
        \tensor{\Lambda(L)}{^0_0} = \frac{1}{2}\Tr[L L^\dagger] > 0,
    \end{equation}
    and hence \(\Lambda(L) \in \Og*(3,1)\). 

    It will be a bit more difficult to prove \(\Lambda(L) \in \SO(3,1)\). To do so, notice that any \(2 \times 2\) complex matrix can be written as
    \begin{equation}
        \tensor{z}{^\mu} \sigma_{\mu},
    \end{equation}
    with complex \(\tensor{z}{^\mu}\). It follows that \(\det(\tensor{z}{^\mu} \sigma_{\mu}) = - \tensor{z}{^\mu}\tensor{z}{_\mu}\). Hence, we can always write \(L \in \SL(2,\Comp)\) as
    \begin{equation}
        L = \tensor{z}{^\mu} \sigma_{\mu}
    \end{equation}
    with \(\tensor{z}{^\mu}\tensor{z}{_\mu} = - 1\). Using this, we notice that
    \begin{equation}
        \tensor{\Lambda(L)}{^\mu_\nu} = \frac{\tensor{z}{^\alpha}\tensor{\bar{z}}{^\beta}}{2}\Tr[\sigma_\alpha \sigma_{\nu} \sigma_\beta \bar{\sigma}^{\mu}].
    \end{equation}
    Using this expression, we can compute the elements of \(\Lambda(L)\) explicitly and compute the determinant by brute force. The result is 
    \begin{equation}
        \det \Lambda(L) = \tensor{z}{^\mu}\tensor{z}{_\mu} \tensor{\bar{z}}{^\nu}\tensor{\bar{z}}{_\nu} = 1,
    \end{equation}
    and hence \(\Lambda(L) \in \SO(3,1)\). We already knew \(\Lambda(L) \in \Og*(3,1)\), and thus we conclude \(\Ran\Lambda \subeq \SO*(3,1)\). One can show that \(\Ran\Lambda = \SO*(3,1)\) by constructing some key elements of the Lorentz group, such as rotations and boosts centered at each Cartesian axis, explicitly.

    Finally, we want to see that \(\Lambda \colon \SL(2,\Comp) \to \SO*(3,1)\) is two-to-one. This can be seen by noticing that 
    \begin{subequations}
        \begin{align}
            \tensor{\Lambda(-L)}{^\mu_\nu} &= \frac{1}{2}\Tr[(-L) \sigma_{\nu} (-L)^\dagger \bar{\sigma}^{\mu}], \\
            &= \frac{1}{2}\Tr[L \sigma_{\nu} L^\dagger \bar{\sigma}^{\mu}], \\
            &= \tensor{\Lambda(L)}{^\mu_\nu},
        \end{align}
    \end{subequations}
    and therefore \(L\) and \(-L\) are both mapped to \(\Lambda(L)\). More rigorously, we find that the kernel of \(\Lambda\), defined by
    \begin{equation}
        \Ker \Lambda = \qty{L \in \SL(2,\Comp) \mid \Lambda(L) = \Eins},
    \end{equation}
    is given by 
    \begin{equation}
        \Ker \Lambda = \qty{-\Eins, \Eins}.
    \end{equation}
    Notice \(\Ker\Lambda = \Z_2\). This also shows \(\Lambda\) maps exactly two elements of \(\SL(2,\Comp)\) to each element of \(\SO*(3,1)\). Not more, not less.

    Bringing everything together, we conclude that \(\SL(2,\Comp)\) is a duplicated copy of \(\SO*(3,1)\). Hence,
    \begin{equation}
        \SL(2,\Comp)/\Z_2 \cong \SO*(3,1).
    \end{equation}
    This is in fact merely an application of the first isomorphism theorem for groups (see, \eg, Theorem A-4.73 on Ref. \citeonline{rotman2015AdvancedModernAlgebraI}).

\section{Brief Review of Linearized Gravity}\label{app: linearized-gravity}
    This appendix briefly reviews linearized gravity on Minkowski spacetime with the goal of fixing notation for the many sections that build upon it. There are numerous references on linearized gravity. It is a standard topic in general relativity textbooks (I particularly notice those by \textcite{carroll2019SpacetimeGeometryIntroduction,wald1984GeneralRelativity}), but some more specialized references are the book by \textcite{maggiore2008TheoryExperiments} and the review article by \textcite{flanagan2005BasicsGravitationalWave}.

    The full metric is written \(\tensor{g}{_a_b} = \tensor{\eta}{_a_b} + \tensor{h}{_a_b}\). Our task is to compute \(\tensor{h}{_a_b}\) given some matter distribution. This will incur in the emission of \glspl{GW}. 

    Define the trace-reversed perturbation \(\tensor{\bar{h}}{_a_b} = \tensor{h}{_a_b} - \frac{1}{2}h\tensor{\eta}{_a_b}\), where \(h = \tensor{h}{_a_b}\tensor{\eta}{^a^b}\). In de Donder gauge\footnote{Also known as Lorenz gauge or harmonic gauge.},
    \begin{equation}\label{eq: de-donder-gauge}
        \tensor{\partial}{_a}\tensor{\bar{h}}{^a^b} = 0,
    \end{equation}
    the Einstein equations become \cite{wald1984GeneralRelativity,maggiore2008TheoryExperiments}
    \begin{equation}
        \tensor{\partial}{_c}\tensor{\partial}{^c}\tensor{\bar{h}}{_a_b} = - 16 \pi \tensor{T}{_a_b}.
    \end{equation}

    \subsection{The Transverse-Traceless Gauge}
        In Cartesian coordinates, we can choose to focus specifically on the spatial components \(\tensor{\bar{h}}{_i_j}\). The remaining components of \(\tensor{\bar{h}}{_\mu_\nu}\) are obtained by employing the de Donder gauge condition \(\tensor{\partial}{_\mu}\tensor{\bar{h}}{^\mu^\nu} = 0\). For simplicity, we can assume that in the vacuum region the gauge is such that \(\tensor{\bar{h}}{_\mu_\nu}\tensor{\eta}{^\mu^\nu} = 0\), \(\tensor{\bar{h}}{_0_i} = 0\), and we make the simplifying assumption that \(\tensor{\bar{h}}{_0_0} = 0\) (this means we are neglecting any constant background field). This is the \glsfmtfull{TT} gauge \cite{maggiore2008TheoryExperiments,carroll2019SpacetimeGeometryIntroduction}. 

        The first point to notice is that, since \(\tensor*{\bar{h}}{^{\text{TT}}_\mu_\nu}\tensor{\eta}{^\mu^\nu} = 0\), it also holds that \(\tensor*{h}{^{\text{TT}}_\mu_\nu}\tensor{\eta}{^\mu^\nu} = 0\), and it holds that \(\tensor*{h}{^{\text{TT}}_\mu_\nu} = \tensor*{\bar{h}}{^{\text{TT}}_\mu_\nu}\).
    
        Next, recall that one can project any symmetric rank-2 spatial tensor to its \gls{TT} component. If we define the projector
        \begin{equation}
            \tensor{P}{_i^j}(\vu{n}) = \tensor{\delta}{_i^j} - \tensor{n}{_i}\tensor{n}{^j},
        \end{equation}
        then we can define
        \begin{equation}
            \tensor{\Lambda}{_i_j^k^l}(\vu{n}) = \tensor{P}{_i^k}(\vu{n})\tensor{P}{_j^l}(\vu{n}) - \frac{1}{2}\tensor{P}{_i_j}(\vu{n})\tensor{P}{^k^l}(\vu{n}).
        \end{equation}
        Then, for a \gls{GW} propagating along the \(\vu{n}\) direction, its \gls{TT} projection is given by \cite{maggiore2008TheoryExperiments}
        \begin{equation}
            \tensor*{h}{^{\text{TT}}_i_j} = \tensor{\Lambda}{_i_j^k^l}(\vu{n}) \tensor*{h}{_k_l}.
        \end{equation}
        The case for other tensors follows in the same manner. For instance,
        \begin{equation}
            \tensor*{\ddot{Q}}{^{\text{TT}}_i_j} = \tensor{\Lambda}{_i_j^k^l}(\vu{n}) \tensor*{\ddot{Q}}{_k_l}.
        \end{equation}
        This algebraic projection should be performed for each Fourier mode \(\vb{k}\).

        \gls{TT} gauge is a particular case of the de Donder gauge. The subsequent expressions can be derived in general de Donder gauge, but it may be useful to have them expressed in the more specific \gls{TT} gauge. This can be done by simply projecting both sides of the equation of interest accordingly. 

    \subsection{Solving the Linearized Einstein Equations}
        It is convenient to work in global Cartesian coordinates. In this case, we write 
        \begin{equation}
            \dal\tensor{\bar{h}}{_\mu_\nu} = - 16 \pi \tensor{T}{_\mu_\nu},
        \end{equation}
        where the d'Alembertian \(\dal\) is given by
        \begin{equation}
            \dal = \tensor{\eta}{^\mu^\nu}\tensor{\partial}{_\mu}\tensor{\partial}{_\nu}.
        \end{equation}
        This means that each Cartesian component of the gravitational field satisfies an independent wave equation with a source. This allows us to write a Green function for the scalar equation and obtain a Green function for the tensorial equation in Cartesian components. 
    
        For the scalar wave equation, we know that a solution to
        \begin{equation}
            \dal G_R(x,y) = \delta^{(4)}(x,y)
        \end{equation}
        in Minkowski spacetime is given by 
        \begin{equation}
            G_R(x,y) = -\frac{\delta(\tensor{x}{^0} - \tensor{y}{^0} - \norm{\vb{x} - \vb{y}})}{4 \pi \norm{\vb{x} - \vb{y}}}\Theta(\tensor{x}{^0} - \tensor{y}{^0}),
        \end{equation}
        with \(\Theta\) standing for the Heaviside step function. This is known as the retarded Green function and it has the property that \(G_R(x,y)\) vanishes for \(x\) outside the future lightcone of \(y\). This topic is well-discussed in books on electrodynamics \cite{lechner2018ClassicalElectrodynamicsModern,wald2022AdvancedClassicalElectromagnetism,zangwill2013ModernElectrodynamics}.
    
        In terms of the Green function, we can write the solution to the linearized Einstein equations in Cartesian coordinates as
        \begin{subequations}\label{eq: h-bar-from-tmunu-retarded-green}
            \begin{align}
                \tensor{\bar{h}}{_\mu_\nu}(x) &= - 16 \pi \int G_R(x,y) \tensor{T}{_\mu_\nu}(y) \dd[4]{y}, \\
                &= 4 \int \frac{\delta(\tensor{x}{^0} - \tensor{y}{^0} - \norm{\vb{x} - \vb{y}}) \tensor{T}{_\mu_\nu}(y)}{\norm{\vb{x} - \vb{y}}}\Theta(\tensor{x}{^0} - \tensor{y}{^0}) \dd[4]{y}, \\
                &= 4 \int \frac{\tensor{T}{_\mu_\nu}\qty(\tensor{x}{^0} - \norm{\vb{x} - \vb{y}}, \vb{y})}{\norm{\vb{x} - \vb{y}}} \dd[3]{y}.
            \end{align}
        \end{subequations}
    
        If the only approximation we were interested in was the linear approximation, we could stop here. Nevertheless, throughout the main text we will be mainly interested in the behavior near infinity, which means we can make a large \(r = \norm{\vb{x}}\) approximation. With this in mind, write \(\vb{x} = r \vu{x}\) and notice that
        \begin{subequations}
            \begin{align}
                \norm{\vb{x} - \vb{y}} &= r\norm{\vu{x} - \frac{\vb{y}}{r}}, \\
                &= r\sqrt{1 - \frac{2 \vu{x} \vdot \vb{y}}{r} + \frac{\norm{\vb{y}}^2}{r^2}}, \\
                &= r - \vu{x} \vdot \vb{y} + \order{\frac{1}{r}}.
            \end{align}
        \end{subequations}
        Using this in \cref{eq: h-bar-from-tmunu-retarded-green} leads to 
        \begin{equation}
            \tensor{\bar{h}}{_\mu_\nu}(t,\vb{x}) = \frac{4}{r} \int \tensor{T}{_\mu_\nu}\qty(t - r + \vu{x} \vdot \vb{y}, \vb{y}) \dd[3]{y} + \order{\frac{1}{r^2}}.
        \end{equation}
        Notice the approximation requires \(r\) to be large when compared to the region in which the stress tensor is non-vanishing.
        
        To proceed from this point, we follow the same techniques used in multipole expansions in electrodynamics. More specifically, I will adapt the discussions in Section 5.3 of Ref. \citeonline{wald2022AdvancedClassicalElectromagnetism}. Let us rewrite the stress tensor in a Taylor series, which will allow us to get rid of the angular dependency. We write
        \begin{equation}
            \tensor{T}{_\mu_\nu}\qty(t - r + \vu{x} \vdot \vb{y}, \vb{y}) = \sum_{n=0}^{+\infty} \frac{1}{n!} \pdv[n]{t} \tensor{T}{_\mu_\nu}\qty(t-r, \vb{y}) \qty(\vu{x} \vdot \vb{y})^n.
        \end{equation}
        Hence, the gravitational field is given by 
        \begin{equation}\label{eq: hmunu-linear-taylor-multipole}
            \tensor{\bar{h}}{_\mu_\nu}(t,\vb{x}) = \frac{4}{r} \sum_{n=0}^{+\infty} \frac{1}{n!} \dv[n]{t} \int \tensor{T}{_\mu_\nu}\qty(t-r, \vb{y}) \qty(\vu{x} \vdot \vb{y})^n \dd[3]{y} + \order{\frac{1}{r^2}}.
        \end{equation}
    
    \subsection{The Quadrupole Approximation}
        Let us now consider the leading term in the Taylor series of \cref{eq: hmunu-linear-taylor-multipole}. This means making an approximation in which \(\tensor{T}{_\mu_\nu}\) varies slowly. We are thus considering 
        \begin{equation}
            \tensor{\bar{h}}{_i_j}(t,\vb{x}) = \frac{4}{r} \int \tensor{T}{_i_j}\qty(t-r, \vb{y}) \dd[3]{y} + \cdots.
        \end{equation}
        Above, I wrote only the spatial components, because we can recover the remaining ones from the de Donder gauge condition (\ref{eq: de-donder-gauge}).

        Following \textcite{carroll2019SpacetimeGeometryIntroduction}, we notice that integration by parts and energy-momentum conservation \(\tensor{\partial}{_\mu}\tensor{T}{^\mu_\nu} = 0\) get us to
        \begin{subequations}
            \begin{align}
                \tensor{\bar{h}}{_i_j}(t,\vb{x}) &= \frac{4}{r} \int \tensor{T}{_i_j}\qty(t-r, \vb{y}) \dd[3]{y}, \\
                &= \frac{4}{r} \int \tensor{\partial}{_k}\qty(\tensor{y}{_i}\tensor{T}{^k_j}\qty(t-r, \vb{y})) \dd[3]{y} - \frac{4}{r} \int \tensor{y}{_i}\tensor{\partial}{_k}\tensor{T}{^k_j}\qty(t-r, \vb{y}) \dd[3]{y}, \\
                &= - \frac{4}{r} \int \tensor{y}{_i}\tensor{\partial}{_k}\tensor{T}{^k_j}\qty(t-r, \vb{y}) \dd[3]{y}, \\
                &= \frac{4}{r} \dv{t} \int \tensor{y}{_i}\tensor{T}{^0_j}\qty(t-r, \vb{y}) \dd[3]{y}.
            \end{align}
        \end{subequations}
        We were able to neglect the total derivative because, by assumption, the stress tensor is only relevant in a small region (this is hidden in the large \(r\) expansion). 

        To proceed, we notice that, since \(\tensor{\bar{h}}{_i_j}\) is knowingly symmetric,
        \begin{subequations}
            \begin{align}
                \tensor{\bar{h}}{_i_j}(t,\vb{x}) &= \frac{4}{r} \dv{t} \int \tensor{y}{_i}\tensor{T}{^0_j}\qty(t-r, \vb{y}) \dd[3]{y}, \\
                &= \frac{2}{r} \dv{t} \int \tensor{y}{_i}\tensor{T}{^0_j}\qty(t-r, \vb{y}) \dd[3]{y} + \frac{2}{r} \dv{t}\int \tensor{y}{_j}\tensor{T}{^0_i}\qty(t-r, \vb{y}) \dd[3]{y} , \\
                &= \frac{2}{r} \dv{t} \int \tensor{\partial}{_k}\qty(\tensor{y}{_i}\tensor{y}{_j}\tensor{T}{^0^k}\qty(t-r, \vb{y})) \dd[3]{y} - \frac{2}{r}\dv{t} \int\tensor{y}{_i}\tensor{y}{_j} \tensor{\partial}{_k}\qty(\tensor{T}{^0^k}\qty(t-r, \vb{y})) \dd[3]{y} , \\
                &= -\frac{2}{r}\dv{t} \int\tensor{y}{_i}\tensor{y}{_j} \tensor{\partial}{_k}\qty(\tensor{T}{^0^k}\qty(t-r, \vb{y})) \dd[3]{y} , \\
                &= \frac{2}{r}\dv[2]{t} \int\tensor{y}{_i}\tensor{y}{_j} \tensor{T}{^0^0}\qty(t-r, \vb{y}) \dd[3]{y} , \\
                &= \frac{2}{r} \left\lfloor\dv[2]{\tensor{Q}{_i_j}}{t}\right\rfloor,
            \end{align}
        \end{subequations}
        where \(\lfloor\cdot\rfloor\) denotes evaluation at retarded Minkowski time \(u=t-r\) and the quadrupole moment tensor \(\tensor{Q}{_i_j}\) is given by
        \begin{equation}\label{eq: def-quadrupole}
            \tensor{Q}{_i_j}(t) = \int \tensor{T}{^0^0}(t,\vb{y}) \tensor{y}{_i} \tensor{y}{_j} \dd[3]{y}.
        \end{equation}
        To keep notation simple, I will write
        \begin{equation}
            \tensor{\ddot{Q}}{_i_j} \equiv \left\lfloor\dv[2]{\tensor{Q}{_i_j}}{t}\right\rfloor
        \end{equation}
        and it should be always understood that \(\tensor{\ddot{Q}}{_i_j}\) is a function of the retarded time \(t-r\) only.
        
        Finally, the power radiated to infinity at retarded time \(u\) per unit solid angle is given by \cite{carroll2019SpacetimeGeometryIntroduction,wald1984GeneralRelativity}
        \begin{equation}\label{eq: power-GWs}
            P = \frac{1}{5} \tensor*{\dddot{Q}}{^{\text{TT}}_i_j}\tensor*{\dddot{Q}}{^{\text{TT}}^i^j}.
        \end{equation}
        The derivation of this equation is a long calculation. Since it will not play a very important role in our discussions, I will omit it.  

\section{Preserving the Strong Carrollian Structure}\label{app: preserving-scs}
    In this appendix, I show how preserving the strong Carrollian structure can affect the \gls{BMS} group. For simplicity, we will from the start ignore the contributions due to Lorentz transformations, since we are interested in knowing whether preserving the strong Carrollian structure is enough to get rid of the supertranslations. This considerably simplifies the calculations, but will still allow us to conclude that the strong Carrollian structure is too strong to define asymptotic symmetry groups. This is due to the fact that, as we shall see, we end up losing the spatial translations with this definition. 

    Since we are working with only supertranslations, we are working from the start with isometries (not confomorphisms) that preserve the strong conformal geometry. Hence, we have that the vector fields \(\tensor{\xi}{^a}\) generating the supertranslations are such that the induced metric \(\tensor{\tilde{h}}{_a_b}\) and the normal vector \(\tensor{n}{^a}\) respect
    \begin{equation}
        \Lie[\xi]\tensor{\tilde{h}}{_a_b} = 0 \qq{and} \Lie[\xi]\tensor{n}{^a} = 0.
    \end{equation}
    To impose the invariance of the strong Carrollian structure, we will also impose that \(\tensor{\xi}{^a}\) should be such that
    \begin{equation}\label{eq: Lie-nabla-0}
        \Lie[\xi] \tensor{\tilde{\nabla}}{_a} = 0.
    \end{equation}
    If we can provide meaning to this equation, it will be a natural condition to impose if we want the strong Carrollian structure to be preserved. While \cref{eq: Lie-nabla-0} may be an odd-looking equation, it could also be interpreted as the Lie derivative of the connection coefficients of \(\tensor{\tilde{\nabla}}{_a}\) at a point---as done in p. 126 of Ref. \citeonline{lovelock1989TensorsDifferentialForms} or Section 4 of Ref. \citeonline{yano1955LieDerivativesIts}.

    Here is how we will define the Lie derivative of the covariant derivative. Firstly, we notice that the Lie derivative can be thought of as a difference between two geometrical objects at a point. The difference between two covariant derivatives is a \((1,2)\)-tensor, so the Lie derivative of the covariant derivative will be a \((1,2)\)-tensor. That being said, we will define \(\Lie[\xi] \tensor{\tilde{\nabla}}{_a}\) by demanding that the formal expression
    \begin{equation}
        \Lie[\xi] (\tensor{\tilde{\nabla}}{_a}\tensor{\psi}{^b}) = (\Lie[\xi] \tensor{\tilde{\nabla}}{_a})\tensor{\psi}{^b} + \tensor{\tilde{\nabla}}{_a} (\Lie[\xi] \tensor{\psi}{^b})
    \end{equation}
    holds. From this formal equation, we can guess the (correct) expression
    \begin{equation}
        (\Lie[\xi] \tensor{\tilde{\Gamma}}{^a_b_c})\tensor{\psi}{^c} = \Lie[\xi] (\tensor{\tilde{\nabla}}{_b}\tensor{\psi}{^a}) - \tensor{\tilde{\nabla}}{_b} (\Lie[\xi] \tensor{\psi}{^a}),
    \end{equation}
    with \(\tensor{\tilde{\Gamma}}{^a_b_c}\) being the connection coefficients of \(\tensor{\tilde{\nabla}}{_a}\). This is Eq. (4.7) of Ref. \citeonline{yano1955LieDerivativesIts}. Using \cref{eq: general-expression-lie-derivative} leads us to 
    \begin{subequations}
        \begin{align}
            (\Lie[\xi] \tensor{\tilde{\Gamma}}{^a_b_d})\tensor{\psi}{^d} &= \tensor{\xi}{^c}(\tensor{\tilde{\nabla}}{_c}\tensor{\tilde{\nabla}}{_b} - \tensor{\tilde{\nabla}}{_b}\tensor{\tilde{\nabla}}{_c})\tensor{\psi}{^a} + \tensor{\psi}{^d}\tensor{\tilde{\nabla}}{_b}\tensor{\tilde{\nabla}}{_d}\tensor{\xi}{^a}, \\
            &= \tensor{\tilde{R}}{_b_c_d^a}\tensor{\xi}{^c}\tensor{\psi}{^d} + \tensor{\psi}{^d}\tensor{\tilde{\nabla}}{_b}\tensor{\tilde{\nabla}}{_d}\tensor{\xi}{^a},
        \end{align}
    \end{subequations}
    where the second step is merely an use of the definition of the Riemann tensor along with some simplification. Notice that \(\tensor{\tilde{R}}{_a_b_c^d}\) is the Riemann tensor associated with \(\tensor{\tilde{\nabla}}{_a}\). This calculation assumes implicitly that the covariant derivative is torsionless.

    We want to impose \cref{eq: Lie-nabla-0}. Thus, we can impose that \(\Lie[\xi] \tensor{\tilde{\Gamma}}{^a_b_c} = 0\). This leads to
    \begin{equation}
        \tensor{\tilde{R}}{_a_c_d^b}\tensor{\xi}{^c} + \tensor{\tilde{\nabla}}{_a}\tensor{\tilde{\nabla}}{_d}\tensor{\xi}{^b} = 0.
    \end{equation}
    To solve this equation, it is convenient to pick \(\tensor{\tilde{\nabla}}{_a} = \tensor{D}{_a}\), the Levi-Civita connection on the sphere. Recall that this choice of connection does lead to a strong Carrollian structure on null infinity. This choice implies that \(\tensor{\tilde{R}}{_a_b_c^d}\) contracted with \(\tensor{n}{^a}\) in any of its indices will vanish. Recalling that the general expression for a vector field generating a supertranslation is (\cref{eq: bms-vector-field})
    \begin{equation}
        \tensor{\xi}{^a} = f \tensor{n}{^a},
    \end{equation}
    where \(f \in \ck[\infty](\Sph^2)\), we get that 
    \begin{equation}
        f\tensor{\tilde{R}}{_a_c_d^b}\tensor{n}{^c} + \tensor{D}{_a}\tensor{D}{_d}(f\tensor{n}{^b}) = 0.
    \end{equation}
    It then follows that 
    \begin{equation}
        \tensor{D}{_a}\tensor{D}{_b}f = 0.
    \end{equation}

    Using stereographic coordinates, we conclude using \cref{eq: christoffel-sphere-stereographic} that this equation corresponds to imposing that
    \begin{subequations}
        \begin{align}
                \tensor{\partial}{_\zeta}\tensor{\partial}{_\zeta}f &= - \frac{2 \bar{\zeta}}{1 + \zeta \bar{\zeta}}\tensor{\partial}{_\zeta}f, \label{eq: scs-1} \\
                \tensor{\partial}{_{\bar{\zeta}}}\tensor{\partial}{_{\bar{\zeta}}}f &= - \frac{2 \zeta}{1 + \zeta \bar{\zeta}}\tensor{\partial}{_{\bar{\zeta}}}f, \label{eq: scs-2} \\
                \tensor{\partial}{_{\zeta}}\tensor{\partial}{_{\bar{\zeta}}} f &= 0, \label{eq: scs-3} \\
                \tensor{\partial}{_{\bar{\zeta}}}\tensor{\partial}{_{\zeta}} f &= 0. \label{eq: scs-4}
            \end{align}
    \end{subequations}

    Denote \(f_{\zeta} = \tensor{\partial}{_{\zeta}}f\) and \(f_{\bar{\zeta}} = \tensor{\partial}{_{\bar{\zeta}}}f\). Then \cref{eq: scs-4} tells us that \(f_{\zeta}\) is a holomorphic function of \(\zeta\), and, equivalently, bears no dependence on \(\bar{\zeta}\). Thus, the \gls{LHS} of \cref{eq: scs-1} does not depend on \(\bar{\zeta}\), but the \gls{RHS} may. This is only possible if both of them vanish. Hence, \(f_{\zeta} = 0\). Similarly, \(f_{\bar{\zeta}} = 0\). We thus get to the system of differential equations
    \begin{equation}
        \left\lbrace
            \begin{aligned}
                &\tensor{\partial}{_{\zeta}} f = 0, \\
                &\tensor{\partial}{_{\bar{\zeta}}} f = 0,
            \end{aligned}
        \right.
    \end{equation}
    which is solved by constant \(f\). 

    Therefore, imposing the preservation of the strong Carrollian structure imposes that the function \(f\) is a constant. On the one hand, this does rule out the supertranslations. On the other hand, it also rules out spatial translations. We thus conclude the preservation of strong Carrollian structure is too strong to be imposed on the asymptotic symmetry group.

\section{Stationary Phase Approximation in the Sphere}\label{app: stationary-phase-approximation-sphere}
    In some points throughout the text we need to compute a few integrals on the sphere at infinity. These integrals typically have the form 
    \begin{equation}\label{eq: oscillatory-integral-sphere}
        \int_{\Sph^2} G(\theta,\phi) e^{-i\omega r (1 - \cos\theta)} \sin\theta \dd{\theta} \dd{\phi}
    \end{equation}
    for large \(r\). 

    These sorts of integrals can be approximated to leading order in \(r\) by employing the stationary phase approximation (see, for example, Theorem 7.7.5 in Ref. \citeonline{hormander2003DistributionTheoryFourier} for a particular case). For an integral over a finite open set \(U \subeq \R^n\), we would have (for large \(r\))
    \begin{equation}
        \int_U F(\vb{x}) e^{i r f(\vb{x})} \dd[n]{x} = \sum_{\vb{x}_0 \in \Sigma} e^{i r f(\vb{x}_0)} \qty[\det(\frac{r f''(\vb{x}_0)}{2 \pi i})]^{-\frac{1}{2}} F(\vb{x}_0) + o\qty(\frac{1}{r^{\frac{n}{2}}}),
    \end{equation}
    where \(f''\) denotes the Hessian of \(f\) and \(\Sigma \subeq U\) is the collection of points with \(\dd{f}|_{\vb{x}_0} = 0\)---\ie, the collections of points in which the phase is stationary. Above, we assume \(F\) and \(f\) to be real functions and for the Hessian of \(f\) to be nonvanishing at \(\Sigma\), among other technical conditions on which we will not focus.

    The key difficulty in \cref{eq: oscillatory-integral-sphere} is the fact that the phase is stationary precisely when its Hessian vanishes, which violates the assumption on the behavior of the Hessian of \(f\). Hence, the stationary phase approximation becomes problematic. 

    To circumvent this issue, we compute the integral in real stereographic coordinates. Define the (real) coordinates \(X\) and \(Y\) on the sphere by 
    \begin{equation}
        X = \cot\frac{\theta}{2} \cos\phi \qq{and} Y = \cot\frac{\theta}{2} \sin\phi,
    \end{equation}
    which are just the real and imaginary parts of the complex stereographic coordinate of \cref{eq: def-stereographic}. These coordinates hold for \(\theta < \pi\).  Define also 
    \begin{equation}
        \tilde{X} = -\tan\frac{\theta}{2} \cos\phi \qq{and} \tilde{Y} = -\tan\frac{\theta}{2} \sin\phi,
    \end{equation}
    which are just the real and imaginary parts of the complex stereographic coordinate of \cref{eq: def-stereographic-antipodal}. These coordinates hold for \(\theta > 0\). We now find that (in the domains in which the expressions make sense)
    \begin{equation}
        1 - \cos\theta = \frac{2}{1 + X^2 + Y^2} = 2 - \frac{2}{1 + \tilde{X}^2 + \tilde{Y}^2},
    \end{equation}
    and 
    \begin{equation}
        \sin\theta \dd{\theta} \dd{\phi} = \frac{4 \dd{X} \dd{Y}}{(1 + X^2 + Y^2)^2} = \frac{4 \dd{\tilde{X}} \dd{\tilde{Y}}}{(1 + \tilde{X}^2 + \tilde{Y}^2)^2}.
    \end{equation}

    and the metric and volume element expressed in terms of them have the same form as for \((X,Y)\).

    Therefore, we found 
    \begin{multline}\label{eq: oscillatory-integral-sphere-stereographic}
        \int_{\Sph^2} G(\theta,\phi) e^{-i\omega r (1 - \cos\theta)} \sin\theta \dd{\theta} \dd{\phi} = \int_{X^2+Y^2 \leq 1} G(X,Y) e^{-\frac{2 i \omega r}{1 + X^2 + Y^2}} \frac{4 \dd{X} \dd{Y}}{(1 + X^2 + Y^2)^2} \\ + \int_{\tilde{X}^2+\tilde{Y}^2 < 1} G(\tilde{X},\tilde{Y}) e^{\frac{2 i \omega r}{1 + \tilde{X}^2 + \tilde{Y}^2} - 2 i \omega r} \frac{4 \dd{\tilde{X}} \dd{\tilde{Y}}}{(1 + \tilde{X}^2 + \tilde{Y}^2)^2}.
    \end{multline}
    Both integrals run over finite sets. This ensures we will not run into problems with a stationary point at infinity.
    
    Notice that 
    \begin{equation}
        \dd\qty(\frac{1}{1 + X^2 + Y^2}) = - \frac{2 X \dd{X}}{1 + X^2 + Y^2} - \frac{2 Y \dd{Y}}{1 + X^2 + Y^2},
    \end{equation}
    which vanishes at the origin\footnote{And would vanish at infinity if we had not restricted the integrals to the unit circle}. The stationary phase approximation yields
    \begin{equation}
        \int_{X^2+Y^2 \leq 1} G(X,Y) e^{-\frac{2 i \omega r}{1 + X^2 + Y^2}} \frac{4 \dd{X} \dd{Y}}{(1 + X^2 + Y^2)^2} = - \frac{2 i \pi e^{-2i\omega r}}{r \omega} G(\theta = \pi) + o\qty(\frac{1}{r})
    \end{equation}
    and
    \begin{equation}
        \int_{\tilde{X}^2+\tilde{Y}^2 < 1} G(\tilde{X},\tilde{Y}) e^{\frac{2 i \omega r}{1 + \tilde{X}^2 + \tilde{Y}^2} - 2 i \omega r} \frac{4 \dd{\tilde{X}} \dd{\tilde{Y}}}{(1 + \tilde{X}^2 + \tilde{Y}^2)^2} = - \frac{2 i \pi}{r \omega} G(\theta = 0) + o\qty(\frac{1}{r}).
    \end{equation}

    Hence, we find that 
    \begin{equation}
        \int_{\Sph^2} G(\theta,\phi) e^{-i\omega r (1 - \cos\theta)} \sin\theta \dd{\theta} \dd{\phi} = - \frac{2 i \pi}{r \omega} G(\theta = 0) - \frac{2 i \pi e^{-2i\omega r}}{r \omega} G(\theta = \pi) + o\qty(\frac{1}{r}).
    \end{equation}

    With a minor change of notation we find
    \begin{equation}\label{eq: stationary-phase-sphere}
        \int_{\Sph^2} G(\vu{q}) e^{-i \omega r (1 - \vu{q} \vdot \vu{x})} \sqrt{\gamma(\vu{q})} \dd[2]{q} = - \frac{2 i \pi}{r \omega} G(\vu{x}) - \frac{2 i \pi e^{-2i\omega r}}{r \omega} G(-\vu{x}) + o\qty(\frac{1}{r}),
    \end{equation}
    where \(\gamma(\vu{q})\) denotes the metric determinant as a function of the unit vector \(\vu{q}\).

\end{appendix}

\printglossary[type=abbreviations,title={Abbreviations}]

\printbibliography[heading=bibintoc,title={References}]
\end{document}